\documentstyle[12pt,epsf,appendix,amssymb]{book}

\advance \topmargin by -\headheight

\advance \topmargin by -\headsep

\evensidemargin \oddsidemargin

  \def\tr{{\hbox{\rm Tr}}}

\def\ie{{\em i.e.}}

\def\ie{\hbox{\it i.e.}}

\def\CC{{\mathchoice
{\rm C\mkern-8mu\vrule height1.45ex depth-.05ex
width.05em\mkern9mu\kern-.05em}
{\rm C\mkern-8mu\vrule height1.45ex depth-.05ex
width.05em\mkern9mu\kern-.05em}
{\rm C\mkern-8mu\vrule height1ex depth-.07ex
width.035em\mkern9mu\kern-.035em}
{\rm C\mkern-8mu\vrule height.65ex depth-.1ex
width.025em\mkern8mu\kern-.025em}}}

\def\RR{{\rm I\kern-2.0pt {\rm R}}}

\def\ZZ{{\rm Z}\kern-3.4pt {\rm Z} \kern2pt}

\def\IB{\relax{\rm I\kern-.18em B}}

\def\ID{\relax{\rm I\kern-.18em D}}

\def\II{\relax{\rm I\kern-.18em I}}

\def\IP{\relax{\rm I\kern-.18em P}}

\def\G{\Gamma}

\def\np{Nucl. Phys.}

\def\prl{Phys. Rev. Lett.}

\newcommand{\beq}{\begin{equation}}

\newcommand{\eeq}{\end{equation}}

\newcommand{\rc}{\nonumber\\}

\newcommand{\bear}{\begin{eqnarray}}

\newcommand{\eear}{\end{eqnarray}}

\def\to{\rightarrow}

\def\tr{{\rm Tr}}

\def\to{\rightarrow}

\newfont{\namefont}{cmr10}

\newfont{\addfont}{cmti7 scaled 1440}

\newfont{\boldmathfont}{cmbx10}

\newfont{\headfontb}{cmbx10 scaled 1728}

\begin{document}
\begin{titlepage}
\begin{center} \Large \bf Holographic flavor in the gauge/gravity duality
\end{center}
\vskip 0.3truein
\begin{center}
\ Daniel Are\'an Fraga
\footnote{PhD Thesis, Universidade de Santiago de
Compostela, Spain (July, 2008). \\
Advisor: Alfonso V. Ramallo}\\
\vspace{0.3in}
Departamento de F\'\i sica de Part\'\i culas, Universidade de
Santiago de Compostela \\and\\
Instituto Galego de F\'\i sica de Altas Enerx\'\i as (IGFAE)\\
E-15782 Santiago de Compostela, Spain\\
\vspace{0.15in}
arean@fpaxp1.usc.es
\vspace{0.3in}
\end{center}
\vskip 1truein
\begin{center}
\bf ABSTRACT
\end{center}

In this paper we review some results on the generalization of the gauge/gravity duality to include fundamental matter by means of probe branes.
We compute the meson spectrum of maximally supersymmetric gauge theories in several dimensions, both in the Coulomb and Higgs branch. We also study the addition of flavor and compute the meson spectrum for non-commutative  theories.
Additionally, we present a thorough search of supersymmetric embeddings of
probe branes in the Klebanov-Witten model.

\smallskip
\end{titlepage}
\setcounter{footnote}{0}
\pagestyle{empty}
\setcounter{page}{0}

\tableofcontents

\pagestyle{headings}

\chapter*{Motivation}
\addcontentsline{toc}{chapter}{Motivation}
\medskip

The work presented in this thesis is related to the AdS/CFT
correspondence, one of the most powerful results of string theory which, as
we will see in this brief historical overview, seems to take one back to
the origins of the theory. 

String theory was born in the late sixties as an attempt to describe the
strong nuclear interaction: the properties of the hadrons of increasing
mass and spin being discovered at that time were well described by
considering them as the modes of a fundamental string.
However, disagreement with experiment and the emergence of QCD led to the
abandonment of this path. Yet at the same time, a new hint of the stringy
nature of gauge theories was provided by 't Hooft
\cite{Hooft}. He showed that in the large $N$ limit the Feynman diagrams
of an $SU(N)$ gauge theory can be organized in a $1/N$ expansion; a
feature that can be mapped to the sum over the genus of the possible
worldsheets in the computation of a string theory amplitude.

One of the basic problems of the string models of the strong interaction,
namely the unavoidable presence of a massless spin two particle in the
spectrum, made string theory a promising candidate for a unified
quantum theory of gravity and the other fundamental forces. Indeed, while
there is no consistent way to describe a quantized theory of gravity in
terms of pointlike particles, string theory not only consistently describes
gravity but requires it and, in addition, gives rise to Yang-Mills gauge
theories with room enough for accommodating the $SU(3)\times SU(2)\times
U(1)$ gauge symmetry of the Standard Model and, in some cases, with chiral
matter as well. Consequently, a great effort was put into the search of a
consistent string theory which culminated with the formulation of string
theories with spacetime supersymmetry. In addition to supersymmetry,
which might soon be found experimentally in the Large Hadron Collider,
string theory requires a ten-dimensional spacetime, reviving
the concept of a space with extra dimensions. This last fact seems to
spoil the expected predictiveness of string theory due to the many
possible compactifications apparently suitable for describing our
universe. A problem that was worsened by the fact that five different
string theories, none of which could be discarded for describing Nature,
were discovered during the eighties.

The uncomfortable multiplicity of theories was solved in the nineties
when it was shown that there is a web of dualities relating all of them.
In fact, each one is a different perturbative limit of an
underlying eleven-dimensional theory known as M-theory \cite{mtheor},
about which little is known apart from the fact that it reduces to
eleven-dimensional supergravity in the low energy limit. In addition, the
huge number of possible compactifications gave rise to the idea of the
landscape \cite{landscape} where our universe is just a point in an
enormous set of string theory vacua.

An important role in establishing the dualities relating the different
theories is played by the solitonic objects known as branes. In
particular, D-branes \cite{Dbranes} are non-perturbative solutions
consisting of hyperplanes where open strings can end.
These open strings describe the dynamics of the branes; the massless modes
of these strings realize a gauge theory on the worldvolume of the brane.
Moreover, as massive objects the branes will curve the space around them
and thus couple to the closed strings describing gravity. In fact,
D-branes can be found as solitonic solutions of supergravity theories which
are the low energy limit of the corresponding string theory. This dual
description points again towards the existence of a duality between gauge
theories and string theory (or, in the large 't Hooft limit,
supergravity). Therefore, twenty years after 't Hooft's work, string
theory was again hinting at the possibility of a description of gauge
theories in terms of strings. Furthermore, the discovery of D-branes led
to the idea of brane-world scenarios, in which our four-dimensional
universe is confined to a D3-brane living in a higher dimensional space.

A precise correspondence between string and gauge theories, known as the
AdS/CFT correspondence, was proposed by Maldacena in 1997 \cite{jm}. As we
will see in the next section, the dual description of D3-branes motivates
the conjecture of an exact duality between the conformal ${\cal N}=4$ Super
Yang-Mills (SYM) $SU(N)$ gauge theory in four dimensions and type IIB
string theory in the ten-dimensional geometry
$AdS_5\times S^5$, namely the direct product of the five-dimensional
Anti-de-Sitter space and a compact five-sphere. This correspondence is
formulated as a holographic relation where string theory living in the
interior of $AdS_5$ encodes the dynamics of a four-dimensional gauge
theory living on the boundary of that space.

An important feature of the correspondence is that it is a
strong-weak coupling duality and, as we will see, one can compute
properties of the gauge theory at strong coupling by performing
calculations in classical supergravity. Though it has undergone many
rigorous tests \cite{MAGOO,n1duality}, this last remarkable property makes
the duality very difficult to be proven exactly. Another drawback is the
fact that ${\cal N}=4$ SYM is very far from being a realistic theory of
nature so there have now been several generalizations to non-conformal and
less (or non) supersymmetric theories. For instance, as we will see in the
last chapter of this thesis, by placing D-branes in a space of reduced
supersymmetry, dual models to
${\cal N}=1$ four-dimensional gauge theories can be constructed
\cite{n1duality}. This has resulted in a better understanding of phenomena
such as confinement or chiral symmetry breaking which are present in QCD.
Another interesting example is that of finite temperature AdS/CFT
which is an extension of the duality consisting in the introduction of
temperature (see \cite{mateosrv} for a review). It is expected that some
properties of the deconfined Quark-Gluon Plasma obtained in the heavy ion
experiments such as those of RHIC could be shared with ${\cal N}=4$ SYM at
finite temperature, a phase where the SYM theory is neither conformal nor
supersymmetric. The temperature is introduced in the correspondence by
considering non extremal D-branes, which are non-BPS D-branes whose
geometry has a horizon. The Hawking temperature of this horizon is
interpreted as the temperature of the dual theory. Finally, it is worth
noting that the original formulation of the correspondence does not
include fundamental degrees of freedom on the gauge theory side.
Therefore, an interesting field of research, which comprises most of the
work presented in this thesis, is the generalization of the AdS/CFT
correspondence to a duality describing gauge theories with fields
transforming in the fundamental representation of the gauge group. All in
all, even though the formulation of an exact string dual to QCD seems
far from our present understanding of string theory, several interesting
features of real-world physics have already been described by means of the
AdS/CFT correspondence and its generalizations.

\subsubsection{About this thesis}

This thesis is based on the papers \cite{conifold,APR,AR,ARR,ARRxl} and has
the following structure:

\begin{itemize}

\item
The first chapter begins with an introduction to the AdS/CFT
correspondence. Next, the generalization of the duality to
include quenched flavors in the gauge theory by means of the addition of
probe branes on the gravity side is described. The chapter ends with a
review of the worldvolume action for D-branes, where we also introduce the
so-called kappa symmetry and its utility to find supersymmetric embeddings
of D-brane probes.

\item
In the second chapter we use the different supersymmetric brane
intersections as a tool for adding flavor to the maximally supersymmetric
gauge theories in several dimensions. By considering the brane of higher
dimensionality as a probe in the background generated by the lower
dimensional brane, one can introduce flavor to the theory living on this
last brane. Generically the fundamental fields will be confined to a
defect given by the intersection of both branes. By studying the
fluctuations of the probes we will compute the meson spectra of the
theories and find a universal behavior of the mass gap.

\item
The third chapter deals with the description of the Higgs branch of the
theories considered in the preceding chapter.
On the gravity side, the Higgs branch corresponds to a deformation of the
embedding and/or the addition of a non-trivial
flux of the worldvolume gauge field of the probe. Again, we characterize
the spectra by analyzing the fluctuations of the probe brane.


\item
In chapter 4 we study the addition of flavor to the non-commutative
deformation of the maximally supersymmetric four-dimensional gauge
theory. We proceed by adding a D7-brane probe to the supergravity dual.
In order to determine the meson spectrum of the dual theory we analyze the
fluctuations of the probe brane. In addition, for the case of mesons
with large spin the spectrum is characterized by considering
the classical dynamics of a rotating string attached to the flavor brane.

\item
Chapter 5 is somewhat different from the first four where we studied
the addition of flavor in different maximally supersymmetric scenarios.
This chapter consists of a study of the supersymmetric embeddings of
D-brane probes in the so-called Klebanov-Witten model: a ten-dimensional
background constructed by placing D3-branes at the tip of the conifold and
conjectured to be dual to a four-dimensional ${\cal N}=1$ superconformal
field theory. We find supersymmetric embeddings of D3-branes dual to
dibaryonic operators in the field theory, configurations with D5-branes
wrapping an internal two-cycle which are suitable for the study of defect
theories in the conifold and embeddings of D7-branes with the right
properties to be considered as flavor branes.

\end{itemize}

\chapter{Introduction}
\label{introduction}
\setcounter{equation}{0}
\medskip

\section{The AdS/CFT Correspondence}
\setcounter{equation}{0}
\medskip

The AdS/CFT correspondence as originally formulated in \cite{jm} (for
a review see \cite{MAGOO,n1duality}) establishes a duality between
four-dimensional
${\cal N}=4$ $SU(N)$ super-Yang-Mills theory and type IIB
string theory on $AdS_5\times S^5$. This duality is based on 
two alternative descriptions of D3-brane stacks. On the one hand, a
D-brane is a solution of the open string sector consisting of a hyperplane
where the open strings can end. On the other hand, the D-branes appear as
non-perturbative solutions of the type II closed string
sector, and at low energy they are solutions of the classical equations
of ten-dimensional type II supergravity.

From the open string point of view let us consider a stack of $N$
D3-branes living in flat ten-dimensional Minkowski space. In the low
energy limit $E\ll 1/ \sqrt{\alpha'}$ (energies much lower than the
string  scale) only the massless modes of the open strings ending on the
branes are present. These realize an ${\cal N}=4$ vector multiplet living
on the four-dimensional worldvolume of the brane. Thus, the low energy
theory on the worldvolume is ${\cal N}=4$ 
$SU(N)$ super-Yang-Mills\footnote{The actual gauge group of the worldvolume theory is
$U(N)$. Up to global issues this is equivalent to an $SU(N)$ theory
times a free $U(1)$ vector multiplet. In the string dual there are no
decoupled modes; the fields in the bulk of $AdS$ describe the $SU(N)$
theory. Nevertheless, there are zero modes living in the region
connecting the near-horizon region ($AdS$) with the bulk which correspond
to the $U(1)$ vector multiplet.} (SYM) whose gauge coupling  is
determined by the string coupling $g_s$ as $g_{YM}^2=4\pi g_s$ (see
section \ref{cp1sscbraneaction}). This low energy limit can be taken by
setting $\alpha'\to0$ while keeping the energy fixed. Then, the 
ten-dimensional Planck length $l_p=G_{10}^{1/8}=
g_s^{1/4}\,\sqrt{\alpha'}$ vanishes, and thus the closed string modes of
the bulk decouple from the brane and give rise to free type II
supergravity. Consequently, the system consists of two decoupled theories:
the gauge theory on the worldvolume of the branes and free gravity in the
bulk.

The same system can be studied as a solution of classical IIB
ten-dimensional supergravity with constant dilaton and an RR
self-dual five-form field strength and metric given by:
\bear
&&ds^2=f(r)^{-{1\over2}}\, dx_{1,3}^2 +
f(r)^{1\over2}\,\left(dr^2+r^2\,d\Omega_5^2\right)\,,\rc\rc
&&g_s\,F_{(5)}=dC_{(4)}+\,{\rm Hodge\,\, dual}\;,\qquad
g_s\,C_{(4)}=f(r)^{-1}\,dx^0\wedge dx^1\wedge dx^2\wedge
dx^3\,,\qquad
\eear
with:
\beq
f(r)=1+{R^4\over r^4}\;,\qquad\quad R^4=4\pi\,g_s\,\alpha'^{\,2}\,N\,.
\label{c1d3sol}
\eeq
Notice that the string coupling constant is given in terms of the
constant value of the dilaton as $g_s=e^{\phi}$.

The decoupling limit is taken by keeping the energy fixed and sending the
string length to zero ($\alpha'\to 0$). In this limit the two kinds of
low energy excitations of the system decouple. First, there are
massless particles propagating in the bulk region which decouple from the
near-horizon region ($r\to 0$): due to the redshift proportional to
$f^{-1/4}$ their wavelength becomes much larger than the size of the
near-horizon region ($\sim R\,$, see below). Likewise, the excitations
living close to the brane cannot leave the near-horizon region. Then, once
again the system consists of two decoupled pieces and, as before, one of
them is free gravity in the bulk. Therefore, it seems natural to identify
the theory living in the near-horizon region with the other decoupled
piece in the open string description, namely the gauge theory living on
the brane.

In order to have a better look at the low energy dynamics in the
near-horizon region ($r\sim 0$) the $\alpha'\to 0$ limit should be taken
while keeping $g_s$ and ${r\over\alpha'}$ fixed. This procedure is known
as the Maldacena or near-horizon limit. In this limit the coupling to the
bulk modes ($\sim g_s\,\alpha'^{\,2}$) vanishes, and the energy of the
string excitations living close to the origin as measured by an observer
at infinity is kept fixed . This last feature can be understood as the
requirement that the generic gauge excitations of the theory living on the
brane at $r=0$ are still present in the low energy limit. Eventually,
taking $r\to 0$ while keeping
${r\over\alpha'}$ fixed and defining the new variable
$U={r\over\alpha'}$, the metric ({\ref{c1d3sol}) becomes:
\beq
ds^2={U^2\over R^2}\, dx_{1,3}^2 + R^2\left({dU^2\over U^2}+
d\Omega_5^2\right)\,,
\label{c1ads5s5metric}
\eeq
where the 1 has been dropped in the harmonic function $f(r)=1+{R^4\over
r^4}$ and the Minkowski coordinates have been rescaled as
$x\to{x\over\alpha'}$. This is the metric of a ten-dimensional
space of the form $AdS_5\times S^5$, where both the
$AdS_5$ and the $S^5$ have the same constant radius $R$.

Finally, the identification outlined above leads  to the conjecture,
stated by Maldacena, that ${\cal N}=4$ $SU(N)$ SYM in 3+1 dimensions is
equivalent to type IIB string theory on $AdS_5\times S^5$. According to
what has been shown, the parameters of the two theories are matched in the
following way:
\bear
4\pi\,g_S=g_{YM}^2={\lambda\over N}\;,\qquad\quad
R^2=\sqrt{g_{YM}^2\,N}\,\alpha'=\sqrt{\lambda}\,\alpha'\,,
\label{c1cpmatch}
\eear
where $\lambda=g_{YM}^2\,N$ is the 't Hooft coupling. In view of these
relations the AdS/CFT correspondence can be shown to be a strong-weak
coupling duality. Indeed, one can treat the gauge theory perturbatively 
when the 't Hooft coupling is small and then:
\beq
\lambda=4\pi\,g_s\,N\ll1\,,
\label{c1gaugpert}
\eeq
whereas one can trust the classical gravity description when the radius
$R$  shared by the $AdS_5$ and $S^5$ is large compared to the string
length, hence:
\beq
{R^2\over\alpha'}=\sqrt{\lambda}\gg1\,.
\label{c1clssgrav}
\eeq
Therefore, for a regime where one side of the duality is weakly
coupled the other one becomes strongly coupled and vice versa.

\subsubsection{The 't Hooft limit}
The correspondence as it has been formulated is known as the strong
version of the duality, for it relates both theories along the whole range
of their coupling constants. However, since it is not known how to
deal with string theory at a generic value of the string coupling, it is
worth considering a milder version of the duality restricted to weakly
coupled string theory. This can be easily seen to be equivalent to
taking the 't Hooft limit on the gauge theory side \cite{Hooft}.
This limit consists of setting $N\to\infty$ while keeping the 't Hooft
coupling $\lambda$ fixed. It corresponds to a topological expansion of
the Feynman diagrams of the field theory where the quantity $1/N$ acts as
the coupling constant (see \cite{Witten:1979kh,mateosrv} for a nice
review). Indeed, by using the double line notation, which consists of
drawing the line associated to a gluon as a pair of lines corresponding
to a fundamental and an antifundamental field, one can associate a Riemann
surface to each Feynman diagram of the theory. By doing so, one can easily
determine the powers of $N$ and $\lambda$ corresponding to each diagram
and it turns out that the expansion of any gauge theory amplitude in terms
of diagrams is given by:
\beq
{\cal A}=\sum_{g=0}^\infty N^\chi\sum_{n=0}^\infty c_{g,n}\,\lambda^n\,,
\label{c1thooftexp}
\eeq
where $\chi$ is the Euler number of the Riemann surface corresponding to
the diagram and $c_{g,n}$ is a constant depending on the specific diagram.
For a compact orientable surface of genus $g$ with no boundaries the Euler
number is given by $\chi=2-2g$. Therefore, in the large $N$ limit with
$\lambda$ fixed the diagrams can be organized in a double series
expansion in powers of $1/N$ and
$\lambda$. Moreover, in the strict $N\to\infty$ limit only the planar
diagrams contribute to the amplitude. Those are the ones associated to a
sphere ($g=0$) and thus scale as $N^2$, while the next kind of diagrams
in the topological expansion, namely the ones corresponding to a torus
($g=1$), scale as $N^0$ and are therefore suppressed by a factor
of $N^{-2}$. 

On the string theory side, taking into account the identifications
(\ref{c1cpmatch}), the expansion (\ref{c1thooftexp}) becomes the genus
expansion of the string theory. 
One should notice that, since $\lambda/N=g_s$,
the gauge theory in the 't Hooft limit is then conjectured to be dual to
classic string theory, this being the milder version of the correspondence
we have mentioned above. Furthermore, the identification of both
expansions implies that $1/N$ corrections at fixed $\lambda$ on the gauge
theory get mapped to $g_s$ corrections on the string theory, while
$1/\lambda$ corrections at each order in $N$ on the gauge theory
correspond to $\alpha'$ corrections on the dual string theory.

\subsubsection{The large $\lambda$ limit}

Even in the classical limit, string theory on generic
curved backgrounds with RR fluxes is not tractable at present.
Accordingly, it is interesting to consider the $\lambda\to\infty$ limit
where type IIB string theory reduces to IIB supergravity. By doing so,
one arrives at the weak version of the correspondence which states that
type IIB ten-dimensional supergravity on $AdS_5\times S^5$ is equivalent
to the large 't Hooft coupling limit of large $N$, ${\cal N}=4$ $SU(N)$
super-Yang-Mills theory. It is for this weak version of the duality that 
most of the checks have been carried out (see \cite{MAGOO,n1duality} and
references therein). 

\begin{figure}
\centerline{\hskip .1in \epsffile{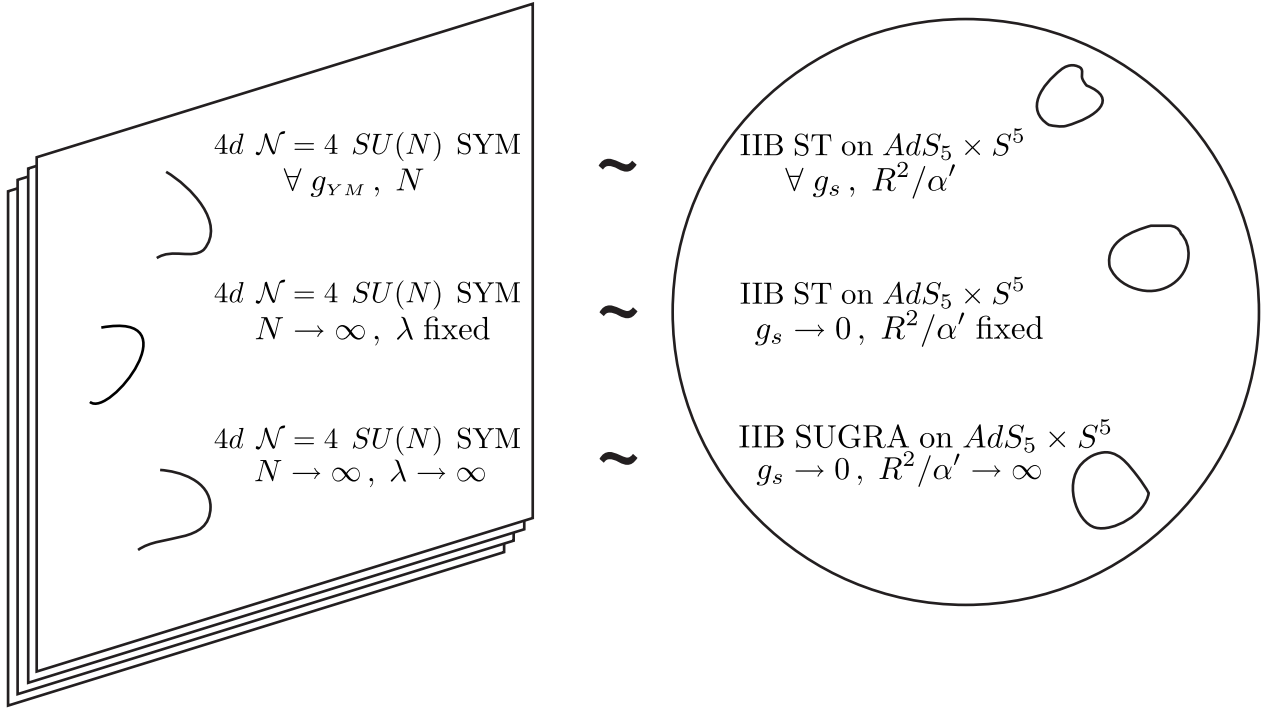}}
\caption{ The three versions of the AdS/CFT correspondence superposed on
a drawing illustrating the two sides of the duality. On the left one can
see a stack of D3-branes hosting open strings, while on the right it is
shown the equivalent $AdS_5$ space where the closed string modes
propagate.}
\label{c1figdual}
\end{figure}

\subsubsection{The global symmetries}

Finally, if both theories are to be equivalent, the global symmetries on
both sides of the duality should match. ${\cal N}=4$ $SU(N)$ SYM in its
superconformal phase is globally invariant under the supergroup
$SU(2,2\,|\,4)$, whose maximal bosonic subgroup is $SO(2,4)\times SO(6)_R$;
the $SO(2,4)$ factor realizing the conformal transformations in 3+1
dimensions and the $SO(6)_R$ being the $R$-symmetry group. Regarding the
fermionic symmetries, the gauge theory is invariant under the
sixteen Poincar\'e supersymmetries generated by four complex Weyl  spinors
$Q^{\,a}_\alpha$ ($a=1,\cdots,4$) transforming in the fundamental of the
$R$-symmetry group, and its complex conjugates $\bar Q_{\dot\alpha\,a}$
transforming in the antifundamental of
$SO(6)_R$. Moreover, these fermionic supercharges do not commute with the
conformal bosonic generators giving rise to sixteen extra fermionic
supercharges referred to as superconformal generators ($S^{\,a}_\alpha$
and $\bar S_{\dot\alpha\,a}$). As for the string theory side, the isometry
group of $AdS_5$ is
$SO(2,4)$, while the isometries of the
$S^5$ form the $SO(6)_R$ group. An array of parallel D-branes in flat
ten-dimensional space preserves 16 of the maximal 32 Poincar\'e
supersymmetries. Yet, near the horizon of a stack of D3-branes, taking the
limit appropriately as described above, the geometry becomes that of
$AdS_5\times S^5$ and an extra set of 16 supersymmetries completes the
isometry group matching the supergroup of the conformal gauge theory.

The ${\cal N}=4$ SYM theory has a non-perturbative duality known as
S-duality. In order to state this invariance it is standard to combine the
real coupling $g_{YM}$ and the YM angle $\theta_{YM}$ into the
following complex coupling\footnote{The lagrangian of a non-abelian YM theory may include a topological term
of the form $\theta_{YM}\,\epsilon_{abcd}\, F^{ab}F^{cd}$.}:
\beq
\tau\equiv {4\pi\, i\over g_{YM}^2}+{\theta_{YM}\over2\pi}\,.
\eeq
Since $\theta_{YM}$ is periodic ($\theta_{YM}=\theta_{YM}+2n\pi$) the gauge
theory is invariant under shifts $\tau\to\tau+1$. In addition, it is
conjectured that
${\cal N}=4$ SYM has a non-perturbative symmetry, namely the
Montonen-Olive duality \cite{MOdual,SYMdualrv}, which takes
$\tau\to-1/\tau$ and exchanges the gauge algebra with its dual, thus
acting as an electric-magnetic duality. In particular, notice that for
$\theta_{YM}=0$ this symmetry reduces to
$g_{YM}\to 1/g_{YM}$ and thus exchanges strong and weak coupling.
The combination of the two symmetries just described forms the S-duality
group $SL(2,{\mathbb Z})$ generated by:
\beq
\tau\longrightarrow\,{a\tau+b\over c\tau+d}\;,\qquad a,b,c,d \in
{\mathbb Z}\;,\quad ad-bc=1\,.
\label{c1sdtransf}
\eeq
The complex coupling can be expressed in terms of string theory quantities
as:
\beq
\tau\equiv {4\pi\, i\over g_{YM}^2}+{\theta_{YM}\over2\pi}={i\over
g_s}+{\chi\over2\pi}\,,
\label{c1cmpcplmap}
\eeq
where $\chi$ is the expectation value of the RR scalar, while as we have
seen $g_s=e^{\phi_\infty}$. This equality follows after supplementing the
relation
$g_{YM}^2=4\pi\,g_s$ with the identification $\theta_{YM}=2\pi\,\chi$.
Indeed, one can easily check that through the Wess-Zumino part of the
action of a D3-brane (see section
\ref{cp1sscbraneaction}) the RR scalar sources a Chern-Simons term ($\sim
F\wedge F$) of the gauge theory living on the worldvolume of the D3,
leading to the relation
$\theta_{YM}=2\pi\,\chi$.

The equality (\ref{c1cmpcplmap}) maps the $SL(2,{\mathbb Z})$
transformations acting on the complex coupling of the gauge theory to the
S-duality symmetry group of type IIB string theory, which acts on
the complex combination of the dilaton and RR scalar expectation values
$\tau=i/g_s+\chi/2\pi$ precisely as (\ref{c1sdtransf}). 
In the eleven-dimensional M-theory interpretation, the ten-dimensional
type IIB string theory vacuum corresponds to M-theory compactified on a
two-torus of complex structure $\tau$ and vanishing area. Therefore, the
$SL(2,{\mathbb Z})$ symmetry, left unbroken by the D3-brane solution,
is the modular group acting on the complex structure of the two-torus 
(for a review see \cite{sdualsch}).
Notice that the subgroup given by the $\tau\to -1/\tau$ transformations
acts, in the $\chi=0$ case, as a strong-weak coupling duality
$g_s\to1/g_s$. It can be seen from the M-theory interpretation that this
transformation interchanges the NSNS and RR two-form potentials (an
electric-magnetic duality as in the SYM case) and consequently
exchanges the fundamental string with the solitonic (RR charged) D-string.

\subsection{Mapping the observables}

As we explained in the previous section, the AdS/CFT correspondence is a
strong-weak coupling duality. Thus, though difficult to be proven exactly,
it becomes a powerful tool which provides a description of the
strong coupling regime of the gauge theory in terms of classical
supergravity. But in order to do that, a precise matching between
observables of the two theories is needed. 

The conformal gauge theory is specified by a complete set of
conformal operators. We will be interested in the gauge invariant
conformal operators polynomial in the canonical fields since these
operators will have a definite conformal dimension and, in addition,
will form finite multiplets of increasing dimension. An special role is
played by the single color-trace operators because out of them the 
multiple trace operators can be constructed using the OPE. In order to
compute correlators for a given operator ${\cal O}$, its generating
functional $\Gamma({\cal O})$ is constructed by adding a source term to
the lagrangian of the CFT:
\beq
e^{\Gamma[{\cal O}]}=\langle\, e^{\int h\,{\cal O}}\,\rangle\,,
\label{c1gen}
\eeq    
where $h=h(x)$, depending on the four Minkowski coordinates, is a source
for the corresponding operator ${\cal O}$.

On the other side of the duality, in the classical string limit, the
closed  string excitations can be described by fields living on
$AdS_5\times S^5$. When compactifying along the $S^5$, these fields can
be decomposed in terms of an infinite basis of spherical harmonics on the
$S^5$, thus giving rise to an infinite tower of five-dimensional modes
receiving contributions to their mass of order $1/R^2$ (recall that $R$ was
the radius of both $AdS_5$ and $S^5$). Therefore, the observables of 
the string theory are a set of five-dimensional fields living on
$AdS_5$, and, for the correspondence to be correct, they have to be
matched to the observables of the gauge theory. 

First, the metric of an $AdS_5$ space with radius $R$ can be written in
the form:
\beq
ds^2={r^2\over R^2} dx_{1,3}^2 + {R^2\over r^2} dr^2 \,,
\label{c1adsmetric}
\eeq
which follows from the $AdS_5$ part of (\ref{c1ads5s5metric}) after
setting $r=\alpha'\,U$ and undoing the rescaling of the Minkowski
variables, \ie \ $x\to\alpha'\,x$. In these variables the boundary
of
$AdS_5$ is at $r\to\infty$ and is isomorphic to four-dimensional Minkowski
space, spanned by the coordinates $(x^0,\cdots x^3)$.
The isometry group of $AdS_5$, \ie \ $SO(2,4)$, reduces on the boundary
to the group of conformal transformations of $\RR^{1,3}$ and so a theory living on the
boundary will be a four-dimensional conformal theory.

Next, let $\hat h(x,r)$ be a scalar field on $AdS_5$ to be
associated to the operator
${\cal O}$ of the gauge theory. Assuming that the interactions take place
in the bulk, and thus the fields become free in the asymptotic region
$r\to \infty$, the equation of motion of $\hat h(x,r)$ is that of a free
scalar when $r\to \infty$, namely $(\Box+m^2)\,\hat h=0$, where $m$ is the
five-dimensional mass of the scalar $\hat h$. As a result, the asymptotic
dependence of
$\hat h$ on the radial coordinate is given by the two independent
solutions:
\beq
\hat h \sim r^{\Delta-4}\;,\qquad\quad \hat h \sim r^{-\Delta}\;,
\qquad (r\to\infty)\,,
\label{c1adsscalar}
\eeq
with $\Delta=2+\sqrt{4+m^2 R^2}$. Hence, the first solution dominates near
the boundary at $r\to\infty$. Then, the solution at the boundary can be
written as:
\beq
\hat h(x,r)\to r^{\Delta-4}\,\hat h_\infty(x)\;,\qquad (r\to\infty)\,,
\label{c15dsclsrc}
\eeq
where $\hat h_\infty(x)$ is a four-dimensional function living on the
boundary  and is defined only up to conformal transformations under
which it has conformal dimension $\Delta-4$ (in units of length)\footnote{
This can be better understood in terms of the variable $z=R^2/r\,$: the
metric (\ref{c1adsmetric}) takes the form ${R^2\over
z^2}(dx_{1,3}^2+dz^2)$ and thus rescalings of $z$ amount to conformal
transformations of the metric of the boundary (which now sits at $z=0$).
Eq. (\ref{c15dsclsrc}) becomes $\hat h(x,z)=z^{4-\Delta}\,\hat h_0(x)$
and, since $\hat h$ is dimensionless, it is clear that $\hat h_0(x)$ has
conformal dimension $\Delta-4$.}. Finally, a map between observables of the
two theories is established \cite{KlebHol,WittenHol} by identifying the
generating functional for the correlators of ${\cal O}$ with the effective
5d action $S_{AdS_5}(\hat h)$  restricted to the solution of the equations
of motion (\ref{c15dsclsrc}) with boundary value $\hat h_\infty(x)$ set to
the value of the source $h(x)$ for the corresponding operator ${\cal O}$,
namely $\hat h_\infty(x) =h(x)$. Then, one can write:
\beq
\langle\, e^{\int h\,{\cal O}}\,\rangle = e^{-S_{AdS_5}(\hat h)}\,,
\label{c1actmap}
\eeq 
where in the supergravity limit ($\lambda\to\infty$) $S_{AdS_5}$
becomes the type IIB supergravity action, while in the full string
theory case it should be replaced by some S-matrix of string theory.
Moreover, in view of the coupling term $\int d^4 x\,h\,{\cal O}$, the
duality relates the quantum numbers of the operator ${\cal O}$ to those
of the dual field $\hat h(x,r)$. For instance, it has just been shown that
the finite boundary value of a massive
scalar field on $AdS_5$, $\hat h_\infty(x)$, has conformal dimension
$\Delta-4$, and therefore, it will couple to an operator of dimension
$-\Delta$. Usually, a conformal operator is labeled by its conformal
weight which is $-1$ times its conformal dimension. Thus, the
correspondence states that a scalar field in $AdS_5$ with mass
$m^2=\Delta(\Delta-4)/R^2$, thus scaling as $r^{\Delta-4}$ at $r\to\infty$, is
dual to a conformal operator ${\cal O}$ with conformal weight\footnote{
Henceforth we will follow the usual convention and
refer to $\Delta$ expressed in mass units as the conformal dimension.}
$\Delta$. Even
though here we have considered the case when the dual field on the string
theory side is a scalar, the correspondence works the same way for fields
of increasing spin. However, the relation between the conformal dimension
of the operator at the boundary and the mass of the five-dimensional field
depends on the spin of the field. We list these  relations for a generic
Anti-de-Sitter space $AdS_{d+1}$ in table \ref{c1massdelta}.
\begin{table}[!h]
\centerline{
\begin{tabular}[b]{|c|c|}   
 \hline
 Field  & $\Delta$  \\ 
\hline 
\rule{0mm}{4.9mm} scalar & ${1\over2}\left(d+\sqrt{d^2+4m^2}\right)$  \\
\hline  
\rule{0mm}{4.9mm} spinor & ${1\over2}\left(d+2|m|\right)$   \\ \hline  
\rule{0mm}{4.9mm} vector & ${1\over2}\left(d+\sqrt{(d-2)^2+4m^2}\right)$  
\\  \hline   
\end{tabular}
}
\caption{Conformal dimension of the dual operator for fields of different
spin in $AdS_{d+1}$. The mass is written in units of $R^2$. The fields
scale as $r^{\Delta-d}$ at $r\to\infty$ so the dual operator must have
conformal weight
$\Delta$.}
\label{c1massdelta} 
\end{table}

The correspondence has been thoroughly checked in the supergravity limit
(see \cite{MAGOO,n1duality}). As was stated, the basic constituents of
the gauge theory are the gauge invariant operators. These operators fall
into multiplets generated by successive applications of the supercharges
$Q$ to a so-called superconformal primary operator:
the conformal supercharges $S$ have dimension $-1/2$ so successively
acting with $S$ on an operator of definite dimension has to result in 0
at some point; otherwise operators of negative dimension would
appear. Hence, there exists an operator of lowest dimension in the
supermultiplet satisfying $[S,{\cal O}]=0\;,\;\; {\cal O}\neq0$ which is
known as the superconformal primary. In addition, for some multiplets at
least one of the supercharges $Q$ commutes with the primary operator, and
therefore, such supermultiplets are shortened. As a consequence,
their dimension is protected from receiving quantum corrections. These
are called chiral multiplets or  BPS multiplets.   

On the $AdS_5\times S^5$ geometry all Kaluza-Klein modes of the
supergravity excitations have been computed and are organized in
five-dimensional ${\cal N}=8$ multiplets. A complete correspondence has
been found between the
${\cal N}=8$ KK short multiplets and the BPS multiplets of ${\cal N}=4$
SYM.

Ultimately, the content of the correspondence  can be summarized by
saying that the five-dimensional dynamics of type IIB string theory in the
interior of $AdS_5$ encodes the 4d gauge theory ${\cal N}=4$ SYM with the
four-dimensional fields sourced by the boundary values of the type IIB
fields.

\subsection{Duality for Dp-branes}
\label{c1scdpdual}
The twofold description of Dp-branes as hyperplanes where the open
strings can end and as a non-perturbative solution of the closed string
sector opens up the possibility of generalizing the correspondence to a
duality between supersymmetric gauge theories in $p+1$
dimensions and string theory on the near-horizon geometry of the
supergravity Dp-brane solutions \cite{IMSY}. Indeed, in the low energy
limit the open strings ending on a  stack of $N$ Dp-branes realize the
degrees of freedom of pure gauge $SU(N)$ SYM in $p+1$ dimensions. By
looking at the low energy action for a Dp-brane (see section
\ref{cp1sscbraneaction}), the coupling constant
$g_{YM}$ can be written as:
\beq
g_{YM}^2=2(2\pi)^{p-2}\,g_s\,(\alpha')^{{p-3}\over2},
\label{c1gymdp}
\eeq
which for $p\neq3$ is dimensionful. Proceeding as before, the system must
be studied in the limit where the massive string modes decouple,
$\alpha'\to0$, as do the open/closed string interactions,
$l_p=g_s^{1/4}\,\sqrt{\alpha'}\to 0\,$; and, at the same time, keeping
$g_{YM}$ finite, which in view of (\ref{c1gymdp}) results in
$g_s\sim(\alpha')^{{3-p}\over2}$ . Hence,
$l_p\sim (\alpha')^{{7-p}\over8}$, so apparently this decoupling limit
holds for Dp-branes with $p<7$. However, we will see that the situation
is more subtle.

From the point of view of the closed string sector, the supergravity
solution for a stack of $N$ Dp-branes is given by (in string frame):
\bear
&&ds^2=f_p(r)^{-{1\over2}}\, dx_{1,p}^2 +
f_p(r)^{1\over2}\,\left(dr^2+r^2\,d\Omega_{8-p}^2\right)\,,\rc\rc
&&e^{-2(\phi-\phi_\infty)}=f_p(r)^{{p-3\over2}}\,,\rc\rc
&&e^{\phi_\infty}\,C_{(p+1)}=(f_p(r)^{-1}-1)\,dx^0\wedge\cdots\wedge
dx^p\,,\rc\rc
&&f_p(r)=1+{R^{7-p}\over r^{7-p}}\;,\qquad
R^{7-p}=2^{5-p}\pi^{5-p\over2}\,g_s\,N\,\Gamma\left({7-p\over2}\right)
(\alpha')^{7-p\over2}\,.
\label{c1dpsol}
\eear
One should notice that for $p\neq3$ the dilaton is not constant anymore, 
and, as a result, the ranges of validity of the different descriptions
of the system will be more complex.

As before, the low energy, near-horizon limit is taken by setting
$\alpha'\to0$ and $r\to0$ while keeping $U=r/\alpha'$ fixed. In this limit the
metric and dilaton of eq. (\ref{c1dpsol}) can be written as:
\bear
&&ds^2=\alpha'\left[\left({U^{7-p}\over
c_p\,g_{YM}^2\,N}\right)^{1\over2}\, dx_{1,p}^2
+\left({c_p\,g_{YM}^2\,N\over U^{7-p}
}\right)^{1\over2}\,\left(dU^2+U^2\,d\Omega_{8-p}^2
\right)\right]\,,\rc\rc
&&e^{-\phi}={(2\pi)^{p-2}\over g_{YM}^2}\left({c_p\,g_{YM}^2\,N\over
U^{7-p}}\right)^{(p-3)\over4}\;,\qquad c_p=2^{6-2p}\,\pi^{9-3p\over2}\,
\Gamma\left({7-p\over2}\right)
\,,
\label{c1dpsolnh}
\eear
in terms of the coupling constant of the gauge theory $g_{YM}^2$, the
number of colors $N$ and the new radial coordinate $U$. The expression of
$e^{-\phi}$ follows after setting $e^{\phi_\infty}=g_s$. Furthermore,
one should notice that $U$  represents an energy scale in the field
theory living on the worldvolume of the branes. Actually, separating one
of the D3s some distance $r$ from the remaining ones corresponds to
giving a VEV to one of the adjoint scalars in the gauge theory. This VEV
is equal to the mass of the string stretching between the branes at $r=0$
and the one at $r$, which is proportional to
${r\over\alpha'}=U$.

In view of the metric given in (\ref{c1dpsolnh}) it becomes obvious that
for $p\neq3$ neither the dilaton nor the curvature are constant. Indeed,
it can be easily seen that the curvature takes the form:
\beq
\alpha'\,{\cal R}\sim \left({U^{3-p}\over
c_p\,g_{YM}^2\,N}\right)^{1\over2}\sim {1\over g_{eff}}\,,
\label{c1dpcurv}
\eeq
where we have introduced the dimensionless effective gauge coupling,
defined in terms of $g_{YM}$ as:
\beq
g_{eff}^2=g_{YM}^2\,N\,U^{p-3}\,.
\label{c1geff}
\eeq
The fact that the curvature is inversely proportional to the effective
gauge coupling shows that the classical supergravity description
applies only for the strong coupling regime and, analogously, the
weakly coupled region of the gauge theory is dual to a strongly curved
geometry where the supergravity description does not hold. So, as for the
D3-brane system, the correspondence is a strong-weak coupling duality.
However, given that now the dilaton also runs with the scale, the range
where the supergravity description holds will get reduced to the region
where $\alpha'\,{\cal R}\ll 1$ and $e^\phi\ll1$, which turns out to be an
intermediate regime of radial distances in the background, or,
equivalently, of energies in the field theory. Indeed, using
(\ref{c1dpcurv}) the condition of small curvature implies $g_{eff}\gg 1$,
and from the form of the dilaton in (\ref{c1dpsolnh}) it follows that
$e^\phi\sim (g_{eff})^{7-p\over2}/N$, so $e^\phi\ll1$ only when
$g_{eff}\ll N^{2\over7-p}$. Hence, the supergravity description applies
in the range:
\beq
1\ll g_{eff}\ll N^{2\over 7-p}\,.
\label{c1dpsugrarange}
\eeq
For $p<3$, when $U\gg1$ the curvature is large and the SUGRA description
is not valid, but the effective coupling runs to $g_{eff}\ll1$ and 
perturbative field theory is applicable on the gauge theory side.
Alternatively, for small values of $U$, both $g_{eff}$ and the dilaton
are large, yet another dual description can be found by up-lifting to
M-theory. For $p>3$ the effective coupling runs in the opposite
direction and the perturbative description holds for the infrared, 
corresponding to the small $U$ region  where the curvature is large.
Furthermore, in the UV ($U\gg1$) both the dilaton and
$g_{eff}$ are large, so a new dual description must be found.

Regarding the isometries, the near-horizon geometry of a stack
of Dp-branes, (\ref{c1dpsolnh}), is invariant under $SO(1,p)\times
SO(9-p)$ for $p\neq3$; thus reflecting the fact that the dual theories
are no longer conformal. Accordingly, the number of
conserved supersymmetries is 16, for the superconformal ones are
broken in this case.
At this point, since the dual field theory is no longer conformal, it is
worth recalling the role of the radial coordinate as the energy scale of 
the dual field-theory. Therefore, taking $U\to\infty$ will amount to
going to the UV regime of the field theory, while $U\sim 0$ corresponds
to the far IR of the theory.

As it will be useful in the following, we shall finish this section
by writing down the near-horizon solution (in string frame) in terms of
the radial coordinate
$r=\alpha'\,U$:
\bear
&&ds^2=\left({r\over R}\right)^{7-p\over2}\,
dx_{1,p}^2 +\left({R\over r}\right)^{7-p\over2}\,\left(dr^2+r^2
\,d\Omega_{8-p}^2\right)\,,\\ \rc
&&e^{-\phi}=\left({R\over r}\right)^{(7-p)(p-3)\over4}\,,
\label{cp1dpsoldiltn}\\ \rc
&&C_{(p+1)}=\left({r\over R}\right)^{7-p}\,dx^0\wedge\cdots\wedge dx^p
\,,
\label{cp1dpsolpot}
\eear
where $R$, which was defined in eq. (\ref{c1dpsol}), can be written as:
\beq
R^{7-p}=c_p\,g_{YM}^2\,N\,(\alpha')^{5-p}\,.
\label{c1Rgym}
\eeq
Notice that when writing the equations (\ref{cp1dpsoldiltn}) and
(\ref{cp1dpsolpot}) we have made the following redefinitions:
\beq
g_s\,e^{-\phi}\to e^{-\phi}\;,\qquad\quad g_s\,C_{(p+1)}\to C_{(p+1)}\,,
\label{cp1gsabsor}
\eeq
which will be implicitly applied in the following chapters.
Moreover, after changing to cartesian coordinates
along the subspace transverse to the worldvolume of the Dp-brane, the
metric takes the form:
\beq
ds^2=\left({r\over R}\right)^{7-p\over2}\,
dx_{1,p}^2 +\left({R\over r}\right)^{7-p\over2}\,d\vec r\cdot d\vec r\,,
\label{cp1dpsolmtrc}
\eeq
where $r^2=\vec r\cdot \vec r$.

%
%
%
%
%
%
%
%
%

\section{Adding flavor to the Correspondence}
\label{cp1flavor}
\setcounter{equation}{0}
\medskip
The original formulation of the duality we have summarized in the
previous section only includes fields in the adjoint representation of
the gauge group: the open strings with both ends on a stack of $N$
Dp-branes realize the adjoint fields of $SU(N)$ SYM in $p+1$
dimensions. Accordingly, the closed string  excitations in the dual
geometry only account for the adjoint fields in the gauge theory. Fields
transforming in the fundamental representation can be introduced in the
brane picture by, for instance, separating one of the branes of the
stack so that the strings between this brane and the remaining ones
will be seen as fundamental fields in the theory on the worldvolume of
the $N-1$ branes. 
This fact suggests that it may be possible to describe a gauge theory
with fundamental degrees of freedom by adding an open string sector to
the geometry of the string theory background dual to the pure gauge SYM
theory.

Shortly after formulating the correspondence, Maldacena studied in
\cite{Wilson} the system consisting of a stack of D3-branes with
fundamental strings stretching to the infinite. These strings introduce
infinitely massive fundamental fields in the theory living on the branes.
The addition of a static quark-antiquark pair was conjectured
to be dual to inserting an open string with its ends on the boundary of
$AdS_5$ in the $AdS_5\times S^5$ background dual to the stack of
D3-branes. 

Alternatively, as was first performed in \cite{KR}, an open
string sector can be added by embedding D-branes in the
supergravity dual. If the number of branes embedded is small compared to
the number of branes generating the background, their backreaction can be
neglected and, accordingly, they can be treated as probes. As we will
describe below, following the philosophy of duality, the fluctuations of
the probe would correspond to open string degrees of freedom connecting
the probe and the background branes and therefore give rise to
fundamental degrees of freedom on the dual gauge theory.

It is important to emphasize here that, by neglecting the backreaction of
the flavor branes on the background, one is describing a gauge
theory with quenched flavors, \ie \ flavors that do not run in internal
loops. Indeed, the flavors are quenched in the strict $N\to\infty$ limit:
the double series expansion (\ref{c1thooftexp}) holds for any pure gauge
Yang-Mills theory (or with matter in the adjoint), but gets modified
by the inclusion of fundamental fields, since the diagrams with
quarks running in internal loops have a different $N$ scaling. By using
the double line notation it is not difficult to see that if we substitute
a gluon running in an internal loop by a quark the diagram loses a free
color line  and thus it has one fewer power of N. However, the
flavor of the quark running in the loop must be summed over, resulting in
an additional power of $N_f$ (where $N_f$ is the number of flavors). In
fact, it is not difficult to see that these diagrams correspond to Riemann
surfaces with boundaries whose Euler number is $\chi=2-2g-b$, with $b$
being the number of boundaries. Then, the topological expansion
(\ref{c1thooftexp}) still holds, but now one must also sum over the number
of boundaries.
In conclusion, the diagrams with quarks running in internal loops are
suppressed by powers of $N_f/N$, and thus they do not contribute to the
amplitude in the limit $N\to\infty$ as long as the number of flavors is
finite. On the string theory side the quenching of the flavors is
reflected on the fact that the corrections due to the presence of $N_f$
flavor branes are of order
$N_f/N$ and thus negligible when $N\to\infty$. This argument will not
apply in the case where both $N$ and $N_f$ are large but $N_f/N$ is
finite, this is the so-called Veneziano limit. In that case, the probe
approximation is not valid anymore and one has to take into account the
backreaction of the flavor branes on the geometry.

\subsubsection{An example: the D3-D5 system}
In general, the probe brane will not share all the
dimensions of the background branes, thereby creating a defect of
some codimension on the worldvolume of the latter. Hence, the strings 
stretching between both kinds of branes give
rise to fundamental fields living on the defect spanned by the probe
brane. Such a setup was first studied in \cite{KR} where it was
considered the orthogonal intersection of a D3- and a D5-brane along two
common spatial dimensions. This intersection is supersymmetric,
preserving $1/2$ of the original 32 supersymmetries.

Let us study the low energy dynamics of the intersection of $N$
D3-branes with $N_f$ D5-branes in the limit where $g_s\to 0$, $N\to
\infty$ with $\lambda=g_s\,N$ fixed, corresponding to the 't Hooft
limit of the gauge theory on the D3s. In the $g_S\,N\ll1$ limit the
correct description of the system is in terms of open strings propagating
on the flat D-branes. In figure \ref{c1d3d5fig} one can see the three
different kinds of open strings present in the D3-D5 intersection: the
3-3 strings realizing the ${\cal N}=4$ vector multiplet living on the
worldvolume of  the D3-branes, the 5-5 strings corresponding to the
adjoint fields of the gauge theory living on the worldvolume of the
D5-brane, and finally, the 3-5 strings restricted to the intersection
and, from the point of view of the theory in the D3s, introducing
fundamental hypermultiplets living on a (2+1)-dimensional defect. In the
usual decoupling limit where
$\alpha'\to 0$ while the energy of the D3 modes (\ie \ the 3-3 strings)
is fixed, the D5-brane excitations, corresponding to the 5-5 strings,
decouple, since its gauge coupling constant vanishes 
($(g_{YM}^{D5})^2\sim g_S\,\alpha'$). Thus, the gauge group $SU(N_f)$ on
the D5 becomes the global flavor symmetry group of the fundamental matter. 

\begin{figure}
\centerline{\hskip .1in \epsffile{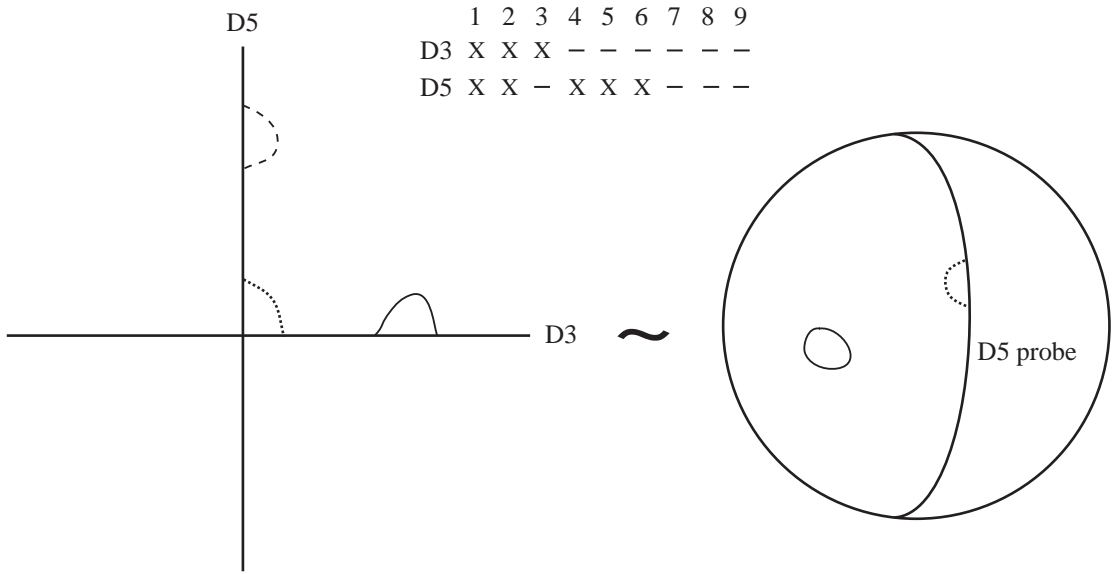}}
\caption{Representation of the duality acting twice on the D3-D5
intersection, whose configuration is shown in the array. In the brane
picture on the left one can see the three different kinds of strings: 3-3,
5-5 (which decouple) and 3-5, these last ones restricted to live on the
(2+1)-dimensional defect. On the right one can see the D5 probe wrapping
an $AdS_4$ submanifold of $AdS_5$.}
\label{c1d3d5fig}
\end{figure}

On the other hand, when $g_S\,N\gg1$, the appropriate description
of the stack of D3-branes is the SUGRA solution which in the decoupling
limit $\alpha'\to 0$ leads to the near-horizon geometry $AdS_5\times
S^5$. Since $N_f\ll N$, the backreaction of the D5-branes on this geometry
is negligible. Furthermore, given that $g_s\to 0$, it happens
that $g_s\,N_f\ll1$ so the description of the D5s as hypersurfaces with
open strings attached still holds. Then, the solution takes the
form of $AdS_5\times S^5$ with $N_f$ D5-brane probes embedded with
open  strings ending on their worldvolume. When the D5s are located at
zero distance from the background D3-branes, they will wrap an
$AdS_4\times S^2$ submanifold of the
$AdS_5\times S^5$ near-horizon geometry. The probes embedded in this way
break half of the supersymmetries of the background (which was maximally
supersymmetric), and so do the fundamental hypermultiplets on the gauge
theory side.

In view of this situation, the authors of
\cite{KR} argued that the AdS/CFT correspondence would act twice on this
system: as usual, the closed string modes propagating through the whole
$AdS_5$ provide a holographic description of the (3+1)-dimensional theory
living on the boundary and, in addition, the fluctuations of the probe
D5-branes wrapping an $AdS_4$ submanifold should describe the physics
confined to the (2+1)-dimensional boundary of this $AdS_4$. Therefore,
the dual field theory will be the (3+1)-dimensional ${\cal N}=4$ SYM
coupled to fundamental fields living on a (2+1)-dimensional defect. These
last ones corresponding to the low energy limit of the strings stretching
between the D3- and D5-branes. The fluctuations of the D5-probe in this
setup will be thoroughly analyzed in section \ref{cp2D3D5}.

This description can be applied to any supersymmetric intersection where
the branes of lower dimension are the ones generating the background, and
thus the system can be used to add flavor to different supergravity
duals \cite{KKW}. Of particular interest is the case of intersections
where the probe branes extend along all the worldvolume directions of the
background branes, since in this case the fundamental fields will not be
confined to a defect in the dual field theory. For the $AdS_5\times S^5$
geometry, dual to four-dimensional ${\cal N}=4$ SYM, the appropriate probe
branes are D7-branes filling all the spacetime directions of the
gauge theory \cite{D3D7}. The construction starts with stacks of $N$
D3- and $N_f$ D7-branes  intersecting along three spatial directions in
flat ten-dimensional space. From the point of view of the theory on the
D3s, the lightest modes of  the 3-7 strings give rise to $N_f$
fundamental hypermultiplets breaking the supersymmetry down to ${\cal
N}=2$. The mass of these fundamentals is proportional to the distance
between the two stacks in the transverse space:
$m_q=L/2\pi\alpha'$. 
On the supergravity side the probe embedding will only preserve $1/4$ of the 32 supersymmetries of $AdS_5\times S^5$, as it must be for the dual of ${\cal N}=2$.
However,  when $L=0$, the probe will wrap an
$AdS_5\times S^3$ submanifold (preserving $1/2$ of the maximal supersymmetry) which suggests  that the dual theory is conformal. Indeed, when the quarks are massless the ${{\cal N}=2}$ theory is classically conformal and, as we have seen, quantum corrections due to the flavors will be of order $N_f/N$ and thus negligible
in the probe limit.

The resulting theory is 4d ${\cal N}=4$ SYM coupled to $N_f$ ${\cal N}=2$
fundamental hypermultiplets. Subsequently, proceeding as before, when
$g_s\,N\gg1$ the  stack of D3-branes can be replaced (in the appropriate
decoupling limit) by the near-horizon $AdS_5\times S^5$ geometry, and if
$N_f\ll N$ the D7-branes can be treated as probes in that geometry. 
As one would expect for a flavor brane introducing an ${\cal N}=2$
fundamental hypermultiplet, the embedding of the probe preserves only one
half of the supersymmetry of the background (in the massless case, when it
wraps an $AdS_5\times S^3$ submanifold of the background). Furthermore,
the  fluctuations of the probe D7s were identified in \cite{KKW} with
mesons in the dual field theory. In \cite{KMMW}, by studying the
Dirac-Born-Infeld action of the probe D7-branes, the mass spectrum of the
complete set of fluctuations was computed. In addition, by looking at their
quantum numbers and UV behavior, the dictionary between the fluctuations
and the corresponding operators of the field theory was established.

%
%
%
%
%
%
%
%
%

\section{The D-brane action}
\label{cp1sscbraneaction}
\setcounter{equation}{0}
\medskip
In the previous section it has been shown how to add flavor to the
AdS/CFT conjecture via D-brane probes. In this context the action describing
the dynamics of these probes, whose fluctuations are dual to the
fundamental degrees of freedom introduced in the dual gauge theory, is of
great importance. Therefore, the low energy bosonic action of a D-brane
will be a fundamental tool throughout this work. Although we will not
derive it here (see ref.
\cite{Dbranes}), it will be useful to write down its form for a generic
Dp-brane in a generic ten-dimensional type II supergravity background.

D-branes are solitonic extended objects where open strings can end.
Consequently, the open strings ending on the brane must satisfy Dirichlet
boundary conditions in the directions orthogonal to the hyperplane. Let us
first recall that in the absence of any such object the massless bosonic
sector of the open string consists of a ten-dimensional vector and,
accordingly, the low energy effective field theory for the open
superstring in flat space with Neumann boundary conditions is ${\cal
N}=1$ SYM. The quantization of the open superstring with Dirichlet
boundary conditions is the same as for the open sector of the type I
superstring. However, the zero modes of the open string along the
directions with Dirichlet boundary conditions are not dynamical and so the
low energy fields corresponding to the massless modes of the  open string
sector do not depend on these directions. Therefore, the low energy
effective theory one gets is the dimensional reduction of the
ten-dimensional one, with the components of the gauge field along
directions with Dirichlet boundary conditions becoming scalars
representing the fluctuations of the D-brane. The hyperplane is then a
dynamical object whose fluctuations correspond to states of the open string
spectrum.

The bosonic action of a Dp-brane can be computed by requiring that the
non-linear sigma model describing the propagation of an open string with
Dirichlet boundary conditions in a general supergravity background is
conformally invariant. The constraints arising for the bosonic fields are
the same as the equations of motion resulting from the following action
(in string frame):
\beq
S_{DBI}=-T_p\int_{\Sigma_{p+1}} d^{p+1}\xi\;e^{-\phi}\sqrt{\,-{\rm
det}\,(\,g+{\cal F}\,)}\,,
\label{cp1DBI}
\eeq
which is known as the Dirac-Born-Infeld (DBI) action. The presence of the
dilaton is due to the fact that the Dp-brane is an object of the open
string spectrum, hence the coupling $g_s^{-1}$ (inside $T_p$ as we will
see below). $T_p$ is the tension of the brane, $\Sigma_{p+1}$ denotes its
worldvolume, and
$g$ is the induced metric on the Dp-brane resulting from the pullback to
the worldvolume of the background metric:
\beq
g_{ab}={\partial\,X^M\over \partial\xi^a}\,{\partial X^N\over\partial
\xi^b}\,G_{MN}\,,
\label{cp1indmetric}
\eeq
where $\xi^a$, $(a=0,\cdots,p)$ are the worldvolume coordinates,
and $X^M$ and $G_{MN}$ are the coordinates and metric of
the background. By making use of the worldvolume and spacetime
diffeomorphism invariance one can go to the so-called ``static gauge''
where the worldvolume of the Dp-brane is aligned with the first $p+1$
spacetime coordinates. In this gauge the embedding of the Dp-brane is
given by the functions $X^m(\xi^a)$, where $X^m$, $(m=1,\cdots,9-p)$ are
the spacetime coordinates transverse to the Dp-brane.
Actually, in this static gauge the induced metric for a Dp-brane in flat
space takes the form:
\beq
g_{ab}=\eta_{ab}+{\partial\,X^m\over \partial\xi^a}\,{\partial X^n\over
\partial\xi^b}\,\delta_{mn}\,.
\label{cp1stindmetric}
\eeq
The two-form ${\cal F}$ appearing in eq. (\ref{cp1DBI}) is given by the
following combination:
\beq
{\cal F}_{ab}=P[B]_{ab}+2\pi\,\alpha'\,F_{ab}\,,
\label{cp1calf}
\eeq
where $F=dA$ is the field strength of the gauge field living on the
worldvolume of the Dp-brane, $P\big[\cdots]$ denotes
the pullback to the worldvolume of the form inside the brackets and $B$ is
the NSNS two-form potential.  Finally, the tension of a Dp-brane can be
computed
\cite{Dbranes} by studying the exchange of a closed string between two
parallel D-branes, resulting in:
\beq
T_p={1\over (2\pi)^p\,g_s\,\alpha'^{\,{p+1\over2}}}\,.
\label{cp1tension}
\eeq
One should notice that the DBI action (\ref{cp1DBI}) not only describes
the massless open string modes given by the worldvolume gauge field $A_a$
and the scalars $X^m$, but also couples the Dp-brane to the massless
closed string modes of the NSNS sector, namely $\phi$, $G_{MN}$ and
$B_{MN}$. Furthermore, as expected, this action is an abelian $U(1)$
gauge theory and when the target space (where the D-brane lives) is flat,
it reduces, to leading order in $\alpha'$, to YM in $p+1$ dimensions with
$9-p$ scalar fields plus higher derivative terms. Indeed, the quadratic
expansion of (\ref{cp1DBI}) for a Dp-brane in flat space reads:
\beq
S_{DBI}\approx-\int_{\Sigma_{p+1}} d^{p+1}\xi\,e^{-\phi}
\left[T_p+{1\over
g_{YM}^2}\left({1\over2}F_{ab}F^{ab}+\partial_a\Phi^m\,
\partial^a\Phi^m\right)\right]\,,
\label{cp1dpYM}
\eeq
where we have applied the usual relation between coordinates and 
fields: $\Phi^m=1/(2\pi\,\alpha')\,X^m$, and the relation (\ref{c1gymdp})
which follows from this expansion\footnote{
In fact, the relation (\ref{c1gymdp}) is such that 
for $N$ coincident Dp-branes the gauge kinetic term of the non-abelian
$SU(N)$ gauge theory living on them takes the standard form:
${2\over4g_{YM}^2}\tr (F_{ab}F^{ab})={1\over4g_{YM}^2}F_{ab}^iF^{ab\;i}$
(since $F_{ab}=F_{ab}^i\,T^i$ and the $SU(N)$ generators $T^i$ satisfy
$\tr (T^i\,T^j)=1/2\,\delta^{ij}$).}.
After taking into account the
fermionic superpartners coming from the fermionic completion of the
action for the super D-brane, the low energy effective theory living on
the worldvolume becomes the  maximally supersymmetric SYM theory in $p+1$
dimensions (for a D-brane in flat space).

In type II superstring theory, the ten-dimensional vacuum is invariant
under ${\cal N}=2$ supersymmetry, but the state containing a D-brane is
invariant only under ${\cal N}=1$, which is the maximum amount of SUSY
preserved by the open strings. Hence, the D-brane is a BPS object and
must carry a conserved charge. The only set of charges with the
appropriate Lorentz properties is given by the antisymmetric RR charges.
Indeed, the worldvolume of a Dp-brane naturally couples to a $(p+1)$-form
RR potential
$C_{(p+1)}$ whose field strength is a $(p+2)$-form. This coupling can
also be motivated by the $g_s^{-1}$ dependence of the D-brane tension
which corresponds to that of an RR soliton. 

By looking again at the amplitude corresponding to the
exchange of a closed string between parallel D-branes, one can see that
it vanishes due to the exact cancellation of the contributions from the
NSNS and RR exchanges. This confirms the fact that the D-branes are BPS
objects and thus the force between them must vanish. In addition, from
the same computation one can extract the tension and charge of the
D-brane. As it should be for a BPS object, the charge of a D-brane is
equal (up to a sign) to its tension given by eq. (\ref{cp1tension}).

Apart from the natural coupling term $\int_{\Sigma_{p+1}} C_{p+1}$, the
action coupling a Dp-brane to the RR potentials must contain additional
terms involving the worldvolume gauge field. By making use of T-duality
one can show that the part of the action of a Dp-brane describing its
coupling to the RR potentials is given by the following Wess-Zumino (WZ)
term:
\beq
S_{\,WZ}=\mu_p\,\int_{\Sigma_{p+1}}\,\sum_{r=0}^{p+1}P\left[C_{(r)}\right]
\wedge e^{\cal F}\,,
\label{cp1WZ}
\eeq
where $\mu_p$ stands for the charge of the Dp-brane and $P$ for the
pullback to its worldvolume. Although the expansion of the
integrand above involves forms of various ranks, the integral only picks
out those proportional to the volume form of the Dp-brane.

Finally, the action of a Dp-brane is given by the sum of the DBI and WZ
terms given by eqs. (\ref{cp1DBI}) and (\ref{cp1WZ}) respectively,
resulting in (in string frame):
\beq
S_{Dp}=-T_p\int_{\Sigma_{p+1}}d^{p+1}\xi\,e^{-\phi}\sqrt{\,-{\rm
det}\,(\,g+{\cal F}\,)}\,\pm\, T_p\int_{\Sigma_{p+1}}\,\sum_{r=0}^{p+1}\,
P\left[C_{(r)}\right]\wedge e^{\cal F}\,,
\label{cp1Dbraneact}
\eeq
where the +(-) sign in front of the second term corresponds to a Dp-brane
(anti Dp-brane).


\subsection{The non-abelian action}
\label{c1ssncact}

We shall now consider the non-abelian generalization of the action
(\ref{cp1Dbraneact}) for the case of  $N$ coincident D-branes.
A remarkable property of D-branes is that the $U(1)$ gauge symmetry of an
individual D-brane is enhanced to a non-abelian $U(N)$ symmetry for $N$
coincident D-branes. When the $N$ parallel D-branes approach each other,
the ground state modes of strings stretching between different D-branes
become massless. These modes carry the appropriate  charges to fill out
$U(N)$ representations and the $U(1)^N$ gauge symmetry of the set of
separated D-branes is enhanced to $U(N)$. 
In this case, the fields $A_a$ and $X^m$ describing the open string degrees
of freedom of the brane become $N\times N$ matrices transforming in the
adjoint of the gauge group. In addition, the background fields typically
depend on the transverse coordinates $X^m$ which have now become
matrices, so the background fields on the worldvolume will possibly
depend (explicitly or implicitly) on the non-abelian fields. Furthermore,
when taking the pullback of the spacetime indices ($M,N,\cdots$) one
now has to use the covariant derivative instead of the usual partial
derivative as we have done in (\ref{cp1indmetric}). For instance, the
pullback of a certain background one-form $H_M$ will now read:
$H_a=H_M\,{\cal
D}_a\,X^M=H_M\left(\partial_a\,X^M+[A_a,X^M]\right)$. 

The non-abelian DBI action can be constructed by introducing the changes
mentioned above in the DBI action of a D9-brane which has no transverse
scalars. In addition, one has to perform a trace over the indices of the
$U(N)$ gauge group. By T-dualizing along $9-p$ coordinates one gets the
non-abelian DBI action for a Dp-brane (in string frame) \cite{simtr,M}:
\beq
\tilde S_{DBI}=-T_p\int_{\Sigma_{p+1}} d^{p+1}\xi\;{\rm Tr}\,
\left[e^{-\phi}\sqrt{\,{\rm -det}\left(\,P[E+E\,(Q^{-1}-\delta)
\,E]_{ab}+2\pi\,\alpha' F_{ab}\right){\rm det}(Q^m_{\,n})}\,\right]\,,
\label{cp1ncDBI}
\eeq
where now we have written the indices over which the determinant is
applied: $a,b$ are worldvolume indices, while $m,n$ run over the
transverse space directions. In addition, we have defined:
\beq
E_{MN}=G_{MN}+B_{MN}\,,\qquad\quad
Q^m_{\,n}=\delta^m_{\,n}+ {i\over2\pi\,\alpha'}\,[X^m,X^p]\,E_{pn}\,.
\eeq
Therefore, the expression $E\,(Q^{-1}-\delta)\,E$ appearing in
(\ref{cp1ncDBI}) reduces to $E_{Mm}\,(Q^{-1}-\delta)^{mn}\,E_{n\,N}$
where the second index of $(Q^{-1}-\delta)^{mn}$ has been raised using
$E^{mn}$. Notice also that in the action (\ref{cp1ncDBI}) appear terms
depending on the commutators of the scalars that one would have lost by
naively substituting the fields in the action for a Dp-brane by their
non-abelian generalizations. In particular, the second determinant
supplies the standard scalar potential of the non-abelian $U(N)$ YM
theory living on the Dp-brane. Indeed, in flat space with vanishing $B$
field, thus setting $G_{MN}=\eta_{MN}$ and $B_{MN}=0$, one can see that:
\beq
\sqrt{{\rm
det}\,(Q^m_{\,n})}=1-{(2\pi\,\alpha')^2\over4}\,[\Phi^m,\Phi^n]\,
[\Phi^m,\Phi^n]+o\left((\Phi^m)^6\right)\,,
\label{cp1ncpot}
\eeq
where again we have made use of the relation
$\Phi^m=1/(2\pi\,\alpha')\,X^m$. Finally, since the action (\ref{cp1ncDBI})
is a highly non-linear functional of the non-abelian fields, a trace
prescription is needed for the gauge indices. It was proposed in
\cite{simtr} that a suitable prescription is a symmetrized trace (STr)
over the gauge indices. This means that the trace is completely symmetric
between all non-abelian expressions, namely $F_{ab}$, ${\cal D}_a\,X^m$
and $[X^m,X^n]$, whose commutators can be consequently ignored while
expanding the action. This suggestion was shown to be consistent with
string scattering amplitudes up to fourth order in the field strength, but
it seems not to be reliable for sixth order and beyond.

As for the Wess-Zumino term coupling the D-brane to the RR fields, the
expression given in eq. (\ref{cp1WZ}) fails to include interactions
involving the commutators of the non-abelian scalar fields. Instead, the
non-abelian generalization consistent with T-duality reads \cite{M}:
\beq
\tilde
S_{\,WZ}=\mu_p\,\int_{\Sigma_{p+1}}\tr\,\bigg(P\bigg[
e^{i\,2\pi\,\alpha'\,{\rm i}_X\,{\rm i}_X}\,
\Big(\sum C_{(r)}\wedge e^{B}\Big)
\bigg]\wedge e^F\bigg)\,,
\label{cp1ncWZ}
\eeq
where ${\rm i}_X$ denotes the interior product by $X^m$ as a vector of
the transverse space. Acting on an $r$-form it reads:
\beq
{\rm i}_X\,C_{(r)}=X^{m}\,C^{(r)}_{\,m\,M_2\cdots M_r}\,dx^{M_2}
\wedge \cdots \wedge dx^{M_r}\,.
\label{cp1intprod}
\eeq
Hence, the exponential $e^{i\,2\pi\,\alpha'\,{\rm i}_X\,{\rm i}_X}$ makes a
non-trivial contribution due to the non-abelian nature of the fields
$X^m$. For instance:
\beq
{\rm i}_{X}\,{\rm i}_{X}\,C^{(2)}=X^m\,X^n\,C^{(2)}_{mn}={1\over2}\,
C^{(2)}_{mn}\,[X^m,X^n]\,.
\label{cp1ncinter}
\eeq
Notice that the integrand in eq. (\ref{cp1ncWZ}) must be evaluated by
considering the expression in each set of brackets in turn, from the
innermost to the outermost. That is to say, first,
$\sum C_{(r)}\wedge e^{B}$ is expanded as a sum of forms
so that only a finite number of terms from the exponential contribute to
the integral. Next, the exponential $e^{i\,2\pi\,\alpha'\,{\rm i}_X\,{\rm
i}_X}$ acts on those terms and, again, only a finite number of terms from
this exponential contribute. Then, the surviving terms act, through a
wedge product, on the expansion of the exponential of the gauge field
strength. Finally, a trace should be performed over the non-abelian
indices. Once again, there are ambiguities when taking the trace of the
terms which are highly non-linear in the non-abelian fields. In \cite{M}
it is adopted the symmetrized trace prescription described above, since it
correctly reproduces the first non-linear interaction terms of the
non-abelian fields obtained from computing string theory amplitudes.

As is well known, a Dp-brane couples not only to the RR potential of
degree $r=p+1$, but it can also couple to the RR potentials of lower
degree $r=p-1,\,p-3,\cdots$ through the interactions with the NSNS
two-form $B$ and the worldvolume field strength $F$. Additionally, in the
non-abelian case one can see from the WZ term (\ref{cp1ncWZ}) that a
Dp-brane can now couple to RR potentials of degree $r=p+3,\,p+5,\cdots$
through additional interactions like (\ref{cp1ncinter}) involving the
commutators of the non-abelian scalars.
This fact gives rise to a new interesting effect which is the D-brane
analog of the dielectric effect (see \cite{M}). Indeed, when a Dp-brane is
placed in a background with a non-trivial RR potential $C_{(r)}$ with
$r>p+2$, new terms will be induced in the scalar potential and there may
appear new minima of the action where the transverse scalars have
non-commuting expectation values. Therefore, the external field may
``polarize" the Dp-branes which would get expanded into a (higher
dimensional) non-commutative worldvolume geometry.

All in all, the non-abelian generalization of the action of a Dp-brane
is the sum of the DBI and WZ terms (\ref{cp1ncDBI}) and (\ref{cp1ncWZ}),
namely:
\bear
&&\tilde S=-T_p\int_{\Sigma_{p+1}}d^{p+1}\xi\;{\rm STr}\left[e^{-\phi}
\sqrt{\,-{\rm det}\left(\,P[E+E\,(Q^{-1}-\delta)\,E]_{ab}+
2\pi\,\alpha' F_{ab}\right){\rm det}(Q^m_{\,n})}\,\right]\pm\rc\rc
&&\qquad\pm\,T_p\int_{\Sigma_{p+1}}{\rm STr}\,\bigg(P\bigg[e^{i\,2\pi\,
\alpha'\, {\rm i}_X\,{\rm i}_X}\,\Big(\sum_{r=0}^{p+1}C_{(r)}\wedge
e^{B}\Big)\bigg]\wedge e^F\bigg)\,,
\label{cp1ncDbraneact}
\eear
where, as in (\ref{cp1Dbraneact}), the +(-) sign in front of the second term corresponds to a Dp-brane
(anti Dp-brane).

\subsection{Kappa symmetry}
\label{cp1sskappasym}

The kappa symmetry is a fermionic gauge
symmetry of the worldvolume theory which is a key ingredient in the
covariant formulation of superstrings and
supermembranes. As we will see, it turns out to be a useful tool to find
supersymmetric embeddings of D-branes in a given background.
Therefore, in this subsection we will briefly introduce the concept of
kappa symmetry and its use to determine the amount of SUSY
preserved by a D-brane probe (see \cite{tcamino} for a nice review).

The bosonic Dp-brane is described by a map $X^m\left(\Sigma_{p+1}\right)$,
where $\Sigma_{p+1}$ refers to the worldvolume of the brane and $X^m$ are
the coordinates of the 10d target space. In order to construct the action
of the super Dp-brane this map is replaced with a supermap\footnote{Only
in this subsection we will use the indices $M,N,\cdots,$ for the
superspace coordinates, while the indices $m,n,\cdots,$ will run over the
bosonic coordinates of the ten-dimensional spacetime.}
$\{Z^M\}=(X^m,\theta^\alpha)$ and, accordingly, the bosonic supergravity
fields with the corresponding superfields. Superspace forms can be
expanded in the coordinate basis $dZ^M$, or alternatively, in the one-form
frame $E^{\underline M}=E^{\underline M}_N\,dZ^N$, where the underlined
indices are flat and $E^{\underline M}_N$ is the supervielbein. Under the
action of the Lorentz group $E^{\underline M}$ decomposes into a vector
$E^{\underline m}$ and a spinor $E^{\underline\alpha}$, which is a
32-component Majorana spinor for IIA superspace and a doublet of chiral
Majorana spinors for IIB superspace. In particular, for flat superspace
one can write:
\beq
E^m=dX^m+\bar\theta\Gamma^m\,d\theta\,,\qquad\quad
E^\alpha=d\theta^\alpha\,,
\label{kpflatbs}
\eeq
where the indices are already flat. The supersymmetric generalization of
the DBI plus WZ terms of the action (see eq. (\ref{cp1Dbraneact})) for a
super Dp-brane reads:
\beq
S=-T_p\int_{\Sigma_{p+1}} d^{p+1}\xi\,\sqrt{-\det (g+\cal{F})}+T_p\int_{\Sigma_{p+1}}
\,\sum_{r=0}^{p+1}\, P\left[C_{(r)}\right]\wedge e^{\cal F}\,,
\label{kpsdpact}
\eeq
with:
\bear
&&g_{ab}=E^{\underline m}_a\,E^{\underline n}_b\,\eta_{\underline m
\underline n}\,,\rc\rc &&{\cal F}_{ab}=F_{ab}-E^{\underline M}_a
E^{\underline N}_b B_{\underline M\underline N}\,,\rc\rc
&&C_{(r)}={1\over r!}\,dZ^{M_1}\wedge\cdots\wedge dZ^{M_r}\,C_{M_1\cdots
M_r}\,,
\label{kpdpfields}
\eear
where the indices $a,b$ run over the worldvolume, $F$ is the field
strength of the worldvolume gauge field $A$ ($F=dA$)\footnote{With
respect to eq. (\ref{cp1Dbraneact}) $F$ has been made dimensionless by
means of the redefinition $2\pi\,\alpha'\,F\to F$.}, $B$ is the NSNS
two-form potential superfield, $C_{(r)}$ is the RR $r$-form gauge potential
superfield and $g_{ab}$ is the induced metric on the worldvolume expressed
in terms of the pullback to the worldvolume of the supervielbein, which is
given by:
\beq
E^{\underline m}_a=\partial_a Z^M\,E^{\underline m}_M\,.
\label{kpvielbpb}
\eeq
In ref. \cite{swedes} the action (\ref{kpsdpact}) was shown to be
invariant under the local fermionic transformations:
\bear
&&\delta_\kappa\,E^{\underline m}=0\,,\rc\rc
&&\delta_\kappa\,E^{\underline \alpha}=(1+\Gamma_\kappa)\kappa\,,
\label{kpkpvar}
\eear
where $\delta_\kappa\,E^{\underline M}=(\delta_\kappa Z^N)
E^{\underline M}_N$. This generalizes to curved backgrounds the kappa
symmetry variations of the flat superspace coordinates:
\beq
\delta_\kappa\,\theta=(1+\Gamma_\kappa)\kappa\,,\qquad\quad\delta_\kappa\,
X^m=\bar\theta\Gamma^m\delta_\kappa\theta\,.
\eeq
One can easily check, reading the supervielbein from (\ref{kpflatbs}),
that these transformations fulfill eq. (\ref{kpkpvar}).

In terms of the induced gamma matrices
$\gamma_{a}=E^{\underline m}_a\,\Gamma_{\underline
m}$, the matrix $\Gamma_\kappa$ takes the form \cite{swedes}:
\beq
\Gamma_{\kappa}={1\over \sqrt{-\det(g+{\cal F})}}\,
\sum_{n=0}^{\infty}\,{(-1)^n\over 2^n n!}\,\gamma^{a_1b_1
\cdots a_n b_n}\,{\cal F}_{a_1 b_1}\,\cdots\,{\cal F}_{a_n b_n}\,
J^{(n)}_{(p)}\,,
\label{kpgammakpdef}
\eeq
where $g$ and ${\cal F}$ are those defined in (\ref{kpdpfields}), and
$J^{(n)}_{(p)}$ is the following matrix:
\beq
J^{(n)}_{(p)} = \cases{(\G_{11})^{n+(p-2)/2}\,\,\G_{(0)} & (IIA)\,,\cr\cr
(-1)^n (\sigma_3)^{n+(p-3)/2}\,\,i\sigma_2 \otimes \G_{(0)} & (IIB)\,,}
\label{kpmtj}
\eeq
with $\Gamma_{(0)}$ being:
\beq
\Gamma_{(0)}={1\over (p+1)!}\,\,\epsilon^{a_1\cdots a_{p+1}}\,\,
\gamma_{a_1\cdots a_{p+1}}\,,
\label{kpGammazero}
\eeq
and $\gamma_{a_1\cdots a_{p+1}}$ denoting the antisymmetrized product of
the induced gamma matrices. In eq. (\ref{kpmtj}) $\sigma_2$ and $\sigma_3$
are Pauli  matrices that act on the two Majorana-Weyl components 
(arranged as a two-dimensional vector) of  the type IIB spinors. As it
will be used several times along this work let us write in a single
expression the form of the kappa symmetry matrix for a Dp-brane in the
type IIB theory:
\beq
\Gamma_{\kappa}={1\over \sqrt{-\det(g+{\cal F})}}\,
\sum_{n=0}^{\infty}\,{(-1)^n\over 2^n n!}\,\gamma^{a_1 b_1\,
\cdots  a_n b_n}\,{\cal F}_{a_1 b_1}\,\cdots\,{\cal F}_{a_n b_n}
(\sigma_3)^{{p-3\over 2}+n}\, (i\sigma_2)\,\Gamma_{(0)}\,.
\label{kpGammakpIIB}
\eeq

What makes this fermionic symmetry a key ingredient in the formulation of
the super Dp-brane is that, upon gauge fixing, it eliminates the extra
fermionic degrees of freedom of the worldvolume theory, guaranteeing the
equality of fermionic and bosonic degrees of freedom on the worldvolume.
As we have seen before the number of bosonic degrees of freedom coming
from the transverse scalars is $9-p$. After adding up the $p-1$ physical
degrees of freedom corresponding to the worldvolume gauge field, the total
number of bosonic degrees of freedom is 8. On the other hand, one has 32
spinors $\theta$ which are cut in half by the equation of motion. We will
see that by gauge fixing the local kappa symmetry the spinorial degrees of
freedom are also halved, thus resulting the expected 8 physical spinors.

As we have said above, one can make use of kappa symmetry when
looking for embeddings of Dp-branes preserving some
supersymmetry \cite{bbs}. Particularly,  we are interested in bosonic
configurations where the fermionic degrees of freedom vanish, \ie \
$\theta =0$, so we only need the variations of
$\theta$ up to linear terms in $\theta$. The supersymmetry plus kappa
symmetry transformations of $\theta$ are:
\beq
\delta\theta=\epsilon+(1+\Gamma_\kappa)\kappa\,,
\label{kpthetatransf}
\eeq
where $\epsilon$ is the SUSY variations parameter. Therefore,
though, generically, these transformations do not leave $\theta$ invariant,
one can choose an appropriate value of $\kappa$ such that
$\delta\theta=0$. Of course, this amounts to gauge fixing the kappa
symmetry. Indeed, let us impose the following gauge fixing condition:
\beq
{\cal P}\,\theta=0\,,
\label{kpkpgaugefix}
\eeq
where ${\cal P}$ is a field independent projector, ${\cal P}^2=1$, so
the non-vanishing components of $\theta$ are given by $(1-{\cal
P})\,\theta$. The condition for preserving the gauge fixing condition
${\cal P}\delta\theta=0$ results in:
\beq
{\cal P}\delta\theta = {\cal P}\epsilon + {\cal P}(1 + \G_{\kappa})\kappa =
0\,,\label{kpunbrsusy}
\eeq
which determines $\kappa=\kappa(\epsilon)$. Then, after gauge fixing the
kappa symmetry, the transformation (\ref{kpthetatransf}) becomes a global
SUSY transformation. The condition of unbroken SUSY for the non-vanishing
components of $\theta$, namely $(1-{\cal P})\,\theta$, reads:
\beq
(1-{\cal P})\,\delta\theta = (1-{\cal P})\,\epsilon + 
(1-{\cal P})(1 + \G_{\kappa})\,\kappa(\epsilon)= \epsilon + 
(1 + \G_{\kappa})\,\kappa(\epsilon)=0\,,
\label{kpunbrsusy2}
\eeq 
where in the last equality we have made use of eq. (\ref{kpunbrsusy}).
Acting on the last equality with $(1-\Gamma_\kappa)$ one gets:
\beq
(1-\Gamma_\kappa)\,\epsilon=0\,.
\label{kpfnalcond}
\eeq  
Therefore, the fraction of supersymmetry preserved by the brane is given
by the number of solutions to this equation. Notice that $\Gamma_\kappa$,
defined in eq. (\ref{kpgammakpdef}), depends on the first derivatives of
the embedding trough the induced metric and the pullback of the $B$
field.  Finally, for backgrounds of reduced supersymmetry, $\epsilon$
(the target space SUSY parameter) must be substituted by the Killing
spinor  of the ten-dimensional geometry where the Dp-brane is embedded.

%
%
%
%
%
%
%
%
%
%

\chapter{Flavor from brane intersections}
\label{cp2dpinter}
\setcounter{equation}{0}
\medskip

\setcounter{equation}{0}
\section{Introduction}
\medskip

It was shown in the last chapter that the so-called AdS/CFT
correspondence is a very useful tool to study strong coupling features of
SYM theories by carrying out classical computations in some supergravity
backgrounds. However, we have seen that in its standard form the
correspondence applies to theories with all their fields transforming in
the adjoint representation of the gauge field. Yet this limitation can be
overcome by studying the dynamics of probe branes embedded in the
appropriate dual backgrounds, as it was described in section
\ref{cp1flavor}. Particularly, in \cite{KMMW} starting from a
supersymmetric intersection of D3- and D7-branes it was possible to
perform a detailed analysis of the spectrum of mesons in the large $N$
limit of ${\cal N}=4$  $SU(N)$ SYM coupled to $N_f$ ${\cal N}=2$
fundamental hypermultiplets when $N_f\ll N$ (in
\cite{Sonnen}-\cite{Apreda:2006bu} different flavor branes and their
spectra  for several backgrounds have been considered, for a review see
\cite{RamRw,Rvdg,Rvjo}). We have argued in the last chapter that the same
procedure can be applied to any supersymmetric intersection where the
branes of lower dimensionality are the ones generating the background and
the others are treated as probes. Therefore, in this chapter we will
generalize those results for the D3-D7 system to a general class of BPS
intersections of two types of branes, both in type II theories and in
M-theory.

In our approach the lower dimensional brane is substituted by the
corresponding near-horizon geometry, while the higher dimensional one
will be treated as a probe. Generically, the addition of the probes to
the supergravity background creates a defect in the gauge theory dual in
which extra hypermultiplets are localized. In the decoupling limit one
sends the string scale $l_s$ to zero keeping the gauge coupling of the
lower dimensional brane fixed. It is straightforward to see that the
gauge coupling of the higher dimensional brane vanishes in this limit
and, as a consequence, the corresponding gauge theory decouples and the
gauge group of the higher dimensional brane becomes the flavor symmetry
of the effective theory at the intersection.

The prototypical example of a defect theory is the one dual to the D3-D5
intersection, described in section \ref{cp1flavor} of the previous
chapter. This system was proposed in ref. \cite{KR} as a generalization
of the usual AdS/CFT correspondence in the
$AdS_5\times S^5$ geometry. Indeed, if the D5-branes are at zero distance
of the D3-branes, they wrap an $AdS_4\times S^2$ submanifold of the 
$AdS_5\times S^5$ background. It was argued in ref. \cite{KR} that the AdS/CFT
correspondence acts twice in this system and, apart from 
the holographic description of the four-dimensional field theory on the boundary of $AdS_5$, the fluctuations of the D5-brane
probe should be dual to the physics confined to the boundary  of $AdS_4$. 

The field theory dual of the D3-D5 intersection corresponds to  ${\cal N}=4$, $d=4$ super
Yang-Mills theory coupled to  ${\cal N}=4$, $d=3$ fundamental hypermultiplets localized
at the defect. In ref. \cite{WFO}  the action of this model in the conformal limit of
zero D3-D5 separation was constructed and a precise
dictionary between operators of the field theory and fluctuation modes of the probe was
obtained (see also refs. \cite{EGK,ST}). We will extend these results to the case in
which the distance between the D3- and D5-branes is non-zero. This non-vanishing
distance breaks  conformal invariance by giving mass to the fundamental
hypermultiplets.  Interestingly, the differential equations for the quadratic
fluctuations can be decoupled and solved analytically, and the
corresponding mass spectra can be given in closed form. These masses
satisfy the degeneracy conditions expected from the structure of the
supersymmetric multiplets found in ref. \cite{WFO}.

The D3-D5 intersection can be generalized to the case of a Dp-D(p+2) BPS
intersection, in which the D(p+2)-brane creates a codimension one defect
in the $(p+1)$-dimensional gauge theory of the Dp-brane. The results
presented here for the D3-D5 system were extended to the generic
Dp-D(p+2) intersection in \cite{AR}. The differential equations of the
fluctuations can also be decoupled in this more general case, and though
they were not analytically solved, the mode structure was disentangled
and the corresponding mass spectra were found by numerical methods.
Furthermore, it was proven in \cite{MyTh} that the  masses of
the fluctuations satisfy degeneracy relations as the ones found here for
the exactly solvable  D3-D5 system.

Another interesting case of defect theory arises from the D3-D3 BPS intersection, in
which the two D3-branes share one spatial dimension. In the conformal limit this
intersection gives rise to a two-dimensional defect in a four-dimensional CFT. 
In this case one
has, in the probe approximation, a D3-brane probe wrapping an  $AdS_3\times S^1$ submanifold of
the  $AdS_5\times S^5$ background. In ref. \cite{CEGK} the spectrum of fluctuations of the
D3-brane probe in the conformal limit was obtained and the corresponding dual fields were
identified (see also \cite{Kirschphd}). Notice that in this intersection
both types of branes have the same dimensionality and the decoupling
argument explained above does not hold anymore.  Therefore, it is more
natural to regard this system as describing two
${\cal N}=4$ four-dimensional theories coupled to each other through a bifundamental
hypermultiplet living on the two-dimensional defect. This fact is reflected in the
appearance of a Higgs branch in the system, in which the two types of D3-branes merge
along some holomorphic curve \cite{CEGK, Kirschphd,Erdmenger:2003kn}. We
will study this system when a non-zero mass is given to the
hypermultiplet. Again, we will be able to solve analytically the
differential equations for the fluctuations and to get the exact mass
spectrum of the model.  This system generalizes to the case of a Dp-Dp
intersection, in which the two Dp-branes have $p-2$ common spatial
directions. For $p\not=3$ the mass spectrum of the different
modes was obtained in \cite{AR} from a numerical integration of the
differential equations of the fluctuations.

The D3-D7 intersection
described above corresponds to a codimension zero ``defect". This
configuration is a particular case of the Dp-D(p+4) BPS intersection  in
which the D(p+4)-brane fills completely the $(p+1)$-dimensional worldvolume
of the Dp-brane and acts as a flavor brane of the corresponding
supersymmetric gauge theory in $p+1$ dimensions. Again, for $p\not=3$ the
mass spectrum was computed in \cite{AR} by means of numerical methods.

This chapter is organized as follows. In section \ref{general} we will
consider a general intersection of two branes of arbitrary
dimensionalities. By placing these two branes at a non-zero distance, and
by imposing a no-force condition on the static configuration, we get an
equation which determines the BPS intersections. Next, we consider
fluctuations of the scalars transverse to both types of branes around the
static BPS configurations. For D-brane probes embedded in $AdS_5\times
S^5$ the corresponding differential equation can be reduced, after a
change of variables, to the hypergeometric differential equation. Thus,
in these cases the form of the fluctuations can be obtained analytically
and the mass spectra of the transverse scalar fluctuations can be found by
imposing suitable boundary conditions in the UV. In the general case the
fluctuation equation can be transformed into the Schr\"odinger equation
for some potential. From the analysis of this potential we can establish
the existence of a discrete spectrum. The mass gap of this spectrum can be
easily read from the fluctuation equation. Remarkably,
for the Dp- Dq-brane intersections the dependence of the mass gap on the
effective coupling constant and the mass of the quark turns out to be
universal. Finally, for the intersections which are not analytically
solvable the use of numerical methods and the WKB approximation to compute
the mass spectra is introduced and some results are presented.

In section \ref{cp2D3D5} we study in detail the complete set of
fluctuations, involving all scalar and vector worldvolume fields, of the
D3-D5 intersection.  In general, these fluctuations are coupled to each
other and one has to decouple them in order  to get a system of
independent equations. The decoupling procedure is actually the same as
in the more general Dp-D(p+2) intersection and is given in detail in
\cite{AR}. In section  \ref{cp2D3D5} we use this procedure to get the
exact mass spectrum of the D3-D5 system. We also recall the
fluctuation/operator dictionary found in ref. \cite{WFO} and check that
the masses we find for the modes are consistent with the arrangement of
the dual operators in supersymmetric multiplets. 

In section \ref{cp2D3D3sc} we perform a complete analysis of the exactly
solvable D3-D3 intersection. In this case the differential equations can
also be decoupled and solved in terms of the hypergeometric function. As
a consequence, the exact mass spectrum can be found and matched with the
fluctuation/operator dictionary established in ref.
\cite{CEGK}. We will also show the appearance of the Higgs branch and how it is
modified by the fact that the hypermultiplet is massive. 

This chapter ends in section \ref{Discussion} where we discuss the results
and comment on open problems and future work. We refer the interested
reader to \cite{AR} where the intersections which are not exactly
solvable are studied; these cases include the Dp-D(p+2), Dp-Dp, Dp-D(p+4) and F1-Dp
intersections of the type II theory, as well as the M2-M2, M2-M5 and M5-M5
intersections of M-theory. The numerical mass spectra and their WKB
estimates were computed for all of them.

%
%
%
%
%
%
%
%
%

\setcounter{equation}{0}
\section{Fluctuations of intersecting branes}
\medskip
\label{general}
\begin{figure}
\centerline{\hskip -.8in \epsffile{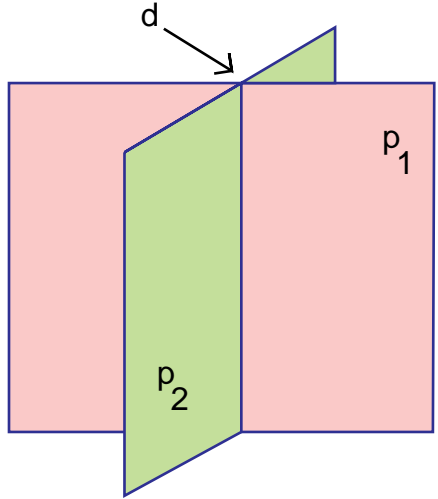}}
\caption{ A general orthogonal intersection of a $p_1$- and $p_2$-brane along $d$
spatial directions.}
\label{intersection}
\end{figure}

Let us consider an orthogonal intersection of a $p_1$-brane and a $p_2$-brane
along $d$ common spatial directions ($p_2\ge p_1$), as depicted in figure
\ref{intersection}.  We shall denote this intersection,
both in type II string theory and M-theory, as  $(d|p_1\perp p_2)$. We shall
treat the lower dimensional $p_1$-brane as a background, whereas the $p_2$-brane
will be considered as a probe. The background metric will be taken as:
\beq
ds^2=\Biggl[{r^2\over R^2}\Biggr]^{\gamma_1}\,\,
(-dt^2+(dx^1)^2\,+\cdots +(dx^{p_1})^2\,)+
\Biggl[{R^2\over r^2}\Biggr]^{\gamma_2}\,\,
d\vec y\cdot d\vec y\,,
\label{cp1metric}
\eeq
where $R$, $\gamma_1$ and $\gamma_2$ are constants that depend on the case
considered, $\vec y=(y^1,\cdots,y^{D-1-p_1})$ with D=10,\,11 and $r^2=\vec y\cdot
\vec y$. In the type II theory the supergravity solution also contains a
dilaton $\phi$, which we will parametrize as:
\beq
e^{-\phi(r)}=\Biggl[\,{R^2\over r^2}\,
\Biggr]^{\gamma_3}\,\,,
\label{dilaton}
\eeq
with $\gamma_3$ being constant (in the case of a background of
eleven-dimensional supergravity we just take $\gamma_3=0$).

Let us now place a $p_2$-brane in this background extended along the directions:
\beq
(\,t,x^1,\cdots,x^d,y^1,\cdots,y^{p_2-d}\,)\,\,.
\label{cp2prwv}
\eeq
We shall denote by $\vec z$ the set of $y$ coordinates transverse to the probe:
\beq
\vec z=(z^1,\cdots, z^{D-p_1-p_2+d-1})\,\,,
\label{cp2trnsv}
\eeq
with $z^m=y^{p_2-d+m}$ for $m=1,\cdots, D-p_1-p_2+d-1$. Notice that the $\vec z$
coordinates are transverse to both background and probe branes. Moreover, we shall
choose spherical coordinates on the
$p_2$-brane worldvolume which is transverse to the $p_1$-brane. If we define:
\beq
\rho^2=(y^1)^2+\cdots+(y^{p_2-d})^2\,,
\label{rho}
\eeq
clearly, one has:
\beq
(dy^1)^2+\cdots+(dy^{p_2-d})^2=d\rho^2+\rho^2
d\Omega^2_{p_2-d-1}\,,
\label{spherical}
\eeq
where $d\Omega^2_{p_2-d-1}$ is the line element of a unit
$(p_2-d-1)$-sphere. Obviously, we are assuming that $p_2-d\ge 2$.

\subsection{BPS intersections}
\label{cp2ssBPS}

\begin{figure}
\centerline{\hskip -.8in \epsffile{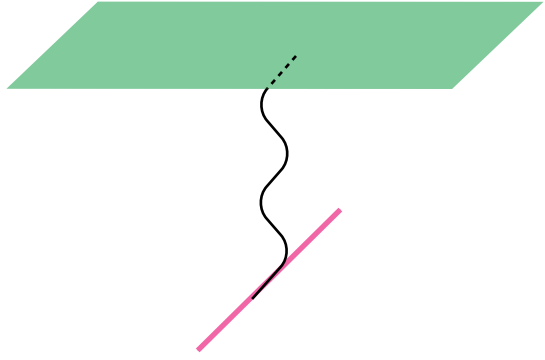}}
\caption{ The two branes of the intersection are separated a finite distance. In the
figure one of the branes is represented as a one-dimensional object. An open string can
be stretched between the two branes.}
\label{separation}
\end{figure}

Let us consider first a configuration in which the probe is located at a
constant value of $|\vec z|$, \ie\ at $|\vec z|=L$ (see figure \ref{separation}). If
$\xi^a$ are a set of worldvolume coordinates, the induced metric on the
probe worldvolume for such a static configuration  will be denoted by:
\beq
ds^2_{I}=\,{\cal G}_{ab}\,d\xi^a d\xi^b\,.
\eeq
In what follows we will use as worldvolume coordinates the cartesian ones
$x^0\cdots x^d$ and the radial and angular variables introduced in eqs. (\ref{rho}) and
(\ref{spherical}). Taking into account that, for an embedding with $|\vec z|=L$, one has 
$r^2=\rho^2\,+\,\vec z^{\,\,2}\,=\,\rho^2\,+\,L^2$, the induced metric can be
written as:
\beq
ds^2_{I}=\Biggl[{\rho^2+L^2\over
R^2}\Biggr]^{\gamma_1}\,(-dt^2+(dx^1)^2+\cdots +(dx^d)^2\,)+
\Biggl[{R^2\over \rho^2+L^2}\Biggr]^{\gamma_2}\,(
d\rho^2+\rho^2 d\Omega^2_{p_2-d-1})\,.
\label{cp2indmetric}
\eeq
The action of the probe is given by the Dirac-Born-Infeld action. In the 
configurations we  study in this section the worldvolume gauge field vanishes  and it
is easy to verify that the lagrangian density reduces to:
\beq
{\cal L}=-e^{-\phi}\,\sqrt{-\det {\cal G}}\,.
\eeq
For a static configuration such as the one with $|\vec z|=L$, the energy
density ${\cal H}$ is just ${\cal H}=-{\cal L}$. By using the explicit form of
${\cal G}$ in (\ref{cp2indmetric}), one can verify that, for the $|\vec
z|=L$ embedding,
${\cal H}$ is given by:
\beq
{\cal H}=\Biggl[{\rho^2+L^2\over
R^2}\Biggr]^{{\gamma_1\over 2}\,(d+1)-{\gamma_2\over 2}\,(p_2-d)-
\gamma_3}\,\rho^{p_2-d-1}\,\sqrt{\det \tilde g}\,,
\eeq
where $\tilde g$ is the metric of the unit $(p_2-d-1)$-sphere. In a BPS
configuration the no-force condition of a supersymmetric intersection requires
that ${\cal H}$ be independent of the distance $L$ between the branes. Clearly, this
can be achieved if the $\gamma_i$-coefficients are related as:
\beq
\gamma_3={\gamma_1\over 2}\,(d+1)-{\gamma_2\over 2}\,(p_2-d)\,.
\label{BPScon}
\eeq
Let us rewrite this last equation as:
\beq
d={\gamma_2\over \gamma_1+\gamma_2}\,p_2+
{2\gamma_3-\gamma_1\over \gamma_1+\gamma_2}\,\,,
\label{BPSrule}
\eeq
which gives the number $d$ of common dimensions of the intersection in terms of the 
parameters $\gamma_i$ of the background and of the dimension $p_2$ of the probe brane.
In the following subsections we shall  consider some particular examples.
\subsubsection{Dp-brane background}
In the string frame, 
the supergravity solution corresponding to a Dp-brane with $p<7$ has been
written in eqs. (\ref{cp1dpsoldiltn}), (\ref{cp1dpsolpot}) and
(\ref{cp1dpsolmtrc}) of the Introduction. It has the form
displayed in eqs. (\ref{cp1metric})\footnote{Notice
that in this equation we are calling $\vec y$ to the vector which was
denoted as $\vec r$ in eq. (\ref{cp1dpsolmtrc}) of the Introduction.}
and (\ref{dilaton}) with $p_1=p$, $R$ given by:
\beq
R^{7-p}=2^{5-p}\,\pi^{{5-p\over 2}}\,\Gamma\Big({7-p\over
2}\Big)\,g_s\,N\, (\alpha')^{{7-p\over 2}}\,\,,
\label{RDp}
\eeq
and with the following  values for the exponents $\gamma_i$:
\beq
\gamma_1=\gamma_2={7-p\over 4}\;,\qquad\quad
\gamma_3={(7-p)(p-3)\over 8}\;.
\label{Dpgammas}
\eeq
Moreover, the Dp-brane solution is endowed with a Ramond-Ramond $(p+1)$-form potential,
whose component along the Minkowski coordinates $x^0\cdots x^p$ can be taken as:
\beq
\Big[C^{(p+1)}\Big]_{x^0\cdots x^p}=
\Biggl[{r^2\over R^2}\Biggr]^{{7-p\over 2}}\,.
\label{cp2CRR}
\eeq

Applying eq. (\ref{BPSrule}) to this background, we get the following relation 
between $d$ and $p_2$:
\beq
d={p_2+p-4\over 2}\,\,.
\eeq
Let us now consider the case in which the probe brane is another D-brane.
As the brane of the background and the probe should live in the same type II theory, 
$p_2-p$ should be even. Since $d\le p$, we are left with the following
three possibilities:
\beq
(p|Dp\perp D(p+4))\,\,,\qquad
(p-1|Dp\perp D(p+2))\,\,,\qquad
(p-2|Dp\perp Dp)\,\,.
\eeq

\subsubsection{Fundamental string background}
In the string frame, 
the metric and dilaton for
the background created by a fundamental string are of the form of eqs.
(\ref{cp1metric}) and (\ref{dilaton}) for:
\beq
\gamma_1=3\;,\qquad \gamma_2=0\;,\qquad
\gamma_3={3\over 2}\;,\qquad R^6=32\pi^2 (\alpha')^3\,g_s^2\,N\,. 
\label{F1back-parameters}
\eeq
In this case one gets from (\ref{BPSrule}) that $d=0$, which corresponds to the following
intersection:
\beq
(0|F1\perp Dp)\,.
\eeq

\subsubsection{M2-brane background}
Our next example is the  geometry created by an M2-brane  in M-theory.
In this case one has:
\beq
\gamma_1=2\;,\qquad
\gamma_2=1\;,\qquad
\gamma_3=0\;,\qquad
R^6=32\pi^2 l_P^{\,6}\,\,N\,,
\label{cp2M2back-parameters}
\eeq
where $l_P$ is the Planck length in eleven dimensions. In this case
eq. (\ref{BPSrule}) becomes:
\beq
d={p_2-2\over 3}\,.
\eeq
Taking $p_2=2,5$ we get the following intersections:
\beq
(0|M2\perp M2)\;,\qquad\qquad (1|M2\perp M5)\,.
\eeq

\subsubsection{M5-brane background}
The background corresponding to an M5-brane in eleven-dimensional
supergravity has:
\beq
\gamma_1={1\over 2}\;,\qquad
\gamma_2=1\;,\qquad
\gamma_3=0\;,\qquad
R^3=\pi\,l_P^{\,3}\,N\,,
\label{M5back-parameters}
\eeq
which leads to:
\beq
d={2p_2-1\over 3}\,.
\eeq
For $p_2=5$ in the previous expression we get the intersection:
\beq
(3|M5\perp M5)\,.
\eeq

\subsection{Fluctuations}
\label{generalfluctuations}

In what follows we will assume that the condition  (\ref{BPScon}) holds. This fact can be
checked for all the particular supersymmetric intersections that will be
analyzed below. 

Let us now study the fluctuations around the $|\vec z|=L$ embedding. Without
loss of generality we can take $z^1=L$, $z^m=0$ ($m>1$) as the unperturbed
configuration and consider a fluctuation of the type:
\beq
z^1=L+\chi^1\;,\qquad\quad z^m=\chi^m\,\,(m>1)\,\,,
\eeq
where the $\chi\,$s are small. The dynamics of the fluctuations is
determined by the Dirac-Born-Infeld lagrangian which, for the
fluctuations of the transverse scalars we study in this section, reduces
to 
${\cal L}\,=\,-e^{-\phi}\sqrt{-\det g}$, where $g$ is the induced metric on the
worldvolume. By expanding this lagrangian and keeping up to
second order terms, one can prove that:
\beq
{\cal L}=-\,{1\over 2}\,\,\rho^{p_2-d-1}\,\,\sqrt{\det \tilde g}\,\,
\Biggl[{R^2\over \rho^2+L^2}\Biggr]^{\gamma_2}\;
{\cal G}^{ab}\,\partial_{a}\chi^m\,\partial_{b}\chi^m\,,
\label{fluct-lag-general}
\eeq
where ${\cal G}^{ab}$ is the (inverse of the) metric
(\ref{cp2indmetric}). The equations of motion derived from this
lagrangian are:
\beq
\partial_{a}\,\Bigg[\,{\rho^{p_2-d-1}\sqrt{\det \tilde g}\over 
(\rho^2+L^2)^{\gamma_2}}\,{\cal G}^{ab}\,\partial_{b}\chi
\,\,\Bigg]=0\,\,,
\label{eom-general}
\eeq
where we have dropped the index $m$ in the $\chi\,$s. Using the explicit
form of the metric elements ${\cal G}^{ab}$, the above equation can be
written as:
\beq
{R^{2\gamma_1+2\gamma_2}\over (\rho^2+L^2)^{\gamma_1+\gamma_2}
}\,\,\partial^{\mu}\partial_{\mu}\,\chi+ {1\over
\rho^{p_2-d-1}}\,\partial_{\rho}\,(\rho^{p_2-d-1}\partial_{\rho}\chi)+
{1\over
\rho^2}\,\nabla^i\nabla_i\,\chi=0\,,
\eeq
where the index $\mu$ corresponds to the directions $x^{\mu}=(t, x^1,\cdots,
x^d)$ and $\nabla_i$ is the covariant derivative on the $(p_2-d-1)$-sphere. In
order to analyze this equation, let us separate variables as:
\beq
\chi=\xi(\rho)\,e^{ikx}\,Y^l(S^{p_2-d-1})\,,
\label{sepvar}
\eeq
where the product $kx$ is performed with the flat Minkowski metric and
$Y^l(S^{p_2-d-1})$ are scalar spherical harmonics which satisfy:
\beq
\nabla^i\nabla_i\,Y^l(S^{p_2-d-1})=-l(l+p_2-d-2)\,\,Y^l(S^{p_2-d-1})\,.
\label{casimir}
\eeq
If we redefine the variables as:
\beq
\varrho={\rho\over L}\;,\qquad
\bar M^2\,=\,-R^{7-p}\,L^{2-2\gamma_1-2\gamma_2}\,k^2\,,
\label{newvariables}
\eeq
the differential equation becomes:
\beq
\partial_{\varrho}\,(\varrho^{p_2-d-1}\partial_{\varrho}\,\xi)+\left[
\bar M^2\,{\varrho^{p_2-d-1}\over (1+\varrho^2)^{\gamma_1+\gamma_2}}-
l(l+p_2-d-2)\varrho^{p_2-d-3}\right]\,\xi=0\,.
\label{cp2fluc}
\eeq

In order to study the solutions of eq. (\ref{cp2fluc}), let us change
variables in such a way that this equation can be written as a Schr\"odinger
equation:
\beq
\partial^2_y\,\psi-V(y)\,\psi=0\,,
\label{Sch}
\eeq
where $V$ is some potential. The change of variables needed to convert eq.
(\ref{cp2fluc}) into (\ref{Sch}) is:
\beq
e^y=\varrho\;,\qquad\quad
\psi=\varrho^{{p_2-d-2\over 2}}\,\xi\,.
\label{Sch-variables}
\eeq
Notice that in this change of variables $\varrho\to\infty$ corresponds to $y\to\infty$,
while the point $\varrho=0$ is mapped into $y=-\infty$. Moreover, 
the resulting potential $V(y)$ takes the form:
\beq
V(y)=\left(l-1+{p_2-d\over 2}\right)^2-\bar M^2\,{e^{2y}\over
(e^{2y}+1)^{\gamma_1+\gamma_2}}\,.
\label{potential}
\eeq

\begin{figure}
\centerline{\hskip -.8in \epsffile{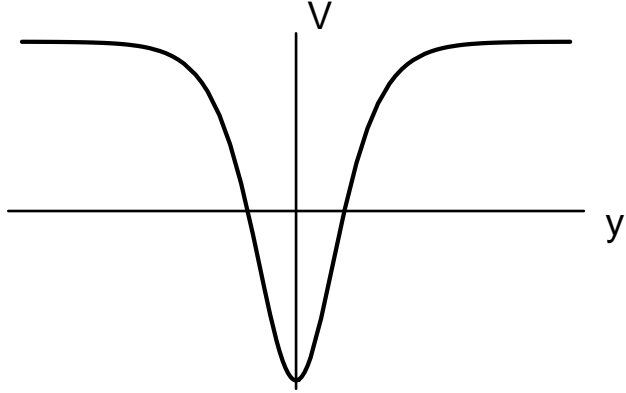}}
\caption{ The Schr\"odinger potential $V(y)$ of eq. (\ref{potential}).}
\label{poten}
\end{figure}

In figure \ref{poten} we have plotted the function $V(y)$. It is 
interesting to notice that, in these new variables, the problem of
finding the mass spectrum can be rephrased as that of finding the values
of $\bar M$ such that a zero-energy level for the potential 
(\ref{potential}) exists. Notice that the classically allowed region in
the Schr\"odinger equation (\ref{Sch}) corresponds to the values of $y$
such that $V(y)\le 0$. We would have a discrete spectrum if this region
is of finite size or, equivalently, if the points
$y=\pm \infty$ are not allowed classically.  As, when $\gamma_1+\gamma_2>1$, one has:
\beq
\lim_{y\to\pm\infty}\,V(y)=\Biggl(l-1+{p_2-d\over 2}\Biggr)^2\,,
\label{generalV-limit}
\eeq
we will have a discrete spectrum for all values of $l\in{\mathbb Z}_+$ if
$p_2-d>2$. Notice that $p_2-d\ge 2$ and when $p_2-d=2$ and $l=0$ the
turning points of the potential $V(y)$ are at $y=\pm \infty$. Moreover,
$V(y)$ has a unique minimum at a value of $y$ given by:
\beq
e^{2y_0}={1\over\gamma_1+\gamma_2-1}\,.
\label{c2potmin}
\eeq

\subsection{The exactly solvable case}
\label{cp2scexactly-solvable}
When $\gamma_1+\gamma_2=2$, the differential equation for the fluctuation can
be solved exactly in terms of a hypergeometric function (see appendix
\ref{hyperchange}). To prove this statement,  let us change variables in eq.
(\ref{cp2fluc}) as follows:
\beq
z=-\varrho^2\,.
\eeq
One can check that  eq. (\ref{cp2fluc}) for $\gamma_1+\gamma_2=2$ is
converted into:
\bear
z(1-z)\,{\partial^2\xi\over \partial z^2}+
{p_2-d\over 2}\,(1-z)\,{\partial\xi\over \partial z}
+\Bigg[\,\,{l(l+p_2-d-2)\over 4}\,(1-z^{-1})-
{\bar M^2\over 4}\,(1-z)^{-1}\,\Bigg]\,\xi=0\,,\rc
\label{hyper}
\eear
which can be reduced to the hypergeometric differential equation. Indeed, 
let us  define $\lambda$ as:
\beq
\lambda\equiv {-1+\sqrt{\,1+\phantom{}\bar M^2\phantom{\big)}}\over
2}\,.
\label{lambda}
\eeq
Notice that eq. (\ref{lambda}) can be easily inverted, namely:
\beq
\bar M^2=4\lambda (\lambda+1)\,.
\label{Mlambda}
\eeq
Then, in terms of the original variable $\varrho$, the solution of eq.
(\ref{hyper}) that is regular when
$\varrho\to 0$ is:
\beq
\xi(\varrho)=\varrho^l\,(\varrho^2+1)^{-\lambda}\,
F(-\lambda,  -\lambda+l-1+{p_2-d\over 2}; l+{p_2-d\over 2};-\varrho^2\,)\,
\,.
\label{scalarhyper}
\eeq
We also want that $\xi$ vanishes when $\varrho\to\infty$. A way to ensure
this is by imposing that:
\beq
-\lambda+l-1+{p_2-d\over 2}=-n\;,
\qquad n=0,1,2,\cdots\,.
\label{quant}
\eeq
When this condition is satisfied the hypergeometric function behaves as $(\varrho^2)^n$
when $\varrho\to\infty$ and $\xi\sim \varrho^{-(l+p_2-d-2)}$ in this limit.
Notice that when $p_2-d=2$ the $l=0$ mode does not vanish at large $\varrho$, in
agreement with our general analysis based on the potential (\ref{potential}). By using
the condition (\ref{quant}) in eq. (\ref{Mlambda}), one gets:
\beq
\bar M^2=4\Bigg(n+l-1+{p_2-d\over 2}\,\Bigg)
\Bigg(n+l+{p_2-d\over 2}\,\Bigg)\,.
\label{cp2exactM}
\eeq
Since in this case $\bar M^2=-R^4L^{-2}k^2$, one gets the following spectrum of
possible values of $k^2=-M^2$:
\beq
M^2={4L^2\over R^4}\,\Bigg(n+l+{p_2-d-2\over 2}\,\Bigg)
\Bigg(n+l+{p_2-d\over 2}\,\Bigg)\,.
\label{exactMass}
\eeq

Let us calculate which intersections satisfy the condition 
$\gamma_1+\gamma_2=2$, needed to reduce the fluctuation equation to the hypergeometric
one. From the list of intersections worked out in subsection
\ref{cp2ssBPS}, it is clear that this exactly solvable cases can only
occur if the background brane is a Dp-brane. Actually, in this
case one must have $\gamma_1=\gamma_2=1$ (see eq. (\ref{Dpgammas})),
which only happens if $p=3$. Thus, the list of exactly solvable
intersections reduces to the following cases:
\beq
(3|D3\perp D7)\,\,,\qquad
(2|D3\perp D5)\,\,,\qquad
(1|D3\perp D3)\,\,.
\label{AdSdefects}
\eeq
Notice that, for the three cases in (\ref{AdSdefects}), $p_2=2d+1$ for $d=3,2,1$.
Therefore, if $dx_{1,d}^2$ denotes the line element for the flat Minkowski space in
$d+1$ dimensions, one can write the induced metric (\ref{cp2indmetric})
as:
\beq
ds^2_{I}={\rho^2+L^2\over R^2}\,dx_{1,d}^2+{R^2\over\rho^2+L^2}\,d\rho^2
+R^2\,{\rho^2\over \rho^2+L^2}\,d\Omega^2_d\,.
\label{AdSindmetric}
\eeq
Moreover, by using the relation between $p_2$ and $d$, one can rewrite
the mass spectra (\ref{exactMass}) of scalar fluctuations for the 
intersections (\ref{AdSdefects}) as:
\beq
M_S^2={4L^2\over R^4}\,\Bigg(n+l+{d-1\over 2}\,\Bigg)
\Bigg(n+l+{d+1\over 2}\,\Bigg)\,.
\label{generalMS}
\eeq

In the three cases in (\ref{AdSdefects}) the background geometry is $AdS_5\times S^5$. 
Moreover, one can see from (\ref{AdSindmetric}) that
the induced metric reduces in the UV limit $\rho\to\infty$ to that of a product space of
the form  $AdS_{d+2}\times S^{d}$. Indeed, the $(3|D3\perp D7)$ intersection is the case
extensively studied in ref. \cite{KMMW} and corresponds in the UV to an 
$AdS_5\times S^3\subset AdS_5\times S^5$ embedding. In this case the D7-brane is a flavor
brane for the ${\cal N}=4$ gauge theory.  The $(2|D3\perp D5)$ intersection represents
in the UV an $AdS_4\times S^2$ defect in $AdS_5\times S^5$. In the conformal limit $L=0$
the corresponding defect CFT has been studied in detail in ref. \cite{WFO} where, in
particular, the fluctuation/operator dictionary was found. In section
\ref{cp2D3D5} we will extend these results to the case in which the brane
separation $L$ is different from zero and we will be able to find
analytical expressions for the complete set of fluctuations. The
$(1|D3\perp D3)$ intersection corresponds in the UV to an
$AdS_3\times S^1$ defect in $AdS_5\times S^5$ which is of codimension two in the gauge
theory directions. The fluctuation spectra and the corresponding field theory dual for
$L=0$ have been analyzed in ref.
\cite{CEGK}. In section \ref{cp2D3D3sc} we will integrate analytically the
differential equations for all the fluctuations of this $(1|D3\perp D3)$
intersection when the D3-brane separation $L$ is non-vanishing.

The $\varrho\to\infty$ limit is simply the high energy regime of the
theory, where the mass of the quarks can be ignored and the theory
becomes conformal. Therefore, the
$\varrho\to\infty$ behavior of the fluctuations should provide us  information about
the conformal dimension $\Delta$ of the corresponding dual operators. 
At this point some subtleties regarding the mapping of observables
(\ref{c1actmap}) between both sides of the correspondence need be
mentioned. In fact, the original mapping was formulated for Euclidean
$AdS$ spaces. In that case there are no normalizable solutions of the
field equations; they diverge either at the origin ($r\to0$) or at the
boundary ($r\to\infty$). When one moves to Lorentzian AdS, normalizable
solutions appear. It was argued in \cite{Kraus} that while the
non-normalizable modes correspond to operator insertions in the boundary
gauge theory, the normalizable ones are identified with the physical
states of those operators. 

In the context of the AdS/CFT correspondence in $d+1$ dimensions, it is
well known that, if the fields are canonically normalized, the
normalizable modes behave at infinity as
$\rho^{-\Delta}$, whereas the non-normalizable ones should behave as
$\rho^{\Delta-d-1}$, which coincides with (\ref{c1adsscalar}) for d=3. In
the case in which the modes are not canonically normalized the behaviors
of both types of modes are of the form 
$\rho^{-\Delta+\gamma}$ and $\rho^{\Delta-d-1+\gamma}$ for some $\gamma$.
Clearly, we can obtain the conformal dimension from the difference
between the exponents. Let us apply this method to  fluctuations which
are given in terms of hypergeometric functions, as in eq.
(\ref{scalarhyper}). Since for large $\varrho$ the hypergeometric
function behaves as:
\beq
F(a_1, a_2;b;-\varrho^2)\approx c_1\, \varrho^{-2a_1}+
c_2\, \varrho^{-2a_2}\;,\qquad (\varrho\to \infty)\,,
\label{asymptotic-hyper}
\eeq
one immediately gets:
\beq
\Delta={d+1\over 2}+a_2-a_1\,.
\label{Delta-hyper}
\eeq
For the scalar fluctuations studied above, one has from the solution (\ref{scalarhyper})
that $a_1=-\lambda$ and $a_2=-\lambda+l+{d-1\over 2}$. By applying eq.
(\ref{Delta-hyper}) to this case, we get the following value for the dimension of the
operator associated to the scalar fluctuations:
\beq
\Delta_S=l+d\,.
\label{generalDeltaS}
\eeq
Notice that the quantization condition (\ref{quant}) selects precisely the normalizable
modes, which behave at large $\varrho$  as $\xi\sim\varrho^{-\Delta_S+1}$. Notice also
that the modes that become constant at $\varrho\to\infty$ correspond to
operators with $\Delta_S=1$ and, therefore, they should not be discarded.
A glance at eq. (\ref{generalDeltaS}) reveals that this situation only
occurs when $d=1$ (\ie\ for the $AdS_3\times S^1$ defect) and $l=0$. This
case will be studied in detail in section \ref{cp2D3D3sc}.

As the hypergeometric function $F(a_1, a_2;b;-\varrho^2)$ is symmetric under
$a_1\leftrightarrow a_2$, it is clear that the roles of $a_1$ and $a_2$ can also be
exchanged in (\ref{Delta-hyper}). If the resulting conformal dimension lies in the
unitarity range $\Delta>0$ (or $\Delta\ge 0$ if $d=1$) we have a second branch of
fluctuations. For the case at hand $\Delta=1-l$ and the unitarity condition requires
generically that $l=0$. This second branch  is selected by
imposing  to the hypergeometric function (\ref{scalarhyper})
the truncation condition $\lambda=n$  for $n=1,2,\cdots$. The resulting spectrum is just
$\bar M^2=4n(n+1)$, $n=1,2,\cdots$. In the rest of this chapter this
second branch of  the fluctuations of the transverse scalars will not be
considered further.

\subsection{WKB quantization}
\label{cp2WKBssc}
The mapping to the Schr\"odinger equation we have performed in section
\ref{generalfluctuations} allows us to apply the semiclassical WKB approximation to
compute the fluctuation spectrum. The WKB method has been very successful
\cite{MInahan,RS} in  the calculation of the glueball spectrum in the context of the
gauge/gravity correspondence \cite{glueball}. The starting point in this calculation is
the WKB quantization rule:
\beq
(n+{1\over 2})\,\pi=\int_{y_1}^{y_2}dy\,\sqrt{-V(y)}\;,
\qquad n\ge 0\,,
\label{WKBquantization}
\eeq
where $n\in{\mathbb Z}$ and $y_1$, $y_2$ are the turning points of the
potential ($V(y_1)=V(y_2)=0$). To evaluate the right-hand side of
(\ref{WKBquantization}) we expand it as a power series in $1/\bar M$ and
keep the leading and subleading terms of this expansion. We obtain  in
this way the expression of $\bar M$ as a function of the principal quantum
number $n$ which is, in principle, reliable for large $n$, although in
some cases it happens to give the exact result. Let us recall the outcome
of this analysis for a general case, following \cite{RS}. With this
purpose,  let us come back to the original variable $\varrho$ and suppose
that we have a differential equation of the type:
\beq
\partial_{\varrho}\,\big(g(\varrho)\,\partial_{\varrho}\,\phi\big)+
\big(\bar M^2\,q(\varrho)+p(\varrho)\big)\,\phi=0\,,
\label{WKBdiffeq}
\eeq
where the functions $g$, $q$ and $p$ behave near $\varrho\approx
0,\infty$ as:
\bear
&&g\approx g_1\varrho^{s_1}\,\,,
\,\,\,\,\,\,\,\,\,\,\,\,\,\,
q\approx q_1\varrho^{s_2}\,\,,
\,\,\,\,\,\,\,\,\,\,\,\,\,\,
p\approx p_1\varrho^{s_3}\,\,,
\,\,\,\,\,\,\,\,\,\,\,\,\,\,{\rm as}\,\,\varrho\to 0\,\,,\rc\rc
&&g\approx g_2\varrho^{r_1}\,\,,
\,\,\,\,\,\,\,\,\,\,\,\,\,\,
q\approx q_2\varrho^{r_2}\,\,,
\,\,\,\,\,\,\,\,\,\,\,\,\,\,
p\approx p_2\varrho^{r_3}\,\,,
\,\,\,\,\,\,\,\,\,\,\,\,\,\,{\rm as}\,\,\varrho\to \infty\,\,.
\eear
The consistency of the WKB approximation requires that $s_2-s_1+2$ and $r_1-r_2-2$ be
strictly positive numbers, whereas $s_3-s_1+2$ and $r_1-r_3-2$ can be
either positive or zero \cite{RS}. In our case (eq. (\ref{cp2fluc})), the
functions
$g(\varrho)$, $q(\varrho)$ and
$p(\varrho)$ are:
\beq
g(\varrho)=\varrho^{p_2-d-1}\;,\qquad
q(\varrho)={\varrho^{p_2-d-1}\over (1+\varrho^2)^{\gamma_1+\gamma_2}}\;,
\qquad
p(\varrho)=-l(l+p_2-d-2)\varrho^{p_2-d-3}\,.
\label{gqp}
\eeq
From the behavior at $\varrho\approx 0$ of the functions written above we
obtain:
\bear
&&g_1=1\,\,,
\,\,\,\,\,\,\,\,\,\,\,\,\,\,\,\,\,\,\,\,\,\,\,\,
\,\,\,\,\,\,\,\,\,\,\,\,\,\,\,\,\,\,\,\,\,\,\,\,
\,\,\,\,\,\,\,\,\,\,\,\,\,\,\,\,\,\,\,\,\,\,\,\,\,\,\,\,
s_1=p_2-d-1\,\,,\rc\rc
&&q_1=1\,\,,
\,\,\,\,\,\,\,\,\,\,\,\,\,\,\,\,\,\,\,\,\,\,\,\,
\,\,\,\,\,\,\,\,\,\,\,\,\,\,\,\,\,\,\,\,\,\,\,\,
\,\,\,\,\,\,\,\,\,\,\,\,\,\,\,\,\,\,\,\,\,\,\,\,\,\,\,\,\,
s_2=p_2-d-1\,\,,\rc\rc
&&p_1=-l(l+p_2-d-2)\,
\,\,,
\,\,\,\,\,\,\,\,\,\,\,\,\,\,\,\,\,\,\,\,\,\,\,\,\,\,\,\,
\,\,\,\,\,\,
s_3=p_2-d-3\,\,.
\label{IR}
\eear
Notice that $s_2-s_1+2=2$ and $s_3-s_1+2=0$ and, thus, we are within the
range of applicability of the WKB approximation. Moreover, from the
behavior at $\varrho\to\infty$  of the functions written in (\ref{gqp}) we
get:
\bear
&&g_2=1\,\,,
\,\,\,\,\,\,\,\,\,\,\,\,\,\,\,\,\,\,\,\,\,\,\,\,
\,\,\,\,\,\,\,\,\,\,\,\,\,\,\,\,\,\,\,\,\,\,\,\,
\,\,\,\,\,\,\,\,\,\,\,\,\,\,\,\,\,\,\,\,\,\,\,\,\,\,\,\,
r_1=p_2-d-1\,\,,\rc\rc
&&q_2=1\,\,,
\,\,\,\,\,\,\,\,\,\,\,\,\,\,\,\,\,\,\,\,\,\,\,\,
\,\,\,\,\,\,\,\,\,\,\,\,\,\,\,\,\,\,\,\,\,\,\,\,
\,\,\,\,\,\,\,\,\,\,\,\,\,\,\,\,\,\,\,\,\,\,\,\,\,\,\,\,
r_2=p_2-d-1-2\gamma_1-2\gamma_2\,\,,\rc\rc
&&p_2=-l(l+p_2-d-2)
\,\,,
\,\,\,\,\,\,\,\,\,\,\,\,\,\,\,\,\,\,\,\,\,\,\,\,\,\,\,\,
\,\,\,\,\,
r_3=p_2-d-3\,\,.
\label{UV}
\eear
Now $r_1-r_2-2=2(\gamma_1+\gamma_2-1)$ and $r_1-r_3-2=0$ and we are also in the
range of applicability of the WKB method if $\gamma_1+\gamma_2>1$. Coming back to the
general case, let us define \cite{RS} the quantities:
\beq
\alpha_1=s_2-s_1+2\;,\qquad\quad
\beta_1=r_1-r_2-2\,,
\label{WKBalphabetauno}
\eeq
and (as $s_3-s_1+2=r_1-r_3-2=0$, see \cite{RS}):
\beq
\alpha_2=\sqrt{(s_1-1)^2-4\,{p_1\over g_1}}\;,\qquad\quad
\beta_2=\sqrt{(r_1-1)^2-4\,{p_2\over g_2}}\;.
\label{alphabeta2}
\eeq
Then, the mass levels for large quantum number $n$ can be written in terms of
$\alpha_{1,2}$ and $\beta_{1,2}$ as \cite{RS}:
\beq
\bar M^2_{WKB}={\pi^2\over \zeta^2}\,(n+1)\,\bigg(n+{\alpha_2\over
\alpha_1}+ {\beta_2\over \beta_1}\bigg)\,,
\label{generalWKBlevels}
\eeq
where $\zeta$ is the following integral:
\beq
\zeta=\int_0^{\infty} d\varrho\,\,\sqrt{{q(\varrho)\over g(\varrho)}}
\;.\label{WKBxidef}
\eeq
In our case $\alpha_{1,2}$ and $\beta_{1,2}$ are easily obtained from the coefficients
written in (\ref{IR}) and (\ref{UV}), namely:
\beq
\alpha_1=2\,\,,
\,\,\,\,\,\,\,\,\,\,\,\,\,\,
\beta_1=2(\gamma_1+\gamma_2-1)\,\,,
\,\,\,\,\,\,\,\,\,\,\,\,\,\,
\alpha_2=\beta_2=2l+p_2-d-2\,.
\label{alphabeta}
\eeq
Moreover, the integral $\zeta$ for our system is given by:
\beq
\zeta=\int_{0}^{\infty}{d\varrho\over 
(1+\varrho^2)^{{\gamma_1+\gamma_2\over 2}}}= 
{\sqrt{\pi}\over 2}\,\,
{\Gamma\Big({\gamma_1+\gamma_2-1\over 2}\Big)\over 
\Gamma\Big({\gamma_1+\gamma_2\over 2}\Big)}\,\,,
\label{zeta}
\eeq
and we get the following WKB formula for the masses:
\beq
\bar M^{WKB}_S=2\sqrt{\pi}\,
{\Gamma\Big({\gamma_1+\gamma_2\over 2}\Big)\over
 \Gamma\Big({\gamma_1+\gamma_2-1\over 2}\Big)}\,
\sqrt{(n+1)\,\Bigg(n+{\gamma_1+\gamma_2\over \gamma_1+\gamma_2-1}\,
\Big(l-1+{p_2-d\over 2}\Big)\Bigg)}\;.
\label{WKBM}
\eeq

\subsection{Numerical computation}
\label{cp2ssgeneralnumeric}

The  formula (\ref{WKBM}) for the masses can be checked numerically by
means of the shooting technique. Notice that the behavior for small $\varrho$ of the
fluctuation $\xi$ (needed when this technique is applied) can be easily obtained. Indeed,
let us try to find a solution of eq. (\ref{cp2fluc}) of the form:
\beq
\xi\sim \varrho^{\gamma}\,\,,
\eeq
and let us neglect the term containing $\bar M^2$ of eq. (\ref{cp2fluc}).
It is immediate to see that $\gamma$ satisfies the equation:
\beq
\gamma^2+(p_2-d-2)\gamma-l(l+p_2-d-2)=0\,\,,
\eeq
whose roots are: 
\beq
\gamma=l,-(l+p_2-d-2)\,\,.
\label{IRroots}
\eeq
Clearly, the solution regular at $\varrho=0$ should correspond to the root $\gamma=l$.
Then, we conclude that near $\varrho=0$ one has:
\beq
\xi\sim \varrho^{l}\;,\qquad (\varrho\approx 0)
\,\,.
\label{cp2generalIR}
\eeq
In order to get the mass levels in the numerical calculation we have to match the 
$\varrho\approx 0$ behavior (\ref{cp2generalIR}) with the behavior for
large $\varrho$. The latter can be easily obtained by using the mapping
written in (\ref{Sch-variables}) to the Schr\"odinger equation 
(\ref{Sch}). Indeed, for $\varrho\to\infty$, or equivalently for
$y\to\infty$, the potential $V(y)$ becomes asymptotically constant and
the wave equation (\ref{Sch}) can be  trivially integrated. Let us call
$V_*\equiv \lim_{y\to \infty} V(y)$, with $V_*>0$. Then, the solutions of (\ref{Sch})
are of the form $\psi\sim e^{\pm \sqrt{V_*}y}$ which, in terms of the original variable 
$\varrho$ are simply $\psi\sim \varrho^{\pm \sqrt{V_*}}$. The actual value of $V_*$ is
given in eq. (\ref{generalV-limit}). Taking into account the relation 
(\ref{Sch-variables}) between $\psi$ and $\xi$, we get that 
$\xi\sim \varrho^{\gamma}$ for $\varrho\to\infty$, where $\gamma$ are exactly the two
values written in eq. (\ref{IRroots}). The so-called $S^l$ modes are characterized by
the following UV behavior:
\beq
\xi\sim \varrho^{-(l+p_2-d-2)}\;,\qquad (\varrho\to\infty)\,\,.
\label{generalUV}
\eeq
In the shooting technique one solves the differential equation for the fluctuations by
imposing the behavior (\ref{cp2generalIR}) at $\varrho\approx 0$ and
then one scans the values of $\bar M$ until the UV behavior
(\ref{generalUV}) is fine tuned. This occurs only for a discrete set of
values of $\bar M$, which determines the mass spectrum we are looking
for. The numerical values obtained for the different intersections  and
their comparison with the WKB mass formula (\ref{WKBM}) are studied in
detail in \cite{AR}. Nevertheless, we include in table \ref{MassDpDp+4}
the values obtained for the Dp-D(p+4) intersections. 
\\
\begin{table}[!h]
\begin{tabular}[b]{|c|c|c|}   
\hline  
\multicolumn{3}{|c|}{$(1|D1\perp D5)$ with $l=0$}\\
\hline
 $n$  & WKB  & Numerical \\ 
\hline   
\ \ 0 & $14.80$  & $14.70$  \\   
\ \ 1 & $49.35$  & $49.22$  \\   
\ \ 2 & $103.63$  & $103.50$  \\   
\ \ 3 & $177.65$  & $177.54$  \\     
\ \ 4 & $271.41$  & $271.30$  \\
\ \ 5 & $384.91$  & $384.80$  \\ 
\hline
\end{tabular}
\qquad
\begin{tabular}[b]{|c|c|c|}   
\hline  
\multicolumn{3}{|c|}{$(2|D2\perp D6)$ with $l=0$}\\
\hline
 $n$  & WKB  & Numerical \\ 
\hline   
\ \ 0 & $11.46$  & $11.34$  \\   
\ \ 1 & $36.67$  & $36.54$  \\   
\ \ 2 & $75.63$  & $75.50$  \\   
\ \ 3 & $128.34$  & $128.20$  \\     
\ \ 4 & $194.80$  & $194.66$  \\
\ \ 5 & $275.01$  & $274.88$  \\ 
\hline
\end{tabular}
\qquad
\begin{tabular}[b]{|c|c|c|}   
\hline  
\multicolumn{3}{|c|}{$(4|D4\perp D8)$ with $l=0$}\\
\hline
 $n$  & WKB  & Numerical \\ 
\hline   
\ \ 0 & $4.31$  & $4.68$  \\   
\ \ 1 & $11.48$  & $11.88$  \\   
\ \ 2 & $21.53$  & $21.94$  \\   
\ \ 3 & $34.45$  & $34.86$  \\     
\ \ 4 & $50.24$  & $50.66$  \\
\ \ 5 & $68.91$  & $69.34$  \\ 
\hline
\end{tabular}
\caption{Values of $\bar M^2$ obtained numerically and with the WKB
method for the scalar modes of the D1-D5, D2-D6 and D4-D8 intersections
for $l=0$.}
\label{MassDpDp+4}
\end{table}

\subsection{Mass gap for Dp-Dq intersections}
\label{cp3massgap}
Before going on to analyze in detail the fluctuations for some
particular intersections in the following sections, we can extract
some interesting information about the spectra from this generic
analysis. Indeed, one should notice that the mass gap of the spectra is
just the inverse of the coefficient relating $\bar M^2$ and $M^2=-k^2$ in
eq. (\ref{newvariables}), namely
$R^{-2\gamma_1-2\gamma_2}\,L^{2\gamma_1+2\gamma_2-2}$, which for the
Dp-brane background reduces to $R^{p-7}\,L^{5-p}$ (see eq.
(\ref{Dpgammas})). It is worth expressing this quantity in terms of the
quark mass
$m_q={L\over2\pi\alpha'}$ and the Yang-Mills coupling constant 
(\ref{c1gymdp}), arriving at:
\beq
R^{p-7}\,L^{5-p}={2^{p-1}\,\pi^{p+1\over2}\over\Gamma\left({7-p\over2}
\right)}\,{m_q^{5-p}\over g_{YM}^2\,N}\,,
\label{cp2mg1}
\eeq
which takes a neat form when written in terms of the effective coupling
constant $g_{eff}(U)$, defined in eq. (\ref{c1geff}) at an energy
scale $U$ equal to the quark mass $m_q$. Therefore, up to a numerical
coefficient, the mass gap of the spectra reads:
\beq
M\sim\sqrt{R^{p-7}\,L^{5-p}}\sim {m_q\over g_{eff}(m_q)}\,.
\label{cp2mg2}
\eeq
This result holds for the complete set of fluctuations of the
different D-brane intersections we will analyze in the subsequent
sections. Indeed, the mass of the different fluctuations will always
be proportional to the reduced mass $\bar M$, as it
happens with the transverse scalar fluctuations analyzed here. 

Another interesting general feature of the Dp-Dq system is the fact that
there is no discrete spectrum when $p\geq5$, that is, when the background
brane is a D5-brane or one of larger dimensionality. In subsection
\ref{generalfluctuations} the equation of motion of the fluctuations was
mapped to a Schr\"odinger equation, and so the task of finding the mass
levels was translated to the problem of finding the masses such that the
potential (\ref{potential}) allows for a zero-energy level. We have seen
in eq. (\ref{generalV-limit}) that the asymptotic values of the
potential are positive, and from eq. (\ref{c2potmin}) it follows that
the potential has a minimum at $y_0$, which, for a Dp-brane background,
is given by
$e^{2y_0}=2/(5-p)$. Hence, for $p\geq5$ there will be no minimum and in
consequence no discrete mass spectrum of the fluctuations.

%
%
%
%
%
%
%
%
%

\setcounter{equation}{0}
\section{Fluctuations of the D3-D5 system}
\medskip
\label{cp2D3D5}
In this section we study in detail the complete set of fluctuations corresponding to the 
$(2|D3\perp D5)$ intersection. The dynamics of the D5-brane probe in the 
$AdS_5\times S^5$ background is governed by the action
(\ref{cp1Dbraneact}), which in this case reduces to:
\beq
S=-\int d^6\xi\,\sqrt{-\det (g+F)}\,+
\int d^6\xi \;P\big[\,C^{(4)}\,\big]\wedge F\,,
\label{cp2DBI-D5}
\eeq
and notice that with respect to (\ref{cp1Dbraneact}) we have made $F$
dimensionless by means of the redefinition $2\pi\,\alpha'\,F\to F$, and,
for simplicity, we have set $T_5=1$.

Let us now find the action for the complete set of quadratic fluctuations around the
static configuration in which the two branes are at a distance $L$ (and the
worldvolume gauge field of the D5 probe is switched off). Recall that in
this embedding the worldvolume metric in the UV is
$AdS_4\times S^2$. As in section 
\ref{general}, let us denote by $\chi$ the scalars transverse to both types of branes
and let us assume that the probe is extended along  $x^1$ and $x^2$. We will  call simply $X$ to the
coordinate 
$x^3$. By expanding up to second order the action (\ref{cp2DBI-D5}), one
gets the following lagrangian for the fluctuations:
\bear
&&{\cal L}=-\rho^2\,\sqrt{\tilde g}\,\left[{1\over 2}\,
{R^2\over \rho^2+L^2}\,
{\cal G}^{ab}\partial_a\chi\partial_b\chi+{1\over 2}\,{\rho^2+L^2\over
R^2}\, {\cal G}^{ab}\partial_a X\partial_b X+{1\over
4}\,F_{ab}F^{ab}\right]-\rc\rc
&&\qquad- 2\,{\rho\over
R^4}\,(\rho^2+L^2)X\,\epsilon^{ij}\,F_{ij}\,,
\label{D3D5-quad-action}
\eear
where $i,j$ are indices of the two-sphere of the worldvolume, $\epsilon^{ij}=\pm 1$
and ${\cal G}_{ab}$ is the induced metric for the static configuration, \ie\ the metric
displayed in eq. (\ref{AdSindmetric}) for $d=2$.

The equation of motion of the scalar $\chi$ is just (\ref{eom-general}) for $p_2=5$,
$d=2$ and $\gamma_2=1$. As shown in subsection
\ref{cp2scexactly-solvable} this equation can be solved exactly in terms
of the hypergeometric  function (\ref{scalarhyper}). Upon imposing the
quantization condition (\ref{quant}) we obtain a tower of normalizable
modes, which we will denote by $S^l$,  given by:
\beq
\xi_S(\varrho)=\varrho^l\,(\varrho^2+1)^{-n-l-{1\over 2}}\,
F(-n-l-{1\over 2},  -n; l+{3\over 2};-\varrho^2\,)\;,
\qquad (l,n\ge 0)\,.
\eeq
Recall from (\ref{generalMS}) and (\ref{generalDeltaS}) that the associated mass
$M_S(n,l)$ and conformal dimension $\Delta_S$ for these scalar modes are given by:
\beq
M_S(n,l)={2L\over R^2}\,
\sqrt{\bigg(n+l+{1\over 2}\bigg)\bigg(n+l+{3\over 2}\bigg)}\,\,,
\qquad\quad \Delta_S=l+2\,.
\label{MSd3d5}
\eeq
Notice that the value found here for $\Delta_S$ is in agreement with the result 
of ref. \cite{WFO}.

As shown in (\ref{D3D5-quad-action}), the scalar $X$ is coupled to the components
$F_{ij}$ of the gauge field strength along the two-sphere. The equation of
motion of $X$ derived from the lagrangian (\ref{D3D5-quad-action}) is:
\beq
R^2\,
\partial_a\,\left[\rho^2\sqrt{\tilde g}\,(\rho^2+L^2)\,
{\cal G}^{ab}\partial_b X\right]-2\rho\,( \rho^2+L^2)\,
\epsilon^{ij}F_{ij}=0\,.
\label{D3D5Xeq}
\eeq
Moreover, the equation of motion of the gauge field is:
\beq
R^4\,
\partial_a\,\Big[\rho^2\sqrt{\tilde g}\,F^{ab}\Big]-4\rho\,
(\rho^2+L^2)\,\epsilon^{bj}\,\partial_j X=0\,,
\label{D3D5Aeq}
\eeq
where $\epsilon^{bj}$ is zero unless $b$ is an index along the two-sphere.

Let us now see how the equations of motion (\ref{D3D5Xeq}) and (\ref{D3D5Aeq}) can be
decoupled and, subsequently, integrated in  analytic form. 
With this purpose in mind, let us see
how one can obtain vector spherical harmonics for the two-sphere from the scalar
harmonics
$Y^l$. Clearly, from a scalar harmonic in $S^2$ we can can construct a vector  by simply
taking the derivative  with respect to the coordinates of the
two-sphere, namely:
\beq
Y_i^l(S^2)\equiv \nabla_i\,Y^l(S^2)\,\,.
\label{S2har}
\eeq
One can check from (\ref{casimir}) that these functions satisfy:
\bear
&&\nabla^i\,Y_i^l={1\over \sqrt{\tilde g}}\,
\partial_i\,\Big[\sqrt{\tilde g}\,\tilde
g^{ij}\,Y_j^l\Big]=-l(l+1)\,Y^l\,,\rc\rc
&&\epsilon^{ij}\,\partial_i\,Y_j^l=0\,.
\eear
Alternatively, we can take the Hodge dual in the sphere and define a new
vector harmonic function $\hat Y_i^l(S^2)$ as:
\beq
\hat Y_i^l(S^2)\,\equiv\,{1\over \sqrt{\tilde g}}\,\tilde g_{ij}\,\epsilon^{jk}\,
\nabla_k\,Y^l(S^2)\,\,.
\label{hatS2har}
\eeq
The $\hat Y_i^l$ vector harmonics satisfy:
\bear
&&\nabla^i\,\hat Y_i^l=0\,,\rc\rc
&& \epsilon^{ij}\,\partial_i\,\hat Y_j^l=l(l+1)\,\sqrt{\tilde
g}\;Y^l\,.
\label{hatYprop}
\eear

Let us analyze the different types of modes, in analogy with the D3-D7 case in
\cite{KMMW}.

\subsection{Type I modes}

We are going to study first the modes which involve the scalar field $X$ and the
components $A_i$ of the gauge field along the two-sphere directions. Generically, the
equations of motion couple   $A_i$ to the other gauge field components $A_{\mu}$ and
$A_{\rho}$. However,  due to the property $\nabla^i\,\hat Y_i^l\,=\,0$ (see eq.
(\ref{hatYprop})), if $A_i$ is proportional
to $\hat Y_i^l$ it does not mix with other components of the gauge field, although it
mixes with the scalar $X$. Accordingly, let us take the ansatz:
\beq
A_{\mu}=0\,\,,\qquad\quad
A_{\rho}=0\,\,,\qquad\quad
A_i=\phi(x,\rho)\,\hat Y_i^l(S^2)\,,
\label{typeIA}
\eeq
while we represent $X$ as:
\beq
X=\Lambda(x,\rho)\,Y^l(S^2)\,.
\label{typeIX}
\eeq
Taking into account that:
\beq
{1\over 2}\,\epsilon^{ij}\,F_{ij}=l(l+1)\,\sqrt{\tilde g}\,\phi\,Y^l\,,
\eeq
one can prove that the equation of motion  of $X$ (eq. (\ref{D3D5Xeq})) becomes:
\bear
&&R^4\,\rho^2\,\partial_{\mu}\partial^{\mu}\Lambda+\partial_{\rho}\Big[
\rho^2\,(\rho^2+L^2)^2\,\partial_{\rho}\,\Lambda\Big]-\rc\rc
&&-l(l+1)\,(\rho^2+L^2)^2\,\Lambda-4l(l+1)\,\rho\,(\rho^2+L^2)\,
\phi=0\,.
\label{D3D5Lambdaeq}
\eear
It can be easily verified that the equations for $A_{\mu}$ and $A_{\rho}$ are
automatically satisfied as a consequence of the relation 
$\nabla^i\,\hat Y_i^l\,=\,0$. Moreover, for $l\not=0$ the equation of motion 
(\ref{D3D5Aeq}) for the gauge field
components along $S^2$ reduces to:
\beq
R^4\,
\partial_{\mu}\partial^{\mu}\phi+\partial_{\rho}\Big[(\rho^2+L^2)^2\,
\partial_{\rho}\,\phi\Big]-l(l+1)\,{(\rho^2+L^2)^2\over\rho^2}\,\phi
-4\rho\,(\rho^2+L^2)\Lambda=0\,.
\label{D3D5phieq}
\eeq
In order to decouple this system of equations, let us follow closely
the steps of ref. \cite{WFO}. First, we redefine the scalar field
$\Lambda$ as follows:
\beq
V=\rho\,\Lambda\,.
\label{VLambda}
\eeq
The system of equations which results after this redefinition can be
decoupled by simply taking suitable linear combinations of the unknown
functions $V$ and $\phi$. This decoupling procedure was used in ref.
\cite{WFO} for the conformal case $L=0$ and, remarkably, it also works for
the case in which the brane separation does not vanish. In ref. \cite{AR}
this method was applied to decouple the fluctuations of the type written
in eqs. (\ref{typeIA}) and (\ref{typeIX}) for the more general
$(p-1|Dp\perp D(p+2))$ intersection.  In ref. \cite{AR} the decoupling
procedure is explained in detail. Here we just need the decoupled
functions, which are:
\bear
&&Z^+=V+ l\,\phi\,,\rc\rc
&&Z^-=V-(l+1)\,\phi\,.
\label{Zpm}
\eear
It is interesting at this point to notice that the $Z^-$ modes only exist for $l\ge 1$
while the $Z^+$ modes make sense for $l\ge 0$. Indeed, the vector harmonic $\hat Y_i^l$
vanishes for $l=0$, since it is the derivative of a constant function (see eq.
(\ref{S2har})). Then, it follows from (\ref{typeIA}) that the vector field vanishes for
these $l=0$ modes and, as a result, the $l=0$ mode of the scalar field is uncoupled. It
is clear from (\ref{Zpm}) that this $l=0$ mode of the field $X$ is just $ Z^+$ for
vanishing $l$. As we have just mentioned, for $l>0$ the equations for $ Z^+$ and 
$ Z^-$ are decoupled. These equations can be obtained by substituting the definition
(\ref{Zpm}) into eqs. (\ref{D3D5Lambdaeq}) and (\ref{D3D5phieq}). In order to
get the corresponding spectra, let us adopt  a plane wave ansatz for $Z^{\pm}$, namely:
\beq
Z^{\pm}=e^{ikx}\,\xi^{\pm}(\rho)\,.
\eeq
Moreover, let us define the reduced variables $\varrho$ and $\bar M$ as in eq.
(\ref{newvariables}), namely $\varrho=\rho/L$ and $\bar M^2=-R^4L^{-2}k^2$. Furthermore,
we define $\lambda$ as in eq. (\ref{lambda}). We will consider separately the $Z^+$ and
$Z^-$ equations.

\subsubsection{$Z^+$ spectra}
By combining appropriately eqs. (\ref{D3D5Lambdaeq}) and (\ref{D3D5phieq}) one can show
that the  equation for $\xi^+$ is indeed decoupled and given by:
\beq
{1\over 1+\varrho^2}\,\partial_{\varrho}\,\Big[(1+\varrho^2)^2\,
\partial_{\varrho}\,\xi^+\Big]+\Bigg[{\bar M^2\over 1+\varrho^2}-(l+1)\,
\big(l+4+{l\over \varrho^2}\big)\Bigg]\,\xi^+=0\,.
\label{xi+eq}
\eeq
Remarkably, eq. (\ref{xi+eq}) can be analytically solved in terms of a
hypergeometric function. Actually, by using the change of variables of
appendix \ref{hyperchange}, one can show that the solution of
(\ref{xi+eq}) which is  regular at $\varrho=0$ is:
\beq
\xi^+(\varrho)=\varrho^{1+l}\,(1+\varrho^2)\,^{-1-\lambda}\,\,
F(-\lambda-1, l+{3\over 2}-\lambda;l+{3\over 2}; -\varrho^2)\,.
\label{xi+expression}
\eeq
By using eq. (\ref{asymptotic-hyper}) one can show that, for large values of the
$\varrho$ coordinate, the function $\xi^+$ written in (\ref{xi+expression}) behaves as:
\beq
\xi^+(\varrho)\,\sim\,c_1\varrho^{l+1}+c_2\varrho^{-l-4}\;,
\qquad (\varrho\to\infty)\,.
\eeq
Clearly, the only normalizable solutions are those for which $c_1=0$. 
This regularity condition at $\varrho=\infty$ can be enforced by means of 
the following quantization condition:
\beq
l+{3\over 2}-\lambda=-n\;,\qquad n=0,1,2,\cdots\,.
\label{I+quantization}
\eeq
The $\xi^+$ fluctuations for which (\ref{I+quantization}) holds will be referred to as
$I_+^l$ modes. Their analytical expression is given by:
\beq
\xi_{I_+}(\varrho)=\varrho^{1+l}\,(\varrho^2+1)^{-n-l-{5\over 2}}\,
F(-n-l-{5\over 2},  -n; l+{3\over 2};-\varrho^2\,)\;,
\qquad (l,n\ge 0)\,,
\eeq
and the corresponding energy levels are:
\beq
M_{I_+}(n,l)={2L\over R^2}\,
\sqrt{\bigg(n+l+{3\over 2}\bigg)\bigg(n+l+{5\over 2}\bigg)}\;. 
\eeq
By using the general expression (\ref{Delta-hyper}) one reaches the 
conclusion that the conformal dimensions of the operators dual to the
$I_+^l$ modes are:
\beq
\Delta_{I_+}=l+4\,\,,
\eeq
which  agrees with the values found in ref. \cite{WFO}.

\subsubsection{$Z^-$ spectra}
The equation for $\xi^-$ can be shown to be:
\beq
{1\over 1+\varrho^2}\,\partial_{\varrho}\,\Big[
(1+\varrho^2\,)^2\,\partial_{\varrho}\,\xi^-\Big]+
\Bigg[\,{\bar M^2\over 1+\varrho^2}-l\,\big(l-3+{l+1\over \varrho^2}
\big)\Bigg]\,\xi^-=0\,,
\eeq
which again can be solved in terms of the hypergeometric function. The solution
regular at $\varrho=0$ is:
\beq
\xi^-(\varrho)=\varrho^{1+l}\,(\varrho^2+1)\,^{-1-\lambda}\,\,
F(-\lambda+1, l-{1\over 2}-\lambda;l+{3\over 2}; -\varrho^2)\,.
\eeq
It is easy to verify by using eq. (\ref{asymptotic-hyper}) that $\xi^-(\varrho)$ has two
possible behaviors at
$\varrho\to
\infty$, namely $\varrho^{-l}$, $\varrho^{l-3}$, where $l\ge 1$.  The former corresponds
to a normalizable mode with  conformal dimension $\Delta=l$, while the latter is
associated to operators with
$\Delta=3-l$. Notice that the existence of these two branches is in agreement with
the results of ref. \cite{WFO}. 

Let us consider first the branch with  $\Delta=l$, which we will refer to as $I_-^l$
fluctuations. One can select these fluctuations by imposing the following
quantization condition:
\beq
l-{1\over 2}-\lambda=-n\;,\qquad n=0,1,2,\cdots\,.
\eeq
The corresponding functions $\xi^-(\varrho)$ are:
\beq
\xi_{I_-}(\varrho)=\varrho^{1+l}\,(\varrho^2+1)\,^{-n-l-{1\over 2}}\,\,
F(-n-l+{3\over 2}, -n;l+{3\over 2}; -\varrho^2)\;,
\qquad (l\ge 1, n\ge 0)\,,
\label{I-sol}
\eeq
and the mass spectrum and conformal dimension are:
\beq
M_{I_-}(n,l)={2L\over R^2}\,
\sqrt{\bigg(n+l-{1\over 2}\bigg)\bigg(n+l+{1\over 2}\bigg)}\;,
\qquad\quad \Delta_{I_-}=l\,.
\eeq

One can check that, indeed, the solution (\ref{I-sol}) behaves as $\varrho^{-l}$
at $\varrho\to\infty$ and, therefore, the associated operator in the conformal limit has
$\Delta=l$ as it should. 

One can select the  branch with $\Delta=3-l$ by requiring that:
\beq
-\lambda+1=-n\;,\qquad n=0,1,2,\cdots\,.
\label{Itildequant}
\eeq
The corresponding functions are:
\beq
\xi_{\tilde I_-}(\varrho)=\varrho^{1+l}\,(\varrho^2+1)\,^{-2-n}\,\,
F(-n,l -n-{3\over 2};l+{3\over 2}; -\varrho^2)\,.
\eeq
Notice that the condition $\Delta=3-l>0$ is only fulfilled for two possible values of
$l$, namely $l=1,2$. We will refer to this branch of solutions as $\tilde I_-^l$
fluctuations. Their mass spectrum is independent of $l$, as follows from eq.
(\ref{Itildequant}).  Thus, one has:
\beq
M_{\tilde I_-}(n,l)={2L\over R^2}\,
\sqrt{\bigg(n+1\bigg)\bigg(n+2\bigg)}\;,\qquad\quad
\Delta_{\tilde I_-}=3-l\;,\qquad (l=1,2)\,.
\label{Itildelevels}
\eeq

\subsection{Type II modes}

Let us consider a configuration with $X=0$ and with the following ansatz for the gauge
fields:
\beq
A_{\mu}=\phi_{\mu}(x,\rho)\,Y^l(S^2)\;,\qquad\quad
A_{\rho}=0\;,\qquad\quad A_i=0\,,
\label{D3D5II}
\eeq
with: 
\beq
\partial^{\mu}\phi_{\mu}=0\,.
\eeq
Due to this last condition one can check that the equations for $A_{\rho}$ and 
$A_i$ are satisfied, while the equation for $A_{\mu}$ yields:
\beq
\partial_{\rho}\big(\rho^2\partial_\rho\phi_{\nu}\big)+R^4\,
{\rho^2\over (\rho^2+L^2)^2}\,\partial_{\mu}\,\partial^{\mu}\phi_{\nu}-
l(l+1)\,\phi_{\nu}=0\,.
\eeq
Expanding in a plane wave basis we get exactly the same equation as for
the scalar fluctuations. Actually, let us represent $\phi_{\mu}$ as:
\beq
\phi_{\mu}=\xi_{\mu}\,e^{ikx}\,\chi(\rho)\,,
\label{D3D5IIbis}
\eeq
where $k^{\mu}\,\xi_{\mu}=0$. The equation for $\chi$ is:
\beq
{1\over \varrho^2}\,\partial_{\varrho}\Big(\varrho^2\partial_{\varrho}\chi
\Big)+\Bigg[{\bar M^2\over (1+\varrho^2)^2}-{l(l+1)\over\varrho^2}\Bigg]
\chi=0\,,
\eeq
where we have already introduced the reduced quantities $\varrho$ and $\bar M$. 
This equation is identical to the one corresponding to the transverse scalars. Thus, the
spectra are the same in both cases. These fluctuations correspond to an operator with
conformal weight $\Delta_{II}=l+2$, in agreement with ref. \cite{WFO}.

\subsection{Type III modes}

These modes have $X=0$ and the following form for the gauge field:
\beq
A_{\mu}=0\;,\qquad\quad
A_{\rho}=\phi(x,\rho)\,Y^l(S^2)\;,\qquad\quad
A_i=\tilde\phi(x,\rho)\,Y^l_i(S^2)\,.
\label{D3D5IIIansatz}
\eeq
Notice that for the gauge field potential written above   the field strength
components $F_{ij}$ along the two-sphere vanish, which ensures that the equation of
motion of $X$ is satisfied for $X=0$. The non-vanishing components of the gauge field
strength are:
\beq
F_{\mu i}=\partial_{\mu}\,\tilde\phi\,Y_i^l\;,\qquad\quad
F_{\mu \rho}=\partial_{\mu}\,\phi\,Y^l\;,\qquad\quad
F_{\rho i}=(\partial_{\rho}\,\tilde\phi-\phi)\,Y_i^l\,.
\eeq
The equation for $A_{\rho}$ becomes:
\beq
R^4\,\rho^2\,\partial_{\mu}\,\partial^{\mu}\phi-l(l+1)\,(\rho^2+L^2)^2\,
(\phi-\partial_{\rho}\tilde\phi\,)=0\,,
\eeq
while the equation for $A_i$ is:
\beq
R^4\,\partial_{\mu}\,\partial^{\mu}\tilde\phi+\partial_{\rho}\,\Big[
(\rho^2+L^2)^2\,(\partial_{\rho}\tilde\phi-\phi)\Big]=0\,.
\eeq
Moreover, the equation for $A_{\mu}$ can be written as:
\beq
\partial_{\mu}\,\Big[l(l+1)\tilde\phi-\partial_{\rho}(\rho^2\phi)\Big]=0\,.
\label{D3D5IIIAmu}
\eeq

Expanding $\phi$ and $\tilde\phi$ in a plane wave basis we can get rid of the $x^{\mu}$
derivative and we can write the following relation between  $\tilde\phi$ and $\phi$:
\beq
l(l+1)\tilde\phi=\partial_{\rho}(\rho^2\phi)\,.
\eeq
For $l\not= 0$, one can use this relation to eliminate $\tilde\phi$ in favor of $\phi$.
The equation of motion of $A_{\rho}$ becomes:
\beq
\partial^2_{\rho}\,(\rho^2\,\phi)-l(l+1)\,\phi+{R^4\,\rho^2\over
(\rho^2+L^2)^2}\,
\partial_{\mu}\,\partial^{\mu}\phi=0\,,
\eeq
which, again,  can be solved in terms of hypergeometric functions. The
equation of motion of $A_i$ is just equivalent to the above equation. To
solve this equation, let us write:
\beq
\phi(x,\rho)=e^{ikx}\,\zeta(\rho)\,.
\eeq
Then, in terms of the reduced variable $\varrho$ the equation for $\zeta$
becomes:
\beq
{1\over \varrho^2}\,\partial^2_{\varrho}\,\Big(\varrho^2\zeta\Big)+
\Bigg[{\bar M^2\over (1+\varrho^2)^2}-{l(l+1)\over\varrho^2}\Bigg]\zeta=0\,,
\label{D3D5IIIeq}
\eeq
where $\bar M$ is the same as for the type I modes. 
Eq. (\ref{D3D5IIIeq}) can be solved in terms of the hypergeometric function as:
\beq
\zeta(\varrho)=\varrho^{l-1}\,(\varrho^2+1)\,^{-\lambda}\,\,
F(-\lambda, l+{1\over 2}-\lambda;l+{3\over 2}; -\varrho^2)\,.
\eeq
The quantization condition and the energy levels are just the same as for the transverse
scalars and the type II modes. At $\rho\to\infty$, $\zeta\sim\rho^{-l-2}$. These
fluctuations correspond to a field with conformal weight $\Delta_{III}=l+2$, as predicted
in \cite{WFO}. 

\subsection{Fluctuation/operator correspondence}
\label{cp2d3d5dict}
\medskip
Let us recall the array corresponding to the D3-D5 intersection:
\beq
\begin{array}{ccccccccccl}
 &1&2&3& 4& 5&6 &7&8&9 & \nonumber \\
D3: & \times &\times &\times &\_ &\_ & \_&\_ &\_ &\_ &     \nonumber \\
D5: &\times&\times&\_&\times&\times&\times&\_&\_&\_ &
\end{array}
\label{D3D5intersection}
\eeq
Before adding the D5-brane we have an $SO(6)$ R-symmetry which
corresponds to the rotation  in the $456789$ directions. The D5-brane
breaks this $SO(6)$  to
$SU(2)_H\times SU(2)_V$, where the $SU(2)_H$  corresponds to  rotations in the 
$456$ directions (which are along the D5-brane worldvolume) and the $SU(2)_V$ is
generated by rotations in the $789$ subspace (which are the directions orthogonal to
both types of branes). 

Let us recall how the ${\cal N}=4$, $d=4$ gauge multiplet decomposes under the 
${\cal N}=4$, $d=3$ supersymmetry. As it is well-known, the ${\cal N}=4$ gauge multiplet
in four dimensions contains a vector $A_{\mu}$ (which has components along the four
coordinates of the D3-brane worldvolume), six real scalars $X^i$ (corresponding to the
directions orthogonal to the D3-brane worldvolume) and four complex Weyl spinors
$\lambda^a$. All these fields are in the adjoint representation of the gauge group. The
$d=4$ vector field $A_{\mu}$ gives rise to a $d=3$ gauge field  $A_{k}$ and to a
scalar field $A_3$. Both types of fields are singlets with respect to the 
$SU(2)_H\times SU(2)_V$ symmetry. Moreover, the adjoint scalars can be arranged in two sets
as:
\beq
X_H\,=\,(X^4,X^5,X^6)\;,\qquad\quad
X_V\,=\,(X^7,X^8,X^9)\,.
\eeq
Clearly, $X_H$ transforms in the $({\bf 3},{\bf 1})$ representation of 
$SU(2)_H\times SU(2)_V$ while $X_V$ does it in the $({\bf 1}, {\bf 3})$. Finally, the
spinors transform in the  $({\bf 2},{\bf 2})$ representation and will be denoted by
$\lambda^{im}$. In addition to the bulk fields, the 3-5 strings introduce a $d=3$ complex
hypermultiplet in the fundamental representation of the gauge field, whose components will
be denoted by $(q^m, \psi^i)$. The bosonic components $q^m$ of this hypermultiplet 
transform in the $({\bf 2},{\bf 1})$ representation, whereas the fermionic ones
$\psi^i$ are in the $({\bf 1}, {\bf 2})$ of $SU(2)_H\times SU(2)_V$. 
The dimensions and quantum numbers of the different fields just discussed are summarized 
in table \ref{tableFieldsD3D5}.

\begin{table}[!h]
\centerline{
\begin{tabular}[b]{|c|c|c|c|}   
 \hline
 Field  & $\Delta$  & $SU(2)_H$ & $SU(2)_V$ \\ 
\hline 
\ \ $A_k$ & $1$  & $0$ &0 \\  \hline  
\ \ $A_3$ & $1$  & $ 0$ & 0 \\ \hline  
\ \ $X_H$ & $1$  & $ 1$ &0 \\  \hline   
\ \ $X_V$ & $1$  & $ 0 $ &1 \\ \hline  
\ \ $\lambda^{im}$ & $3/2$  & $ 1/2 $ &1/2 \\ \hline  
\ \ $q^m$ & $1/2$  & $ 1/2 $ &0 \\ \hline  
\ \ $\psi^i$ & $1$  & $ 0 $ &1/2 \\ \hline 
\end{tabular}
}
\caption{Quantum numbers and dimensions of the fields of the D3-D5 intersection.}
\label{tableFieldsD3D5} 
\end{table}

Let us now determine the quantum numbers of the different fluctuations of the D3-D5
system. We will denote by $S^l$ the scalar fluctuations and by $I_+^l$ and $I_-^l$  the
two types of vector fluctuations of type I. The $I_-^l$ fluctuations which we will
consider from now on are those corresponding to the operator of dimension $\Delta=l$.
Moreover, the modes of type II and III correspond to fluctuations of the vector gauge field
in $AdS_4$ and will be denoted collectively by $V^l$. In all cases, $l$ corresponds to the
quantum number of the spherical harmonics in the $456$ directions and, thus, it can be
identified with the isospin of the $SU(2)_H$ representation. Moreover, it is clear that
the scalar modes are fluctuations in the $789$ directions and therefore are in the vector
representation of $SU(2)_V$, while the other fluctuations are singlets under $SU(2)_V$.
With all this data and with the values of the dimensions determined previously, we can
fill the values displayed in  table \ref{tableModesD3D5}.

\begin{table}[!h]
\centerline{
\begin{tabular}[b]{|c|c|c|c|}   
 \hline
 Mode  & $\Delta$  & $SU(2)_H$ & $SU(2)_V$ \\ 
\hline 
\ \ $S^l$ & $l+2$  & $l\ge 0$ &1 \\  \hline  
\ \ $I_+^l$ & $l+4$  & $l\ge 0$ & 0 \\ \hline  
\ \ $I_-^l$ & $l$  & $l\ge 1$ &0 \\  \hline   
\ \ $V^l$ & $l+2$  & $l\ge 0 $ &0 \\ \hline  
\end{tabular}
}
\caption{Quantum numbers and dimensions of the modes of the D3-D5 intersection.}
\label{tableModesD3D5} 
\end{table}

Let us now recall our results for the mass spectra. 
The mass of the scalar fluctuations $M_S(n,l)$ is given in eq. (\ref{MSd3d5}).
The masses of the other modes are given in terms of $M_S(n,l)$ as:
\beq
M_{I_+}(n,l)=M_S(n,l+1)\;,\qquad
M_{I_-}(n,l)=M_S(n,l-1)\;,\qquad
M_{V}(n,l)=M_S(n,l)\,.
\label{massrelations}
\eeq

Let us now match, following ref. \cite{WFO}, the fluctuation modes with composite
operators  of the ${\cal N}=4$, $d=3$ defect theory by looking at the dimensions 
and $SU(2)_H\times SU(2)_V$ quantum numbers of these two types of objects. Let us consider
first the fluctuation mode with the lowest dimension, which according to our previous
results is $I_-^1$. This mode is a triplet of $SU(2)_H$ and a singlet of $SU(2)_V$ and has
$\Delta=1$. There is only one operator with these characteristics. Indeed, let us define
the following operator:
\beq
{\cal C}^I\equiv \bar q^m \sigma_{mn}^I\, q^n\,\,,
\eeq
where the $\sigma^I$ are Pauli matrices. This operator has clearly the same dimension and
$SU(2)_H\times SU(2)_V$ quantum numbers as the mode $I_-^1$. Therefore, we have the
identification \cite{WFO}:
\beq
I_-^1\sim {\cal C}^I\,\,.
\eeq
Since all the other operators dual to fluctuation modes will have
larger conformal dimension, ${\cal C}^I$ must be the lowest chiral
primary. Hence, by acting with the supersymmetry generators we can
obtain the other operators in the same multiplet as
${\cal C}^I$. The bosonic ones are \cite{WFO}:
\bear
&&{\cal E}^A=\bar\psi^i\,\sigma_{ij}^A\,\psi^j+2\bar q^m
X_V^{Aa}T^a q^m\,,\rc\rc
&&J_B^k=i\bar q^m D^k q^m-i(D^kq^m)^{\dagger}\,q^m+
\bar\psi^i\,\rho^k \psi^i\,,
\eear
where the $T^a$ are the matrices of the gauge group and the $\rho^k$ are Dirac matrices in
$d=3$. Notice that these two operators have dimension $\Delta=2$. Moreover, 
${\cal E}^A$ transforms in the $({\bf 1}, {\bf 3})$ representation of $SU(2)_H\times
SU(2)_V$, whereas $J_B^k$ is an $SU(2)_H\times SU(2)_V$ singlet. It is
straightforward to find the modes that have these same quantum numbers
and dimension. Indeed, one gets that:
\beq
S^0\sim {\cal E}^A\,\,,\qquad\quad
V^0\sim J_B^k\,\,.
\eeq
Notice that our mass spectrum is consistent with these identifications, since according to
eq. (\ref{massrelations}),  the fluctuations $I_-^1$, $S^0$ and $V^0$ have all the same
mass spectrum, namely $M_s(n,0)$. 

Let us next consider the modes corresponding to higher values of $l$. Following ref.
\cite{WFO} we define the operator:
\beq
{\cal C}_l^{I_0\cdots I_l}\,\equiv\,
{\cal C}^{(I_0}\,X_H^{I_1}\cdots X_H^{I_l)}\,\,,
\eeq
where the parentheses stand for the traceless symmetrization of the indices. The operator
${\cal C}_l$ has dimension $\Delta=l+1$, is a singlet of $SU(2)_V$ and transforms in the
spin
$l+1$ representation of $SU(2)_H$. It is thus a natural candidate to be identified with
the mode $I_-^{l+1}$, namely:
\beq
I_-^{l+1}\sim {\cal C}_l^{I_0\cdots I_l}\,\,.
\eeq
The mode $I_+^0$ has been identified in \cite{WFO} with a four-supercharge descendant of
the second-floor chiral primary ${\cal C}_1^{IJ}$. Notice that our mass spectra supports
this identification since $M_{I_+}(n,0)\,=\,M_{I_-}(n,2)$. Actually, our results are
consistent with having the modes $I_-^{l+1}$, $S^l$, $V^l$ and $I_+^{l-1}$ in the same
massive supermultiplet for $l\ge 1$ and with the identification of this supermultiplet
with the one obtained from the chiral primary ${\cal C}_l$, \ie:
\beq
(I_-^{l+1}, S^l, V^l, I_+^{l-1})\sim ({\cal C}_l,\cdots)\,\,,\qquad
(l\ge 1)\,\,.
\label{identf}
\eeq
As a check notice that the four modes on the left-hand side of eq. (\ref{identf}) have the
same mass spectrum, namely $M_S(n,l)$. Moreover, 
$\Delta(S^l)=\Delta(V^l)=\Delta(I_-^{l+1})+1$ and 
$\Delta(I_+^{l-1})=\Delta(I_-^{l+1})+2$, which is in agreement with the fact that the
supercharge has dimension $1/2$. 

%
%
%
%
%
%
%
%
%

\setcounter{equation}{0}
\section{Fluctuations of the D3-D3 system}
\label{cp2D3D3sc}
\medskip
Let us analyze in this section the modes of the $(1|D3\perp D3)$ intersection. In the
probe approximation we are considering the equations of motion of these fluctuation
modes are obtained from the Dirac-Born-Infeld action
(see eq. (\ref{cp1Dbraneact})) of a D3-brane in the $AdS_5\times S^5$
background. This action is given by:
\beq
S=-\int d^4\xi\,\sqrt{-\det (g+F)}\,+\int
d^4\xi\;P\big[\,C^{(4)}\,\big]\,,
\label{DBID3-D3}
\eeq
where, again, we have set $T_3=1$ for simplicity and redefined
$2\pi\,\alpha'\,F\to F$. We want to expand the action (\ref{DBID3-D3})
around the static configuration in which the two branes are separated a
distance
$L$.  Recall from section \ref{general} that the induced metric on the
worldvolume of the D3-brane probe of such a configuration is just the one
written in eq. (\ref{AdSindmetric}) for $d=1$, which reduces to
$AdS_3\times S^1$ in the UV limit. As in section \ref{general}, let us
denote by $\chi$ the scalars transverse to both types of branes. In this
case the defect created by the probe has codimension two in the Minkowski
directions of $AdS_5\times S^5$. Let us assume that the D3-brane probe is
extended along the Minkowski coordinate $x^1$ and let us define:
\beq
\lambda_1=x^2\;,\qquad\quad
\lambda_2=x^3\,.
\eeq 
With these notations, the lagrangian for the quadratic fluctuations can be readily
obtained from (\ref{DBID3-D3}):
\bear
&&{\cal L}=-{\rho\over 2}\,{R^2\over \rho^2+L^2}\,{\cal G}^{ab}
\partial_a\chi\partial_b\chi-{\rho\over 2}\,{\rho^2+L^2\over R^2}\,
{\cal G}^{ab}\partial_a \lambda_i\partial_b \lambda_i-{\rho \over 4}
\,F_{ab}F^{ab}+\rc\rc
&&\;\qquad
+{(\rho^2+L^2)^2\over R^4}
 \,\epsilon^{ij}\partial_{\rho} \lambda_i\,\partial_{\varphi} \lambda_j\,,
\label{D3D3quadraticL}
\eear
where $i,j=1,2$ and ${\cal G}_{ab}$ is  the metric (\ref{AdSindmetric})
for $d=1$. The equation of motion of $\chi$ derived from
(\ref{D3D3quadraticL}) is just the one studied in section \ref{general},
namely (\ref{eom-general}) for $p_2=3$, $d=1$ and $\gamma_2=1$.
Separating variables as in (\ref{sepvar}) we arrive at an equation which
can be analytically solved in terms of hypergeometric functions. The
corresponding solution has been written in eq. (\ref{scalarhyper}). After
imposing the truncation condition (\ref{quant})  we obtain the so-called
$S^l$ modes, whose explicit expression is:
\beq
\xi_S(\varrho)=\varrho^l\,(\varrho^2+1)^{-n-l}\,
F(-n-l,  -n; l+1;-\varrho^2\,)\,.
\label{D3D3xiS}
\eeq
Notice that in this case the harmonics are just exponentials of the type
$e^{i\l\varphi}$, where $\varphi$ is just the angular coordinate of the $S^1$ circle.
Then, the quantum number $l$ can take also negative values. The modes written in
(\ref{D3D3xiS}) are those which are regular at $\varrho=0$ for
non-negative $l$. When $l<0$ one can get regular modes at the origin by
using the second solution of the hypergeometric function. The result is
just (\ref{D3D3xiS}) with $l$ changed by $-l$. However, since the scalar
field
$\chi$ whose fluctuation we are analyzing is real, changing $l$ by $-l$
in $e^{i\l\varphi}$ makes no difference and we can restrict ourselves to
the case $l\ge 0$.  The mass spectrum and associated conformal dimensions
of the fluctuations  (\ref{D3D3xiS}) are:
\beq
M_S(n,l)={2L\over R^2}\,\sqrt{\bigg(n+l\bigg)\bigg(n+l+1\bigg)}\;,\qquad
\quad \Delta_S=l+1\,,
\label{MSd3d3}
\eeq
where $n\ge 0$, except for the case $l=0$ where $n\ge 1$. Notice that for $n=l=0$ the
function $\xi_S(\varrho)$ is just constant. Moreover,  $M_S$ vanishes in this case and
thus we can take the solution $\chi$ to be also independent of the Minkowski
coordinates. This constant zero mode corresponds just to changing the value of the
distance $L$ and should not be considered as a true fluctuation. Therefore, we shall
understand that $n\ge 0$ in (\ref{D3D3xiS}) and (\ref{MSd3d3}), except in the case $l=0$
where $n\ge 1$.

\subsection{Scalar fluctuations}
Let us now study the fluctuations of the $\lambda_i$ scalars. Notice that these fields
are coupled through the Wess-Zumino term in (\ref{D3D3quadraticL}). Actually, 
the equations of motion for the $\lambda_i\,$s derived from
(\ref{D3D3quadraticL}) are:
\beq
R^2\,\partial_a\,\left[\rho\,(\rho^2+L^2)\,
{\cal G}^{ab}\partial_b \lambda_i\right]-4\rho\,(\rho^2+L^2)\,\epsilon^{ij}
\,\partial_{\varphi} \lambda_j=0\,.
\eeq
To solve these equations let us introduce the reduced variables 
$\bar M^2\,=\,-R^4L^{-2}k^2$, $\varrho=\rho/L$ and 
let us expand the $\lambda_i\,$s in modes as:
\beq
\lambda_i=e^{ikx}\,e^{-il\varphi}\,\xi_i(\rho)\;,\qquad i=1,2\,.
\label{D3D3separation}
\eeq
Then, the functions $\xi_i(\rho)$ satisfy the coupled equations:
\beq
{1\over \varrho (1+\varrho^2)}\,\partial_{\varrho}\,\Big[\varrho\,
(1+\varrho^2)^2\,\partial_{\varrho}\,\xi_i\Big]+
\Bigg[{\bar M^2\over 1+\varrho^2}-l^2\,\big(1+{1\over \varrho^2}\big)
\Bigg]\,\xi_i+4\,i\,l\,\epsilon^{ij}\xi_j=0\,.
\eeq
In order to diagonalize this system of equations, let us define the following complex
function:
\beq
w=\xi_1+ i\,\xi_2\,,
\label{D3D3w-def}
\eeq
which satisfies the differential equation:
\beq
{1\over \varrho (1+\varrho^2)}\,\partial_{\varrho}\,\Big[\varrho\,
(1+\varrho^2)^2\,\partial_{\varrho}\,w\Big]+\Bigg[{\bar M^2\over 1+
\varrho^2}-l^2\,\big(1+{1\over \varrho^2}\big)\Bigg]\,w+ 4l\,w=0\,.
\label{D3weqn}
\eeq
Equation  (\ref{D3weqn}) can be solved in terms of a hypergeometric function, namely:
\beq
w^{(1)}=\varrho^{l}\,(1+\varrho^2)\,^{-\lambda-1}\,\,
F(-\lambda+1, l-\lambda-1;l+1; -\varrho^2)\,,
\label{w1}
\eeq
where $\lambda$ is related to $\bar M$ as in  (\ref{lambda}). 
Notice that $w^{(1)}$ is regular at $l=0$ for $l\ge 0$. Actually, since $l$ can be
negative in this case  one can also consider the second solution of the hypergeometric
equation, which is:
\beq
w^{(2)}=\varrho^{-l}\,(1+\varrho^2)\,^{-\lambda-1}\,\,
F(-\lambda-1, -\lambda-l+1;1-l; -\varrho^2)\,.
\label{w2}
\eeq
By applying eq. (\ref{Delta-hyper}) to the present case, we obtain that 
the conformal dimension of a fluctuation of the type 
(\ref{hyper}) is just $\Delta=l-1$ or $\Delta=3-l$. Actually, 
it is straightforward to verify that the solutions of the differential equation 
(\ref{D3weqn}) present two different behaviors at $\varrho\to \infty$, namely 
$\varrho^{-l}$ and $\varrho^{l-4}$. The first behavior corresponds to an operator with
$\Delta=l-1$, while $\Delta=3-l$ is the dimension of an operator whose dual fluctuation
behaves as $\varrho^{l-4}$ for large $\varrho$. In the following we will refer to the
fluctuations with
$\Delta=l-1$ as
$W_+^l$, while those with
$\Delta=3-l$ will be denoted by $W_-^l$. These two branches
\footnote{For $l=2$ both UV behaviors coincide and   there is no distinction between
the two branches. In this case there are solutions of the fluctuation equation which
behave as $\varrho^{-2}\log \varrho$ for large $\varrho$.  }
will be studied
separately in their unitarity range
$\Delta\ge 0$ by finding the truncations of the hypergeometric series of 
$w^{(1)}$ and $w^{(2)}$ with the appropriate behavior at large $\varrho$.

\subsubsection{$W_+^l$ fluctuations}
Let us consider the solution $w^{(1)}$ in eq. (\ref{w1}) with the following truncation
condition:
\beq
l-\lambda-1=-n\;,\qquad\quad n=0,1,2,\cdots\,.
\eeq
The resulting solution is:
\beq
w=\varrho^{l}\,(1+\varrho^2)\,^{-l-n}\,\,
F(2-l-n, -n;l+1; -\varrho^2)\,.
\label{W+sol}
\eeq
One can check easily that $w\sim\varrho^{-l}$ for large $\varrho$ if $l=1, n=0$ or $l\ge
2, n\ge 0$. Notice that $\Delta=l-1$ in this case and the unitarity range is $l\ge 1$.
The mass spectrum becomes:
\beq
M_{W_+}(n,l)={2L\over R^2}\,\sqrt{\Big(n+l-1\Big)\Big(n+l\Big)}\;,\qquad
\quad \Delta_{W_+}=l-1\;,\qquad (l\ge 1)\,.
\label{D3D3W+mass}
\eeq
The  $l=1$ fluctuation  is a special case. Indeed, in this case one has
$\Delta=0$ and, as $n$ must vanish,  $M$ is zero, \ie\ we
have a massless mode despite of the fact that we have introduced a mass scale by
separating the branes. As argued in ref. \cite{CEGK} this is related to the appearance of
the Higgs branch on the field theory side. Let us look closer at this $l=1$, $n=0$ mode.
In this case
$M=\lambda=0$ and  the hypergeometric function is just equal to one. Thus:
\beq
w\sim {\rho\over \rho^2+L^2}\,\,,
\label{higgsbranch}
\eeq
where we have reintroduced the constant $L$. In particular, for large
$\rho$:
\beq
w \approx {c\over \rho}\,\,,\qquad (\rho \to \infty)\,\,,
\eeq
where $c$ is a constant. Let us consider the solution in which $k=0$ (which certainly has 
$M=0$). This solution does not depend on the coordinates $(t,x^1)$. Let us introduce the
$\varphi$ dependence and define the following two complex variables:
\beq
\Lambda\equiv \lambda_1+i\lambda_2=x^2+ix^3\,\,,\qquad\quad
Y=\rho \, e^{i\varphi}\,.
\eeq
Then, the modes we are studying satisfy for large $\rho$:
\beq
\Lambda Y\approx c\,\,,\qquad (\rho \to \infty)\,,
\eeq
which is just the holomorphic curve of the Higgs branch found in ref. \cite{CEGK} by
looking at the vanishing of the F-terms of the SUSY theory. 

It is worth to stress here the difference between the $l=1$ solution
(\ref{higgsbranch}) and the fluctuations (\ref{W+sol}) for $l>1$. Indeed,
in the latter case we get a full tower of solutions, depending on the
excitation number $n$, whereas for $l=1$ we have only the single function
(\ref{higgsbranch}). Moreover, the mass spectra (\ref{D3D3W+mass}) is
simply related to the one corresponding to the transverse scalar
fluctuations $S^l$ only for $l>1$ (see subsection \ref{cp2D3D3dict}).
One can regard (\ref{higgsbranch}) as a non-trivial solution in which the
D3-brane probe is deformed at no cost along the directions of the
worldvolume of the D3-brane of the background. We will confirm this in the
next chapter, when studying the Higgs branch of this defect theory, by
showing that a D3-brane embedded along such a curve preserves one half
of the supersymmetry of the background.

It is also interesting to point out that the differential equation (\ref{D3weqn}) can be
solved by taking $w=\rho^{-l}$ and $M=0$. This fact can be checked directly from eq.
(\ref{D3weqn}) or by taking $\lambda\to 0$ in the solution (\ref{w2}). Notice, however
that this solution is not well-behaved at $\rho \to 0$ for $l\ge 1$, contrary to what
happens to the function written in eq. (\ref{higgsbranch}). A second solution with $M=0$
can be obtained by putting $\lambda=0$ in (\ref{w1}). This solution is regular at
$\rho\to 0$. However, for $l>1$ the hypergeometric function which results from taking 
$\lambda=0$ in (\ref{w1}) contains logarithms for $L\not=0$ and its interpretation in
terms of a holomorphic curve is unclear to us.

\subsubsection{$W_-^l$ fluctuations}
The allowed range of values of $l$ for the fluctuations $W_-^l$ is $l\le 3$. We have 
found a discrete tower of states only for $l\le 1$. As in the previous subsection, $l=1$ is
special. In this case the solutions regular at $\varrho=0$ which decrease as
$\varrho^{-3}$ for large 
$\varrho$ are:
\beq
w=\varrho\,(1+\varrho^2)\,^{-n-1}\,\,
F(1-n, -n;2; -\varrho^2)\,\,,\qquad (l=1)\,,
\eeq
where $n\ge 1$. Indeed, this solution is just (\ref{W+sol}) for $l=1$ and $n\ge 1$.
Moreover, for $l\le 0$ the solutions which behave as $\varrho^{l-4}$ for large $\varrho$
can be obtained by putting $-\lambda-l+1\,=\,-n$, with $n\ge 0$, on the solution
$w^{(2)}$ of eq.  (\ref{w2}). One gets:
\beq
w=\varrho^{-l}\,(1+\varrho^2)\,^{-2-n+l}\,\,
F(-2-n+l, -n;1-l; -\varrho^2)\,\,,\qquad (l\le 0)\,.
\eeq
The mass spectrum for $l\le 1$ can be written as:
\beq
M_{W_-}(n,l)={2L\over R^2}\,\sqrt{\Big(n+1-l\Big)\Big(n+2-l\Big)}\;,\qquad 
\quad \Delta_{W_-}=3-l\;,\qquad (l\le 1)\,,
\eeq
where it should be understood that $n\ge 1$ for $l=1$ and $n\ge 0$ otherwise.

\subsection{Vector fluctuations}
We will now study the fluctuations of the worldvolume gauge field. We will try to
imitate the discussion of section \ref{cp2D3D5} for the D3-D5 system.
Obviously, the analogue of the type I modes does not exists for a
one-dimensional sphere. Let us analyze the spectra of the other two types
of modes. 
\subsubsection{Type II modes}
Let us consider the ansatz:
\beq
A_{\mu}=\xi_{\mu}\,\phi(\rho)\,e^{ikx}\,e^{-il\varphi}\;,\qquad\quad
A_{\rho}=0\;,\qquad\quad A_{\varphi}=0\,,
\label{D3D3typeII}
\eeq
with $\xi_{\mu}$ being a constant vector such that $k^{\mu}\xi_{\mu}=0$. In terms of
the reduced variables $\varrho$ and $\bar M$, the equation for
$\phi(\rho)$ is:
\beq
\partial_{\varrho}\,\Big[\,\varrho\,\partial_{\varrho}\phi\,\Big]+
\Bigg[\,{\varrho\over (\varrho^2+1)^2}\,\bar M^2-{l^2\over
\varrho}\,\Bigg]\,\phi=0\,\,,
\eeq
which is the same as for the transverse scalars $\chi$. Therefore, the mass spectrum of
these type II vector modes is just the same as in (\ref{MSd3d3}).

\subsubsection{Type III modes}
We now adopt the ansatz:
\beq
A_{\mu}=0\;,\qquad\quad
A_{\rho}=\phi(\rho)\,e^{ikx}\,e^{-il\varphi}\;,\qquad\quad
A_{\varphi}=\tilde\phi(\rho)\,e^{ikx}\,e^{-il\varphi}\,. 
\label{D3D3typeIII}
\eeq
The equation for $A_{\rho}$ is:
\beq
il\partial_{\rho}\tilde\phi-l^2\phi+
M^2\,R^4\,{\rho^2\over (\rho^2+L^2)^2}\,\phi=0\,\,,
\eeq
while the equation for $A_{\varphi}$ yields:
\beq
\partial_{\rho}\left[{(\rho^2+L^2)^2\over \rho}
\left(\partial_{\rho}\tilde\phi+il\phi\right)\right]+
{M^2\,R^4\over \rho}\,\tilde\phi=0\,.
\eeq
Finally, the equation for $A_{\mu}$ gives a relation between $\phi$ and $\tilde\phi$,
namely:
\beq
\rho\,\partial_{\rho}\Big(\rho\phi\Big)=il\tilde\phi\,.
\label{D3D3phi-tildephi}
\eeq
For $l\not=0$ we can use (\ref{D3D3phi-tildephi}) to eliminate $\tilde\phi$ in favor of
$\phi$. The remaining equations reduce to the following equation for $\phi$:
\beq
\partial_{\varrho}\,\Big[\varrho\,\partial_{\varrho}(\varrho\,
\phi)\Big]+\Bigg[{\varrho^2\over (\varrho^2+1)^2}\,\bar M^2-l^2
\Bigg]\,\phi=0\,,
\label{D3D3-typeIII-phi}
\eeq
where we have already introduced the reduced variables $\varrho$ and $\bar M$. 
The solution of (\ref{D3D3-typeIII-phi}) regular at $\varrho=0$ is:
\beq
\phi=\varrho^{l-1}\,(1+\varrho^2)\,^{-\lambda}\,\,
F(-\lambda, l-\lambda;l+1; -\varrho^2)\,,
\eeq
with  $\lambda$ being the quantity defined in (\ref{lambda}). 
By imposing the quantization condition:
\beq
l-\lambda=-n\,,\qquad\quad n=0,1,\cdots\,,
\eeq
we get a tower of fluctuation modes which behaves as $\rho^{-l-1}$ when $\rho\to\infty$.
The corresponding   mass levels and conformal dimensions are:
\beq
M_V(n,l)={2L\over R^2}\,\sqrt{\Big(n+l\Big)\Big(n+l+1\Big)}\,\,,
\qquad\quad \Delta_V=l+1\,\,,
\eeq
which again coincide with the results obtained for the scalar modes. 
\subsection{Fluctuation/operator correspondence}
\label{cp2D3D3dict}
\medskip
The array corresponding to the D3-D3 intersection is:
\beq
\begin{array}{ccccccccccl}
 &1&2&3& 4& 5&6 &7&8&9 & \nonumber \\
D3: & \times &\times &\times &\_ &\_ & \_&\_ &\_ &\_ &     \nonumber \\
D3': &\times&\_&\_&\times&\times&\_&\_&\_&\_ &
\end{array}
\label{D3D3intersection}
\eeq
where the $D3'$ is the probe brane. First of all, let us discuss the isometries of this
configuration. Clearly, the addition of the brane probe breaks the $SO(6)$ symmetry
corresponding to rotations in the 456789 directions to $SO(4)\times U(1)_{45}$, where the
$SO(4)\approx SU(2)\times SU(2)$ factor is generated by rotations in the 6789 subspace and
the $U(1)_{45}$ corresponds to rotations in the plane spanned by coordinates $4$ and $5$.
In addition we have an extra $U(1)_{23}$ generated by the rotations in the $23$ plane. 

The field content of the defect theory can be obtained by reducing the ${\cal N}=4$,
$d=4$  gauge multiplet  down to two dimensions and by adding the corresponding $3-3'$
sector \cite{CEGK}. The resulting theory has $(4,4)$ supersymmetry in $d=2$. In
particular, two of the six $d=4$ adjoint chiral scalar superfields give rise to a field
$Q$ whose lowest component (which we will denote by $q$) describes the fluctuations of
the D3 in the directions $4$ and $5$. This field $q$ is a singlet of $SO(4)$ and 
$U(1)_{23}$ and is charged under $U(1)_{45}$. The strings stretched between the D3 and
the $D3'$ give rise to two chiral multiplets $B$ and $\tilde B$ which are
fundamental and antifundamental with respect to the gauge group. The
lowest components of $B$ and
$\tilde B$ are two scalar fields $b$ and $\tilde b$ which are singlets under $SO(4)$ and
are  charged under $U(1)_{23}$ and $U(1)_{45}$. Moreover, the fermionic components of
$B$ and $\tilde B$ can be arranged in two $SU(2)$ multiplets $\psi^+$ and $\psi^-$ which
are neutral with respect to the two $U(1)\,$s and charged under one of the
two $SU(2)\,$s of the decomposition $SO(4)\approx SU(2)\times SU(2)$. The
dimensions and quantum numbers of the different fields just discussed are
summarized in table
\ref{tableFieldsD3D3}.
\begin{table}[!h]
\centerline{
\begin{tabular}[b]{|c|c|c|c|c|}   
 \hline
\rule{0mm}{4.5mm} Field  & $\Delta$  & $SO(4)$ & $U(1)_{23}$ &$U(1)_{45}$ \\ 
\hline 
  \rule{0mm}{5.5mm} $q$ & $1$  & $(0,0)$ & $0$ & $1$ \\  \hline  
\rule{0mm}{5.5mm}  $b, \tilde b$ &$0$  &  $(0,0)$  & $-1/2$ & $1/2$ \\ \hline  
\rule{0mm}{5.5mm} $\psi^+$ & $1/2$  & $(1/2, 0)$ &$0$&$0$ \\  \hline   
\rule{0mm}{5.5mm} $\psi^-$ & $1/2$  & $(0, 1/2)$ &$0$&$0$ \\ \hline  
\end{tabular}
}
\caption{Quantum numbers and dimensions of the fields of the D3-D3 intersection.}
\label{tableFieldsD3D3} 
\end{table}

In order to establish the fluctuation/operator dictionary in this case, let
us determine the quantum numbers of the different fluctuations. The
fluctuations in the directions $6789$ (which are transverse to both types
of D3-branes) will be denoted by $S^l$. Clearly, they transform in the
$(1/2,1/2)$ representation of $SO(4)$ and are neutral under 
$U(1)_{23}$. Moreover, since the rotations of the 45 plane are just those
along the one-sphere of the probe worldvolume, the integer $l$ is just the
charge under $U(1)_{45}$. The  $w$ coordinate parametrizes the 23 plane.
Let us denote by $W_+^l$ and $W_-^l$ to the two branches of $w$
fluctuations. The $W_{\pm}^l\,$s are $SO(4)$ singlets and are charged
under both $U(1)_{23}$ and $U(1)_{45}$.  As in the D3-D5 case, the
fluctuations of types II and III correspond, in the conformal limit, to
the components of a vector field in $AdS_3$ and will be denoted by $V^l$.
They are singlets under $SO(4)$ and $U(1)_{23}$ . Table
\ref{tableModesD3D3} is filled in with the dimensions and quantum numbers
of the different fluctuations.

\begin{table}[!h]
\centerline{
\begin{tabular}[b]{|c|c|c|c|c|}   
 \hline
\rule{0mm}{4.5mm} Mode  & $\Delta$  & $SO(4)$ & $U(1)_{23}$ &$U(1)_{45}$ \\ 
\hline 
  \rule{0mm}{5.5mm}$S^l$ & $l+1$   & $(1/2,1/2)$ & $0$ & $l\ge 0$ \\  \hline  
\rule{0mm}{5.5mm}  $W_+^l$ &$l-1$  &  $(0,0)$  & $-1$&$l\ge 1$ \\ \hline  
\rule{0mm}{5.5mm} $W_-^l$ & $3-l$  & $(0, 0)$ &$-1$&$l\le 1$ \\  \hline   
\rule{0mm}{5.5mm} $V^l$ & $l+1$  & $(0, 0)$ &$0$&$l\ge 0$ \\ \hline  
\end{tabular}
}
\caption{Quantum numbers and dimensions of the modes of the D3-D3 intersection.}
\label{tableModesD3D3} 
\end{table}

The mass spectrum $M_S(n,l)$ of the fluctuations $S^l$ has been written in eq.
(\ref{MSd3d3}). The masses of the other modes can
be written in terms of $M_S(n,l)$ as:
\bear
&&M_{W_+}(n,l)=M_S(n,l-1)\;,\qquad(l\ge 2),\rc
&&M_{W_-}(n,l)=M_S(n,1-l)\;,\qquad(l\le 1),\rc
&&M_{V}(n,l)=M_S(n,l)\;,\qquad\qquad\,\,\,(l\ge 0).
\label{D3massrelations}
\eear
Notice that the relation between $M_S$ and $M_{W_-}$ is consistent with the absence of
the
$n=0$ mode in the $S^0$ and $W_-^1$ fluctuations.

In order to relate the different fluctuations to composite operators of the defect theory
let us define following ref. \cite{CEGK} the operator ${\cal B}^l$ for $l\ge 1$ as:
\beq
{\cal B}^l\equiv \tilde b\, q^{l-1}\,b\,.
\eeq
Notice that ${\cal B}^l$ has the same dimension and quantum numbers as the fluctuation 
$W_+^l$. Similarly, the operator ${\cal G}^l$ for $l\le 1$, defined as:
\beq
{\cal G}^l\equiv D_-\tilde b\,q^{\dagger 1-l}\, D_+b+
D_+\tilde b\,q^{\dagger 1-l}\, D_-b\,,
\eeq
where $D_\pm=D_0\,\pm\,D_1$, has the right properties to be identified
with the dual of the fluctuations $W_-^l$. Moreover, to define the
operator dual to $S^l$ we have to build a vector of $SO(4)$. The natural
objects to build  an operator of this sort are the spinor fields
$\psi^{\pm}$. Indeed, such an operator can be written as:
\beq
{\cal C}^{\mu l}\equiv\sigma_{ij}^{\mu}\,\left(\epsilon_{ik}\bar\psi_k^{+}
\,q^l\psi_j^{-}+\epsilon_{jk}\bar\psi_k^{-}\,q^l \psi_i^{-}\right)\,,
\eeq
where $l\ge 0$ and 
$\mu$ is an $SO(4)$ index. Therefore, according to the proposal of ref.
\cite{CEGK}, we have:
\beq
S^l\sim {\cal C}^{\mu l}\;,\qquad\quad
W_+^l\sim {\cal B}^l\;,\qquad\quad
W_-^l\sim {\cal G}^l\,.
\eeq
As argued in ref. \cite{CEGK}, the fluctuations $W_+^l$ for $l=1$ are dual to the field 
${\cal B}^1=\tilde b b$, which parametrizes the classical Higgs branch of the theory,
whereas for higher values of $l$ these fluctuations correspond to other holomorphic
curves. Moreover, the field ${\cal C}^{\mu l}$ is a BPS primary and 
${\cal G}^{1-l}$ is a two supercharge descendant of this primary. Notice that this is
consistent with our relation (\ref{D3massrelations}) between the masses of the $S^l$ and
$W_-^l$ fluctuations. Lastly, the fluctuations $V^l$ are dual to two-dimensional vector
currents. The dual operator at the bottom of the Kaluza-Klein tower is a global $U(1)$
current ${\cal J}_B^{M}$:
\beq
V^0\sim {\cal J}_B^{M}\,.
\eeq
The expression of ${\cal J}_B^{M}$ has been given in ref. \cite{CEGK}, namely:
\def\lrD{\buildrel \leftrightarrow\over D}
\beq
{\cal J}_B^{M}\equiv\psi_i^{\alpha}\,\rho_{\alpha\beta}^M\,\psi_i^{\beta}+
i \bar b \lrD{}^M b+i \tilde b\,\lrD{}^M\,\bar{\tilde b}\;,\qquad
(M=0,1)\,,
\eeq
where $\alpha,\beta=+, -$ and $\rho^M$ are Dirac matrices in two dimensions. 

%
%
%
%
%
%
%
%
%

\setcounter{equation}{0}
\section{Discussion}
\medskip
\label{Discussion}

In this chapter we have studied the fluctuation spectra of brane probes
in the near-horizon background created by a stack of other branes. In the
context of the generalization of the gauge/gravity correspondence
proposed in refs. \cite{KR,KKW} these fluctuations are dual to open
strings stretching between the two types of branes of the intersection
and can be identified, on the field theory side of the correspondence,
with composite operators made up from hypermultiplets in the fundamental
representation of the gauge group.

In particular, we have analyzed the set of systems given by  the BPS
intersections of two types of branes, considering the brane of higher
dimensionality as a probe brane in the background sourced by (a stack of)
the lower dimensional brane. By allowing a finite separation between the
probe and the origin of the holographic direction we introduce an explicit
mass scale in the problem, which is related to the mass of the
hypermultiplet. We have been able to give a unified description of the
transverse fluctuations of the brane probe and, through an analysis of the
equivalent Schr\"odinger problem, we have found that these fluctuations
have a discrete mass spectrum. This spectrum has been computed
analytically for the cases where the background brane is a D3-brane. For
the rest of the systems the use of numerical computations and WKB
estimates has been described (more detailed results were presented in
\cite{AR}). 

Interestingly, the mass gap of the spectra for the different Dp- Dq-brane
intersections, including the exactly solvable ones (for $p=3$), turns
out to be universal when written in terms of the quark mass and the
effective dimensionless coupling of the dual theory, namely $M\sim
m_q/g_{eff}(m_q)$, where $m_q$ is the quark mass and $g_{eff}$
stands for the dimensionless effective Yang-Mills coupling constant.

We have thoroughly studied the cases of D5- and D3-brane probes moving in
the $AdS_5\times S^5$ geometry, corresponding respectively to the D3-D5 and
D3-D3 intersections. In these two cases, if the brane reaches the origin of
the holographic coordinate,  it wraps an $AdS_{d+2}\times S^d$ ($d=2,1$)
submanifold of the  $AdS_5\times S^5$ background and the corresponding
field theory duals are defect conformal field theories with a fundamental
hypermultiplet localized at the defect. The  spectra of conformal
dimensions and the precise mapping between probe fluctuations and
operators of the dual defect theory for the D3-D5 and D3-D3 intersections
were obtained in refs. \cite{WFO} and \cite{CEGK} respectively. 
After giving mass to the fundamentals as described above the system is no
longer conformal and develops a mass gap. 
We have studied the whole set of fluctuations for these configurations and,
remarkably, the corresponding differential equations can be solved in
terms of the hypergeometric function and the mass spectra of the
fluctuations can be obtained analytically.  These mass spectra display
some degeneracies which are consistent with the structure of the
supermultiplets found in refs. \cite{WFO,CEGK} for the corresponding dual
operators. An extension of this complete analysis of the D3-D5 and
D3-D3 systems to the case of a supersymmetric Dq-brane probe in the
background of a Dp-brane (for $p\not=3$) was presented in \cite{AR}.

One of the limitations of our approach is the fact that we have worked in
the probe approximation, thus neglecting the backreaction of the branes on
the geometry. Actually, for some cases the supergravity solution
representing the localized intersection has been constructed
\cite{Beyond}-\cite{Gomis:2006cu}, but these backgrounds are
rather complicated and it is not easy to extract information about the
gauge theory dual. It might be more fruitful to follow the approach
recently proposed in
\cite{CNP} for the
${\cal N}=1$ theories (see also \cite{unqchKW,unqchKS}), where the
supergravity action is supplemented with the action of the brane, which
has been conveniently smeared,  and  a solution of the equations of motion
for the supergravity plus brane system is found. In this approach the
adjoint (color) degrees of freedom are represented by fluxes, whereas the
fundamental (flavor) fields are generated by branes.

Despite its limitations, we think that this work provides a non-trivial
realization of the holographic idea (see \cite{KOBS} for a rigorous
treatment)  and it will be  interesting to look at some generalizations of
our results. In particular, it has been shown that the string dual to the
Higgs branch of gauge theories can be constructed by recombining color and
flavor branes (see, for instance, \cite{Giveon:1998sr}). Therefore, in the
next chapter, by appropriately deforming the different kinds of Dp-
Dq-brane intersections studied in this chapter, we will study  the Higgs
branch of the (defect) dual theories.

It is also of great interest to look at defect theories with reduced
supersymmetry. One of such theories is obtained by embedding a
D5-brane in the $AdS_5\times T^{1,1}$ geometry, the so-called
Klebanov-Witten model \cite{KW}. The precise form of the embedding in this
case can be found in chapter \ref{cpSSprobes}, where we will study the
different supersymmetric embeddings of probe branes in $AdS_5\times T^{1,1}$.
Actually, the $T^{1,1}$ space can be substituted by any Sasaki-Einstein space
\cite{Yamaguchi}, as illustrated in \cite{CEPRV,Canoura:2006es} for the
$Y^{p,q}$ and $L^{a,b,c}$ manifolds respectively. The supersymmetric defects
in the Maldacena-N\'u\~nez background \cite{MN,CV} have been obtained in
ref. \cite{CPR}.

\vskip .8cm

\medskip

\begin{subappendices}

\setcounter{equation}{0}
\section{Change of variables for the exact spectra}
\medskip
\label{hyperchange}
Let us consider an equation of the type:
\beq
z(1-z)\,\phi''+(\alpha+\beta z)\,\phi'+
[\,\gamma+\delta\,z^{-1}+\epsilon\,(1-z)^{-1}\,]\,\phi=0\,,
\label{hypereq}
\eeq
where $\phi=\phi(z)$, the prime denotes derivative with respect to $z$ and $\alpha, \beta,
\gamma,\delta$ and $\epsilon$ are constants. The solution of this equation can be written
as:
\beq
\phi(z)=z^{f}\,(1-z)^{g}\,P(z)\,\,,
\eeq
where $P(z)$ satisfies the hypergeometric equation:
\beq
z(1-z)\,P''+[c-(a+b+1)z]\,P'-ab\,P=0\,.
\eeq
The values of the exponents $f$ and $g$ are:
\bear
&&f={1-\alpha+\lambda_1\,\sqrt{(\alpha-1)^2-4\delta}\,\over 2}\,,\rc\rc
&&g={\alpha+\beta+1+\lambda_2\,\sqrt{(\alpha+\beta+1)^2-4\epsilon}\over 2}
\,,
\label{fyg}
\eear
where $\lambda_1$ and $\lambda_2$ are signs which can be chosen by convenience. Moreover,
$a$, $b$ and $c$ are given by:
\bear
&&a=f+g-{1+\beta+\sqrt{4\gamma+(\beta+1)^2}\over 2}\,,\rc\rc
&&b=f+g-{1+\beta-\sqrt{4\gamma+(\beta+1)^2}\over 2}\,,\rc\rc
&&c=\alpha+2f\,.
\eear
There are two  solutions for $P(z)$ in terms of the hypergeometric function. The first
one is:
\beq
P(z)=F(a,b;c;z)\,.
\eeq
The second solution is:
\beq
P(z)=z^{1-c}\,F(a-c+1, b-c+1; 2-c;z)\,.
\eeq

\end{subappendices}

%
%
%
%
%
%
%
%
%
%

\chapter{Holographic flavor on the Higgs branch}
\setcounter{equation}{0}
\section{Introduction}
\label{cp3introsc}
\medskip

In this chapter we will study the holographic dual to the Higgs branch of
the different gauge theories with fundamentals dealt with in the previous
chapter. There, by considering different SUSY intersections of branes in
type II SUGRA and M-theory, we constructed the holographic dual to gauge
(defect) theories with fundamental matter in several spacetime
dimensions. In the brane picture both stacks of branes intersected
orthogonally: in this configuration the dual theory is in its Coulomb
phase, parametrized by the positions of the background branes in the
space transverse to both the background and flavor branes.
Nevertheless, one could have more involved situations such as Higgs
phases. On the field theory side the Higgs phase corresponds to having
non-zero VEVs of quark fields. As it is well-known (see e.g.
ref. \cite{Giveon:1998sr}) the Higgs branch of gauge theories can be
realized in string theory by recombining color and flavor branes. This
recombination can be described in two different and complementary ways.
From the point of view of the flavor brane (the so-called macroscopic
picture) the recombination is achieved by a non-trivial embedding of the
brane probe in the background geometry and/or by a non-trivial  flux of
the worldvolume gauge field. On the other hand, 
the description of the recombination from the point of view of the color
brane defines  the microscopic picture. In most of the cases this
microscopic picture can be regarded as a dielectric effect \cite{M}, in
which a set of color branes gets polarized into a higher-dimensional
flavor brane. Interestingly, the microscopic description of the Higgs
branch allows a direct relation with the field theory analysis and the
micro-macro matching is essential to understand how gauge theory
quantities are encoded in the configuration of the flavor brane.

Let us briefly review what is known about how to realize the Higgs phase
of theories dual to the Dp- Dq-brane intersections, especially when
$p=3$, since in this case the dual theories are better studied. 
The Higgs phase of the D3-D7 intersection was studied in \cite{EGG} (see
also ref. \cite{Guralnik:2004ve}). There it was proposed that, from the
point of view of the D7-brane, one can realize  a (mixed Coulomb-)Higgs
phase of the D3-D7 system by switching on an instanton configuration of
the worldvolume gauge field of the D7-brane. This instantonic gauge
field lives on the directions of the D7-brane worldvolume that are
orthogonal to the gauge theory directions. The size of the instanton has
been identified in \cite{EGG,Guralnik:2004ve} with the VEV of the quark
fields. The meson  spectra depends on this size and  was shown  to
display, in the limit of infinite instanton size, an spectral flow
phenomenon.

It would be interesting to find a similar deformation of the D3-D5
intersection giving rise to the Higgs phase of the dual defect theory.
Let us recall that, in the probe approximation, the D5-brane probe
is wrapping an $AdS_4\times S^2$ defect in the $AdS_5\times S^5$
geometry dual to the D3-branes (if the transverse distance between both
branes is not zero this is only true in the UV). By switching on a
magnetic field along the $S^2$, one can still have a supersymmetric
intersection if the D5-brane is appropriately bent along the D3-brane
\cite{ST}, which corresponds to a different $AdS_4\times S^2\subset
AdS_5\times S^5$ embedding. Furthermore, the worldvolume flux induces
D3-brane charge to the D5-brane probe reflecting the fact that some of
the D3-branes of the background recombine with the probe D5-brane.
Therefore, this configuration is the ideal candidate for realizing the
Higgs phase of the dual theory.

As for the D3-D3 intersection, its dual defect conformal field theory
was studied in ref. \cite{CEGK}, where the Higgs branch was identified 
with a particular holomorphic embedding of the probe D3-brane in the
$AdS_5\times S^5$ geometry. This embedding was shown to correspond to the
vanishing of the $F$- and $D$-terms in the dual superconformal field
theory (see also refs.
\cite{Kirschphd,Erdmenger:2003kn}).

In this chapter we will generalize the results of refs. \cite{EGG} and
\cite{CEGK} for the D3-D7 and D3-D3 systems to any Dp-D(p+4) and Dp-Dp
intersection respectively. Furthermore, we will study the dual theory for
the D3-D5 intersection with flux and show that it corresponds to the Higgs
branch of the defect theory studied in \cite{WFO} and dealt with in
section \ref{cp2D3D5}. We will also extend our analysis to the generic
Dp-D(p+2) intersection. 

As we have seen in the last chapter, each type of intersection is dual to
a defect hosting a field theory living inside a bulk gauge theory.
Therefore, we can  label each case by the codimensionality of the defect.
We will see that, generically, all of them behave in a similar way, in the
sense that the Higgs phase is achieved by adding extra worldvolume flux
to the flavor brane. For all the cases we will study the meson spectra
and, except for the codimension zero case, it will become continuous and
gapless. Let us recall that we will be working in the limit where the
brane introducing the defect is considered as a probe in the background
generated by the other kind of branes.

We begin in section \ref{cp3codimzr} by analyzing the codimension zero
defect, which corresponds to the Dp-D(p+4) intersection. We first study
the field theory of the D3-D7 system, where we identify a mixed
Coulomb-Higgs branch which is obtained by imposing the vanishing of both
the $F$- and $D$-terms. This branch is characterized by a non-zero
commutator of the adjoint fields of ${\cal N}=4$ SYM which, from the
point of view of the flavor brane, corresponds to having a non-vanishing
flux of the worldvolume gauge field along the directions orthogonal to
the color brane. We will then describe such non-commutative scalars by
using the Myers action for a dielectric D3-brane and we will argue that,
macroscopically, this configuration can be described in terms of a
D7-brane with a self-dual instantonic gauge field. From this matching
between the D3- and D7-brane descriptions we will be able to extract the
relation proposed in ref. \cite{EGG} between the VEV of the quark field
and the size of the instanton. Afterwards we perform the computation of
the meson spectrum of the general Dp-D(p+4) systems, which in this case
remains discrete. We estimate the value of the mesonic mass gap as a
function of the instanton size. For large instantons this gap is
independent of the size, in agreement with the spectral flow found in
ref. \cite{EGG}, while for small instantons the mass gap is proportional
to the size of the instanton and vanishes in the zero-size limit. 

In section \ref{cp3codim1sc} we discuss  the codimension one  defects,
whose most prominent  example is the D3-D5 intersection. We begin by
studying the D3-D5 intersection with flux. The worldvolume gauge field
has the  non-trivial effect of inducing D3-brane charge in the D5-brane
worldvolume, which motivates the subsequent description of the system in 
terms of D3-branes expanded to a D5-brane due to dielectric effect
\cite{M}. Next, we will analyze the field theory
dual and show that the vacuum conditions of the dielectric theory can be
mapped to the $F$- and $D$-flatness constraints of the dual gauge
theory, justifying the identification with the Higgs phase. The Higgs
vacua of the field theory involve a non-trivial dependence of the
defect fields on the coordinate transverse to the defect. In the
supergravity side this is mapped to a bending of the flavor brane, which
is actually required by supersymmetry (see \cite{ST}). Subsequently,
after an straightforward generalization of the supergravity analysis to any
Dp-D(p+2) intersection, we study the fluctuations and show that the
spectrum of mesonic bound states is continuous and gapless. The reason is
that the IR theory is modified because of the non-trivial profile of the
flavor brane, so that in the Higgs phase, instead of having an effective
$AdS\times S$ worldvolume for the flavor brane,  one has Minkowski space,
thus losing the KK-scale which would give rise to a discrete spectrum.

In section \ref{cp3m2m5sc} we study a close relative to the Dp-D(p+2)
intersection,  namely the M2-M5 intersection in M-theory. In this case,
we see that we can dissolve M2-branes by turning on  a three-form flux 
on the M5-brane worldvolume and introducing some bending of the
M5-brane. The supersymmetry of this configuration is explicitly
confirmed in appendix \ref{sam2m5kappa} by looking at the kappa symmetry
of the embedding.  For this setup a microscopic description is not at
hand, since it would involve an action for coincident M2-branes which is
not known at present. 

Section \ref{cp3codim2sc} is devoted to the analysis of  the codimension
two defects, which correspond to the Dp-Dp intersections. This case, as
anticipated in 
\cite{CEGK,Erdmenger:2003kn}, is somehow different, since the Higgs
phase is realized by the choice of a particular embedding of the probe
Dp-brane with no need of extra flux. This case is rather particular
since, as we will show, the profile can be an arbitrary holomorphic curve
in suitable coordinates, although only one of them gives the desired
Higgs phase, while the rest remain unidentified. 

We then finish in section \ref{cp3scdiscuss} where we discuss our results
and comment on possible extensions of this work.

%
%
%
%
%
%
%
%
%

\setcounter{equation}{0}
\section{The codimension zero defect}
\label{cp3codimzr}

Let us start considering the D3-D7 intersection, where the D3-branes
are fully contained in the D7-branes as shown in the following array:
\beq
\begin{array}{ccccccccccl}
 &1&2&3& 4& 5&6 &7&8&9 & \nonumber \\
D3: & \times &\times &\times &\_ &\_ & \_&\_ &\_ &\_ &     \nonumber \\
D7: &\times&\times&\times&\times&\times&\times&\times&\_&\_ &
\end{array}
\label{cp3D3D7array}
\eeq

Clearly, the D3-D7 string
sector gives rise to extra fundamental matter living in the $3+1$ common
directions. It can be seen that the dual gauge theory is an
${\cal N}=2$ SYM theory in $3+1$ dimensions obtained by adding
$N_f$ ${\mathcal{N}}=2$ fundamental hypermultiplets to the ${\mathcal{N}}=4$
SYM theory. We can further break the classical conformal invariance of
the theory by adding a mass term for the quark hypermultiplets. The
lagrangian is given by:
\bear
\label{LFT}
&&{\cal L}= {1\over g_{YM}^2} \int d^2 \theta d^2 \bar\theta\,\left[\tr
\,\big(\Phi_I^{\dagger}\,e^{2V} \Phi_I\big) + Q_i^\dagger\,e^V Q^i +
\tilde Q_i\,e^{-V} \tilde Q^{i\dagger}\right]+\rc\rc
&&\;\;\quad +\,{1\over8\pi}\,{\rm Im} \left[\tau \int d^2 \theta\, \tr
\left({\cal W}^\alpha\,{\cal W}_\alpha\right)\right] +{1\over
g_{YM}^2}\left[ \int d^2
\theta\,W +\int d^2 \bar\theta\,\bar W\right]
\,,
\eear
where $\tau=4\pi i/ g_{YM}^2+\theta_{YM}/2\pi$ and the superpotential of
the theory is given by: 
\begin{equation}
W=\tilde{Q}_i(m+\Phi_3)Q^i+{1\over 3}\,\epsilon^{IJK}\,{\rm Tr}\,\left[
\Phi_I\Phi_J\Phi_K\right]\,.
\end{equation}
In eq. (\ref{LFT}) we are working in ${\cal N}=1$ language, where 
$Q_i$, ($\tilde{Q}_i$) $i=1,\cdots, N_f$ are the chiral (antichiral)
superfields in the fundamental hypermultiplets, while $\Phi_I$ $(I=1,2,3)$
are the three adjoint hypermultiplets of ${\mathcal{N}}=4$ SYM.
We will denote the (scalar) bottom components of the superfields by
lowercase letters. The complex scalars $\phi_I$ correspond to the
transverse scalars of the D3-brane: $\phi_1=X^1+iX^{2}\,$,
$\phi_2=X^3+iX^{4}\,$ and
$\phi_3=X^5+iX^{6}\,$ where  $X^I$ ($I=1,\cdots,6$) is the  scalar which
corresponds to the direction $I+3$ in the array (\ref{cp3D3D7array}).
It is worth  mentioning that an identity matrix in color space
is to be understood to multiply the mass parameter of the quarks  $m$.

We are interested in getting the classical SUSY vacua of this theory, which
can be obtained by imposing the corresponding $D$- and $F$-flatness
conditions that follow from the lagrangian (\ref{LFT}). Let us start
by imposing the vanishing of the
$F$-terms corresponding to the quark hypermultiplets, which amounts to set:

\begin{equation}
{\mathbf \tilde q}_i(\phi_3+m)=0\,\,\,,\qquad\quad
(\phi_3+m){\mathbf q}^i=0\,\, .
\label{Phi3}
\end{equation}
These equations can be satisfied by taking $\phi_3$ as:

\beq
\phi_3=\left( \begin{array}{cccccc}
\tilde{m}_1& & & & & \\
 &\ddots& & & \\
  & & \tilde{m}_{N-k}& & &\\
   & & &-m& &\\
   & & & &\ddots& \\
   & & & & & -m
   \end{array}
   \right)\,\, ,
\label{Phi3sol}
 \eeq
where the number of $m\,$s is $k$ and, thus,  in order to 
have $\phi_3$ in the Lie algebra of $SU(N)$, 
one must have $\Sigma_{j=1}^{N-k}\tilde{m}_j=km$.  This choice of $\phi_3$
lead us to take ${\mathbf q}^i$ and ${\mathbf \tilde q}_i$ as:
\beq
\begin{tabular}{c c}
${\mathbf \tilde q}_i=\left( 0\cdots
0,\tilde{q}^1_i\cdots,\tilde{q}^{k}_{i}
\right)\,,$ & $\qquad {\mathbf q}^i=\left(\begin{array}{c} 
0\\ \vdots\\0\\ q^i_1\\ \vdots\\ q^i_{k} 
\end{array}\right)\,.$
\end{tabular}
\label{QQtilde-sol}
\eeq
Indeed, it is trivial to check that the values of $\phi_3$, ${\mathbf
\tilde q}_i$ and ${\mathbf q}^i$ displayed in eqs. (\ref{Phi3sol}) and
(\ref{QQtilde-sol}) solve eq. (\ref{Phi3}).  Since the quark VEV in this
solution has some components which are zero and others that are different
from zero, this choice of vacuum leads to a mixed Coulomb-Higgs phase.

The vanishing of the $F$-terms associated to the adjoint 
scalars gives rise to:
   
\begin{equation}
  [\phi_1,\phi_3]=[\phi_2,\phi_3]=0\ ,
\end{equation}
 together with the equation:  
\begin{equation}
\label{F3}
{\mathbf q}^i\,{\mathbf \tilde q}_i+[\phi_1,\phi_2]=0\ .
\end{equation}
In (\ref{F3}) ${\mathbf q}^i\,{\mathbf \tilde q}_i$ denotes a matrix in
color space of components ${\mathbf q}^i_a {\mathbf \tilde q}_i^b$. For a
vacuum election as in eq. (\ref{QQtilde-sol}) we can restrict ourselves
to the lower
$k\times k$ matrix block, and we can write eq. (\ref{F3}) as:
\begin{equation}
\label{F}
q^i\tilde{q}_i+[\phi_1,\phi_2]=0\ ,
\end{equation}
where now, and it what follows,  it is understood that $\phi_1$ and
$\phi_2$ are $k\times k$ matrices. 

Eq. (\ref{F}) contains an important piece of
information since it shows that a non-vanishing VEV of the quark fields $q$
and $\tilde q$ induces a non-zero commutator of the adjoint fields
$\phi_1$ and $\phi_2$. Therefore, in the Higgs branch, some scalars
transverse to the D3-brane are necessarily non-commutative. Notice that
$\phi_1$ and $\phi_2$ correspond precisely to the directions transverse
to the D3-brane which lie on the worldvolume of the D7-brane  (\ie\ they
correspond to the directions $4,\cdots, 7$ in the array
(\ref{cp3D3D7array})). This implies that the description of this
intersection from the point of view of the D7-branes must involve a
non-trivial configuration of the worldvolume gauge field components of
the latter along the directions $4,\cdots,7$. We will argue in the next
subsection that this configuration corresponds to switching on an
instantonic flux along these directions.

In order to match the field theory vacuum with our brane description we
should also be able to reproduce the $D$-flatness condition arising from the
lagrangian (\ref{LFT}). Assuming that the quark fields ${\mathbf q}$ and
${\mathbf \tilde q}$ are only non-vanishing on the lower $k\times k$
block, we can write this condition as:
\begin{equation}
|q^i|^2-|\tilde{q}_i|^2+[\phi_1,\phi_1^{\dagger}]+[\phi_2, 
\phi_2^{\dagger}]=0\,.
\label{cp3dteq}
\end{equation}
The constraints (\ref{F}) and (\ref{cp3dteq}), together with the condition
$[\phi^I,\phi^3]=0$,  define the mixed Coulomb-Higgs phase of the theory.

\subsection{Gravity dual of the mixed Coulomb-Higgs phase}
\label{MacroD3D7}

As it is well-known, there is a one-to-one correspondence between 
the Higgs phase of ${\cal{N}}=2$ gauge theories and the moduli space of
instantons (\cite{MRD1, MRD2, W}). This comes from the fact that the $F$-
and $D$-flatness conditions can be directly mapped into the ADHM
equations (see \cite{Tong} for a review). Because of this map, we can
identify the Higgs phase of the gauge theory with the space of 4d
instantons. In the context of string theory, an ${\cal N}=2$ theory can
be engineered by intersecting Dp- with D(p+4)-branes over a
$(p+1)$-dimensional space. In particular, if we consider the D3-D7
system, the low energy effective lagrangian is precisely given by
(\ref{LFT}). In this context, the higgsing of the theory amounts to adding
some units of instantonic DBI flux in the subspace transverse to the D3
but contained in the D7, which provides a natural interpretation of the
Higgs phase-ADHM equations map.

Let us analyze this in more detail. Suppose we have $N$ D3-branes
 and $N_f$ D7-branes. In the field theory limit in which we take
$\alpha'$ to zero but keeping fixed the Yang-Mills coupling of the theory
on the D3s, the gauge dynamics on the D7-brane is decoupled. Then, the
$SU(N_f)$ gauge symmetry of the D7-brane is promoted to a global $SU(N_f)$
flavor symmetry on the effective theory describing the system, which is
${\cal N}=4$ SYM plus $N_f$ ${\mathcal{N}}=2$ hypermultiplets arising
from the D3-D7 strings; and whose lagrangian is the one written in
(\ref{LFT}). The gravity dual of this theory would be obtained by replacing
the branes by their backreacted geometry and taking the appropriate low
energy limit. However, in the limit in which $N_f\ll N$ we can consider
the D7-branes as probes in the  near-horizon geometry created by the
D3-branes, namely $AdS_5\times S^5$. The form of the metric and the
four-form RR potential\footnote{
We are writing only the electric part of $C_4$ since it will be the only
one contributing in the following.}
can be obtained, respectively, from
eqs. (\ref{cp1dpsolmtrc}) and (\ref{cp1dpsolpot}) of the Introduction just
by setting $p=3$:
\bear
&&ds^2={r^2\over R^2}\,dx_{1,3}^2 +{R^2\over r^2}\,d\vec r\cdot d\vec
r\,,\rc\rc
&&C_{(4)}=\left({r^2\over R^2}\right)^2\,dx^0\wedge\cdots\wedge dx^3\,.
\label{cp3d3sol}
\eear
This metric can be written in a form more suitable four our purposes.
Let us split the six-dimensional vector $\vec r$ parametrizing the space
transverse to the worldvolume of the D3-branes as: $\vec r=(\vec y,
\vec z)$, where $\vec y$ has four components corresponding to the
directions $4,\cdots,7$ in the array (\ref{cp3D3D7array}). Consequently,
$\vec z=(z^1,z^2)$ stands for the coordinates $8,9$ in
(\ref{cp3D3D7array}) parametrizing the space transverse to both the
D3- and D7-branes. After defining $\rho^2=\vec y\cdot\vec y$, so that
$r^2=\rho^2+\vec{z}^{\,2}$, the metric in eq. (\ref{cp3d3sol}) can be
rewritten as:
\beq
\label{bckgr}
ds^2=\frac{\rho^2+\vec{z}^{\,2}}{R^2}\,dx^2_{1,3}
+\frac{R^2}{\rho^2+\vec{z}^{\,2}}\left(d\vec{y}^{\,2}+d\vec{z}^{\,2}
\right)\,.
\eeq

The non-abelian DBI action (\ref{cp1ncDBI}) for a stack
of $N_f$ D7-branes with commutative transverse scalars (hence $Q=
{{\rm 1}\kern-4.5pt {\rm 1}}$ in (\ref{cp1ncDBI})) is given by\footnote{
Notice that, with our notations, $F_{ab}$ is dimensionless and,
therefore, the relation between $F_{ab}$ and the gauge potential $A$ is
$F_{ab}=\partial_a A_b-\partial_b A_a+{1\over 2\pi\alpha'}\, [A_a,A_b]$,
whereas the gauge covariant derivative is $D_a=\partial_a+{1\over
2\pi\alpha'}\, A_a$. Eq. (\ref{DBI-D7-general}) follows from
(\ref{cp1ncDBI}) after doing $2\pi\,\alpha' F\to F$.} :
\beq
S_{DBI}^{D7}=-T_7\,\int\,d^8\xi\;e^{-\phi}\;
{\rm Str}\left[\sqrt{-\det\, (\,g+F\,)}\,\right]\,.
\label{DBI-D7-general}
\eeq
Let us assume that
we take $\xi^a\,=\,(x^{\mu}\,,\,y^i\,)$ as worldvolume coordinates and
that we consider a D7-brane embedding in which $|\vec{z}|=L$, where $L$
represents the constant transverse separation between the two stacks of
D3- and D7- branes. Notice that this transverse separation will give a
mass 
$L/2\pi\alpha'$ to the D3-D7 strings, which corresponds to the quark mass
in the field theory dual.  For
an embedding with $|\,\vec z\,|=L$, the induced metric takes the form:
\beq
g_{x^{\mu} x^{\nu}}={\rho^2+L^2\over R^2}\,\eta_{\mu\nu}\;,
\qquad\quad
g_{y^{i} y^{j}}={R^2\over \rho^2+L^2}\,\,\delta_{ij}\,.
\label{inducedgD3-D7}
\eeq
Let us now assume that the worldvolume field strength $F$ has non-zero
entries only along the directions of the $y^i$ coordinates and let us denote
$F_{y^iy^j}$ simply by $F_{ij}$. 
Then, after using
eq. (\ref{inducedgD3-D7}) and the fact that the dilaton is trivial for the
$AdS_5\times S^5$ background, the DBI action (\ref{DBI-D7-general}) takes the
form: 
\begin{equation}
S_{DBI}^{D7}\,=-T_{7}\int d^4x\,d^4y\;{\rm Str}\,\Bigg\{
\sqrt{\,\det\Bigg(\delta_{ij}+\Bigg(\frac{\rho^2+L^2}{R^2}\Bigg)
F_{ij}\Bigg)}\,\Bigg\}\ .
\label{DBI-D3D7-reduced}
\end{equation}
The matrix appearing on the right-hand side of eq. (\ref{DBI-D3D7-reduced}) is
a $4\times 4$ matrix whose entries are $SU(N_f)$ matrices. However, inside
the symmetrized trace such matrices can be considered as commutative numbers.
Actually, we will evaluate the determinant in (\ref{DBI-D3D7-reduced}) by
means of the following identity. Let $M_{ij}=-M_{ji}$ be a $4\times 4$
antisymmetric matrix. Then, one can check that:
\beq
\det (1+M)=1+{1\over 2}\,M^2+{1\over 16}\,(\,{}^*M\,M\,)^2\,,
\label{matrix-identity}
\eeq
where $M^2$ and ${}^*M\,M$ are defined as follows:
\beq
M^2\equiv M_{ij}\,M_{ij}\;,\qquad\quad
{}^*M\,M\equiv {}^*M_{ij}\,M_{ij}\,,
\label{MM}
\eeq
and ${}^*M$ is defined as the following matrix: 
\beq
{}^*M_{ij}\,=\,{1\over 2}\,\epsilon_{ijkl}\,M_{kl}\,.
\label{*M}
\eeq
When the $M_{ij}$ matrix is self-dual  (\ie\ when ${}^*M=M$), the three terms
on the right-hand side of (\ref{matrix-identity}) build up a perfect square.
Indeed, one can check by inspection that, in this case, one has:
\beq
\det (1+M)\Big|_{{\rm self-dual}}=\left(1+{1\over 4}\,M^2\right)^2\,.
\label{detM(s-d)}
\eeq

Let us apply these results to our problem. First of all, 
by using 
(\ref{matrix-identity}) one can rewrite eq. (\ref{DBI-D3D7-reduced}) as:
\begin{equation}
S_{DBI}^{D7}=-T_{7}\int d^4x\,d^4y\;{\rm Str}\,\Bigg\{
\sqrt{1+{1\over 2}\,\Bigg(\frac{\rho^2+L^2}{R^2}\Bigg)^2 F^2
+\frac{1}{16}\Bigg(\frac{\rho^2+L^2}{R^2}\Bigg)^{4}
\Big({}^*FF\Big)^2}\,\Bigg\}\,.
\label{DBI-D3D7-explicit}
\end{equation}

Let us now consider the Wess-Zumino(WZ) piece of the worldvolume action
given in eq. (\ref{cp1ncWZ}). For a stack of $N_f$ D7-branes with
commutative transverse scalars in the
$AdS_5\times S^5$ background this action reduces to:
\beq
S_{WZ}^{D7}={T_{7}\over 2}\int
{\rm Str}\,\left[P[\,C^{(4)}\,]\wedge F\wedge F\right]\,.
\label{WZactionD7}
\eeq
By using the same set of coordinates as in
(\ref{DBI-D3D7-reduced}), and the explicit expression of $C^{(4)}$ (see eq.
(\ref{cp3d3sol})), one can rewrite $S_{WZ}^{D7}$ as:
\begin{equation}
S_{WZ}^{D7}=T_{7}\int d^4x\,d^4y\;{\rm Str}\, \Bigg\{\frac{1}{4}\Bigg(
\frac{\rho^2+L^2}{R^2}\Bigg)^{2}\,\,{}^*FF\Bigg\}\,.
\label{c3WZexp}
\end{equation}

Let us now consider a configuration in which the worldvolume gauge field is
self-dual in the internal $\RR^4$ of the worldvolume spanned by the $y^i$
coordinates which, as one can check, satisfies the equations of motion of
the D7-brane probe.  For such an instantonic gauge configuration
${}^*F=F$, where 
${}^*F$ is defined as in eq. (\ref{*M}). As in eq. (\ref{detM(s-d)}), 
when $F={}^*F$ the DBI action 
(\ref{DBI-D3D7-explicit}) contains the square root of a perfect square and
we can write:
\beq
S_{DBI}^{D7}({\rm self}-{\rm dual})=-T_7\,
\int d^4x\,d^4y\;{\rm Str}\, \Bigg\{1+\frac{1}{4}\,
\Bigg(\frac{\rho^2+L^2}{R^2}\Bigg)^{2}\,\,{}^*FF\,\Bigg\}\,.
\label{SDBI-sd}
\eeq
Moreover, by comparing eqs. (\ref{c3WZexp}) and (\ref{SDBI-sd}) one
readily realizes that the WZ action cancels against the second term of
the right-hand side of eq. (\ref{SDBI-sd}). To be more explicit, once we
assume the instantonic character of $F$, the full action for a self-dual
configuration is just:
\beq
S^{D7}({\rm self}-{\rm dual})=-T_7\int d^4x\,d^4y\;
{\rm Str}\, \big[1\big]=-T_7\,N_f \int d^4x\,d^4y\,.
\label{totalaction}
\eeq
Notice that in the total action (\ref{totalaction}) the transverse distance
$L$ does not appear.  This no-force condition is an explicit manifestation
of the SUSY of the system. Indeed, the fact that the DBI action is a square
root of  a perfect square is required for supersymmetry, and actually can be
regarded as the saturation of  a BPS bound.

In order to get a proper interpretation of the role of the instantonic gauge
field on the D7-brane probe, let us recall that for self-dual configurations
the integral of the Pontryagin density ${\cal P}(y)$ is quantized for
topological reasons. Actually, with our present normalization of $F$, 
${\cal P}(y)$ is given by:
\beq
{\cal P}(y)\,\equiv\,\frac{1}{16\pi^2}\,{1\over (2\pi\alpha')^2}\,\,
{\rm tr}\,\left[{}^*FF\right]\,,
\label{Pontryagin}
\eeq
and, if $k\in{\mathbb Z}$ is the instanton number, one has:
\begin{equation}
\int d^4y\,\,{\cal P}(y)\,
\,= \,k\ .
\label{instanton-number}
\end{equation}
A worldvolume gauge field satisfying
(\ref{instanton-number}) is inducing $k$ units  of D3-brane charge into the
D7-brane worldvolume along the subspace spanned by the Minkowski coordinates
$x^{\mu}$. To verify this fact, let us rewrite the WZ action
(\ref{WZactionD7}) of the D7-brane as:
\begin{equation}
S_{WZ}^{D7}={T_{7}\over 4}\int d^4x\,d^4y\;C_{x^0x^1x^2x^3}^{(4)}\;
{\rm tr}\,\left[{}^*FF\right]=T_3\int d^4x\,d^4y\;
C_{x^0x^1x^2x^3}^{(4)}\,{\cal P}(y)\,,
\label{D3induced}
\end{equation}
where we have used (\ref{Pontryagin}) and the relation
$T_3=(2\pi)^4\,(\alpha')^2\,T_7$ between the tensions of the D3- and
D7-branes (see eq. (\ref{cp1tension})). If $C_{x^0x^1x^2x^3}^{(4)}$ does
not depend on the coordinate
$y$, we can integrate over $y$ by using eq. (\ref{instanton-number}),
namely:
\beq
S_{WZ}^{D7}=k\,T_3\int d^4x\;C_{x^0x^1x^2x^3}^{(4)}\,.
\label{D3charge}
\eeq
Eq. (\ref{D3charge}) shows that the coupling of the D7-brane with
$k$ instantons in the worldvolume to the RR potential $C^{(4)}$ of the
background is identical to the one corresponding to $k$ D3-branes, as claimed
above.  It is worth to remark here that the existence of these instanton
configurations relies on the fact that we are considering $N_f>1$ flavor
D7-branes, \ie\ that we have a non-abelian worldvolume gauge theory.

\subsection{A microscopic interpretation of the D3-D7 intersection with
flux}
\label{cp3microD3D7}

The fact that the D7-branes carry $k$ dissolved D3-branes on them opens up
 the possibility of a new perspective on the system, which could be regarded
not just from the point of view of the D7-branes with dissolved D3s, but
also from the point of view of the dissolved D3-branes which expand due to
dielectric effect \cite{M} to a transverse fuzzy $\RR^4$. To see this,
let us assume that we have a stack of $k$ D3-branes
in the background given by (\ref{bckgr}). These D3-branes are extended along
the four Minkowski coordinates $x^{\mu}$, whereas the transverse coordinates  
$\vec{y}$ and $\vec{z}$ must be regarded as the matrix scalar fields $Y^i$ and
$Z^j$,  taking values in the adjoint representation of  $SU(k)$. Actually, we
will assume in what follows that the $Z^j$ scalars are abelian, as it
corresponds to a configuration in which the D3-branes are localized (\ie\
not polarized) in the  space transverse to  the D7-brane.

The dynamics of a stack of coincident D3-branes is determined by the
non-abelian action written in eq. (\ref{cp1ncDbraneact}), which is the sum
of a Dirac-Born-Infeld and a Wess-Zumino part and is also known as the
dielectric action. For a D3-brane the non-abelian DBI term (\ref{cp1ncDBI})
takes the form:
\beq
S_{DBI}^{D3}=-T_3\,\int d^4\xi\; {\bf Str}\,\left\{
\sqrt{-\det\left[ P[G+G(Q^{-1}-\delta)G]_{ab}\,\right]}\;
\sqrt{\det Q}\,\right\}\,.
\label{cp3dielectricBI}
\eeq
In eq. 
(\ref{cp3dielectricBI})  $G$ is the background metric, 
${\bf Str}(\cdots)$
represents the symmetrized trace over the $SU(k)$ indices 
and $Q$ is a matrix which depends on
the commutator of the transverse scalars (see below)\footnote{This matrix
$Q$ should not be confused with the fundamental superfields $Q_i$ and
$\tilde Q_i$.}. The Wess-Zumino term (\ref{cp1ncWZ}) for a
D3-brane in the
$AdS_5\times S^5$ background under consideration reduces to:
\beq
S_{WZ}^{D3}=T_{3}\,\int d^4\xi \;{\bf Str}\,
\left[P\left[C^{(4)}\right]\right]\,.
\label{cp3dielectricWZ}
\eeq
As we are assuming that only the $Y$ scalars are non-commutative, 
the  only elements of the matrix $Q$ appearing in (\ref{cp3dielectricBI})
that differ from those of the  unit matrix are
given by:
\beq
Q_{y^iy^j}=\delta_{ij}+{i\over 2\pi\alpha'}\,[Y^i,Y^k]\,G_{y^ky^j}\,.
\eeq
By using the explicit form of the metric elements along the $y$ coordinates 
(see eq. (\ref{bckgr})), one can rewrite $Q_{y^iy^j}$ as:
\beq
Q_{y^iy^j}=\delta_{ij}+{i\over 2\pi\alpha'}\,
{R^2\over \hat r^{\,2}}\,[Y^i, Y^j]\,,
\eeq
where $\hat r^{\,2}$ is the matrix:
\beq
\hat r^{\,2}=\left(Y^i\right)^2+Z^2\,.
\eeq
Let us now define the matrix $\theta_{ij}$ as: 
\beq
i\theta_{ij}\equiv {1\over 2\pi\alpha'}\,[Y^i,Y^j]\,.
\label{thetaij}
\eeq
It follows from this definition that $\theta_{ij}$ is antisymmetric in the
$i,j$ indices and, as  a matrix of $SU(k)$, is hermitian:
\beq
\theta_{ij}=-\theta_{ji}\;,\qquad\quad
\theta_{ij}^{\dagger}=\theta_{ij}\,.
\eeq
Moreover, in terms of $\theta_{ij}$, the matrix $Q_{y^iy^j}$ can be
written as:
\beq
Q_{y^iy^j}=\delta_{ij}-{R^2\over \hat r^{\,2}}\,\theta_{ij}\,.
\eeq
Using these definitions, we can write the DBI action
(\ref{cp3dielectricBI}) for the dielectric 
D3-brane in the $AdS_5\times S^5$ background as:
\beq
S_{DBI}^{D3}= -T_3\int d^4x\;{\bf Str}
\left[\left({\hat r^{\,2}\over R^2}\right)^2\,
\sqrt{\det\left(\delta_{ij}-{R^2\over \hat
r^{\,2}}\,\theta_{ij}\right)}\;\right]\,,
\label{DBI-dielectricD3}
\eeq
where we have chosen the Minkowski coordinates $x^{\mu}$ as our set of
worldvolume coordinates for the dielectric D3-brane.  Similarly, the WZ term
can be written as:
\beq
S_{WZ}^{D3}=T_{3}\int d^4x \;{\bf Str}\,
\left[\left({\hat r^{\,2}\over R^2}\right)^2\right]\,.
\label{SWZ-D3}
\eeq
Let us now assume that the matrices $\theta_{ij}$ are self-dual with respect
to the $ij$ indices, \ie\ that ${}^*\theta\,=\,\theta$. Notice that, in terms
of the original matrices $Y^i$, this is equivalent to the condition:
\begin{equation}
\label{selfduality}
[Y^i,Y^j]=\frac{1}{2}\epsilon_{ijkl}[Y^k,Y^l]\ . 
\end{equation}
Moreover, the self-duality condition implies that there are three independent
$\theta_{ij}$ matrices, namely:
\beq
\theta_{12}=\theta_{34}\;,\qquad\quad
\theta_{13}=\theta_{42}\;,\qquad\quad
\theta_{14}=\theta_{23}\,.
\label{sd-explicit}
\eeq

The description of the D3-D7 system from the perspective of the color
D3-branes should match the field theory analysis performed at the beginning
of this section. In particular, the $D$- and $F$-flatness conditions of
the adjoint fields in the Coulomb-Higgs phase of the ${\cal N}=2$ SYM
with flavor should be the same as the ones satisfied by the transverse
scalars of the dielectric D3-brane. In order to check this fact,  let
us define the following complex combinations of the $Y^i$ matrices:
\beq
2\pi\alpha'\,\phi_1\equiv{Y^1+iY^2\over \sqrt{2}}\;,
\qquad\quad
2\pi\alpha'\,\phi_2\equiv{Y^3+iY^4\over \sqrt{2}}\,,
\label{Phi12}
\eeq
where we have introduced the factor $2\pi\alpha'$ to take into account the
standard relation between coordinates and scalar fields in string theory. 
We are going to identify $\phi_1$ and $\phi_2$ with the adjoint scalars
of the field theory side.  To verify this identification, let us compute the
commutators of these matrices and match them with the ones obtained from
the $F$-flatness conditions of the field theory analysis.  From the
definitions (\ref{thetaij}) and  (\ref{Phi12}) and the self-duality
condition (\ref{sd-explicit}),  it is straightforward to check that:
\bear
&&[\phi_1,\phi_2]=-{\theta_{23}\over 2\pi\alpha'}
+i{\theta_{13}\over 2\pi\alpha'}\,,\rc\rc
&&[\phi_1,\phi_1^{\dagger}]=[\phi_2,\phi_2^{\dagger}]=
{\theta_{12}\over 2\pi\alpha'}\,.
\eear
By comparing with the results of the field theory analysis
(eqs. (\ref{F}) and (\ref{cp3dteq})), we get the
following identifications between the $\theta\,$s and the vacuum
expectation values of the matter fields:
\beq
q^i\tilde{q}_i={\theta_{23}\over 2\pi\alpha'}
-i{\theta_{13}\over 2\pi\alpha'}\;,\qquad\quad
|\tilde{q}_i|^2-|q^i|^2={\theta_{12}\over \pi\alpha'}\,.
\label{q-theta}
\eeq
Moreover, from the point of view of this dielectric description, the
$\phi_3$ field in  the field theory is proportional to  $Z^1+iZ^2$. Since
the stack of branes is localized in that directions, $Z^1$ and $Z^2$ are
abelian and clearly we have that $[\phi_1,\phi_3]=[\phi_2,\phi_3]=0$,
thus matching the last $F$-flatness condition for the adjoint field
$\phi_3$.

It is also interesting to relate the present microscopic description of
the D3-D7 intersection, in terms of a stack of dielectric D3-branes, to the
macroscopic description of subsection \ref{MacroD3D7}, in terms of the
flavor D7-branes. With this purpose in mind, let us compare the actions of
the D3- and D7-branes. First of all, we notice that, when the matrix $\theta$
is self-dual, we can use eq. (\ref{detM(s-d)}) and write the DBI action
(\ref{DBI-dielectricD3}) as:
\beq
S_{DBI}^{D3}({\rm self}-{\rm dual})=-T_3\int d^4x\;{\bf Str}\,
\left[\left({\hat r^{\,2}\over R^2}\right)^2+{1\over 4}\,\theta^2\right]\,.
\label{SD3-sd}
\eeq
Moreover, by inspecting eqs. (\ref{SWZ-D3}) and (\ref{SD3-sd}) we discover
that the WZ action cancels against the first term of the right-hand side of
(\ref{SD3-sd}), in complete analogy to what happens to the D7-brane. Thus, one
has:
\begin{equation}
S^{D3}({\rm self}-{\rm dual})=-\,{T_3\over 4}\int d^{4}x\; {\bf Str}
\left[\theta^2\,\right]=-\pi^2\,T_7\,(\,2\pi\alpha'\,)^2
\int d^{4}x\;{\bf Str}\left[\theta^2\right]\,,
\label{completeactionD3}
\end{equation}
where, in the last step, we have rewritten the result in terms of the tension
of the D7-brane. Moreover, an important piece of information is obtained by
comparing the WZ terms of the D7- and D3-branes (eqs. (\ref{D3induced}) and
(\ref{SWZ-D3})). Actually, from this comparison we can establish a map
between matrices in the D3-brane description and functions of the $y$
coordinates in the D7-brane approach. Indeed, let us suppose that $\hat f$
is a $k\times k$ matrix and let us call $f(y)$ the function to which  $\hat
f$ is mapped. It follows from the identification between the D3- and
D7-brane WZ actions that the mapping rule is:
\begin{equation}
{\bf Str} [\hat f]\;\Rightarrow\; \int d^4y\;{\cal P}(y)\,f(y)\,,
\label{micro-macro}
\end{equation}
where the kernel ${\cal P}(y)$ on the right-hand side of (\ref{micro-macro})
is the Pontryagin density defined in eq. (\ref{Pontryagin}). Actually, the
comparison between both WZ actions tells us that the matrix $\hat r^2$ is
mapped to the function $\vec y^{\,2}+\vec z^{\,2}$. Notice also that, when
$\hat f$ is the unit
$k\times k$ matrix and $f(y)=1$, both sides of (\ref{micro-macro}) are 
equal to the instanton number $k$ (see eq. (\ref{instanton-number})). Another
interesting piece of information comes from comparing the complete actions
of the D3- and D7-branes. It is clear from (\ref{completeactionD3}) and
(\ref{totalaction}) that:
\beq
(\,2\pi\alpha'\,)^2\,{\bf Str} \left[\theta^2\right]\;
\Rightarrow\; \int d^4y\;{N_f\over \pi^2}\,.
\label{theta-map}
\eeq
By comparing eq. (\ref{theta-map}) with the general relation
(\ref{micro-macro}), one gets the function that corresponds
to the matrix $\theta^2$, namely:
\beq
(2\pi\alpha')^2\,\theta^2\;\Rightarrow\;{N_f\over\pi^2 \,{\cal P}(y)}\,.
\label{theta-instanton}
\eeq
Notice that $\theta^2$ is a measure of the non-commutativity of the adjoint
scalars in the dielectric approach, \ie\ is a quantity that characterizes the
fuzziness of the space transverse to the D3-branes. Eq.
(\ref{theta-instanton}) is telling us that this fuzziness is related to the 
(inverse of the) Pontryagin density for the macroscopic D7-branes. Actually,
this identification is reminiscent of the one found in ref. \cite{SW} between
the non-commutativity parameter and the NSNS $B$ field in the string
theory realization of non-commutative geometry. Interestingly, in our
case the commutator matrix $\theta$ is related to the VEV of the matter
fields $q$ and
$\tilde q$ through the $F$- and $D$-flatness conditions (\ref{F}) and
(\ref{cp3dteq}).  Notice that eq. (\ref{theta-instanton}) implies that
the quark VEV is somehow related to the instanton density on the flavor
brane. In order to make this correspondence more precise, let us consider
the one-instanton configuration of the $N_f=2$ gauge theory on the
D7-brane worldvolume. In the so-called singular gauge, the $SU(2)$ gauge
field is given by:
\begin{equation}
{A_i\over 2\pi\alpha'}=2i\Lambda^2
\frac{\bar{\sigma}_{ij}\,\, y^j}{\rho^2(\rho^2+\Lambda^2)}\,,
\label{instanton-potential}
\end{equation}
where $\rho^2=\vec y\cdot\vec y$,
$\Lambda$ is a constant (the instanton size) and
the matrices $\bar{\sigma}_{ij}$ are defined as:
\beq
\bar{\sigma}_{ij}={1\over 4}\,\,\left(\bar\sigma_i\,\sigma_j-
\bar\sigma_j\,\sigma_i\right)\;,\qquad\quad
\sigma_i=(i\vec \tau\,,\,1_{2\times 2}\,)\,,\qquad\quad
\bar\sigma_i=\sigma_i^{\dagger}=
(-i\vec \tau\,,\,1_{2\times 2})\,.
\label{sigmaij}
\eeq
In (\ref{sigmaij}) the $\vec \tau\,$s are the Pauli matrices. Notice that
we are using a convention in which the $SU(2)$ generators are hermitian
as a consequence of the relation  
$\bar{\sigma}_{ij}^{\dagger}\,=\,-\bar{\sigma}_{ij}$.  The non-abelian
field strength $F_{ij}$ for the gauge potential  $A_i$ in 
(\ref{instanton-potential}) can be easily computed, with the result:
\beq
{F_{ij}\over 2\pi\alpha'}=-{4i\Lambda^2\over (\rho^2+\Lambda^2)^2}\;
\bar{\sigma}_{ij}-{8i\Lambda^2\over \rho^2 (\rho^2+\Lambda^2)^2}
\left(y^i\,\bar{\sigma}_{jk}-y^j\,\bar{\sigma}_{ik}\right)\,y^k\,.
\eeq
Using the fact that the matrices $\bar{\sigma}_{ij}$ are anti self-dual one
readily verifies that $F_{ij}$ is self-dual. Moreover, 
one can prove that:
\beq
{F_{ij}\,F_{ij}\over (\,2\pi\alpha'\,)^2}={48\Lambda^4\over
(\rho^2+\Lambda^2)^4}\,,
\label{F2-inst}
\eeq
which gives rise to the following instanton density:
\beq
{\cal P}(y)={6\over \pi^2}\,\,{\Lambda^4\over
(\rho^2+\Lambda^2)^4}\,.
\eeq
As a check one can verify that eq. (\ref{instanton-number}) is satisfied with
$k=1$.

Let us now use this result in (\ref{theta-instanton}) to get some qualitative
understanding of the relation between the Higgs mechanism in field theory and
the instanton density in its holographic description. For simplicity we will
assume that all quark VEVs are proportional to some scale $v$, \ie\ that:
\beq
q, \tilde q\,\sim\,v\,.
\eeq
Then, it follows from (\ref{q-theta}) that:
\beq
\theta\sim\alpha'\,v^2\,,
\eeq
and, by plugging this result in (\ref{theta-instanton}) one arrives at the
interesting relation:
\begin{equation}
v\sim \frac{\rho^2+\Lambda^2}{\alpha'\Lambda}\;.
\label{holoVEV}
\end{equation}
Eq. (\ref{holoVEV}) should be understood in the holographic sense, \ie\
$\rho$ should be regarded as the energy scale of the gauge theory.
Actually, in the far IR ($\rho\approx 0$) the  relation   (\ref{holoVEV})
reduces to:
\begin{equation}
v\sim\frac{\Lambda}{\alpha'}\,,
\label{v-Lambda}
\end{equation}
which, up to numerical factors, is precisely the relation between the quark
VEV and the instanton size that has been  obtained in \cite{EGG}. Let us now 
consider  the full expression  (\ref{holoVEV})  for $v$. 
For any finite non-zero $\rho$ the quark VEV $v$ is non-zero. Indeed, in both
the  large and small instanton limits  $v$ goes to infinity.
However, in the far IR a subtlety arises, since there the quark VEV goes to
zero in the small instanton limit. This region should be clearly singular,
because a zero quark VEV would mean to unhiggs the theory, which would lead
to the appearance of extra light degrees of freedom.

To finish this subsection, let us notice that
the dielectric effect  considered here is not triggered by the 
influence of any external field other than the metric background. In this
sense it is an example of a purely gravitational dielectric effect, as in
\cite{gravitationaldielectriceffect}.

\subsection{Fluctuations in Dp-D(p+4) with flux}
\label{cp3sscDp4flucts}

So far we have seen how we can explicitly map the Higgs phase of the field 
theory to the instanton moduli space in the D7-brane picture through the
dielectric description. In this section we will concentrate on the
macroscopic description and we will consider fluctuations around the
instanton configuration. These fluctuations should correspond to mesons in
the dual field theory.

Since we have a similar situation for all the Dp-D(p+4) intersections,
namely a one to one correspondence between the Higgs phase of the
corresponding field theory and the moduli space of instantons in 4
dimensions, in this section we will work with the general  Dp-D(p+4) system.
Both the macroscopic and the microscopic analysis of the previous section
can be extended in a straightforward manner to the general case, so we will
briefly sketch  the macroscopic computation to set notations, and
concentrate on the fluctuations. The metric, dilaton, and RR
$(p+1)$-form potential of the supergravity solution corresponding to a
stack of $N$ Dp-branes were written in eqs.
(\ref{cp1dpsolmtrc}), (\ref{cp1dpsoldiltn}) and (\ref{cp1dpsolpot}) of
the Introduction. 

Next, we will separate again the $\vec{r}$ coordinates appearing in
(\ref{cp1dpsolmtrc}) in two sets, namely $\vec r=(\vec y,
\vec z)$, where $\vec y$ has four components, and we will denote
$\rho^2=\vec y\cdot\vec y$. As $r^2=\rho^2+\vec{z}^{\,2}$, the
metric in (\ref{cp1dpsolmtrc}) can be written as:
\begin{equation}
\label{bckgrDp}
ds^2=\left(\frac{\rho^2+\vec{z}^{\,2}}{R^2}\right)^{\alpha}dx_{1,p}^2
+\left(\frac{R^2}{\rho^2+\vec{z}^{\,2}}\right)^{\alpha}(d\vec{y}^{\,2}
+d\vec{z}^{\,2})\;,\qquad\alpha={7-p\over4}\,.
\label{cp3dpdp+4bckgr}
\end{equation}
In this background we will consider a stack of $N_f$ D(p+4)-branes
extended along $(x^{\mu}, \vec y)$ at fixed distance $L$ in the
transverse space spanned by the $\vec z$ coordinates (\ie\ with $|\,\vec
z\,|=L$). If
$\xi^a=(x^{\mu}, \vec y)$ are the worldvolume coordinates, the
action of a probe D(p+4)-brane is:
\bear
&&S^{D(p+4)}=-T_{p+4}\int d^{p+5}\xi\;e^{-\phi}\,
{\rm Str}\left[\sqrt{-\det\left(g+F\right)}\right]+\rc\rc
&&\qquad\qquad\qquad\qquad
+\,{T_{p+4}\over 2}\int {\rm Str}\left[P\left(C^{(p+1)}\right)\wedge
F\wedge F\right]\,,
\label{SD(p+4)}
\eear
where $g$ is the induced metric and $F$ is the $SU(N_f)$ worldvolume gauge
field strength. In order to write $g$  more compactly, let us define the
function $h$ as follows:
\beq
h(\rho)\equiv\left(\frac{R^2}{\rho^2+L^{2}}\right)^{\alpha}\,.
\label{h}
\eeq
Then, one can write the non-vanishing elements of the induced metric as:
\beq
g_{x^{\mu}\,x^{\nu}}={\eta_{\mu\nu}\over h}\;,
\qquad\quad g_{y^{i}\,y^{j}}=h\,\delta_{ij}\,.
\eeq
Let us now assume that the only non-vanishing components of the worldvolume
gauge field $F$ are those along the $y^i$ coordinates. Following the same
steps as in subsection \ref{MacroD3D7}, the action for the D(p+4)-brane probe
can be written as:
\bear
&&S^{D(p+4)}=-T_{p+4}\int d^4x\,d^4y\;{\rm Str}\Bigg[
\sqrt{\,1+{1\over 2}\,\left(\frac{\rho^2+L^2}{R^2}\right)^{2\alpha} F^2
+\frac{1}{16}\left(\frac{\rho^2+L^2}{R^2}\right)^{4\alpha}
\Big(\,{}^*FF\,\Big)^2}-\rc\rc
&&\qquad\qquad\qquad\qquad\qquad\qquad\qquad
-\,\frac{1}{4}\left(\frac{\rho^2+L^2}{R^2}\right)^{2\alpha}\,{}^*FF
\Bigg]\,,
\label{DBI-DpDp+4-explicit}
\eear
where $F^2$ and ${}^*FF$ are defined as in eqs. (\ref{MM}) and (\ref{*M}). If,
in addition, $F_{ij}$ is self-dual, one can check that the equations of
motion of the gauge field are satisfied and, actually, 
there is a cancellation between the DBI
and WZ parts of the action (\ref{DBI-DpDp+4-explicit}) 
generalizing  (\ref{totalaction}), namely:
\begin{equation}
S^{D(p+4)}({\rm self}-{\rm dual})=-T_{p+4}\int{\rm Str}[1]=
-N_{f}\,T_{p+4}\int d^{p+1}x\int d^4y\,.
\end{equation}

We turn now to the analysis of the fluctuations around the self-dual
configuration and the computation of the corresponding meson spectrum 
for this
fluxed Dp-D(p+4) intersection.  We will not compute the whole set of
excitations. Instead, we will focus on the fluctuations of the worldvolume
gauge field, for which we will  write: 
\beq
A=A^{inst}+a\,,
\eeq
where $A^{inst}$ is the gauge potential corresponding to a self-dual gauge
field strength $F^{inst}$ and 
$a$ is the fluctuation. The total field strength $F$ reads:
\begin{equation}
F_{ab}=F^{inst}_{ab}+f_{ab}\,,
\end{equation}
with $f_{ab}$ being given by:
\beq
f_{ab}=\partial_{a}a_{b}-\partial_{b}a_{a}+{1\over 2\pi\alpha'}\,
[A^{inst}_{a},a_{b}]+{1\over 2\pi\alpha'}\,
[a_{a},A^{inst}_{b}]+{1\over 2\pi\alpha'}\,[a_{a},a_{b}]\,,
\eeq
\noindent where the  indices $a$, $b$  run now over all the worldvolume
directions. Next, let us expand the action  (\ref{SD(p+4)}) in powers of the
field $a$ up to second order. With this purpose in mind,  we rewrite the
square root in the DBI action as:
\beq
\sqrt{-\det\left(g+F^{inst}+f\right)}=\sqrt{-\det\left(g+F^{inst}\right)}
\;\sqrt{\det\left(1+X\right)}\,,
\label{cp3DetXp4}
\eeq
where $X$ is the matrix:
\beq
X\equiv\left(g+F^{inst}\right)^{-1}\;f\,.
\label{defX}
\eeq
We will expand the right-hand side of (\ref{cp3DetXp4}) in powers of $X$
by using the equation\footnote{The trace used in eqs.
(\ref{detX-expansion}) and (\ref{TrX-TrX2}) should not be confused with
the trace over the $SU(N_f)$ indices.}:
\beq
\sqrt{\det\,(1+X)}=1+{1\over 2}\,{\rm Tr}\, X-{1\over 4}\,
{\rm Tr}\, X^2+ {1\over 8}\left({\rm Tr}\, X\right)^2+o(X^3)\,.
\label{detX-expansion}
\eeq
To apply this expansion to our problem we need to know previously the value
of $X$, which has been defined in eq. (\ref{defX}). Let us denote by ${\cal
G}$ and 
${\cal J}$ to the symmetric and antisymmetric part of the inverse 
of $g+F^{inst}$, \ie:
\beq
\left(g+F^{inst}\right)^{-1}={\cal G}+{\cal J}\,.
\eeq
One can easily compute the matrix elements of ${\cal G}$, with the
result:
\beq
{\cal G}^{\mu\nu}=h\,\eta^{\mu\nu}\;,\qquad\quad
{\cal G}^{ij}={h\over H}\,\delta_{ij}\,,
\eeq
where $h$ has been defined in (\ref{h}) and
the function $H$ is given by:
\beq
H\equiv h^2+{1\over 4}\left(F^{inst}\right)^2\,.
\label{H-def}
\eeq
Moreover, the non-vanishing elements of ${\cal J}$ are:
\beq
{\cal J}^{ij}=-{F^{inst}_{ij}\over H}\,.
\eeq
Using these results one can easily obtain the expression of $X$ and
the traces of its powers appearing on the right-hand side of
(\ref{detX-expansion}), which are given by:
\bear
&&{\rm Tr}\, X={1\over H}\,F^{inst}_{ij}\,f_{ij}\,,\rc\rc
&&{\rm Tr}\, X^2=-h^2\,f_{\mu\nu}\,f^{\mu\nu}-{2h^2\over H}\,
f_{i\mu}\,f^{i\mu}-{h^2\over H^2}\,f_{ij}\,f^{ij}+
{1\over H^2}\, F^{inst}_{ij}\,F^{inst}_{kl}\,f^{jk}\,f^{li}\,.
\label{TrX-TrX2}
\eear
By using these results we get, 
after a straightforward computation, the action up to quadratic order in the 
fluctuations, namely:
\begin{eqnarray}
S^{D(p+4)}&=&-T_{p+4}\int\, {\rm Str}\,
\Bigg\{ 1+\frac{H}{4}f_{\mu\nu}f^{\mu\nu}+\frac{1}{2}\,f_{i\mu}f^{i\mu}+
\frac{1}{4H}f_{ij}f^{ij}+\rc\rc
&+&\frac{1}{8h^2H}(F^{ij}f_{ij})^2-\frac{1}{4h^2H}\,F^{ij}F^{kl}f_{jk}f_{li}
-\frac{1}{8h^2}\,\,f_{ij}f_{kl}\epsilon^{ijkl}\Bigg\}\,,
\end{eqnarray}
where  we are dropping the superscript in the instanton field strength.

From now on we will assume again that $N_f=2$ and that the unperturbed
configuration is the one-instanton $SU(2)$ gauge field written in eq. 
(\ref{instanton-potential}). As in ref. \cite{EGG}, we will concentrate on the
subset of fluctuations
for which $a_i=0$, \ie\ on those for which the fluctuation field $a$ has
non-vanishing  components only along the Minkowski directions.  However, we
should impose this ansatz at the level of the equations of motion in order
to ensure the consistency of the truncation. Let us consider  first  the
equation of motion for $a_i$, which 
after imposing $a_i=0$ reduces to:
\beq
D_i\,\partial^{\mu}\,a_{\mu}=0\,.
\label{ai-eom}
\eeq
Moreover,  the equation for $a_{\mu}$ when $a_i=0$ becomes: 
\beq
H\,D^{\mu}\,f_{\mu\nu}+D^i\,f_{i\nu}=0\,,
\label{amu-eq}
\eeq
where now $H$ is given in (\ref{H-def}), with $\big(\,F^{inst}\,\big)^2$ as
in (\ref{F2-inst}).  Eq. (\ref{ai-eom}) is solved by requiring:
\beq
\partial^{\mu}\,a_{\mu}=0\,.
\label{transv}
\eeq
Using this result, eq. (\ref{amu-eq}) can be written as:
\beq
H\,\partial^{\mu}\partial_{\mu}\,a_{\nu}+\partial_i\partial_i\,a_{\nu}+
\partial^i\left[{A_i\over 2\pi\alpha'}\,,a_{\nu}\right]+
\left[{A_i\over 2\pi\alpha'}\,,\partial_ia_{\nu}\right]+
\left[{A_i\over 2\pi\alpha'}\,,\left[{A_i\over 2\pi\alpha'}\,,a_\nu
\right]\right]=0\,.
\label{fluc-eom-DpDp+4}
\eeq
Let us now adopt the following ansatz for $a_{\mu}$:
\beq
a_{\mu}^{(l)}=\xi_{\mu}(k)\,f(\rho)\,\,e^{ik_{\mu}x^{\mu}}\,\tau^l\,,
\label{amu-ansatz}
\eeq
where $\tau^l$ is a Pauli matrix. 
This ansatz solves eq. (\ref{transv}) provided the following transversality
condition is fulfilled:
\beq
k^{\mu}\,\xi_{\mu}=0\,.
\eeq
Moreover, one can  check that, for this ansatz, one has:
\bear
&& \partial^i\,\left[A_i\,,a_{\nu}^{(l)}\right]=
\left[A_i\,,\partial_ia_{\nu}^{(l)}\right]=0\,,\rc\rc
&&\left[{A_i\over 2\pi\alpha'}\,,\left[\,{A_i\over
2\pi\alpha'}\,,a_\nu^{(l)}\right]\right]=
-\frac{8\Lambda^4}{\rho^2(\rho^2+\Lambda^2)^2}\,
\xi_{\nu}(k)\,f(\rho)\,e^{ik_{\mu}x^{\mu}}\,\tau^l\,.
\eear
Let us now use these results in eq. (\ref{fluc-eom-DpDp+4}). Denoting 
$M^2=-k^2$ (which will be identified with the mass of the meson in the dual
field theory) and using eq. (\ref{F2-inst}) to compute the function $H$
(see eq.  (\ref{H-def})), one readily reduces (\ref{fluc-eom-DpDp+4}) to the
following second order differential equation for the function $f(\rho)$
of the ansatz (\ref{amu-ansatz}):
\begin{equation}
\label{e1}
\left[\frac{R^{4\alpha}M^2}{(\rho^2+L^2)^{2\alpha}}
\left(1+\frac{12(2\pi\alpha')^2\Lambda^4}
{R^{4\alpha}}\frac{(\rho^2+L^2)^{2\alpha}}{(\rho^2+\Lambda^2)^4}\right)
-\frac{8\Lambda^4}{\rho^2(y^2+\Lambda^2)^2}+\frac{1}{\rho^3}
\partial_\rho(\rho^3\partial_\rho)\right]f=0\ .
\end{equation}
In order to analyze eq. (\ref{e1}), let us introduce a new radial variable
$\varrho$ and a reduced mass $\bar M$, which are related to $\rho$ and $M$ as:
\beq
\rho=L\varrho\;,\qquad\quad \bar{M}^2=R^{7-p}L^{p-5}M^2\,.
\eeq
Moreover, it is interesting to rewrite the fluctuation equation in terms of
field theory quantities. Accordingly, let us introduce the 
quark mass $m_q$ and its VEV $v$ as follows:
\beq
m_q={L\over 2\pi\alpha'}\;,\qquad\quad
v={\Lambda\over 2\pi\alpha'}\,.
\eeq
Notice that the relation between $v$ and the instanton size $\Lambda$ is
consistent with our analysis of subsection \ref{cp3microD3D7} (see eq.
(\ref{v-Lambda})) and with the proposal of ref. \cite{EGG}. 
Let us also recall the definitions of the Yang-Mills
coupling $g_{YM}$ and the effective dimensionless coupling $g_{eff}(U)$
at the energy scale $U$ given, respectively, in eqs. (\ref{c1gymdp}) and
(\ref{c1geff}) of the Introduction. Then, the equation (\ref{e1}) can
be rewritten as:
\bear
\label{fluctuationsDp+4}
&&\Bigg[\frac{\bar{M}^2}{(1+\varrho^2)^{2\alpha}}\left(1+
c_{p}(v,m_q)\,\frac{(1+\varrho^2)^{2\alpha}}{(\varrho^2+
(\frac{v}{m_q})^2)^4}\right)-\left(\frac{v}{m_q}\right)^4\frac{8}
{\varrho^2(\varrho^2+(\frac{v}{m_q})^2)^2}+\rc\rc
&&+\frac{1}{\varrho^3}\,\partial_{\varrho}(\varrho^3\partial_{\varrho})
\Bigg]f=0\,,
\eear
where $c_{p}(v,m_q)$ is defined as:
\beq
c_{p}(v,m_q)\equiv\frac{12\cdot 2^{p-1}\pi^{\frac{p+1}{2}}}
{\Gamma(\frac{7-p}{2})}\,{v^4\over g_{eff}^2(m_q)\,m_q^4}\,.
\label{cp}
\eeq
Notice that everything conspires to absorb the powers of $\alpha'$ and give
rise to  the effective coupling at the 
quark mass in $c_{p}(v,m_q)$.

The equation (\ref{fluctuationsDp+4}) differs in the $\bar{M}$ term  from the
one obtained in
 \cite{EGG}, where  the term proportional to
$c_{p}(v,m_q)$ is absent. We would like to point out that in order to arrive
to (\ref{fluctuationsDp+4}) we expanded up to quadratic order in the
fluctuations and we have kept all orders in the instanton field. The extra
factor compared to that in (\cite{EGG}) comes from the fact that, for a
self-dual worldvolume gauge field, the unperturbed DBI action actually
contains the square root of a perfect square, which can be evaluated exactly
and shows up in the lagrangian of the fluctuations. 
This extra term is proportional to the inverse of
the effective Yang-Mills coupling. In order to trust the supergravity
approximation the effective Yang-Mills coupling should be large, which would
suggest that the effect of this term is indeed negligible. We will see
however that in the region of small $v/m_q$ the full term is actually
dominating in the IR region and determines the meson spectrum. 
In addition, in order to ensure the validity of
the DBI approximation, we should have slowly varying gauge fields, which
further imposes that $F\wedge F$ should be much smaller than $\alpha'$.

In order to study the fluctuation equation (\ref{fluctuationsDp+4}) one
can proceed along the lines of section \ref{generalfluctuations} and
convert the equation into a Schr\"odinger equation. By applying the change
of variables (\ref{Sch-variables}) with $p_2=7$, $d=3$ and calling $z$
to the new variable, denoted by $y$ in eq. (\ref{Sch-variables}), one
arrives at a Schr\"odinger equation of the form (\ref{Sch}) with a
potential given by:
\bear
&&V(z)=1+\left(\frac{v}{m_q}\right)^4\frac{8}
{\left(e^{2z}+\big(\frac{v}{m_q}\big)^2\right)^2}-\rc\rc
&&\qquad\qquad\qquad-\bar M^2\,{e^{2z}\over \left(e^{2z}+1
\right)^{{7-p\over 2}}}\,\left[1+c_{p}(v,m_q)\,
{\left(e^{2z}+1\right)^{{7-p\over 2}}\over 
\left(e^{2z}+(\frac{v}{m_q})^2\right)^4}\right]\,.
\label{Sch-pot}
\eear
Notice that the reduced mass $\bar M$ is just a parameter in $V(z)$.
Actually, in these new variables the problem of finding the mass spectrum can
be rephrased as that of finding the values of $\bar M$  that allow
a zero-energy level for the potential (\ref{Sch-pot}). By using the
standard techniques in quantum mechanics one can convince oneself that such 
solutions exist. Indeed, the potential (\ref{Sch-pot}) is strictly positive
for 
$z\to\pm\infty$ and has some minima for finite values of $z$. The actual
calculation of the mass spectra must be done by means of numerical
techniques. A key ingredient in this approach is the knowledge of the
asymptotic behavior of the solution when $\varrho\to 0$ and
$\varrho\to\infty$. This behavior can be easily obtained from the form of
the potential $V(z)$ in (\ref{Sch-pot}). Indeed, for $\varrho\to\infty$, or
equivalently for $z\to +\infty$, the potential $V(z)\to 1$, and the
solutions of the Schr\"odinger equation behave as $\psi\sim e^{\pm z}$
which, in terms of the original variables, corresponds to $f={\rm
constant},\,\,\varrho^{-2}$.  Similarly for $\varrho\to0$ (or
$z\to-\infty$) one gets that 
$f=\varrho^{2},\,\,\varrho^{-4}$. Thus, the IR and UV
behaviors of the fluctuation are:
\bear
&&f(\varrho)\approx a_1\,\varrho^2+a_2\,\varrho^{-4}\;,
\qquad (\varrho\to 0)\,,\rc\rc
&&f(\varrho)\approx b_1\,\varrho^{-2}+b_2\,\,,
\qquad (\varrho\to \infty)\,.
\label{UVIRbehavior}
\eear
The normalizable solutions are those that are regular at $\varrho\approx 0$
and decrease at $\varrho\approx \infty$. Thus they correspond to having
$a_2=b_2=0$ in (\ref{UVIRbehavior}).
Upon applying a shooting technique, we can determine the values of $\bar M$
for which such normalizable solutions exist. Notice that $\bar M$ depends
parametrically on the quark mass $m_q$ and on its VEV $v$. In general, for
given values of $m_q$ and $v$, one gets a tower of discrete values of  
$\bar M$. In figure \ref{massinstanton} we have plotted the values of the
reduced mass for the first level, as a function of the quark VEV. For
illustrative purposes we have included the values obtained with the
fluctuation equation of ref. \cite{EGG}. As anticipated above, both results
differ significantly in the region of small $v$ and coincide when
$v\to\infty$. Actually, when $v$ is very large we recover the spectral flow
phenomenon described in \cite{EGG}, \ie\ $\bar M$ becomes independent of the
instanton size and equals the mass corresponding to a higher Kaluza-Klein
mode on the worldvolume sphere. However, we see that when $v/m_q$ goes to
zero, the masses of the associated fluctuations also go to zero. Actually,
this limit is pretty singular. Indeed, it corresponds to the small instanton
limit, where it is  expected that the moduli space of instantons becomes
effectively non-compact and  that extra massless degrees of
freedom show up in the spectrum.

\begin{figure}
\centerline{\hskip -.1in \epsffile{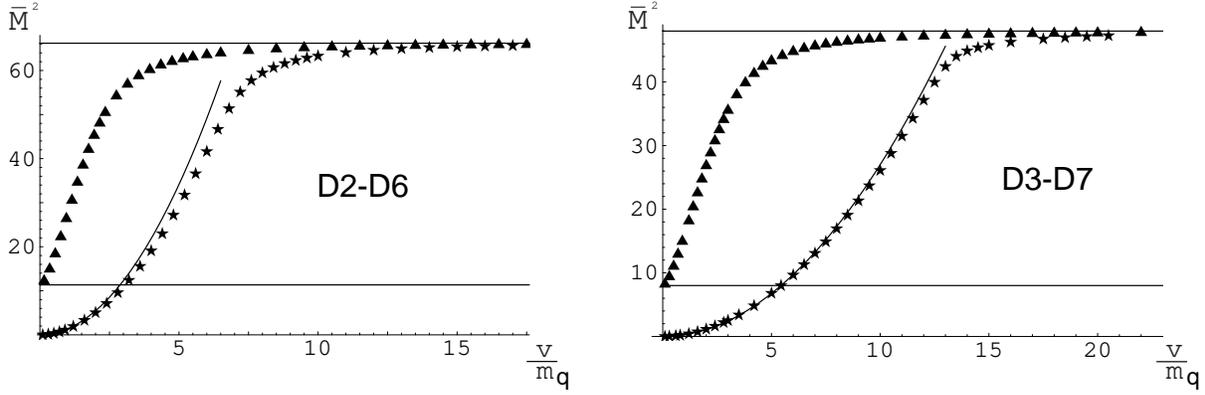}}
\caption{In this figure we  plot the numerical masses for the
first level as a function of 
the instanton size for both the full equation (with stars) and for the
equation obtained in \cite{EGG} (with solid triangles). The quark mass
$m_q$ is such that $g_{eff}(m_q)=1$. The solid line
corresponds to the WKB prediction (\ref{WKB-smallv}) for small $v$. The plot
on the left (right) corresponds to the D2-D6 (D3-D7) intersection.}
\label{massinstanton}
\end{figure}

It turns out that the mass levels for small $v$ are nicely represented
analytically by means of the WKB approximation for the Schr\"odinger
problem. The WKB method was presented in subsection \ref{cp2WKBssc} of the
previous chapter and applying straightforwardly the steps introduced there
to the fluctuation equation (\ref{fluctuationsDp+4}) we obtain the
following expression for the WKB values of
$\bar M$:
\beq
\bar M_{WKB}^2={\pi^2\over \zeta^2}
\,(n+1)\,\left(n+3+{2\over 5-p}\right)\,,
\label{MWKB}
\eeq
where $\zeta$ is the following integral:
\beq
\zeta=\int_{0}^{+\infty} d\varrho\,\sqrt{{1\over (1+
\varrho^2)^{{7-p\over 2}}}+{c_{p}(v,m_q)\over\left[\big({v\over m_q}
\big)^2+\varrho^2\right]^4}}\,.
\label{zetaWKB}
\eeq
Let us evaluate analytically $\zeta$ when $v$ is small. First of all, as can
be easily checked, we notice that, when $v$ is small,  the second term under
the square root in (\ref{zetaWKB}) behaves as:
\beq
{1\over \left[\big({v\over m_q}\big)^2+\varrho^2\right]^2}\approx{\pi\over
2}\,\left({m_q\over v}\right)^3\,\delta(\varrho)\;,
\qquad{\rm as}\;\; v\to 0\,.
\label{deltaVEV}
\eeq
Then, one can see that this term dominates the integral defining $\zeta$
around $\varrho\approx 0$ and, for small $v$, one can approximate $\zeta$
as:
\beq
\zeta\approx {\sqrt{c_{p}(v,m_q)}\over 2}
\int_{-\epsilon}^{\epsilon}{d\varrho\over \left[\big({v\over m_q}\big)^2+
\varrho^2\right]^2}+\int_0^{+\infty}\,{d\varrho\over (1+
\varrho^2)^{{7-p\over 4}}}\;,
\eeq
where $\epsilon$ is a small positive number and we have used the fact that
the function in (\ref{zetaWKB}) is an even function of $\varrho$. 
Using (\ref{deltaVEV}), one can
evaluate $\zeta$ as:
\beq
\zeta\approx{\pi\over 4}\,\left({m_q\over v}\right)^3\,
\sqrt{c_{p}(v,m_q)}+{\sqrt{\pi}\over 2}\,
{\Gamma\Big({5-p\over 4}\Big)\over \Gamma\Big({7-p\over 4}\Big)}\,.
\label{zeta-approx}
\eeq
Clearly, for $v\to 0$, we can neglect the last term in (\ref{zeta-approx}).
Using the expression of $c_{p}(v,m_q)$ (eq. (\ref{cp})), we arrive at:
\beq
\zeta\,\approx\,{\sqrt{3}\,\cdot 2^{{p-3\over 2}}\,\pi^{{p+5\over 4}}\over
\sqrt{\Gamma\Big({7-p\over 2}\Big)}}\,\,
{m_q\over g_{eff}(m_q)\, v}\,,
\eeq
and plugging this result in (\ref{MWKB}), we get the WKB mass of the ground
state ($n=0$) for small $v$:
\beq
\bar M_{WKB}^2\approx{(17-3p)\,\Gamma\Big({5-p\over 2}\Big)\over
3\cdot 2^{p-2}\,\pi^{{p+1\over 2}}}\,\,\,\,
\Bigg(\,{g_{eff}(m_q)\, v\over m_q}\,\Bigg)^2\,.
\label{WKB-smallv}
\eeq
Thus, we predict that $\bar M^2$ is a quadratic function of $v/m_q$ with the
particular coefficient given on the right-hand side of (\ref{WKB-smallv}). 
In figure \ref{massinstanton} we have represented by a solid line the value
of $\bar{M}$ obtained from eq. (\ref{WKB-smallv}). We notice that, for small
$v$, this equation nicely fits the values obtained by the numerical
calculation.

Let us now study the dependence of the mass gap as a function of the quark
mass $m_q$ and the quark VEV $v$. First of all, we notice that the relation
between the reduced mass $\bar M$ and the mass $M$ can be rewritten in terms
of the quark mass $m_q$ and the dimensionless coupling constant
$g_{eff}(m_q)$ as:
\beq
M\,\propto {m_q\over g_{eff}(m_q)}\,\,\bar M\,.
\label{M-barM}
\eeq
For large $v$ the reduced mass $\bar M$ tends to a value independent of both
$m_q$ and $v$. Thus, the meson mass $M$ depends only on $m_q$ in a
holographic way, namely:
\beq
M\,\sim {m_q\over g_{eff}(m_q)}\;,
\qquad (v\to\infty)\,.
\eeq
Notice that this dependence on $m_q$ and $g_{eff}(m_q)$ is exactly the
same as in the unbroken symmetry case, although the numerical coefficient
is different from that found in section \ref{general}.  On the contrary,
for small $v$, after combining eq. (\ref{M-barM}) with  the WKB result
(\ref{WKB-smallv}), we get that the mass gap depends linearly on
$v$ and is independent on the quark mass $m_q$:
\beq
M\sim v\,,\qquad (v\to 0)\,,
\eeq
and, in particular, the mass gap disappears in the limit $v\to 0$, which
corresponds to having a zero-size instanton. 

%
%
%
%
%
%
%
%
%

\setcounter{equation}{0}
\section{The codimension one defect}
\label{cp3codim1sc}
In this section we will study the Higgs branch of the codimension one
defect theories dual to the Dp-D(p+2) intersection. As before, we will
first study the $p=3$ case, whose dual field theory action was
constructed in \cite{WFO} and we will generalize our results to the
generic intersection later. Following a different order to the previous
section, we will begin by introducing the D3-D5 intersection with flux,
namely the macroscopic description of the gravity dual.

Let us first recall the array describing the D3-D5 intersection we have
dealt with in the previous chapter:
\beq
\begin{array}{ccccccccccl}
 &1&2&3& 4& 5&6 &7&8&9 & \nonumber \\
D3: & \times &\times &\times &\_ &\_ & \_&\_ &\_ &\_ &     \nonumber \\
D5: &\times&\times&\_&\times&\times&\times&\_&\_&\_ &
\end{array}
\label{cp3D3D5intersection}
\eeq
Once again, as the number of D5s is much lower than the number of
D3s, the near-horizon region becomes $AdS_5\times S^5$ whose metric
was written in eq. (\ref{cp3d3sol}), where the Minkowski dimensions
correspond to the first three directions in (\ref{cp3D3D5intersection}),
which are parallel to the D3-brane. In addition, the D3-brane background
is endowed with an RR four-form potential written in eq.
(\ref{cp3d3sol}).  Proceeding in an analogous way to the D3-D7
intersection, we shall split the vector $\vec r$ of eq. (\ref{cp3d3sol})
as $\vec r=(\vec y, \vec z)$, where now $\vec y=(y^1,y^2,y^3)$ and 
$\vec z=(z^1,z^2,z^3)$ corresponding, respectively, to the directions
$4,\,5,\,6$ and $7,\,8,\,9$ of the array (\ref{cp3D3D5intersection}).
Then, the metric (\ref{cp3d3sol}) can be written in the following form:
\beq
ds^2={\rho^2+\vec z^{\,\,2}\over R^2}\,dx_{1,3}^2+
{R^2\over \rho^2+\vec z^{\,\,2}}\left(d\rho^2+\rho^2 d\Omega_2^2+d\vec z
\cdot d\vec z\right)\,,
\label{cp3ads5s5metric}
\eeq
where $\rho^2=\vec y\cdot\vec y\,$, and we have changed to spherical
coordinates along the subspace parametrized by $\vec y$, namely
$(dy^1)^2+(dy^2)^2+(dy^3)^2=d\rho^2+\rho^2\,d\Omega_2^2\,$,
with $d\Omega_2^2=\,d\theta^2+\sin^2\theta d\varphi^2$, corresponding to 
a unit two-sphere.

We are interested in studying the dynamics of a D5-brane probe embedded in
the $AdS_5\times S^5$ geometry. The action of a D5-brane is the sum of a
Born-Infeld and a Wess-Zumino term written in eq. (\ref{cp2DBI-D5}) of
the previous chapter.  According to the array (\ref{cp3D3D5intersection})
and the redefinition of coordinates we have just described, let us take as
worldvolume coordinates $\xi^a=(x^0, x^1, x^2,\rho,\theta,\varphi)$. The
embedding  of the D5-brane probe is then specified by  the values of
$x^3$ and $\vec z$ as functions of the $\xi^a$s. We will consider static
embeddings in which $|\vec z|$ is a fixed constant, namely
$|\vec z|=L$. The simplest of such embeddings is the one in which the
coordinate $x^3$ is also a constant and the worldvolume gauge field $F$
vanishes. This configuration was studied extensively in section
\ref{cp2D3D5} of the previous chapter. The probe D5 introduces a defect
along the gauge theory directions consisting of a flat wall located
at $x^3=$constant. The induced worldvolume metric, eq.
(\ref{cp2indmetric}) with the corresponding coefficients, is, for large
$\rho$, of the form
$AdS_4\times S^2$. We shall now generalize this D5-brane probe embedding
by switching on a magnetic field $F$ along the two-sphere of the
worldvolume. Concretely, let us assume that $F$ is given by:
\beq
F=q\,{\rm Vol}\,(S^2)\equiv{\cal F}\,,
\label{cp3wvflux}
\eeq
where $q$ is a constant and ${\rm Vol}\,(S^2)$ is the volume form of the 
worldvolume two-sphere.  To understand the implications of having a
magnetic flux across the worldvolume $S^2$, let us look at the form of
the Wess-Zumino term, namely:
\beq
S_{WZ}\sim\int_{S^2} F\int P[\,C^{(4)}\,]\sim q\,x'\,,
\label{cp3WZbending}
\eeq
where  $x\equiv x^3$ and the prime denotes the derivative  with respect
to the radial coordinate $\rho$. It is clear by inspecting the right-hand
side of eq. (\ref{cp3WZbending}) that the worldvolume flux acts as a
source of  a non-trivial dependence of $x$ on the coordinate $\rho$.
Actually, assuming that $x$ only depends on $\rho$, the action of the
probe takes the form:
\beq
S_{D5}=-4\pi\,T_{5}\int d^3x\,d\rho\left[\rho^2\,
\sqrt{1+{(\rho^2+L^2)^2\over R^4}\,x'^{\,2}}\;\sqrt{1+{(\rho^2+L^2)^2\over
R^4}\,{q^2\over \rho^4}}-{(\rho^2+L^2)^2\over R^4}\,q\,x'\right]\,,
\label{cp3effe-action}
\eeq
where we have assumed that $\vec z$ is constant ($|\,\vec z\,|=L$) and we
have integrated over the coordinates of the two-sphere. The
Euler-Lagrange equation for $x(\rho)$ derived from (\ref{cp3effe-action})
is quite involved. However, there is a simple first order equation for
$x(\rho)$ which solves this equation \cite{ST}, namely:
\beq
x'(\rho)={q\over \rho^2}\,.
\label{cp3first-order}
\eeq
Actually, the first order equation (\ref{cp3first-order}) is a BPS
equation required by supersymmetry, as can be verified by checking the
kappa symmetry of the embedding \cite{ST}. The integration of eq.
(\ref{cp3first-order}) is straightforward:
\beq
x(\rho)=x_0-{q\over \rho}\,,
\label{cp3bending}
\eeq
where $x_0$ is a constant. The dependence on $\rho$ of the right-hand
side of eq. (\ref{cp3bending}) represents the bending of the D5-brane
profile required by supersymmetry when there is a non-vanishing flux of
the worldvolume gauge field. Notice also that now the probe is located at
a fixed value of $x$ only at the asymptotic value $\rho\to\infty$, whereas
when $\rho$ varies the $D5$-brane fills one-half of the worldvolume of
the D3-brane (\ie\ $x^3\le x_0$ for $q>0$). 

It is also interesting to study the modifications of the induced metric
introduced by the bending. Actually, when $q\not=0$  this induced metric
takes the form:
\beq
{\cal G}_{ab}\,d\xi^a d\xi^b={\rho^2+L^2\over R^2}\,dx^2_{1,2}+
{R^2\over \rho^2+L^2}\left[\left(1+{q^2\over R^4}\,{(\rho^2+L^2)^2
\over \rho^4}\right)d\rho^2+\rho^2\,d\Omega_2^2\right]\,.
\label{cp3ind-met-flux}
\eeq
It can be readily verified from (\ref{cp3ind-met-flux}) that the UV metric
at $\rho\to\infty$ (or, equivalently, when the D3- and D5-branes are at
zero distance $L$) takes the form:
\beq
AdS_4(R_{eff})\times S^2 (R)\,,
\label{cp3UVind-met-flux}
\eeq
where the radius of the $AdS_4$ changes from its fluxless value $R$ to 
$R_{eff}$, with the latter given by:
\beq
R_{eff}=\left(1+{q^2\over R^4}\right)^{{1\over 2}}\,R\,.
\label{cp3Reff}
\eeq
Notice that the radius of the $S^2$ is not affected by the flux, as is
clear from (\ref{cp3ind-met-flux}).

One can understand the appearance of this UV metric as follows.
Let us suppose that we have an $AdS_5$ metric of the form:
\beq
ds^2_{AdS_5}={\rho^2\over R^2}\,dx^2_{1,3}+{R^2\over \rho^2}\,d\rho^2\,.
\label{cp3AdSmetric}
\eeq
Let us now change variables from $(\rho, x^3)$ to new
coordinates $(\varrho, \eta)$, as follows:
\beq
x^3=\bar x-{\tanh\eta\over \varrho}\;,\qquad\quad
\rho=R^2\varrho\cosh\eta\,,
\label{cp3chngvblsd3d5}
\eeq
where $\bar x$ is a constant. 
It can be easily seen that the $AdS_5$ metric (\ref{cp3AdSmetric})
in the new variables takes the form:
\beq
ds^2_{AdS_5}=R^2\left(\cosh^2\eta\,ds^2_{AdS_4}+d\eta^2\right)\,,
\label{cp3foliation}
\eeq
where $ds^2_{AdS_4}$ is the metric of $AdS_4$ with unit radius, given by:
\beq
ds^2_{AdS_4}=\varrho^2\,dx^2_{1,2}+{d\varrho^2\over\varrho^2}\,.
\eeq
Eq. (\ref{cp3foliation}) shows clearly the foliation of $AdS_5$ by 
$AdS_4$ slices with $\eta={\rm constant}$. The effective radius of the 
$AdS_4$ slice depends on the value of $\eta$ as follows:
\beq
R_{eff}=R\cosh\eta\,.
\label{cp3sliceradius}
\eeq
It can be straightforwardly checked by using the change of variables
(\ref{cp3chngvblsd3d5}) with $\bar x=x_0$ that our embedding
(\ref{cp3bending}) corresponds to one of these $AdS_4$ slices with a
constant  value of $\eta$ given by:
\beq
\eta=\eta_q=\sinh^{-1}\left({q\over R^2}\right)\,.
\eeq
Moreover, one can verify that the  $AdS_4$ radius $R_{eff}$ of eq. 
(\ref{cp3sliceradius}) reduces to the expression given in (\ref{cp3Reff})
when
$\eta=\eta_q$.

The worldvolume gauge field (\ref{cp3wvflux}) is constrained by a flux
quantization condition \cite{Flux} which, with our notations, reads:
\beq
\int_{S^2}\,F={2\pi k\over T_f}\,\,,\qquad
k\in {\mathbb Z}\,\,,\qquad T_f={1\over 2\pi\alpha'}\,.
\label{cp3fluxquantization}
\eeq
It is now immediate to conclude that the condition
(\ref{cp3fluxquantization}) restricts the constant $q$ to be of the form:
\beq
q=k\pi\alpha'\,,
\label{cp3q-k}
\eeq
where $k$ is an integer.

\subsection{Microscopic interpretation} 
\label{cp3scd3d5micro}

The presence of a worldvolume flux as in (\ref{cp3wvflux}) induces,
through the Wess-Zumino term, a D3-brane charge, proportional to
$\int_{S^2}\,F$, on the D5-brane. For this reason, it is not surprising
that the D5-brane configuration just described admits a microscopic
description in terms of a bound state of coincident D3-branes. Actually, 
the integer $k$ of the quantization condition (\ref{cp3fluxquantization})
has the interpretation of the number of D3-branes that build up the
D5-brane. The dynamics of a stack of coincident D3-branes is determined
by the non-abelian action written in eq. (\ref{cp1ncDbraneact}). It is
the sum of a Born-Infeld and a Wess-Zumino part written in eqs.
(\ref{cp3dielectricBI}) and (\ref{cp3dielectricWZ}).

Let us now choose $x^0,x^1,x^2$ and $\rho$ as our set of worldvolume
coordinates of the D3-branes. Moreover, we shall introduce new
coordinates 
$Y^I(I=1,2,3)$ for the two-sphere of the metric (\ref{cp3ads5s5metric}).
These new coordinates satisfy $\sum_I\,Y^I\,Y^I\,=\,1$ and
the line element $d\Omega_2^2$ is given by:
\beq
d\Omega_2^2=\sum_I\,dY^I\,dY^I\;,\qquad\sum_I\,Y^I\,Y^I=1\,.
\label{cp3ncys}
\eeq
We will assume that the $Y^I$s are the only non-commutative scalars.
They will be represented by $k\times k$ matrices.  In this case the 
matrix $Q$ appearing in eq. (\ref{cp3dielectricBI}) is given by:
\beq
Q_J^I=\delta_{J}^{I}+{i\over 2\pi\alpha'}\,[Y^I,Y^K]\,G_{KJ}\,.
\eeq
Actually, we shall adopt an ansatz in which the  $Y^I$s are constant
and given by:
\beq
Y^I={J^I\over \sqrt{C_2(k)}}\,,
\label{cp3d5Yansatz}
\eeq
where the $k\times k$ matrices $J^I$ correspond to the $k$-dimensional
irreducible representation of the $SU(2)$ algebra:
\beq
[J^I,J^J]=2i\,\epsilon_{IJK}\,J^K\,,
\label{cp3Jcommutator}
\eeq
and $C_2(k)$ is the quadratic Casimir of the $k$-dimensional irreducible
representation of  $SU(2)$ ($C_2(k)=k^2-1$). Therefore, the $Y^I$ scalars
parametrize a fuzzy two-sphere. Moreover, let us assume that we consider
embeddings in which the scalars $\vec z$ and $x^3$ are commutative and
such that $|\vec z|=L$ and $x^3=x(\rho)$ (a unit $k\times k$ matrix is
implicit). With these conditions, as the metric (\ref{cp3ads5s5metric})
does not mix the directions of the two-sphere with the other coordinates,
the matrix $Q^{-1}-\delta$ does not contribute to the first square root
on the right-hand side of (\ref{cp3dielectricBI}) and we get:
\beq
\sqrt{-\det\big[ P[G]\,\big]}={\rho^2+L^2\over R^2}\,\,
\sqrt{1+{(\rho^2+L^2)^2\over R^4}\,x'^{\,2}}\,.
\eeq
Moreover, by using the ansatz (\ref{cp3d5Yansatz}) and the commutation
relations (\ref{cp3Jcommutator}) we obtain that, for large $k$, the second
square root appearing in (\ref{cp3dielectricBI}) can be written as:
\beq
{\rm Str}\Bigg[\sqrt{\det Q}\Bigg]\approx {R^2\over \pi\alpha'}\,\,
{\rho^2\over \rho^2+L^2}\,\,\sqrt{1+{(\rho^2+L^2)^2\over R^4}\,
{(k\pi\alpha')^2\over \rho^4}}\,.
\eeq

Using these results, the Born-Infeld part of the D3-brane action
in this large $k$ limit takes the form:
\beq
S_{BI}^{D3}=-{T_3\over \pi\alpha'}\,\,\int\,d^3x\,d\rho\,
\rho^2\,\sqrt{1+{(\rho^2+L^2)^2\over R^4}\,x'^{\,2}}\,\,
\sqrt{1+{(\rho^2+L^2)^2\over R^4}\,{q^2\over \rho^4}}\,,
\label{cp3d5microBI}
\eeq
where we have already used (\ref{cp3q-k}) to write the result in terms of
$q$.  Due to the relation $T_3\,=\,4\pi^2\,\alpha'\,T_5$ between the
tensions of the D3- and D5-branes, one checks by inspection that the
right-hand side of (\ref{cp3d5microBI}) coincides with the Born-Infeld
term of the D5-brane action (\ref{cp3effe-action}). Notice also that the
quantization integer $k$ in (\ref{cp3fluxquantization}) is identified with
the number of D3-branes.  Moreover, the Wess-Zumino term
(\ref{cp3dielectricWZ}) becomes:
\beq
S_{WZ}^{D3}=kT_3\,\int\,d^3x\,d\rho\,\,\,{(\rho^2+L^2)^2\over R^4}\,x'\,.
\label{microWZ}
\eeq
The factor $k$ in (\ref{microWZ}) comes from the trace of the unit
$k\times k$ matrix. By comparing (\ref{microWZ}) with the Wess-Zumino
term of the  macroscopic action  (\ref{cp3effe-action}) one readily
concludes that they  coincide because of the relation $4\pi
qT_5\,=\,kT_3$, which follows easily from eqs. (\ref{cp1tension}) and
(\ref{cp3q-k}).

\subsection{Field theory analysis} 
\label{cp3scd3d5fieldth} 

In this subsection we will analyze the configuration described above from
the point of view of the field theory at the defect which, from now on, we
shall assume that it is located at $x^3=0$. Recall that the defect arises
as a consequence of the impurity created on the worldvolume of the
D3-branes  by the D5-brane which intersects with them according to the
array (\ref{cp3D3D5intersection}).  We are interested in analyzing, from
the field theory point of view, the configurations in which some fraction
of the D3-branes end on the D5-brane and recombine with it at the defect
point $x^3=0$, realizing in this way a (mixed Coulomb-)Higgs branch of
the defect theory. 

The field theory dual to the D3-D5 intersection 
has been worked out by DeWolfe \textit{et al.} in ref. \cite{WFO} and its
field content was described in subsection \ref{cp2d3d5dict} of the
previous chapter. As a reminder, we list the bosonic fields of the defect
theory in table \ref{cp3d3d5fields}, where 
\begin{table}[!h]
\centerline{
\begin{tabular}{|c|c|c|c|}
\hline 
Field & $SU(N)$ & $SU(2)_H$ & $SU(2)_V$\\
\hline
$A_{\mu}$ & adjoint & singlet & singlet\\
\hline
$\phi_H^I$ & adjoint & vector & singlet\\
\hline
$\phi_V^A$ & adjoint & singlet & vector\\
\hline
$q$ & fundamental & doublet & singlet\\
\hline
$\bar{q}$&fundamental& doublet & singlet\\
\hline
\end{tabular}
}
\caption{Bosonic field content of the gauge theory dual to the D3-D5
intersection.}
\label{cp3d3d5fields}
\end{table}
now we are denoting the six adjoint scalars of the theory by $\phi_H^I$
and $\phi_V^A$. They are related to the coordinates transverse to the
D3-brane $X_H=(X^4, X^5, X^6)$ and $X_V=(X^7, X^8, X^9)$ (corresponding
to the directions $4,\,5,\,6$ and $7,\,8,\,9$, respectively, in the array
(\ref{cp3D3D5intersection})) through the usual relation:
\beq
X_H^I=2\pi\alpha'\,\phi^I_H\;,\qquad\quad
X_V^A=2\pi\alpha'\,\phi^A_V\,,
\label{cp3d5adj}
\eeq
between coordinates and scalar fields.

Assuming that only the fields $\phi_H$, $\phi_V$, $q$ and $\bar q$
are non-vanishing, the defect action for this theory has a potential term
which can be written as \cite{WFO}:
\begin{eqnarray}
S_{\rm defect}=&-&\frac{1}{g^2}\int d^3x\,\left[
 \bar{q}^m\,(\phi_V^A)^2\,q^m+\frac{i}{2}\epsilon_{IJK}\bar{q}^m
\sigma_{mn}^I\,[\phi_H^J,\phi_H^K]\,q^n\right]-\rc\rc
&-&\frac{1}{g^2}\int d^3x\left[\bar{q}^m\sigma_{mn}^I\partial_3\,
\phi_H^I\,q^n+\frac{1}{2}\delta(x_3)(\bar{q}^m
\sigma_{mn}^IT^aq^n\,)^2\right]\,,
\label{cp3actiondefect}
\end{eqnarray}
where the integration is performed over the $x^3=0$ three-dimensional
submanifold and $g$ is the Yang-Mills coupling constant. In the
supersymmetric configurations we are looking for, the potential term must
vanish. Let us cancel the contribution of $\phi_V$ to the right-hand side 
of (\ref{cp3actiondefect}) by requiring that: 
\beq
\phi_V\,q=0\,.
\eeq
We can insure this property by taking $q$ as:
\beq
q=\left(\begin{array}{c}0 \\ \vdots\\ 0\\ 
\alpha_1\\ \vdots \\ \alpha_k\end{array}\right)\,,
\label{cp3d5qvev}
\eeq
and by demanding that  $\phi_V$ is of the form:
\beq
\phi_V=\pmatrix{A&0\cr 0&0}\,,
\eeq
where $A$ is an $(N-k)\times (N-k)$ traceless matrix. Moreover, we shall
take $\phi_V$, $q$ and $\bar q$ constant, which is enough to guarantee
that their kinetic energy vanishes. Notice that the scalars $\phi_V$
correspond to the directions $7,\,8,\,9$ in the array
(\ref{cp3D3D5intersection}), which are orthogonal to both the D3- and
D5-brane. Having $\phi_V\not=0$ is equivalent to taking $|\vec
z|=L\not=0$ in both the macroscopic and microscopic supergravity
approaches of previous sections and corresponds to a non-zero
value of the mass of the hypermultiplets (see the first term in the
defect action (\ref{cp3actiondefect})).

Let us now consider the configurations of $\phi_H$ with vanishing energy.
First of all we will impose that $\phi_H$ is a matrix whose only
non-vanishing entries are in the lower $k\times k$ block. In this way the
mixing terms of $\phi_V$ and $\phi_H$ cancel. Moreover, assuming that
$\phi_H$ only depends on the coordinate $x^3$, the surviving terms in the
bulk action are \cite{WFO}:
\beq
S_{\rm bulk}=-{1\over g^2}\int d^4x\;{\rm Tr}\,\left[\,
{1\over 2}\,\,(\partial_3\phi_H^I)^2-{1\over 4}\,\,
[\phi_H^I, \phi_H^J]^2\right]\,,
\label{cp3DWbulkaction}
\eeq
where the trace is taken over the color indices. It turns out that the
actions (\ref{cp3actiondefect}) and (\ref{cp3DWbulkaction}) can be
combined in such a way that their sum can be written as an integral over
the four-dimensional spacetime of the trace of
a square. In order to write this
expression, let  us  define the matrix
$\alpha^I\,=\,\alpha^{Ia}\,\,T^a$, where the $T^a$s are the generators
of the gauge group and the $\alpha^{Ia}$s are defined as the following
expression bilinear in $q$ and $\bar q$:
\beq
\alpha^{Ia}\equiv \bar{q}^m\,\sigma_{mn}^I\,T^a\,q^n\,.
\eeq
It is now straightforward to check that the sum of (\ref{cp3actiondefect})
and (\ref{cp3DWbulkaction}) can be put as:
\beq
S_{\rm defect}+S_{\rm bulk}=-{1\over 2g^2}\int d^4x\;{\rm Tr}\,
\left[\partial_3\phi_H^I+{i\over 2}\,\epsilon_{IJK}\,[\,\phi_H^J,
\phi_H^K\,]+\alpha^I\,\delta(x^3)\right]^2\,,
\label{cp3D5actionsquare}
\eeq
where we have used the fact that
$\epsilon_{IJK}\,{\rm Tr}\,\Big(\,\partial_3\phi_H^I\,
[\,\phi_H^J, \phi_H^K\,]\,\Big)$ is a total derivative with respect to
$x^3$ and, thus, can be dropped if we assume that $\phi_H$ vanishes at 
$x^3=\pm\infty$. It is now clear from (\ref{cp3D5actionsquare}) that we
must require the Nahm equations \cite{Neq}: 
\beq
\partial_3 \phi_H^I+{i\over 2}\,\epsilon_{IJK}
[\,\phi_H^J, \phi_H^K\,]+\alpha^I\,\delta(x^3)=0\,.
\label{cp3Nahm}
\eeq
(For a nice review of the Nahm construction in string theory see 
\cite{Tong}). 

Notice that when $\alpha^I$ vanishes, eq. (\ref{cp3Nahm}) admits the
trivial solution $\phi_H=0$. On the contrary, if the fundamentals $q$ and
$\bar q$ acquire a non-vanishing vacuum expectation value
as in (\ref{cp3d5qvev}),  $\alpha^I$  is
generically non-zero and the solution of (\ref{cp3Nahm}) must be
non-trivial. Actually, it is clear from (\ref{cp3Nahm}) that  in this case
$\phi_H$ must blow up at $x^3=0$, which shows how a non-vanishing vacuum
expectation value of the fundamentals acts as a source for the brane
recombination in the Higgs branch of the theory. Let us check these facts
more explicitly by solving (\ref{cp3Nahm}) for $x^3\not=0$, where the
$\delta$-function term is zero. We shall adopt the ansatz:
\beq
\phi_H^I(x)=f(x)\,\phi_0^I\,,
\eeq
where $x$ stands for $x^3$ and $\phi_0^I$ are constant 
matrices.  The differential
equation (\ref{cp3Nahm}) reduces to: 
\beq
{f'\over f^2}\,\,\phi_0^I+{i\over 2}\,\epsilon_{IJK}\,[\,\phi_0^J\,,
\phi_0^K\,]=0\,,
\label{cp3sourcelessNahm}
\eeq
where the prime denotes derivative with respect to $x$. 
We shall solve this equation by first putting:
\beq
\phi_0^I={1\over \sqrt{C_2(k)}}\,\pmatrix{0&0\cr 0&J^I}\,,
\eeq
where the $J^I$ are matrices in the $k$-dimensional irreducible
representation of the $SU(2)$ algebra, which satisfy the commutation
relations (\ref{cp3Jcommutator}),  and we have normalized the
$\phi_0^I\,$s such that $\phi_0^I\phi_0^I$ is the unit matrix in
the $k\times k$ block.  By using this representation of the
$\phi_0^I\,$s,  eq. (\ref{cp3sourcelessNahm}) reduces to:
\beq
{f'\over f^2}={2\over \sqrt{C_2(k)}}\,,
\eeq
which can be immediately integrated, resulting in:
\beq
f=-{\sqrt{C_2(k)}\over 2 x}\,.
\eeq
For large $k$, the quadratic Casimir $C_2(k)$ behaves as $k^2$ and this
equation reduces to:
\beq
f=-{k\over 2 x}\,.
\label{cp3Nahmsol}
\eeq
Next, by taking into account the standard relation between coordinates
and scalar fields (\ref{cp3d5adj}) and the fact that
$\rho^2\,\equiv\,X_H^I\,X_H^I$, one immediately gets the following
relation between $\rho$ and $f$:
\beq
\rho=2\pi\alpha' f\,,
\eeq
and the solution (\ref{cp3Nahmsol}) of the Nahm equation can be written
as:
\beq
\rho=-{\pi k\alpha'\over x}\,,
\eeq
which, if we take into account the quantization condition (\ref{cp3q-k}), 
is just our embedding (\ref{cp3bending}) for $x_0=0$. 
As expected, $\rho$ blows up at
$x=0$, while its dependence for $x\not=0$ gives rise to the same bending
as in the brane approach. Notice also that, in this field theory
perspective, the integer $k$ is the rank of the gauge theory subgroup in
which the Higgs branch of the theory is realized, which corresponds to
the number of D3-branes that recombine into a D5-brane.

\subsection{Fluctuations in Dp-D(p+2) with flux} 
\label{cp3Dp+2fluct}

We have just seen that one can enter the Higgs branch of the gauge theory
dual to the D3-D5 intersection by switching on a magnetic two-form flux
along the worldvolume of the D5-brane. As in section
\ref{cp3sscDp4flucts} we shall now go back to the macroscopic description
and study the fluctuations around the probe embedding with flux. As for
the fluxless embedding ($q=0$), studied in chapter \ref{cp2dpinter}, this
configuration can be easily generalized to all the Dp-D(p+2)
intersections. Therefore, in this section we will work directly in the
generalized Dp-D(p+2) system, since both the macroscopic and microscopic
description of the last subsections can be extended to this generic case.
In fact, we will first sketch the generalization of the macroscopic
description and then, after briefly commenting on the generalization of
the microscopic interpretation, study the fluctuations around the
generic embedding with flux.

Let us begin by presenting the generalization of the array
(\ref{cp3D3D5intersection}) for the generic Dp-D(p+2) intersection along
$p-1$ dimensions:
\beq
\begin{array}{ccccccccccl}
 &1&\cdots&p-1& p& p+1&p+2 &p+3&\cdots&9 & \nonumber \\
Dp: & \times &\cdots &\times &\times &\_ & \_&\_ &\cdots &\_ &     \nonumber
\\ D(p+2): &\times&\cdots&\times&\_&\times&\times&\times&\cdots&\_ &
\end{array}
\label{cp3dpdp+2array}
\eeq
As described above, we will study embeddings of the D(p+2)-brane probe in
the background generated by a stack of $N$ Dp-branes. The
fluxless configuration where the D(p+2)-brane is located at a fixed value
of the coordinate $x^p$ and separated a fixed distance in the transverse
space was studied in detail in ref. \cite{AR}. Now we are interested in a
generalization of the embedding just studied for the D3-D5 case where a
magnetic gauge field is switched on along the worldvolume of the
D(p+2)-brane, and in consequence, this brane gets bent along the $x^p$
direction.

The near-horizon metric of a stack of Dp-branes given by eq.
(\ref{cp1dpsolmtrc}) can be rewritten in a more useful form,
as we did in eq. (\ref{cp3ads5s5metric}) for the $p=3$
case. Indeed, let us split again the $\vec{r}$ coordinates (which
generate the space transverse to the Dp-branes) in two sets, namely
$\vec r=(\vec y,
\vec z)$, where
$\vec y$ has three components (corresponding to the directions $\,p+1,
\;p+2$ and $p+3$ in the array (\ref{cp3dpdp+2array})), and we will define
$\rho^2=\vec y\cdot\vec y$. Taking into account that
$r^2=\rho^2+\vec{z}^{\,2}$ and changing again to spherical coordinates
along the subspace spanned by $\vec y$, the metric (\ref{cp1dpsolmtrc})
can be written as:
\begin{equation}
ds^2=\Bigg(\frac{r^2}{R^2}\Bigg)^{\alpha}\,dx_{1,p}^2+
\Bigg(\frac{R^2}{r^2}\Bigg)^{\alpha}\left(d\rho^2+\rho^2d\Omega_2^2+
d\vec{z}^{\,2}\right)\,,
\label{cp3dpdp+2bckgr}
\end{equation}
where $\alpha$ was defined in eq. (\ref{cp3dpdp+4bckgr}) and
$d\Omega_2^2=d\theta^2+\sin^2\theta\,d\varphi^2$. Notice that the
coordinates
$(\rho,\theta,\varphi)$ parametrize the directions
$p+1$, $p+2$ and $p+3$ in (\ref{cp3dpdp+2array}).

We shall now consider  a D(p+2)-brane probe in this background. Its
action (given by eq. (\ref{cp1Dbraneact})) reads:
\beq
S^{D(p+2)}=-\,T_{p+2}\int d^{p+3}\xi\,e^{-\phi}\,
\sqrt{-\det (g+F)}+T_{p+2}\int P\left[C^{(p+1)}\right]\wedge F\,.
\label{cp3dp+2act}
\eeq
According to the array (\ref{cp3dpdp+2array}), let us take in what follows
$\xi^a=(x^0,x^1\cdots,x^{p-1},\rho,\theta,\varphi)$ as worldvolume
coordinates. Moreover, we will assume that there exists   a constant
separation on the transverse space,  $\vec{z}^{\,2}=L^2$, which gives 
mass to the quarks,  and we will switch on a  magnetic worldvolume field
on the internal $S^2$ given by eq. (\ref{cp3wvflux}). As anticipated
above, in order to solve the equations of motion of the probe, we have to
consider a non-trivial transverse field $x^p\,=\,x(\rho)$. Moreover, since nothing depends on the
internal $S^2$, upon integration over this compact manifold, it is
straightforward to see that the action reads:
\bear
&& S^{D(p+2)}=-4\pi T_{p+2}\int d^{p} x\,d\rho\,
\Bigg[\rho^2\sqrt{1+\left(\frac{\rho^2+L^2}{R^2}
\right)^{2\alpha}x'^2}\;\sqrt{1+\left(\frac{\rho^2+L^2}{R^2}
\right)^{2\alpha}\frac{q^2}{\rho^4}}\,-\rc\rc
&&\qquad\qquad\qquad\qquad\qquad\qquad\quad
-q\left(\frac{\rho^2+L^2}{R^2}\right)^{2\alpha}x'\Bigg]\,.
\label{cp3dp+2explaction}
\eear
One can check that the Euler-Lagrange equation for $x(\rho)$ derived from
(\ref{cp3dp+2explaction}) is solved if one imposes the first order
equation (\ref{cp3first-order}), which is easily integrated yielding
(\ref{cp3bending}). Therefore, the bending of the D(p+2)-brane along the
$x^p$ direction is of the same form for any Dp-D(p+2)
intersection, namely eq. (\ref{cp3bending}).
For this configuration, the two square roots in (\ref{cp3dp+2explaction})
become equal and there is a cancellation between the WZ and (part of) the
DBI term.  Then, the energy for such a brane,
which is nothing but minus the lagrangian since our configuration is static,
reduces to:
\beq
E=4\pi T_{p+2}\int \rho^2\,,
\label{cp3dp+2onshl}
\eeq
where,  as in the Dp-D(p+4) case, the distance  $L$ does not explicitly
appear,  displaying the supersymmetry properties of the configuration.
Indeed, one can verify as in \cite{ST} that the condition 
(\ref{cp3first-order}) is a BPS equation that can be derived from the
kappa symmetry of the probe and that the energy (\ref{cp3dp+2onshl})
saturates a BPS bound. In addition, the worldvolume gauge field flux $F$
of  eq. (\ref{cp3wvflux}) on  the internal $S^2$ has the non-trivial
effect of inducing Dp-brane  charge in the D(p+2)-brane worldvolume. 
This follows easily: inserting the explicit form of $F$
given in eq. (\ref{cp3wvflux}) in the WZ term of the action
(\ref{cp3dp+2act}) and integrating along the internal $S^2$ one arrives
at:
\beq
k\, T_p\int\, P[C^{p+1}]\,,
\eeq
after taking into account the flux quantization condition
(\ref{cp3fluxquantization}) and using the relation $T_{p+2}\,(2\pi
)^2\,\alpha'=T_p$. Notice that now the integration is  over $p+1$
dimensions, and also that $q$ is related to $k$ as in eq. (\ref{cp3q-k}).

The fact that the introduction of the flux induces Dp-brane charge
along the D(p+2)-brane worldvolume opens the possibility of describing the
system in terms of Dp-branes polarized to D(p+2)-branes, as it was done in
section \ref{cp3scd3d5micro} for the $p=3$ case. Indeed, let us take as
worldvolume coordinates of the Dp-branes the set $(x^0,\cdots,
x^{p-1},\rho)$ and consider the other
coordinates in the metric (\ref{cp3dpdp+2bckgr}) as scalar fields which,
in general, are non-commutative. Furthermore, we can introduce new
coordinates  $Y^I(I=1,2,3)$ for the two-sphere of the metric
(\ref{cp3dpdp+2bckgr}). We can assume that they are the only
non-commutative scalars and take them to be the same as in the $p=3$
case, thus defined by eqs. (\ref{cp3ncys}), (\ref{cp3d5Yansatz}) and
(\ref{cp3Jcommutator}). With these assumptions it is easy to evaluate the
dielectric action for the Dp-brane in the large $k$ limit as it was
done for $p=3$, and one easily finds agreement with the macroscopic
action (\ref{cp3dp+2explaction}) once the relation (\ref{cp3q-k}) between
$k$ and $q$ is taken into account.

Let us  now study the
fluctuations around the Dp-D(p+2) intersection with flux described above.
Without loss of generality we can take the unperturbed configuration as
$z^1=L$, $z^m=0$ $(m>1)$.  Next, let us consider a fluctuation of the type:
\bear
&&z^1=L+\chi^1\;,\qquad\quad z^m=\chi^m\,,\qquad
(m=2,\cdots, 6-p)\,,\rc\rc
&&x^p={\cal X}+x\;,\qquad\quad  F={\cal F}+f\,,
\eear
where the bending ${\cal X}$ and the worldvolume gauge field ${\cal F}$ are
given by eqs. (\ref{cp3bending}) and  (\ref{cp3wvflux}) respectively, and
we assume that $\chi^m$, $x$  and $f$ are small. The induced metric on the
D(p+2)-brane worldvolume can be written as:
\beq
g={\cal G}+g^{(f)}\,,
\eeq
with ${\cal G}$ being the induced metric of the unperturbed configuration:
\beq
{\cal G}_{ab}\,d\xi^a\,d\xi^b=h^{-1}\,dx_{1,p-1}^2+h\left[\left(1+
{q^2\over\rho^4 h^2}\right)d\rho^2+\rho^2\,d\Omega_2^2\right]\,,
\eeq
where $h=h(\rho)$ is the function defined in (\ref{h}). Moreover, $g^{(f)}$
is the part of $g$ that depends on the derivatives of the fluctuations,
namely:
\beq
g^{(f)}_{ab}={q\over \rho^2 h}\,\left(\delta_{a\rho}\,\partial_{b}\,x+
\delta_{b\rho}\,\partial_{a}\,x\right)+{1\over h}\,\partial_{a}\,x\,
\partial_{b}\,x+h\,\partial_{a}\,\chi^m\,\partial_{b}\,\chi^m\,.
\eeq
Generically, apart from these terms in the derivatives
of the fluctuations, $g^{(f)}$ should include terms proportional to the
fluctuations themselves. However, it is not difficult to check that these
do not contribute to the action at quadratic order, so that there is no
potential for these fields as expected from BPS considerations\footnote{One
can check this from the fact that the combination $\det\,(\tilde
g+{\cal F})$, with $\tilde g=g-g^{(f)}$ (\ie\ the induced metric minus the
terms in derivatives of the fluctuations) does not depend on $r$ and thus
is independent of the fluctuations.}. Let us next rewrite the Born-Infeld
determinant as:
\beq
\sqrt{-\det\,(g+ F)}=\sqrt{-\det \left({\cal G}+{\cal
F}\right)}\,\sqrt{\det\,(1+X)}\,,
\label{cp3detXp2}
\eeq
where the matrix $X$ is given by:
\beq
X\equiv\left({\cal G}+{\cal  F}\right)^{-1}\,\left(g^{(f)}+f\right)\,.
\label{cp3matrixX}
\eeq
We shall evaluate the right-hand side of (\ref{cp3detXp2}) by expanding
it  in powers of $X$ by means of eq. (\ref{detX-expansion}). In order to
evaluate more easily the trace of the powers of $X$ appearing on the
right-hand side of this equation, let us separate the symmetric and
antisymmetric part in the inverse of the matrix ${\cal G}\,+\,{\cal  F}$:
\beq
\left(\,{\cal G}+{\cal  F}\,\right)^{-1}=\hat{\cal G}^{-1}+{\cal J}\,,
\eeq
where:
\beq
\hat{\cal G}^{-1}\equiv{1\over ({\cal G}+{\cal  F})_S}\;,\qquad
{\cal J}\equiv{1\over ({\cal G}+{\cal  F})_A}\,.
\eeq
Notice that $\hat{\cal G}$ is just  the open string metric which, for the
case at hand, is given by:
\beq
\hat{\cal G}_{ab}\,d\xi^a\,d\xi^b=h^{-1}\,dx_{1,p-1}^2+h\,\left(\,1+
{q^2\over \rho^4 h^2}\,\right)\left(\,d\rho^2+\rho^2\,d\Omega_2^2\,
\right)\,.
\label{cp3openstrmetric}
\eeq
Moreover,  the antisymmetric matrix ${\cal J}$ takes the form:
\beq
{\cal J}^{\theta\varphi}=-{\cal J}^{\varphi\theta}=
-{1\over \sqrt{\tilde g}}\,\,{q\over q^2+\rho^4\,h^2}\,,
\eeq
where $\theta, \varphi$ are the standard polar coordinates on $S^2$ and 
$\tilde g=\sin^2\theta$ is the determinant of its round metric. It is now
straightforward to show that:
\beq
\tr \,X=h\,\,\hat{\cal  G}^{ab}\,\partial_a\chi^m\,\partial_b\chi^m+
{1\over h}\,\hat{\cal G}^{ab}\,\partial_a x\,\partial_b x+
{q\over q^2+\rho^4\,h^2}\,\left[2\rho^2\,\partial_\rho x+
{\epsilon^{ij}f_{ij}\over\sqrt{\tilde  g}}\right]\,,
\eeq
while, up to quadratic terms in the fluctuations,  $\tr\, X^2$ is given by:
\bear
&&\tr\, X^2=-f_{ab}f^{ab}+{2\over h}\,\,{q^2\over q^2+\rho^4h^2}\,
\left[\hat{\cal G}^{ab}\,\partial_a x\,\partial_b x+
\hat{\cal G}^{\rho\rho}\,(\partial_\rho x)^2\right]+\rc\rc
&&\qquad\qquad
+{q^2\over \left(q^2+\rho^4h^2\right)^2}\left[{1\over 2\tilde g}
\big(\epsilon^{ij}f_{ij}\big)^2-4\rho^2 \,{\epsilon^{ij}
\partial_i x\,f_{j\rho}\over\sqrt{\tilde g}}\right]\,,
\eear
where the indices $i,j$ refer to the directions along the $S^2$ and
$\epsilon^{ij}=\pm 1$.  Using these results one can readily compute the DBI
term of the lagrangian density. Dropping constant global factors that do not
affect the equations of motion, one gets:
\bear
&&{\cal L}_{DBI}=-\rho^2\,\sqrt{\tilde g}\Bigg[1+{q^2\over\rho^4\,
h^2}+{h\over 2}\left(1+{q^2\over \rho^4 h^2}\right)\,\hat{\cal
G}^{ab}\,\partial_a\chi^m\,\partial_b\chi^m+\rc\rc
&&\qquad\qquad\qquad\qquad
+{1\over 2 h}\,\hat{\cal G}^{ab}\,\partial_a x\,\partial_b x+
{1\over4}\left(1+{q^2\over \rho^4 h^2}\right)\,f_{ab}f^{ab}\Bigg]+\rc\rc
&&\qquad\qquad\qquad\qquad
+{A(\rho)\over 2}\,x\,\epsilon^{ij}\,f_{ij}-{q\sqrt{\tilde g}\over h^2}
\,\partial_{\rho}x-{q\over 2\rho^2 h^2}\,\epsilon^{ij}\,f_{ij}\,,
\eear
where the indices $a,b$ are raised with the open string metric $\hat{\cal G}$
and
$A(\rho)$ is the following function:
\beq
A(\rho)\equiv{d\over d\rho}\left[
{q^2\over h^2\left(q^2+\rho^4h^2\right)}\right]\,.
\eeq
To get the above expression of ${\cal L}_{DBI}$ we have integrated by 
parts and made use of
the Bianchi identity for the gauge field fluctuation:
\beq
\epsilon^{ij}\,\partial_i f_{j\rho}+{\epsilon^{ij}\over 
2}\,\partial_\rho f_{ij}=0\,.
\eeq
Similarly, the WZ term can be written as:
\beq
{\cal L}_{WZ}=\sqrt{\tilde g}\,{q^2\over \rho^2
h^2}+\sqrt{\tilde g}\, {q\over h^2}\,\partial_{\rho}x
+{q\over 2\rho^2 h^2}\,\epsilon^{ij}\,f_{ij}+
{\partial_{\rho}h\over h^3}\,x\epsilon^{ij}\,f_{ij}\,.
\eeq
By combining ${\cal L}_{DBI}$ and ${\cal L}_{WZ}$ and dropping the term
independent of the fluctuations,  we get that the  total lagrangian density
is given by:
\bear
&&{\cal L}=-\rho^2\,\sqrt{\tilde g}\,\Bigg[{h\over 2}\,
\left(1+{q^2\over \rho^4 h^2}\right)\,\hat{\cal G}^{ab}\,\partial_a
\chi^m\,\partial_b\chi^m+{1\over 2 h}\,\hat{\cal G}^{ab}\,\partial_a x\,
\partial_b x+\rc\rc
&&\qquad\qquad +{1\over 4}\left(1+{q^2\over \rho^4 h^2}\right)
f_{ab}f^{ab}\Bigg]-{C(\rho)\over 2}\,x\,\epsilon^{ij}f_{ij}\,.
\label{cp3dp+2flag}
\eear
In eq. (\ref{cp3dp+2flag}), and in what follows, the function
$C(\rho)$ is given by:
\beq
C(\rho)\equiv{d\over d\rho}\left[{\rho^4\over q^2+\rho^4h^2}\right]\,.
\label{cp3C}
\eeq
As it is manifest from (\ref{cp3dp+2flag}), the transverse scalars $\chi$
do not couple  to other fields, while the scalar $x$ is coupled to the
fluctuations $f_{ij}$ of the gauge field strength along the two-sphere. For
the fluxless case $q=0$ these equations were solved in ref. \cite{AR}.
There  it was shown that they give rise to a discrete meson mass spectrum,
which can be  computed numerically and, in the case of the D3-D5
intersection, analytically (see section \ref{cp2D3D5}). Let us examine
here the situation for
$q\not=0$.  The equation of motion of the transverse scalars $\chi$ that
follows from  (\ref{cp3dp+2flag}) is:
\beq
\partial_a\left[\sqrt{\tilde g}\,\rho^2 h\left(1+{q^2\over \rho^4 h^2}
\right)\hat{\cal G}^{ab}\,\partial_b\chi\right]=0\,.
\label{cp3chi-fluct}
\eeq
By using the explicit form of the open string  metric $\hat{\cal G}^{ab}$ (eq.
(\ref{cp3openstrmetric})), we can rewrite (\ref{cp3chi-fluct}) as:
\beq
\partial_{\rho}\,\left(\rho^2\partial_\rho\chi\right)+\left[
\rho^2h^2+{q^2\over \rho^2}\right]\,\partial^{\mu}\partial_{\mu}\,\chi+
\nabla^i\nabla_i\,\chi=0\,.
\eeq
Let us separate variables and
write the scalars in terms of the eigenfunctions of the laplacian in the
Minkowski and sphere parts of the brane geometry as:
\beq
\chi=e^{ikx}\, Y^l(S^2)\,\xi(\rho)\,\, ,
\eeq
where the product $kx$ is performed with the Minkowski metric and $l$ is the
angular momentum on the $S^2$. 
The fluctuation equation for the function $\xi$  is:
\beq
\partial_{\rho}\,(\,\rho^2\,\partial_{\rho}\,\xi\,)+
\left\{\left[R^{4\alpha}\,\frac{\rho^2}{(\rho^2+L^2)^{2\alpha}}+
\frac{q^2}{\rho^2}\right]M^2-l(l+1)\right\}\xi=0\,, 
\label{cp3radial-chi-fluct}
\eeq
where $M^2=-k^2$ is the mass of the meson. When the distance $L\not=0$ and 
$q=0$ eq.  (\ref{cp3radial-chi-fluct})
can be reduced to eq. (\ref{cp2fluc}) of the previous chapter which gives
rise to a set of normalizable solutions that occur for a discrete set of
values of $M$. However, the situation changes drastically when  the flux
is switched on. Indeed, let us  consider  the equation 
(\ref{cp3radial-chi-fluct}), when $L, q\not=0$, in the IR, \ie \ when
$\rho$ is close to zero. In this case, for small values of $\rho$, eq. 
(\ref{cp3radial-chi-fluct})  reduces to:
\beq
\partial_{\rho}\left(\rho^2\partial_{\rho}\,\xi\right)+\left[{q^2\,M^2
\over\rho^2}-l(l+1)\right]\xi\,=0\,,\qquad(\rho\approx 0)\,.
\label{cp3IRfluc}
\eeq
Eq.  (\ref{cp3IRfluc}) can be solved in terms of Bessel functions,
namely:
\beq
\xi={1\over \sqrt{\rho}}\,\,J_{\pm (l+{1\over 2})}\,\Bigg({qM\over \rho}
\Bigg)\,,\qquad(\rho\approx 0)\,.
\label{cp3Bessel}
\eeq
Near $\rho\approx 0$ the Bessel function (\ref{cp3Bessel}) oscillates
infinitely as:
\beq
\xi\approx e^{\pm i{qM\over \rho}}\,,\qquad(\rho\approx 0)\,.
\label{cp3IRfluct-behavior}
\eeq
The behavior (\ref{cp3IRfluct-behavior}) implies that the spectrum of
$M$ is  continuous and gapless. Actually, one can
understand this result by rewriting the function (\ref{cp3Bessel}) in
terms of the coordinate $x^p$ by using (\ref{cp3bending}). Indeed,
$\rho\approx 0$ corresponds to large $|x^p|$ and $\xi(x^p)$ can be
written in this limit as  a simple plane wave:
\beq
\xi\approx e^{\pm iMx^p}\,,\qquad(\,|x^p|\to\infty\,)\,.
\label{cp3planewave}
\eeq
Thus, the fluctuation spreads out of the defect locus at fixed $x^p$,
reflecting the fact that the bending has the effect of recombining, rather
than intersecting, the Dp-branes with the D(p+2)-branes. We
can understand this result by looking at the IR form of the open string
metric (\ref{cp3openstrmetric}), which, in view of (\ref{cp3dp+2flag}),
is the metric governing the dynamics of the fluctuations. One gets:
\beq
\hat{\cal G}_{ab}\,d\xi^a\,d\xi^b\approx{L^{2\alpha}\over R^{2\alpha}}
\left[dx_{1,p-1}^2+q^2\left({d\rho^2\over \rho^4}+{1\over \rho^2}\,
d\Omega_2^2\right)\right]\,,\qquad(\rho\approx 0)\,.
\label{cp3IRmetric}
\eeq
By changing variables from $\rho$ 
to  $u=q/\rho$, this metric can be written as:
\beq
{L^{2\alpha}\over R^{2\alpha}}\left[dx_{1,p-1}^2+ du^2+u^2\,
d\Omega_2^2\right]\,,
\label{cp3IRMinkowski}
\eeq
which is nothing but the $(p+3)$-dimensional Minkowski space and thus one
naturally expects to get plane waves  as in (\ref{cp3planewave}) as
solutions of the fluctuation equations. This fact is generic for all the
fluctuations of this system. Notice that the other fields in the
Lagrangian (\ref{cp3dp+2flag}) are coupled. However, as it was shown in
\cite{ARRxl}, they can be easily decoupled by generalizing the method
applied in \cite{AR} (and in section \ref{cp2D3D5} for the $p=3$ case) for
the fluxless case. Furthermore, as it happened there, the decoupled
fluctuation equations can be mapped to that satisfied by the scalars
$\chi$.  Thus, we conclude that the full mesonic mass spectrum is
continuous and gapless, as a consequence of the recombination of the color
and flavor branes induced by the worldvolume flux.

%
%
%
%
%
%
%
%
%

\setcounter{equation}{0}
\section{Codimension one defects in M-theory}
\label{cp3m2m5sc}

We will consider now a close relative in M-theory of the 
Dp-D(p+2) intersections, namely the M2-M5 intersection along one common
spatial dimension. The corresponding array is:
\beq
\begin{array}{cccccccccccl}
 &1&2&3& 4& 5&6 &7&8&9 &10 \nonumber \\
M2: & \times &\times &\_ &\_ &\_ & \_&\_ &\_ &\_ &\_    
\nonumber
\\ M5: &\times&\_&\times&\times&\times&\times&\_&\_&\_ &\_
\end{array}
\label{cp3M2M5intersection}
\eeq
Since this configuration can be somehow thought as the
uplift of the D2-D4 intersection  to eleven dimensions, 
we  expect a behavior similar to the one studied in section
\ref{cp3codim1sc}. Indeed, notice that the M5-brane induces a codimension
one defect in the M2-brane worldvolume. As in the previous examples we will
treat the highest dimensional brane (\ie\ the M5-brane) as a probe in the
background created by the lower dimensional object, which in this case is the
M2-brane. The near-horizon metric of the M2-brane background of
eleven-dimensional supergravity is:
\begin{equation}
ds^2=\frac{r^4}{R^4}dx_{1,2}^2+\frac{R^2}{r^2}d\vec{r}^{\,2}\,\, ,
\label{cp3M2metric}
\end{equation}
where $R$ is constant, $dx_{1,2}^2$ represents the Minkowski metric in the
directions $x^0, x^1, x^2$ of the M2-brane worldvolume and $\vec r$ is an
eight-dimensional vector transverse to the M2-brane. The metric
(\ref{cp3M2metric}) is the one of the $AdS_4\times S^7$ space, where the
radius of the $AdS_4$ ($S^7$) factor is $R/2$ ($R$). The actual value of
$R$ for a stack of $N$ coincident M2-branes was written in eq.
(\ref{cp2M2back-parameters}) of the previous chapter. In addition, this
background is endowed with a three-form potential $C^{(3)}$, whose
explicit expression is:
\beq
C^{(3)}={r^6\over R^6}\,\,dx^0\wedge dx^1\wedge dx^2\,.
\label{cp3C3M2}
\eeq

The dynamics of the M5-brane probe is governed by the so-called PST action
\cite{PST}. In the PST formalism the worldvolume fields are a three-form
field strength $F$ and an auxiliary scalar $a$. This action is given by
\cite{PST}:
\bear
S\,&=&\,T_{M5}\int d^6\xi\left[-\sqrt{-{\rm det}(g_{ij}+\tilde H_{ij})}+
{\sqrt{-{\rm det}g}\over 4\partial a\cdot\partial a}\,
\partial_i a\,(\star H)^{ijk}\,H_{jkl}\,\partial^l a\right]+\rc\rc
&&+T_{M5}\int\left[P[C^{(6)}]+{1\over 2}\,F\wedge\,P[C^{(3)}]\right]\,,
\label{cp3pstact}
\eear
where $T_{M5}=1/(2\pi)^5\,l_p^6$ is the tension of the M5-brane, $g$ is the
induced metric and $H$ is the following combination of the worldvolume
gauge field $F$ and the pullback of the three-form $C^{(3)}$:
\beq
H=F-P[C^{(3)}]\,.
\label{cp3m2hfield}
\eeq
Moreover, the field ${\tilde H}$ is defined as follows:
\beq
{\tilde H}^{ij}={1\over 3!\,\sqrt{-{\rm det}\,g}}\,
{1\over \sqrt{-(\partial a)^2}}\,
\epsilon^{ijklmn}\,\partial_k\,a\,H_{lmn}\,,
\label{cp3m2htild}
\eeq
and the worldvolume indices in (\ref{cp3pstact}) are lowered with the
induced metric $g_{ij}$. 

In order to study the embedding of the M5-brane in the M2-brane background,
let us introduce a more convenient set of coordinates. Let us split the
vector $\vec r$ as $\vec r=\,(\vec y,\vec z)$, where $\vec y=(y^1,\cdots,
y^4)$ is the position vector along the directions $3,\,4,\,5,\,6$  in the
array (\ref{cp3M2M5intersection}) and $\vec z=(z^1,\cdots, z^4)$
corresponds to the directions $7,\,8,\,9$ and $10$. Obviously, if
$\rho^2=\vec y\cdot
\vec y$, one has that 
$\vec r^{\,2}=\rho^2+\vec z^{\, 2}$ and 
$d\vec r^{\,2}=d\rho^2+\rho^2\,d\Omega_3^2+d\vec{z}^{\,2}$, where 
$d\Omega_3^2$ is the line element of a three-sphere. Thus, the metric
(\ref{cp3M2metric}) becomes:
\begin{equation}
ds^2=\frac{\big(\,\rho^2+\vec z^{\,2}\,\big)^2}{R^4}\,\,dx_{1,2}^2
+\frac{R^2}{\rho^2+\vec z^{\,2}}\,\left(\,d\rho^2+\rho^2d\Omega_3^2
+d\vec{z}^{\,2}\,\right)\,.
\end{equation}
We will now choose $x^0$, $x^1$, $\rho$ and the three angular coordinates
that parametrize $d\Omega_3^2$ as our worldvolume coordinates $\xi^i$.
Moreover, we will assume that the vector $\vec z$ is constant and we will
denote its modulus by $L$, namely:
\beq
|\,\vec z\,|=L\,.
\eeq
To specify completely the embedding of the M5-brane we must give  the form
of the remaining scalar $x^2$ as a function of the worldvolume coordinates.
As for the Dp-D(p+2) intersection, the simplest embedding is that with
$x^2=$ constant where the M5-brane introduces a defect consisting of a
flat wall located at a certain value of $x^2$ in the gauge theory living
on the M2-brane worldvolume. This configuration was studied in \cite{AR}
where a discrete mass spectrum for the transverse scalar fluctuations was
found. Analogously to what we have done in section
\ref{cp3codim1sc}, we will consider here a more complex embedding where
the M5-brane probe is tilted along $x^2$. For simplicity we will assume
that
$x^2$ only depends on the radial coordinate $\rho$, \ie\ that:
\beq
x^2=x(\rho)\,.
\eeq
Moreover, we will switch on a magnetic field $F$ along the three-sphere of
the M5-brane worldvolume, in the form:
\begin{equation}
F=q\,{\rm Vol}\,(\,S^3\,)\,\, ,
\label{cp3M5flux}
\end{equation}
where $q$ is a constant and ${\rm Vol}\,(\,S^3\,)$ is the volume form of the
worldvolume three-sphere. Notice that the induced metric for this
configuration is given by:
\beq
g_{ij}\,d\xi^i d\xi^j={\left(\rho^2+L^2\right)^2\over
R^4}\,\,dx^{2}_{1,1}
+{R^2\over \rho^2+L^2}\left\{\left(1+{\left(\rho^2+L^2\right)^3\over
R^6}\,(x')^2\right)d\rho^2+\rho^2\,d\Omega_3^2\right\}\,.
\label{cp3inducedmetricM5}
\eeq

In order to write the PST action for our ansatz we must specify the value of
the PST scalar $a$. As pointed out in ref. \cite{PST} the field $a$ can be
eliminated by gauge fixing, at the expense of losing manifest covariance.
Here we will choose a gauge such that the auxiliary PST scalar is:
\beq
a=x_1\,.
\eeq
It is now straightforward to prove that the only non-vanishing component of
the field $\tilde H$ is:
\beq
\tilde H_{x^0\rho}=-{i\over R^4}\,{\left(\rho^2+L^2\right)^2\over \rho^3}
\left(1+{\left(\rho^2+L^2\right)^3\over R^6}\,(x')^2\right)^{{1\over 2}}
\,q\,.
\label{cp3M2tildeHexp}
\eeq
Using these results we can write the PST action (\ref{cp3pstact}) as:
\bear
&&S=-2\pi^2\,T_{M5}\int d^2x\,d\rho\Bigg[\rho^3\,
\sqrt{1+{\left(\rho^2+L^2\right)^3\over R^6}\,(x')^2}\;
\sqrt{1+{\left(\rho^2+L^2\right)^3\over R^6}\,{q^2\over \rho^6}}+\rc\rc
&&\qquad\qquad\qquad\qquad\qquad\qquad\qquad\qquad\qquad
+{\left(\rho^2+L^2\right)^3\over R^6}\,q\,x'\Bigg]\,.
\label{cp3PSTansatz}
\eear
Let ${\cal L}$ be the lagrangian density for the PST action, which we can
take as given by the expression inside the brackets in
(\ref{cp3PSTansatz}).  Since $x$ does not appear explicitly in the
action,  one can immediately write a first integral of the equation of
motion of
$x(\rho)$, namely:
\beq
{\partial {\cal L}\over \partial x'}={\rm constant}\,.
\label{PSTcyclic}
\eeq
By setting the constant on the right-hand side of (\ref{PSTcyclic}) equal
to zero, this equation reduces to a simple first order equation for 
$x(\rho)$,
\ie:
\begin{equation}
x'=-\frac{q}{\rho^3}\,,
\label{cp3BPSx}
\end{equation}
which can be immediately integrated to give:
\begin{equation}
x(\rho)=\bar x+\frac{q}{2\rho^2}\,,
\label{cp3M2x-explicit}
\end{equation}
where $\bar x$ is a constant. Notice that the flux parametrized by $q$ 
induces a bending of the M5-brane, which is characterized by the non-trivial
dependence of $x$ on the holographic coordinate $\rho$. Actually, when the
first order eq. (\ref{cp3BPSx}) holds, the two square roots in
(\ref{cp3PSTansatz}) are equal and there is a cancellation with the last
term in (\ref{cp3PSTansatz}). Indeed, the  on-shell action takes the
form:
\beq
S=-2\pi^2T_5\int d^2x\,d\rho \,\rho^3\,\, ,
\eeq
which is  independent of the M2-M5 distance $L$. This is usually a signal of
supersymmetry and, indeed, we will verify in appendix \ref{sam2m5kappa}
that the embeddings in which the flux and the bending are related as in 
(\ref{cp3BPSx}) are kappa symmetric. Thus, eq. (\ref{cp3BPSx}) can be
regarded as the first order BPS equation required by supersymmetry.
Notice also that the three-form flux (\ref{cp3M5flux}) induces M2-brane
charge in the M5-brane worldvolume, as it is manifest from the form of
the PST action (\ref{cp3pstact}). In complete analogy with the Dp-D(p+2)
system, we can interpret the present M-theory configuration in terms of
M2-branes that recombine with the M5-brane. Moreover, in order to gain
further insight on the effect of the bending, let us rewrite the induced
metric (\ref{cp3inducedmetricM5}) when the explicit form of $x(\rho)$
written in eq. (\ref{cp3M2x-explicit}) is taken into account. One gets:
\beq
g_{ij}\,d\xi^i d\xi^j={\left(\rho^2+L^2\right)^2\over
R^4}\,dx^{2}_{1,1}+{R^2\over \rho^2+L^2}\left\{\left(1+{q^2\over R^6}\,
{\left(\rho^2+L^2\right)^3\over\rho^6}\right)d\rho^2+\rho^2\,
d\Omega_3^2\right\}\,.
\label{cp3AdS3-M5metric}
\eeq
From (\ref{cp3AdS3-M5metric}) one readily notices that the UV induced
metric at 
$\rho\to\infty$ (or, equivalently when the M2-M5 distance L is zero) takes
the form $AdS_3 (R_{eff}/2)\times S^3 (R)$, where the $AdS_3$ radius 
$R_{eff}$ depends on the flux as:
\beq
R_{eff}=\Bigg(\,1+
{q^2\over R^6}\,\Bigg)^{1\over 2}\,R\,.
\label{cp3M2Reff}
\eeq
Therefore, our M5-brane is wrapping an  $AdS_3$ submanifold of the  $AdS_4$
background. Actually, there are infinite ways of embedding an  $AdS_3$ within
an $AdS_4$ space and the bending of the probe induced by the flux is
selecting one particular case of these embeddings. In order to shed light on
this, let us suppose that we have an $AdS_4$ metric of the form:
\beq
ds^2_{AdS_4}={\rho^4\over R^4}\,\,dx^2_{1,2}+
{R^2\over \rho^2}\,\,d\rho^2\,.
\label{cp3AdS4metric}
\eeq
Let us now change variables from $(x^{0,1}, x^2, \rho)$ to 
$(\hat x^{0,1}, \varrho, \eta)$, as follows:
\beq
x^{0,1}=2\,\hat x^{0,1}\,,\qquad\quad
x^2=\bar x+{2\over \varrho}\,\tanh\eta\,,\qquad\quad
\rho^2={R^3\over 4}\,\varrho\cosh\eta\,,
\label{cp3M2changevariables}
\eeq
where $\bar x$ is a constant. In these new variables the $AdS_4$ metric
(\ref{cp3AdS4metric}) can be written as a foliation by $AdS_3$ slices,
namely:
\beq
ds^2_{AdS_4}={R^2\over 4}\,
(\,\cosh^2\eta\,ds^2_{AdS_3}+d\eta^2\,)\,,
\label{cp3M2foliation}
\eeq
where $ds^2_{AdS_3}$ is given by:
\beq
ds^2_{AdS_3}=\varrho^2\,
\big(\,-(d\hat x^{0})^2+(d\hat x^{1})^2\,\big)+
{d\varrho ^2\over \varrho^2}\,.
\eeq
Clearly the $AdS_3$ slices in (\ref{cp3M2foliation}) can be obtained by
taking
$\eta={\rm constant}$. The radius of such $AdS_3$ slice is $R_{eff}/2$, with:
\beq
R_{eff}=R\,\,\cosh\eta\,.
\label{cp3M2sliceradius}
\eeq
Moreover, one can verify easily by using the change of variables
(\ref{cp3M2changevariables})  that our embedding
(\ref{cp3M2x-explicit}) corresponds to one of such 
$AdS_3$ slices with:
\beq
\eta=\eta_q=\sinh^{-1}\,\Big({q\over R^3}\Big)\,.
\eeq
Furthermore, one can check that the $AdS_3$ radius $R_{eff}$ of eq. 
(\ref{cp3M2sliceradius}) reduces to (\ref{cp3M2Reff}) when
$\eta=\eta_q$.

\subsection{Fluctuations}

Let us now study the fluctuations of the M2-M5 intersection. For simplicity 
we will focus on the fluctuations of 
the transverse scalars which, without loss of generality,  
we will parametrize as:
\beq
z^1=L+\chi^1\,,\qquad\quad z^m=\chi^m\,,\qquad (m=2,\cdots,4)\,.
\eeq
Let us substitute this ansatz in the PST action and keep 
up to second order terms in the fluctuation $\chi$. As the calculation is
very similar to the one performed in subsection \ref{cp3Dp+2fluct},
we skip the details and give the final result for the effective lagrangian of
the fluctuations, namely:
\beq
{\cal L}=-\rho^3\,\sqrt{\tilde g}\,\,{R^2\over \rho^2+L^2}\left[1+
{q^2\over R^6}\,{\left(\rho^2+L^2\right)^3\over\rho^6}\right]
\hat{\cal G}^{ij}\,\partial_i\chi\,\partial_j\chi\,,
\label{cp3lag-fluc-M2M5}
\eeq
where $\tilde g$ is the determinant of the round metric of the $S^3$ and
$\hat{\cal G}_{ij}$ is the following effective metric on the M5-brane
worldvolume:
\beq
\hat{\cal G}_{ij}\,d\xi^i\,d\xi^j={\left(\rho^2+L^2\right)^2\over R^4}\,
dx^{2}_{1,1}+{R^2\over \rho^2+L^2}\left(1+{q^2\over R^6}
{\left(\rho^2+L^2\right)^3\over \rho^6}\right)\left(d\rho^2+\rho^2\,
d\Omega_3^2\right)\,.
\label{cp3effect-metricM2M5}
\eeq
Notice the close analogy with the Dp-D(p+2) system studied in subsection
\ref{cp3Dp+2fluct}. Actually, (\ref{cp3effect-metricM2M5}) is the
analogue of the open string metric in this case. The equation of motion for
the scalars can be derived straightforwardly from the lagrangian density 
(\ref{cp3lag-fluc-M2M5}). For $q=0$ this equation was integrated in ref.
\cite{AR} where the meson mass spectrum was also computed. This fluxless
spectrum is discrete and displays a mass gap. As happened with the
codimension one defects in type II theory studied in section
\ref{cp3codim1sc}, the situation changes drastically when $q\not=0$. To
verify this fact let us study the form of the effective metric
(\ref{cp3effect-metricM2M5}) in the UV ($\rho\to\infty$) and in the IR
($\rho\to 0$). After studying this metric when $\rho\to\infty$, one easily
concludes that the UV is of the form $AdS_3 (R_{eff}/2)\times S^3
(R_{eff})$, where $R_{eff}$ is just the effective radius with flux written
in (\ref{cp3M2Reff}). Thus, the effect of the flux in the UV is just a
redefinition of the $AdS_3$ and
$S^3$ radii of the  metric governing the fluctuations. On the contrary,
for $q\not=0$ the behavior of this metric in the  IR  changes
drastically with respect to the fluxless case. Indeed, for $\rho\approx
0$ the metric (\ref{cp3effect-metricM2M5}) takes the form:
\beq
{L^{4}\over R^{4}}\left[dx_{1,1}^2+q^2\left({d\rho^2\over \rho^6}+{1\over
\rho^4}\,d\Omega_2^2\right)\right]\;,\qquad(\rho\approx 0)\,.
\label{IR-M2M5metric}
\eeq
Notice the analogy of (\ref{IR-M2M5metric}) with the IR metric
(\ref{cp3IRmetric}) of the Dp-D(p+2) defects. Actually, the IR limit of
the equation of motion of the fluctuation can be integrated, as in
(\ref{cp3Bessel}),  in terms of Bessel functions, which for
$\rho\approx 0$ behave as plane waves of the form $e^{\pm i Mx}$, where $x$
is the function (\ref{cp3M2x-explicit}). Notice that $\rho\approx 0$
corresponds to large  $x$ in  (\ref{cp3M2x-explicit}). Thus, the
fluctuations spread out of the defect and oscillate infinitely in the IR
and, as a consequence, the mass spectrum is continuous and gapless. In
complete analogy with the Dp-D(p+2) with flux, this is a consequence of
the recombination of the M2- and M5-branes and should be understood
microscopically in terms of dielectric multiple M2-branes polarized into
an M5-brane, once such an action is constructed. 

%
%
%
%
%
%
%
%
%

\setcounter{equation}{0}
\section{The codimension two defect}
\label{cp3codim2sc}

We now analyze the codimension two defect, which can be engineered in type
II string  theory as a Dp-Dp intersection over $p-2$ spatial dimensions.
We will consider a single Dp$'$-brane intersecting a stack of
$N$ Dp-branes, according to the array:
\beq
\begin{array}{ccccccccccl}
 &1&\cdots&p-2& p-1& p&p+1 &p+2&\cdots&9 & \nonumber \\
Dp: & \times &\cdots &\times &\times &\times & \_&\_ &\cdots &\_ &    
\nonumber
\\ Dp\,': &\times&\cdots&\times&\_&\_&\times&\times&\cdots&\_ &
\end{array}
\label{cp3DpDpintersection}
\eeq
In the
limit of large $N$ we can think of the system as a probe Dp$'$-brane in
the near-horizon geometry of the Dp-branes given by (\ref{cp1dpsolmtrc}).
It is clear from the array (\ref{cp3DpDpintersection}) that the
Dp$'$-brane produces a defect of codimension two  in the field theory
dual to the stack of Dp-branes. The defect field theory dual to the
D3-D3 intersection was studied in detail in ref. \cite{CEGK} (see also
\cite{Kirschphd}).  Notice also that this same D3-D3
intersection was considered in \cite{Gukov:2006jk} in connection with
the surface operators of ${\cal N}=4$ super Yang-Mills theory, in the
context of the geometric Langlands program. 

In order to describe the dynamics of the Dp$'$-brane probe, let us relabel the 
$x^{p-1}$ and $x^{p}$ coordinates appearing in the metric
(\ref{cp1dpsolmtrc}), and corresponding to the directions $p-1$ and $p\,$
in (\ref{cp3DpDpintersection}), as:
\beq
\lambda^1\equiv x^{p-1}\,,
\qquad\quad
\lambda^2\equiv x^{p}\,.
\eeq
Moreover, we will split the coordinates $\vec r$ transverse to the Dp-branes as
$\vec r\,=\,(\vec y\,,\vec z)$, where $\vec y\,=\,(y^1,y^2)$ corresponds
to the $p+1$ and $p+2$ directions in (\ref{cp3DpDpintersection}) and $\vec
z\,=\,(z^1,\cdots, z^{7-p})$  to the remaining transverse coordinates.
With this split of coordinates the background metric (\ref{cp1dpsolmtrc})
reads:
\beq
ds^2=\left[{\vec y^{\,2}+\vec z^{\,2}\over
R^2}\right]^{\alpha}\left(dx_{1,p-2}^2+d \vec\lambda^{\,2}\right)
+ \left[{R^2\over\vec y^{\,2}+\vec z^{\,2}}\right]^{\alpha}
\left(d\vec y^{\,2}+d \vec z^{\,2}\right)\,,
\label{cp3Dpmetric}
\eeq
where $dx_{1,p-2}^2$ is the Minkowski metric in the coordinates
$x^0,\cdots, x^{p-2}$ and $\alpha$ has been defined in
(\ref{cp3dpdp+4bckgr}).

\subsection{Supersymmetric embeddings}

As we have shown in section \ref{cp1sskappasym}, the supersymmetric
configurations of a D-brane probe in a given  background are those for
which the following condition is satisfied \cite{bbs}:
\beq
\Gamma_{\kappa}\,\epsilon=\epsilon\,,
\label{cp3Gammakappa}
\eeq
where $\epsilon$ is a Killing spinor of the background and
$\Gamma_{\kappa}$ is a  matrix which depends on the
embedding of the probe. In particular, $\Gamma_{\kappa}$ depends on
the induced gamma matrices which are defined as:
\beq
\gamma_{a}=\partial_{a}\,X^M\,E^{\underline N}_{M}\,\Gamma_{\underline
N}\,,
\label{cp3inducedgamma}
\eeq
where $a$ is a worldvolume index, $X^M$ are the coordinates of the
ten-dimensional spacetime and $E^{\underline N}_{M}$ are the coefficients
appearing in the expression of the frame one-forms in terms of the
differentials of the coordinates. Recall that the embedding of a D-brane
is given by the functions $X^M(\xi^a)$, where $\xi^a$ are the coordinates
of the worldvolume of the brane.

To study the embeddings of the Dp$'$-brane probe in the background  
(\ref{cp3Dpmetric}), let us consider
$\xi^a\,=\,(x^0,\cdots,x^{p-2},y^1,y^2)$ as worldvolume coordinates. In
this approach $\vec \lambda$ and $\vec z$ are scalar fields that
characterize the embedding. Actually, we will restrict ourselves to the
case in which $\vec
\lambda$ depends only on the $\vec y$ coordinates (\ie\ 
$\vec\lambda\,=\,\vec\lambda(\,\vec y\,)$) and the transverse separation 
$|\,\vec z\,|$ is constant, \ie\ $|\,\vec z\,|=L$. The induced
gamma matrices $\gamma_{x^\mu}$ ($\mu=0,\cdots, p-2$) and $\gamma_{y^i}$ ($i=1,2$) 
can be computed from eq. (\ref{cp3inducedgamma}), with the result:
\bear
&&\gamma_{x^\mu}=\left[{\rho^2+L^2\over R^2}\right]^{{\alpha\over 2}}\,
\Gamma_{x^\mu}\,,\rc\rc
&&\gamma_{y^i}=
\left[{R^2\over \rho^2+L^2}\right]^{{\alpha\over 2}}\,\Gamma_{y^i}+
\left[{\rho^2+L^2\over R^2}\right]^{{\alpha\over 2}}\,
\left[\partial_i\lambda^1\,\Gamma_{\lambda^1}+\partial_{i}\lambda^2
\,\Gamma_{\lambda^2}\right]\,,
\label{DpDpinducedgammas}
\eear
where $\partial_i\equiv \partial_{y^i}$ and, as before, we have 
defined $\rho^2=\vec y\cdot\vec y$.   To simplify matters, let us assume
that $p$ is odd so that we will be working on the type IIB theory. The
general expression of the kappa symmetry matrix $\Gamma_{\kappa}$ is
written in eq. (\ref{kpGammakpIIB}). For the present case
this matrix reads:
\beq
\Gamma_{\kappa}={1\over \sqrt{-\det(g)}}
\left[{\rho^2+L^2\over R^2}\right]^{{(p-1)\alpha\over 2}}\,
(\sigma_3)^{{p-3\over 2}}\,(i\sigma_2)\,
\Gamma_{x^0\cdots x^{p-2}}\,\,\gamma_{y^1y^2}\,.
\label{GammakappaDpDp}
\eeq
The antisymmetrized product $\gamma_{y^1y^2}$ can be straightforwardly 
computed from the expression of the $\gamma_{y^i}$ matrices in
(\ref{DpDpinducedgammas}). One gets:
\bear
&&\left[{\rho^2+L^2\over R^2}\right]^{\alpha}\,
\gamma_{y^1y^2}=\Gamma_{y^1y^2}+
\left[{\rho^2+L^2\over R^2}\right]^{2\alpha}\,
\left(\partial_1\lambda^1\,\partial_2\lambda^2-
\partial_1\lambda^2\,\partial_2\lambda^1\right)\,
\Gamma_{\lambda^1\lambda^2}+\rc\rc
&&+\left[{\rho^2+L^2\over R^2}\right]^{\alpha}\,
\left[\partial_2\lambda^1\,\Gamma_{y^1\lambda^1}+
\partial_2\lambda^2\,\Gamma_{y^1\lambda^2}-
\partial_1\lambda^1\,\Gamma_{y^2\lambda^1}-
\partial_1\lambda^2\,\Gamma_{y^2\lambda^2}\right]\,.
\label{cp3gammay1y2}
\eear
Let us now use this expression to fulfill the kappa symmetry condition
(\ref{cp3Gammakappa}). For a generic value of $p$ the Killing spinors of
the Dp-brane background satisfy the projection condition:
\beq
(\sigma_3)^{{p-3\over 2}}\,\, (i\sigma_2)\,
\Gamma_{x^0\cdots x^{p-2}}\,\Gamma_{\lambda^1\lambda^2}\,
\epsilon=\epsilon\,.
\label{cp3Dp-projection}
\eeq
In addition, we will also impose the projection corresponding to the Dp$'
$- brane probe, namely:
\beq
(\sigma_3)^{{p-3\over 2}}\,\, (i\sigma_2)\,
\Gamma_{x^0\cdots x^{p-2}}\,\Gamma_{y^1y^2}\,\epsilon=\epsilon\,.
\label{cp3Dp'projection}
\eeq
Notice that (\ref{cp3Dp-projection}) and (\ref{cp3Dp'projection}) are
compatible,  as it should for a supersymmetric intersection. Moreover,
they can be combined to give:
\beq
\Gamma_{y^1y^2}\,\epsilon=\Gamma_{\lambda^1\lambda^2}\,\epsilon\,,
\label{cp3DpDpcondition}
\eeq
which implies that:
\bear
&&\left[{\rho^2+L^2\over R^2}\right]^{\alpha}\,
\gamma_{y^1y^2}\,\epsilon=\left\{1+
\left[{\rho^2+L^2\over R^2}\right]^{2\alpha}
\left(\partial_1\lambda^1\,\partial_2\lambda^2-
\partial_1\lambda^2\,\partial_2\lambda^1\right)\right\}
\Gamma_{\lambda^1\lambda^2}\,\epsilon+\rc\rc
&&\qquad
+\,\left[{\rho^2+L^2\over R^2}\right]^{\alpha}\,
\left[(\partial_2\lambda^1+\partial_1\lambda^2)
\Gamma_{y^1\lambda^1}+(\partial_2\lambda^2-\partial_1\lambda^1)
\Gamma_{y^1\lambda^2}\right]\,\epsilon\,.
\label{secondgammay1y2}
\eear
We can now use this result to compute $\Gamma_{\kappa}\,\epsilon$, 
where $\Gamma_{\kappa}$ is given in eq. (\ref{GammakappaDpDp}). By using
the condition (\ref{cp3Dp-projection}) one easily gets that the terms of
the first line of the right-hand side of (\ref{secondgammay1y2}) give
contributions proportional to the identity matrix, while those on the
second line of (\ref{secondgammay1y2}) give rise to terms containing
matrices that do not act on $\epsilon$ as the identity unless we impose
some extra projections which would reduce the amount of preserved
supersymmetry. Since we do not want this to happen, we require that the
coefficients of
$\Gamma_{y^1\lambda^1}$ and $\Gamma_{y^1\lambda^2}$ in
(\ref{secondgammay1y2}) vanish, \ie:
\beq
\partial_1\lambda^1=\partial_2\lambda^2\,,\qquad\quad
\partial_2\lambda^1=-\partial_1\lambda^2\,.
\label{cp3CR}
\eeq
Notice that eq. (\ref{cp3CR}) is nothing but the Cauchy-Riemann
equations.  Indeed, let us define the following complex combinations of
worldvolume coordinates and scalars
\footnote{The complex worldvolume coordinate $Z$ should not be confused with the real transverse scalars $\vec z$. Notice also that $\rho^2=|Z|^2$. }:
\beq
Z=y^1+iy^2\,,\qquad\quad W=\lambda^1+i\lambda^2\,.
\eeq
In addition, if we define the holomorphic and antiholomorphic derivatives as:
\beq
\partial={1\over 2}\,(\partial_1-i\partial_2)\;,\qquad\quad
\bar\partial={1\over 2}\,(\partial_1+i\partial_2)\,,
\eeq
then  (\ref{cp3CR}) can be written as:
\beq
\bar\partial\,W=0\,,
\eeq
whose general solution is an arbitrary holomorphic function of $Z$, namely:
\beq
W=W(Z)\,.
\eeq
It is also straightforward to check that for these holomorphic embeddings the induced metric 
takes the form:
\beq
\left[{\rho^2+L^2\over R^2}\right]^{\alpha}\,dx^2_{1,p-2}+
\left[{R^2\over
\rho^2+L^2}\right]^{\alpha}\left[1+\left[{\rho^2+L^2\over
R^2}\right]^{2\alpha}\,\partial W\bar\partial \bar W\right]dZ\,d\bar Z\,,
\label{DpDp-ind-metric-holo}
\eeq
whose determinant is:
\beq
\sqrt{-\det (g)}=\left[{\rho^2+L^2\over R^2}\right]^{{(p-3)\alpha\over
2}}\left[1+\left[{\rho^2+L^2\over R^2}\right]^{2\alpha}\,\partial W\bar
\partial\bar W\right]\,.
\eeq
Using this result one can easily verify that the condition $\Gamma_{\kappa}\epsilon=\epsilon$ is indeed satisfied. Moreover, 
for  these holomorphic embeddings the DBI lagrangian density
(see eq. (\ref{cp1DBI})) takes the form:
\beq
{\cal L}_{DBI}=-T_p\,e^{-\phi}\,\sqrt{-\det (g)}=-T_p
\left[1+\left[{\rho^2+L^2\over R^2}\right]^{2\alpha}\,
\partial W\bar\partial \bar W\right]\,,
\label{cp3DpDpLDBI}
\eeq
where we have used the value of $e^{-\phi}$ for the Dp-brane background 
displayed in eq. (\ref{cp1dpsoldiltn}). On the other hand, from the form
of the RR potential 
$C^{(p+1)}$ written in (\ref{cp1dpsolpot}) one can readily check that,
for these holomorphic embeddings,  the WZ piece of the lagrangian (eq.
(\ref{cp1WZ})) can be written as:
\beq
{\cal L}_{WZ}=T_p\left[{\rho^2+L^2\over R^2}\right]^{2\alpha}\,
\partial W\bar\partial \bar W\,,
\eeq
and cancels against the second term of ${\cal L}_{DBI}$ (see
eq. (\ref{cp3DpDpLDBI})). Thus, once again, the on-shell action is
independent of the distance $L$, a result which is a consequence of
supersymmetry and holomorphicity.

Notice that, from the point of view of supersymmetry, any holomorphic
 curve $W(Z)$ is allowed. Obviously, we could have $W=$ constant. In this
case the probe sits at a particular constant point of its transverse
space and does not recombine with branes of the background. If, on the
contrary, $W(Z)$  is not constant, Liouville theorem ensures us that it
cannot be bounded in  the whole complex plane. The points at which $|W|$
diverge are spikes of the probe profile,  and one can interpret them  as
the points where the probe and background branes merge. Observe that, as
opposed to the other cases studied in this chapter, the non-trivial
profile of the embedding is not induced by the addition of any
worldvolume field. Thus,  we are not dissolving any further charge in
the probe brane and a dielectric interpretation is not possible now. 

The field theory dual  for the $p=3$ system has been worked out 
in refs.  \cite{CEGK,Kirschphd}, and we have described
its field content in subsection \ref{cp2D3D3dict} of the previous chapter.
It corresponds to two ${\cal N}=4$ four-dimensional theories coupled to
each other through a two-dimensional defect that hosts a bifundamental
hypermultiplet. The Coulomb branch is realized by taking the embedding
$W=$ constant: the configuration thoroughly analyzed in section
\ref{cp2D3D3sc}. Alternatively, one can seek
for a Higgs branch arising from the corresponding $D$- and $F$-flatness
conditions of the supersymmetric defect theory.  Actually, it was shown in
\cite{CEGK, Kirschphd} that this Higgs branch corresponds to the embedding
$W=c/Z$, where $c$ is a constant. Interestingly, only for these embeddings
the induced UV metric is of the form $AdS_3\times S^1$. Indeed, one can
check that the metric (\ref{DpDp-ind-metric-holo}) for $p=3$ (and then
$\alpha=1$) and for the profile $W=c/Z$ reduces in the UV to that of the
$AdS_3\times S^1$ space, where the two factors have the same radius
$R_{eff}=\sqrt{1+{c^2\over R^4}}\,\,R$. Thus, as in the M2-M5 intersection
of section \ref{cp3m2m5sc}, the constant $c$ parametrizes the particular
$AdS_3\times S^1$ slice of the $AdS_5\times S^5$ space that is occupied
by our D3-brane probe.

\subsection{Fluctuations of the Dp-Dp intersection}

Let us now study the fluctuations around the previous configurations.  
We will concentrate on the fluctuations of the scalars transverse to both
types of branes, \ie\ those along the $\vec z$ directions. Let $\chi$ be
one of such fields. Expanding the action up to quadratic order in the
fluctuations it is easy to see that the lagrangian density for  $\chi$ is:
\beq
{\cal L}=-\left[{R^2\over \rho^2+L^2}\right]^{\alpha}
\left[1+\left[{\rho^2+L^2\over R^2}\right]^{2\alpha}\,
\partial W\bar\partial \bar W\right]\,{\cal G}^{ab}\,\partial_a\chi\,
\partial_b\chi\,,
\eeq
where ${\cal G}_{ab}$ is the induced metric (\ref{DpDp-ind-metric-holo}).
Let us parametrize the complex variable $Z$  in terms of polar
coordinates  as $Z=\rho\, e^{i\theta}$ and let us separate variables  in
the fluctuation equation as:
\beq
\chi=e^{ikx}\,e^{il\theta}\,\,\xi(\rho)\,,
\eeq
where the product $kx$ is performed with the Minkowski metric of the defect. 
If $M^2=-k^2$, the equation of motion for the radial
 function $\xi(\rho)$ takes the form:
\beq
\left[\left[{R^2\over \rho^2+L^2}\right]^{2\alpha}\left[1+
\left[{\rho^2+L^2\over R^2}\right]^{2\alpha}\,
\partial W\bar\partial \bar W\right]M^2-{l^2\over\rho^2}+{1\over\rho}\,
\partial_{\rho}\left(\rho\,\partial_\rho\right)\right]\xi(\rho)=0\,.
\label{cp3DpDpflucteq}
\eeq
For $W=$ constant, eq. (\ref{cp3DpDpflucteq}) was studied in \cite{AR}
(and, for $p=3$, in section \ref{cp2D3D3sc}), where it was shown that it
gives rise to a mass gap and a discrete spectrum of $M$. As in the case
of the codimension one defects, this conclusion changes completely when
we go to the Higgs branch. Indeed, let us consider the embeddings with
$W\sim 1/Z$. One can readily prove that for
$\rho\to\infty$ the function $\xi(\rho)$ behaves as 
$\xi(\rho)\sim c_1\rho^{l}+c_2\rho^{-l}$, 
which is exactly the same behavior as in the 
$W=$ constant case.  However, in the opposite limit $\rho\to 0$ 
the fluctuation equation can be solved in terms of Bessel functions which
oscillate infinitely as $\rho\to0$. Notice that, for our Higgs branch
embeddings, $\rho\to 0$ means $W\to\infty$ and, therefore, the
fluctuations are no longer localized at the defect, as it happened in the
cases of the Dp-D(p+2) and M2-M5 intersections. Thus we conclude that,
also in this case, the mass gap is lost and the spectrum is continuous. 

%
%
%
%
%
%
%
%
%

\section{Discussion}
\label{cp3scdiscuss}
In this chapter we have studied the holographic description of the Higgs
branch of a large class of theories with fundamental matter.  These 
theories are embedded in string theory as supersymmetric systems of
intersecting branes. The strings joining both kinds of  branes 
give rise to bifundamental matter confined to the
intersection, which once the suitable field theory limit is taken,
becomes fundamental matter with a flavor symmetry.

The general picture that emerges from our results is that the Higgs phase
is realized by recombining both types of intersecting branes. From the
point of view of the higher dimensional flavor brane the recombination
takes place when a suitable embedding is chosen and/or some flux of the
worldvolume gauge field is switched on. This  flux is
dissolving color brane charge in the flavor branes and, thus,  it is
tempting to search for a microscopic description from the point of view
of those dissolved branes. Indeed, we have seen that the vacuum
conditions of the dielectric description (when this description is
available) match exactly the $F$- and $D$-flatness constraints that give 
rise to the Higgs phase on the field theory side, which gives support to
our holographic description of the Higgs branch.

The first case studied was the Dp-D(p+4) intersection, where the flavor 
D(p+4)-brane fills completely the worldvolume directions of the color 
Dp-brane. Following \cite{EGG}, we argued that the holographic description
of the Higgs branch of this system corresponds to having a self-dual
gauge field along the directions of the worldvolume of the D(p+4)-brane
that are orthogonal to the Dp-brane. To confirm this statement we have
worked out in detail the microscopic description of this system and we
have computed the meson mass spectra as a function of the quark VEV.

We also analyzed other intersections that are dual to gauge theories
containing defects of non-vanishing codimension. The paradigmatic example
of these theories is the Dp-D(p+2) system, where a detailed microscopic
description can be found. Other cases include the M2-M5 intersection in
M-theory as well as the Dp-Dp system, which gives rise to a codimension
two defect. In this latter case the field theory limit does
not decouple the flavor symmetry, so we actually have a $SU(N)\times SU(M)$
theory. In addition, the profile of the intersection is only constrained to
be holomorphic in certain coordinates, but is otherwise unspecified. In
any case, it turns out that conformal invariance in the UV is preserved only
for two particular curves, which can be shown to correspond to the Coulomb
and Higgs phases (see \cite{CEGK}). In all these non-zero codimension defect
theories we studied the meson spectrum and we have shown that it is
continuous and  that the mass gap is lost. The reason behind this result is
the fact that, due to the recombination of color and flavor branes in the
Higgs branch, the defect can spread over the whole bulk, which leads to an
effective Minkowski worldvolume metric in the IR for the flavor brane.
This implies the loss of a KK scale coming from a compact manifold and,
therefore, the disappearance  of the discrete spectrum. Notice that the 
case of the Dp-D(p+4) system  is different,
since in this case  the defect fills the whole color brane and there is no
room for spreading on the Higgs branch.

Also the Dp-Dp case deserves special attention, since it behaves in a
completely different manner to all the other intersections. As we
already mentioned  the intersection profile is not uniquely
fixed by supersymmetry. However,  just for two of all the possible embeddings
we recover conformal invariance in the UV. While one of them corresponds to
the Coulomb phase, the other corresponds to the Higgs phase. It should be
stressed that in this case there is no need for extra flux to get the Higgs
phase, which in this sense is purely geometrical. The other important
difference is that in this case the field
theory limit does not decouple any of the gauge symmetries. Then, our fields
will be bifundamentals under the gauge group on each Dp-brane. Taking into
account the relation with the surface operators in gauge theories
\cite{Gukov:2006jk},  it would be interesting to gain more understanding of
this system.

Let us now discuss some of the possible extensions of our work. 
As in the previous chapter, our analysis has been performed in the probe
approximation so it would be interesting to go beyond it and find the way
in which the backreacted geometry encodes some of the phenomena that we
have uncovered in the probe approximation. In fact, the backreacted geometry
corresponding to the D3-D5 intersection was found in refs.
\cite{Lunin:2006xr,Gomis:2006cu}. Also, it would be interesting to see if
one can apply the smearing procedure proposed in \cite{CNP} (see
also \cite{unqchKW,unqchKS}) to find a solution of the equations of motion
of the gravity plus branes systems studied in this chapter.

Another problem of great interest is trying to describe holographically
(even in the probe approximation) the Higgs branch of theories with less
supersymmetry. The most obvious case to look at   would be that of branes
intersecting on the conifold, such as the D3-D7 systems in the
Klebanov-Witten model \cite{KW} and its generalizations. Actually, the
supersymmetric D3-D5 intersections with flux on the conifold will be
analyzed in chapter \ref{cpSSprobes}, while for more general
Sasaki-Einstein cones they were studied in refs.
\cite{CEPRV,Canoura:2006es}. These configurations are the analogue of the
ones analyzed in section
\ref{cp3codim1sc}, and it would be desirable to find its field theory
interpretation.

\vskip .8cm
\medskip

\begin{subappendices}

\section{Supersymmetry of the M2-M5 intersection}
\label{sam2m5kappa}
\setcounter{equation}{0}
\medskip

In this appendix we will show that the M2-M5 intersections with flux
studied in section \ref{cp3m2m5sc} are supersymmetric. We will verify this
statement by looking at the kappa symmetry of the M5-brane embedding, which
previously requires the knowledge of the Killing spinors of the background.
In order to write these spinors in a convenient way, 
let us rewrite the $AdS_4\times S^7$ near-horizon metric
(\ref{cp3M2metric}) of the M2-brane background as:
\beq
ds^2={r^4\over R^4}\,\,dx_{1,2}^2+{R^2\over r^2}\,dr^2+
R^2\,d\Omega_7^2\,,
\eeq
where $d\Omega_7^2$ is the line element of a unit seven-sphere and 
$R$ is given in eq. (\ref{cp2M2back-parameters}). In what follows we
shall represent  the metric of $S^7$  in terms of polar coordinates
$\theta^1,\cdots \theta^7$:
\beq
d\Omega_7^2=(d\theta^1)^2+\sum_{k=2}^7\,\,
\Bigg(\,\prod_{j=1}^{k-1}\,(\sin\theta^j)^2\,\Bigg)\,(d\theta^k)^2\,.
\eeq
Moreover, we shall consider the vielbein:
\bear
&&e^{x^\mu}={r^2\over R^2}\,\,dx^\mu\;,\qquad
(\mu=0,1,2)\,,\rc\rc
&&e^{r}={R\over r}\,\,dr\,,\rc\rc
&&e^{\theta^i}=R\,\,\Bigg(\,\prod_{j=1}^{i-1}\,\sin\theta^j\,\Bigg)\,
d\theta^i\;,\qquad (i=1,\cdots, 7)\,,
\eear
where, in the last line, it is understood that for $i=1$ the product is
absent. 

The Killing spinors of this background are obtained by solving the equation 
$\delta\psi_M\,=\,0$, where the supersymmetric variation of the gravitino
in  eleven-dimensional supergravity is given by:
\beq
\delta\psi_M=D_M\,\epsilon+{1\over 288}\,
\Bigg(\,\Gamma_{M}^{\,\,N_1\cdots N_4}-8\delta_{M}^{N_1}\,
\Gamma^{\,\,N_2\cdots N_4}\,\Bigg)\,\epsilon\,\,
F^{(4)}_{N_1\cdots N_4}\,.
\label{gravitino}
\eeq
In eq. (\ref{gravitino}) $F^{(4)}=dC^{(3)}$, where $C^{(3)}$ has been written 
in eq. (\ref{cp3C3M2}). In order to write  equation (\ref{gravitino})
more explicitly,  let us define the matrix:
\beq
\Gamma_*\equiv\Gamma_{x^0x^1x^2}\,.
\eeq
Notice that $\Gamma_*^2=1$. From the equations
$\delta\psi_{x^\alpha}=\delta\psi_{r}\,=0$ we get the value of the
derivatives of $\epsilon$ with respect to the $AdS_4$ coordinates, namely:
\bear
&&\partial_{x^\alpha}\,\epsilon=-{r^2\over R^3}\,\,
\Gamma_{x^\alpha r}\,\big(\,1\,+\Gamma_*\,)\,,\rc\rc
&&\partial_{r}\,\epsilon=-{1\over r}\,\Gamma_*\,\epsilon\,.
\label{AdS4-dependence}
\eear
First of all, 
let us solve the first equation in (\ref{AdS4-dependence}) by taking 
$\epsilon=\epsilon_1$ with $\Gamma_*\epsilon_1=-\epsilon_1$, where 
$\epsilon_1$ is independent of the Minkowski coordinates $x^\alpha$.  The second
equation in (\ref{AdS4-dependence}) fixes the dependence of $\epsilon$ on 
$r$, which is:
\beq
\epsilon_1=r\,\eta_1\;,\qquad\quad \Gamma_*\eta_1=-\eta_1\,,
\eeq
where $\eta_1$ only depends on the coordinates of the $S^7$. 

Let us now find a second solution of eq. (\ref{AdS4-dependence}),  given
by the ansatz:
\beq
\epsilon_2=\big(f(r)\,\Gamma_r+g(r)\,x^{\alpha}\,\Gamma_{x^\alpha}
\big)\,\eta_2\,,
\eeq
where $f(r)$ and $g(r)$ are functions to be determined and
$\eta_2$  is a spinor independent of $x^\alpha$ and $r$. By plugging 
this ansatz in
(\ref{AdS4-dependence}) we get the conditions:
\bear
&&g(r)=-{2r^2\over R^3}\,f(r)\,,\qquad\quad 
\Gamma_*\eta_2=-\eta_2\,,\rc\rc
&&f'(r)=-{f(r)\over r}\,,\qquad\quad\,\,\,\,\,\,
g'(r)={g(r)\over r}\,,
\eear
which can be immediately integrated, giving rise to the following spinor:
\beq
\epsilon_2=\Big(\,{1\over r}\,\,\Gamma_r-{2r\over R^3}\,\,
x^{\alpha}\,\Gamma_{x^{\alpha}}\,\Big)\,\eta_2\;,\qquad\quad
\Gamma_*\eta_2=-\eta_2\,,
\eeq  
where $\eta_2=\eta_2(\theta)$. Then, a general Killing spinor of
$AdS_4\times S^7$ can be written as $\epsilon_1+\epsilon_2$, namely as:
\beq
\epsilon=r\,\eta_1(\theta)+
\Big(\,{1\over r}\,\,\Gamma_r-{2r\over R^3}\,\,
x^{\alpha}\,\Gamma_{x^{\alpha}}\,\Big)\,\eta_2(\theta)\;,\qquad\quad
\Gamma_*\eta_i(\theta)=-\eta_i(\theta)\,.
\label{asd4S7KS}
\eeq

The dependence of the $\eta_i\,$s on the angle $\theta^1$ can be 
determined from the
condition $\delta\psi_{\theta^1}=0$, which reduces to:
\beq
\partial_{\theta^1}\,\epsilon=
-{1\over 2}\,\,\Gamma_{r\theta^1}\,\Gamma_*\,\epsilon\,.
\label{thetaonedep}
\eeq
It can be checked that eq. (\ref{thetaonedep}) gives rise to the following
dependence of the spinor  $\eta_i$ on the angle $\theta^1$:
\beq
\eta_i=e^{{\theta^1\over 2}\,\Gamma_{r\theta^1}}\,\,\tilde \eta_i\,,
\eeq
where $\tilde \eta_i$ does not depend on $\theta^1$. Similarly, one can get the
dependence of the $\eta_i\,$s on the other angles of the seven-sphere. The
result can be written as:
\beq
\eta_i(\theta)=U(\theta)\,\hat\eta_i\,,
\label{eta-hateta}
\eeq
with $\hat\eta_i$ being  constant spinors such that 
$\Gamma_*\hat\eta_i=-\hat\eta_i$
and $U(\theta)$ is the following rotation
matrix:
\beq
U(\theta)=e^{{\theta^1\over 2}\,\Gamma_{r\theta^1}}\,\,
\prod_{j=2}^7\,e^{{\theta^{j}\over 2}\,
\Gamma_{\theta^{j-1}\,\,\theta^{j}}}\,.
\label{Utheta}
\eeq
Notice that   $\epsilon$ depends on two arbitrary constant spinors $\hat\eta_1$
and  $\hat\eta_2$ of sixteen components  each one  and, thus, this background
has the maximal number of supersymmetries, namely thirty-two. 

\subsection{Kappa symmetry}
\medskip
The number of supersymmetries preserved by the M5-brane probe is
the number of independent solutions of the equation
$\Gamma_{\kappa}\epsilon=\epsilon$, where 
$\epsilon$ is one of the Killing spinors (\ref{asd4S7KS}) and 
$\Gamma_{\kappa}$ is the kappa symmetry matrix of the PST
formalism \cite{PST, APPS}.
In order to write the expression of this matrix, let
us define the following quantities:
\beq
\nu_p\,\equiv {\partial_p a\over \sqrt{-(\partial a)^2}}\,,
\qquad\quad
t^m\equiv{1\over 8}\,
\epsilon^{m\,n_1n_2p_1p_2q}\,\tilde H_{n_1n_2}\,\tilde
H_{p_1p_2}\,\nu_q\,.
\label{csiete}
\eeq
Then, the  kappa symmetry matrix is:
\beq
\Gamma_{\kappa}=-{\nu_m\gamma^m\over \sqrt{-{\rm det} (g+\tilde H)}}
\left[\gamma_n t^n+{\sqrt{-g}\over 2}\,\gamma^{np}\,\tilde H_{np}
+{1\over 5!}\gamma_{i_1\cdots i_5}\,\epsilon^{i_1\cdots i_5n}\nu_n
\right]\,.
\label{cocho}
\eeq
In eq. (\ref{cocho}) $g$ is the induced metric on the worldvolume, 
$\gamma_{i_1i_2\cdots }$  are antisymmetrized
products of the worldvolume Dirac matrices 
$\gamma_i=\partial_iX^M\,E^{\underline N}_M\,
\Gamma_{\underline N}$ and the indices are raised with the inverse of $g$.

We shall consider here the embedding with $L=0$, which corresponds to
having massless quarks. In the polar coordinates we are using for the $S^7$
this corresponds to taking:
\beq
\theta^1=\cdots=\theta^4={\pi\over 2}\,.
\label{M5embedding}
\eeq
Moreover, we shall denote the three remaining angles of the $S^7$ as 
$\chi^i\equiv \theta^{4+i}$, $(i=1,2,3)$. We will describe the M5-brane embeddings
by means of the following set of worldvolume coordinates:
\beq
\xi^i=(x^0,x^1,r,\chi^1,\chi^2,\chi^3)\,,
\eeq
and we will assume that: 
\beq
x\equiv x^2=x(r)\,.
\eeq
The induced metric for such embedding is given by
(\ref{cp3inducedmetricM5}) with $L=0$ and
$\rho=r$, namely:
\beq
g_{ij}\,d\xi^i\,d\xi^j={r^4\over R^4}\,dx_{1,1}^2+
{R^2\over r^2}\,\Big(\,1+\,{r^6\over R^6}\,x'^2\,\Big)\,dr^2+
R^2\,d\Omega_3^2\,.
\eeq
The induced Dirac matrices for this embedding  are:
\bear
&&\gamma_{x^{\mu}}={r^2\over R^2}\,\,\Gamma_{x^{\mu}}\,,
\qquad (\mu=0,1)\,,\rc\rc
&&\gamma_{r}={R\over r}\,\,
\big(\Gamma_r+{r^3\over R^3}\,\,x'\,\,\Gamma_{x^2}\,\big)\,,\rc\rc
&&\gamma_{\Omega_3}\,\equiv\gamma_{\chi^1\,\chi^2\,\chi^3}=
R^3\,\sqrt{\tilde g}\,\Gamma_{\Omega_3}\,,
\eear
where $\sqrt{\tilde g}=\sin^2\chi^1\,\sin\chi^2$ and 
$\Gamma_{\Omega_3}\equiv\Gamma_{\chi^1\,\chi^2\,\chi^3}$. Notice also that:
\bear
&&\gamma^{x^0}=-{R^2\over r^2}\,\,\Gamma_{x^0}\;,
\qquad\quad\gamma^{x^1}={R^2\over r^2}\,\,\Gamma_{x^1}\,,\rc\rc
&&\gamma^{r}={r\over R}\,\,\Big(\,1+\,{r^6\over R^6}\,x'^2\,\Big)^{-1}\,
\Big(\,\Gamma_r+{r^3\over R^3}\,\,x'\,\,\Gamma_{x^2}\,\Big)\,.
\eear
We will also assume that we have switched on a magnetic worldvolume gauge
field $F$, parametrized as in (\ref{cp3M5flux}) in terms of a flux
number $q$.  Moreover, in the gauge $a=x^1$, the only non-vanishing
component of $\nu_p$ is:
\beq
\nu_{x^1}=-i\,{r^2\over R^2}\,,
\eeq
and one can check that the only non-vanishing component of $\tilde H$  is
$\tilde H_{x^0r}$, whose expression is given by (\ref{cp3M2tildeHexp})
with $L=0$ and
$\rho=r$. Furthermore, the worldvolume vector $t^m$ defined in eq.
(\ref{csiete}) is zero and one can verify that the kappa symmetry matrix
$\Gamma_{\kappa}$ takes the form:
\beq
\Gamma_{\kappa}=
{\sqrt{\tilde g}\,r^3\over \sqrt{-{\rm det} (g+\tilde H)}}\,\,\Bigg(\,
{q\over  R^3}+\Gamma_{\Omega_3}\,\Bigg)\,\,\Bigg(\,
\Gamma_{rx^2}+{r^3\over R^3}\,x'\,\Bigg)\,\,\Gamma_*\,.
\eeq
For the embeddings we are interested in the function $x(r)$ is given by:
\beq
x=\bar x\,+{q\over 2}\,{1\over r^2}\,,
\label{M5profile}
\eeq
where $\bar x$ is a constant (see eq. (\ref{cp3M2x-explicit})). In order
to express the form of $\Gamma_{\kappa}$ for these embeddings, let us
define the matrix ${\cal P}$ as:
\beq
{\cal P}\equiv\Gamma_{x^2 r}\,\Gamma_{\Omega_3}\,.
\eeq
Notice that ${\cal P}^2=1$. Moreover, the kappa symmetry matrix can be
written as:
\beq
\Gamma_{\kappa}=-{1\over 1+{q^2\over R^6}}\left({\cal P}+{q^2\over R^6}+
{q\over R^3}\,\,\Gamma_{x^2 r}\left(1-{\cal P}\right)\right)\,\Gamma_*\,.
\label{Gkappa-embedding}
\eeq
Let us represent the Killing spinors $\epsilon$ on the M5-brane worldvolume 
as is eq. (\ref{asd4S7KS}).  By using the explicit function 
$x(r)$ written  in eq. (\ref{M5profile}), one gets:
\beq
\epsilon={1\over r}\,\,
\Big(\, \Gamma_{r}\eta_2-{q\over  R^3}\,\Gamma_{x^2}\,\eta_2\,\Big)+
r\,\Big(\,\eta_1-{2 \bar x\over R^3}\,\Gamma_{x^2}\,\eta_2\,\Big)-
{2r\over R^3}\,\,x^p\,\Gamma_{x^p}\,\eta_2\,,
\label{KS-onthe wv}
\eeq
where the index $p$ can take the values $0,1$ and
we have organized the right-hand side of (\ref{KS-onthe wv}) according to
the different dependences on $r$ and $x^p$. By substituting
(\ref{Gkappa-embedding}) and (\ref{KS-onthe wv}) into the equation 
$\Gamma_{\kappa}\epsilon=\epsilon$ and comparing the terms on the two sides of
this equation that have the same dependence on the coordinates, one gets the
following three equations:
\bear
&&\left({\cal P}+1\right)\eta_2=0\,,\rc\rc
&&\left[1-{q\over  R^3}\,\Gamma_{x^2r}\right]\left({\cal P}-1\right)
\left(\eta_1-{2\bar x\over R^3}\,\Gamma_{x^2}\,\eta_2\right)=0\,,\rc\rc
&&\left[{q\over  R^3}\,\Gamma_{x^2r}-1\,\right]
\Gamma_{x^p}\left({\cal P}+1\right)\eta_2=0\,.
\label{kappasystem}
\eear
In order to solve these equations, let us classify the sixteen spinors
$\eta_1$ according to their ${\cal P}$-eigenvalue as:
\beq
{\cal P}\,\eta_1^{(\pm) }=\pm \eta_1^{(\pm) }\,.
\label{M5conditionone}
\eeq
Notice that ${\cal P}$ and $\Gamma_*$ commute and, then, the condition of
having well-defined ${\cal P}$-eigenvalue is perfectly compatible with having
negative  $\Gamma_*$-chirality. We can now solve the system
(\ref{kappasystem}) by taking
$\eta_2=0$ (which solves the first and third equation) and choosing $\eta_1$ to be
one of the eight spinors $\eta_1^{(+)}$
of positive ${\cal P}$-eigenvalue. Thus, this solution of (\ref{kappasystem}) is:
\beq
\eta_1=\eta_1^{(+)}\,,\qquad\quad
\eta_2=0\,.
\eeq
Another  set of solutions corresponds to taking spinors $\eta_1^{(-)}$ of negative 
${\cal P}$-eigenvalue and a spinor $\eta_2$ related to $\eta_1^{(-)}$ as follows:
\beq
\eta_2={R^3\over 2\bar x}\,\,\Gamma_{x^2}\,\,\eta_1^{(-)}\,.
\label{second-kappa-spinor}
\eeq
Notice that the second equation in (\ref{kappasystem}) is automatically satisfied.
Moreover, as $[{\cal P},\Gamma_{x^2}\,]=0$, the spinor $\eta_2$ in
(\ref{second-kappa-spinor}) has negative ${\cal P}$-eigenvalue and, therefore, the
first and third equation in the system (\ref{kappasystem})  are also satisfied. 

In order to complete the proof of the supersymmetry of our M2-M5 configuration we
must verify that the kappa symmetry conditions found above can be fulfilled at all
points of the M5-brane worldvolume. Notice that, when evaluated for the embedding 
(\ref{M5embedding}), the spinors $\eta_{1,2}$ depend on the angles $\chi^i$ that
parametrize the $S^3\subset S^7$. To ensure that the conditions
(\ref{M5conditionone}) and (\ref{second-kappa-spinor}) can be imposed at all
points of the $S^3$ we should be able to translate them into some algebraic
conditions for the constant spinors $\hat\eta_i$. Recall (see eq.
(\ref{eta-hateta})) that the spinors $\eta_i$ and $\hat \eta_i$ are related by
means of the matrix 
$U(\theta)$. Let us denote by $U_{*}(\chi)$ the rotation matrix restricted to the
worldvolume, \ie:
\beq
U_{*}(\chi)\equiv U(\theta)_{\big|\theta^1=\cdots =\theta^4={\pi\over
2}}\,.
\eeq
Moreover, let us define $\hat{\cal P}$ as the result of conjugating the matrix 
${\cal P}$ with the rotation matrix $U_{*}(\chi)$:
\beq
\hat{\cal P}\,\equiv U_{*}(\chi)^{-1}\,\,{\cal P}\,\,U_{*}(\chi)\,.
\eeq
By using (\ref{Utheta}) a simple calculation shows that $\hat{\cal P}$ is
the following constant matrix:
\beq
\hat{\cal P}=\Gamma_{x^2\theta^4}\,\Gamma_{\Omega_3}\,.
\eeq
Moreover, from the definition of $\hat{\cal P}$ it follows that:
\beq
{\cal P}\,\eta_1^{(\pm) }=\pm \eta_1^{(\pm) }\qquad
\Longleftrightarrow\qquad
\hat{\cal P}\,\hat\eta_1^{(\pm) }=\pm \hat\eta_1^{(\pm) }\,.
\eeq
Therefore, the condition (\ref{M5conditionone}) for $\eta_1$ is equivalent to
require that the corresponding constant spinor $\hat\eta_1$ be an eigenstate of
the constant matrix $\hat{\cal P}$.  Finally, as $[U_{*}, \Gamma_{x^2}]=0$, eq. 
(\ref{second-kappa-spinor}) is equivalent to the following condition, to be
satisfied by the constant spinors $\hat \eta_1$ and $\hat\eta_2$:
\beq
\hat\eta_2={R^3\over 2\bar x}\,\,\Gamma_{x^2}\,\,\hat\eta_1^{(-)}\,.
\label{second-constant-spinor}
\eeq
Taken together, these results prove that the kappa symmetry condition
$\Gamma_{\kappa}\,\epsilon\,=\,\epsilon$ can be imposed at all points of the
worldvolume of our M5-brane embedding and that this configuration is
${1\over 2}$-supersymmetric. 

\end{subappendices}

%
%
%
%

%
%
%
%
%
%
%
%
%
%

\chapter{Flavoring the non-commutative gauge theories}
\label{cpncflavor}
\setcounter{equation}{0}
\section{Introduction}
\label{cpncscintro}
\medskip

The aim of this chapter is to study the addition of
flavor degrees of freedom to the non-commutative deformation of
the maximally supersymmetric four-dimensional gauge theory through its
supergravity dual. As in the two previous chapters we will consider the
addition of probe branes to the supergravity background. Once again, their
fluctuations are conjectured to be dual to the fundamental degrees of
freedom realized by the strings joining these probe branes and those
generating the background. As before, we will neglect the backreaction of
the probes on the background, thus restricting our study to the addition
of quenched flavors.

The supersymmetric gauge theory dual to the supergravity background
we will be working with is a gauge theory living on a spacetime which has
two spatial coordinates that are non-commutative.  Such non-local
theories have a long story (see ref. \cite{NCFT} for a review) and have
been  the object of intense study in recent years after the discovery 
that they can be obtained from limits of string theory and M-theory.
Indeed, by considering string theory in  the presence of a Neveu-Schwarz
$B$ field, and performing a suitable limit one ends up  with a gauge
theory on a non-commutative space
\cite{NCString, SW}.

In the context of the gauge/gravity correspondence it is quite 
natural to have a supergravity
solution dual to  the non-commutative gauge theory for large $N$ and 
strong 't Hooft coupling.
Actually, as pointed out in ref. \cite{MR}, this background can be 
obtained from the decoupling
limit of the type IIB supergravity solution \cite{Bound} representing 
a stack of non-threshold
bound states of  D3- and D1-branes (for a similar analysis in more general brane
setups see ref. \cite{AOSJ}). The corresponding 
ten-dimensional metric breaks
four-dimensional Lorentz invariance since it distinguishes between 
the coordinates of the
non-commutative plane and the other two Minkowski coordinates.
As expected, this solution has a non-vanishing Neveu-Schwarz
$B$ field directed along the non-commutative directions, as well as 
two Ramond-Ramond potentials
and a running dilaton.  This background contains  a parameter 
$\Theta$ which corresponds to the non-commutative deformation of the
corresponding gauge theory. For  any non-vanishing value of
$\Theta$ the solution preserves sixteen supersymmetries, while for 
$\Theta=0$ it reduces to the maximally supersymmetric $AdS_5\times S^5$
background. 

To add flavor to the (D1,D3) background we will consider a D7-brane 
probe. First of all, we will make use of  kappa symmetry to determine
the static configuration of  the probe that preserves supersymmetry. It
turns out that this embedding is the same as the one considered in the
commutative case \cite{KMMW}. This is the general embedding
corresponding to orthogonally intersecting branes described in section
\ref{general}, particularized to the D3-D7
intersection. However, the metric induced on the worldvolume is 
different now and, in addition, the brane captures a non-vanishing
Neveu-Schwarz $B$ field on its worldvolume. Next, we  will study the
fluctuations of the scalar and worldvolume gauge field around the static 
configuration. Some of the scalar fluctuation modes are coupled to the
worldvolume electric field. The  corresponding equations of motion are
coupled  and break Lorentz invariance. However, we will be  able to find
a set of decoupled differential equations, which can be analyzed by
means of  several techniques.

One way to extract information from the decoupled equations of the 
fluctuations is by transforming them into a Schr\"odinger equation,
proceeding as we did in section \ref{generalfluctuations} for the generic
fluctuations analyzed in chapter \ref{cp2dpinter}.
The analysis of the Schr\"odinger equation can shed light on the
qualitative  behavior of the spectrum of the fluctuation modes. It turns
out that, in order to have a  discrete spectrum for some of the modes, the
momentum along some directions must be bounded from  above, the bound
being a function of the non-commutativity parameter
$\Theta$. Actually, for any  non-vanishing value of $\Theta$, the tower
of energy levels of some modes is cut off from  above and, in some
window of values  $\Theta_1<\Theta<\Theta_2$ of the non-commutativity
parameter,  this upper bound is so low that these modes disappear from
the discrete spectrum. We have  determined the value of $\Theta_{1,2}$
both by using the WKB approximation and by solving  numerically the
differential equation of the fluctuations by means of the shooting
technique.  Moreover, if $\Theta$ is large enough the spectrum of
fluctuations reduces to the one corresponding  to the commutative theory,
a fact that is reminiscent of the Morita equivalence of algebras in  the
non-commutative torus (see \cite{NCFT} for a review and further
references).

In order to have a complementary picture of the meson spectrum we 
have studied the classical
dynamics of a rotating string whose ends are attached to the D7 
flavor brane. The profile of the
string can be obtained by solving the Nambu-Goto equations of motion 
in the background under
consideration with appropriate boundary conditions. We will verify 
that, as compared to the
string in the $AdS_5\times S^5$ geometry, the non-commutative 
deformation results in a tilting
of the string. The energy spectrum has a Regge-like behavior for low 
angular momentum, with an
asymptotic Regge slope which, remarkably,  is the same as in the 
commutative case. For large
angular momentum the energy obtained is that corresponding to two 
non-relativistic masses bound
by a Coulomb potential. We will confirm these behaviors by an 
explicit calculation of the static
potential energy from a hanging string attached to the flavor brane.
As an aside, we also consider a moving hanging string and briefly comment
on some effects related to the breaking of Lorentz symmetry.

This chapter is organized as follows. In section \ref{cpncscback} we will
review the  main features of the supergravity dual of the non-commutative
gauge theory. In section \ref{cpncscprobe} we  will start to study the
dynamics of a D7-brane probe in this background. First, we will 
determine the static supersymmetric embedding of the probe and then, we
will study  its  fluctuations. The corresponding meson spectrum will be
obtained in section \ref{cpncscspectrum}. In  section \ref{cpncscstrings}
we will perform a semiclassical analysis of a rotating string attached
to the flavor brane in the non-commutative background. The calculation
of the corresponding static potential energy is done in section
\ref{cpncschangstrings}. Finally, in section \ref{cpncscconcl} we summarize
our results and draw some  conclusions.

%
%
%
%
%
%
%
%
%

\setcounter{equation}{0}
\section{The SUGRA dual of non-commutative gauge theories}
\label{cpncscback}
\medskip
In this section we will review the main features of the gravity dual 
of non-commutative gauge
theories obtained in ref.  \cite{MR} (see also ref. \cite{Bound}). 
This supergravity background
is obtained by taking the decoupling limit of the solution 
corresponding to the non-threshold
bound state of a D3-brane and a D1-brane and contains a $B$ field 
along two of the spatial
worldvolume directions of the D3-brane.  The string frame metric 
takes the form:
\beq
ds^2={r^2\over R^2} \left[dx_{0,1}^2+h\,dx^2_{2,3}\right]+{R^2\over
r^{2}}
\left[(dy^1)^2\,+\cdots\,+(dy^6)^2\right]\,,
\label{MRmetric}
\eeq
where $R^4=4\pi g_s N(\alpha')^2$ with $N$ being the number of 
D3-branes. The radial coordinate
$r$ is given by  $r^2=(y^1)^2\,+\cdots+\,(y^6)^2$ and the function $h$ is:
\beq
h={1\over 1+\Theta^4r^4}\,,
\eeq
with $\Theta$ being a constant.  The dilaton is given by:
\beq
e^{2\phi}=h\,.
\label{MRdilaton}
\eeq
Notice that the function $h$ distinguishes in the
metric the non-commutative plane $x^2x^3$ from the two other 
Minkowski directions. Moreover,
this same function is responsible of the running of the dilaton in 
eq. (\ref{MRdilaton}).
Obviously, when $\Theta=0$ the dilaton is constant and  we recover 
the $AdS_5\times S^5$ metric.
When $\Theta\not=0$ this background is dual to a gauge theory in 
which the coordinates $x^2$ and
$x^3$ do not commute, being $[x^2, x^3]\sim\Theta^2$.

This supergravity solution
also contains an NSNS three-form $H$, an RR three-form $F^{(3)}$ and a 
self-dual RR
five-form $F^{(5)}$. The expressions of the three-forms $H$ and $F^{(3)}$ are:
\bear
&&H=-{\Theta^2\over R^2}\,\partial_{y^i}\,[r^4h]\,dx^2\wedge dx^3\wedge
dy^i\,,\rc\rc
&&F^{(3)}=4\Theta^2\,{r^2\over R^2}\,\, y^i\,
dx^0\wedge dx^1\wedge dy^i\,,
\eear
while the RR five-form $F^{(5)}$ can be represented as
\beq
F^{(5)}=\hat F^{(5)}+{}^*\hat F^{(5)}\,,
\eeq
with $\hat F^{(5)}$ being given by:
\beq
\hat F^{(5)}=4\,{r^2\over R^4}\,h\,y^i\,dx^0\wedge\cdots \wedge
dx^3\wedge dy^i\,.
\eeq
These forms\footnote{When writing the dilaton and the RR forms of the background we
have applied the redefinitions (\ref{cp1gsabsor}).} can be represented in
terms of the NSNS $B$ field and the RR potentials $C_{(2)}$ and $C_{(4)}$
as follows:
\beq
H=dB\,,\qquad\quad
F^{(3)}=dC^{(2)}\,,\qquad\quad
F^{(5)}=dC^{(4)}-H\wedge C^{(2)}\,.
\eeq
The explicit form of these potentials is:
\bear
&&B=-\Theta^2\,{r^4\over R^2}\,h\,dx^2\wedge dx^3\,,\rc\rc
&&C^{(2)}=\Theta^2\,{r^4\over R^2} \,dx^0\wedge dx^1\,,\rc\rc
&&C^{(4)}\,=\hat C^{(4)}+\,\tilde C^{(4)}\,,
\label{potentials}
\eear
where:
\beq
\hat C^{(4)}=h\,{r^4\over R^4}\,\,dx^0\wedge\cdots \wedge dx^3\,,
\label{C4}
\eeq
and $\tilde C^{(4)}$ is a potential for ${}^*\hat F^{(5)}$, \ie\ ,
${}^*\hat F^{(5)}=d\tilde C^{(4)}$.
Notice that, as expected, the $B$ field has only components
along the non-commutative plane $x^2x^3$, which approach a constant 
as $r$ becomes large (\ie\
for large energies in the gauge theory). However, for large $r$ the 
metric changes drastically
with respect to the $AdS_5\times S^5$ geometry, since the  $x^2x^3$ 
directions collapse in this
limit.

The background described above preserves sixteen supersymmetries. The 
corresponding
Killing spinors have been obtained in ref. \cite{GG}. In the natural
frame for the coordinate system we are using, namely:
\beq
e^{x^{0,1}}={r\over R}\,dx^{0,1}\,,\qquad\quad
e^{x^{2,3}}={r\over R}\,h^{{1\over 2}}\,dx^{2,3}\,,\qquad\quad
e^{y^{i}}={R\over r}\,dy^{i}\,,
\eeq
they can  be represented as:
\beq
\epsilon=e^{-{\beta\over 2}\Gamma_{x^2x^3}\sigma_3}\,\,\tilde\epsilon\,,
\eeq
where $\beta$ is an angle determined by the equations \cite{GG}:
\beq
\cos\beta=h^{{1\over 2}}\,,\qquad\quad
\sin\beta=\Theta^2\,r^2\,h^{{1\over 2}}\,,
\label{MRbeta}
\eeq
and $\tilde\epsilon$ is the spinor satisfying the projection:
\beq
\Gamma_{x^0\cdots x^3}\,(i\sigma_2)\,\tilde\epsilon=-\tilde\epsilon\,.
\label{tildeprojection}
\eeq
 From the above equations it is immediate to prove that:
\beq
\Gamma_{x^0\cdots x^3}\,(i\sigma_2)\,\epsilon=-\,
e^{\beta\,\Gamma_{x^2x^3}\sigma_3}\,\,\epsilon\,.
\label{MRprojection}
\eeq

%
%
%
%
%
%
%
%
%

\setcounter{equation}{0}
\section{D7-brane probe}
\label{cpncscprobe}
\medskip
We are interested in configurations in which the D7-brane fills the
spacetime directions $x^0\cdots x^3$ of the gauge theory. Therefore, it
is quite natural to choose the following set of  worldvolume coordinates:
\beq
\xi^a=(x^0,\cdots,x^3,y^1,\cdots,y^4)\,,
\eeq
and consider embeddings in which the remaining two coordinates $y^5$ 
and $y^6$ depend on the $\xi^a$s. Notice that the norm of the vector
$(y^5, y^6)$ determines  the distance between the
D7-brane and the D3-branes of the background. We will consider first 
the case in which this norm is
constant along the worldvolume of the D7-brane and, in general, different
from zero. By looking at the  kappa symmetry condition of the
probe \cite{bbs}, we will show that these 
configurations are supersymmetric.
Next, we will study in detail the fluctuations around these 
embeddings and we will be able to
determine the corresponding meson spectrum.

\subsection{Kappa Symmetry}
In section \ref{cp1sskappasym} it has been shown how to look for
supersymmetric embeddings of D-brane probes via kappa symmetry.
One has to look for solutions of the equation (\ref{kpfnalcond}), \ie \
$\Gamma_\kappa\,\epsilon=\epsilon$, where $\epsilon$ is the killing
spinor of the background and $\Gamma_\kappa$ is a matrix depending on
the embedding whose form (for Dp-branes in type IIB theory) can be read
from eq. (\ref{kpGammakpIIB}).

Let us now consider  the  embeddings in which $y^5$ and $y^6$ are 
constant and the worldvolume
gauge  field $F$ is zero.  The induced metric for this configuration
will be denoted by $g^{(0)}$ and the value of ${\cal F}$ in this case is:
\beq
{\cal F}^{(0)}=-P[B]=\Theta^2\,{r^4\over 
R^2}\,\,h\,dx^2\wedge dx^3\,.
\label{MRinducedB}
\eeq
After a short calculation one can prove that for this embedding the
matrix $\Gamma_{(0)}$ (defined in eq. (\ref{kpGammazero})) is:
\beq
\Gamma_{(0)}=h\,\Gamma_{x^0\cdots x^3}\,\Gamma_{y^1\,\cdots \,y^4}\,.
\eeq
Moreover, by using the expression of $g^{(0)}$, ${\cal F}^{(0)}$ and 
$h$, one gets:
\beq
-\det \big(\,g^{(0)}+{\cal F}^{(0)}\big)=h\,.
\eeq
Then, substituting these results in the expression of $\Gamma_{\kappa}$ (eq.
(\ref{kpGammakpIIB})), one arrives at:
\beq
\Gamma_{\kappa}=h^{{1\over 2}}\,\Gamma_{y^1\,\cdots \,y^4}
\left[\Gamma_{x^0\cdots x^3}\,(i\sigma_2)+\Theta^2\,r^2\,\Gamma_{x^0x^1}\,
\sigma_1\right]\,,
\eeq
which can be rewritten as:
\beq
\Gamma_{\kappa}=\Gamma_{y^1\,\cdots \,y^4}\,
e^{-\beta\,\Gamma_{x^2x^3}\sigma_3}\,\Gamma_{x^0\cdots x^3}\,(i\sigma_2)\,,
\eeq
where $\beta$ is the angle defined in eq. (\ref{MRbeta}).

 From this expression  and the equation (\ref{MRprojection}) satisfied 
by the Killing spinors, it
is clear that the condition $\Gamma_{\kappa}\epsilon=\epsilon$ is
equivalent to:
\beq
\Gamma_{y^1\,\cdots \,y^4}\,\epsilon=-\epsilon\,,
\eeq
which, taking into account that $[\Gamma_{y^1\,\cdots \,y^4}, 
\Gamma_{x^2x^3}]=0$,
can be put as:
\beq
\Gamma_{y^1\,\cdots \,y^4}\,\tilde\epsilon=-\tilde\epsilon\,.
\eeq
Notice that this condition is compatible with the one written in eq.
(\ref{tildeprojection}). Thus, this embedding with $y^5$ and $y^6$ 
constant, vanishing
worldvolume gauge field  and induced
$B$ field   as in eq. (\ref{MRinducedB}), preserves eight 
supersymmetries and is
$1/4$-supersymmetric.

\subsection{Fluctuations}

Let us now consider fluctuations of the scalars  $\vec y\equiv 
(y^5,y^6)$ and of the worldvolume
gauge field $A_m$ around the configuration with $\vec 
y^{\,\,2}=(y^5)^2+(y^6)^2=L^2$ and $A_m=0$. The lagrangian of the D7-brane probe can be
read from eq. (\ref{cp1Dbraneact}). The Dirac-Born-Infeld term reads:
\beq
{\cal L}_{DBI}=-e^{-\phi}\,\sqrt{-\det(g+{\cal F})}\,,
\label{BI}
\eeq
while the Wess-Zumino part takes the form:
\beq
{\cal L}_{WZ}={1\over 2}\,P[C^{(4)}]\wedge {\cal F} \wedge {\cal F}+
{1\over 6}\,P[C^{(2)}]\wedge {\cal F} \wedge {\cal F}\wedge {\cal F}\,.
\label{cpncWZ}
\eeq
Let us analyze first the DBI term. We shall expand the
determinant in (\ref{BI})  up to quadratic terms in the fluctuations. With
this  purpose in mind let us put:
\beq
g=g^{(0)}+g^{(1)}\,,\qquad\quad
{\cal F}={\cal F}^{(0)}+{\cal F}^{(1)}\,,
\eeq
where $g^{(0)}$ is the induced metric of the unperturbed configuration,
${\cal F}^{(0)}=-P[B]$ and:
\beq
g^{(1)}_{mn}={R^2\over r^2}\,\partial_{m}\vec y\cdot 
\partial_{n}\vec y\,,\qquad\quad
{\cal F}^{(1)}=F\,.
\label{cpncflucts}
\eeq
In principle, apart from this second order term in the derivative
of the fluctuations, $g^{(1)}$ should include terms proportional to the
fluctuations themselves. However, it is not difficult to check that these
do not contribute to the action at quadratic order, so that there is no
potential for these fields as expected from BPS considerations\footnote{
This can be checked from the fact that the combination
$e^{-\phi}\,\sqrt{\det\,(\tilde g)}$
with $\tilde g=g-g^{(1)}$ (\ie\ the induced metric minus the
terms in derivatives of the fluctuations) does not depend on $r$ and thus
is independent of the fluctuations.}. The DBI determinant can be written
as:
\beq
\sqrt{-\det(g+{\cal F})}=\sqrt{-\det \big(\,g^{(0)}+{\cal 
F}^{(0)}\,\big)}\,
\sqrt{\det(1+X)}\,,
\label{cpncdetX}
\eeq
where the matrix $X$ is given by:
\beq
X\equiv\bigg(\,g^{(0)}+{\cal F}^{(0)}\,\bigg)^{-1}\,\,
\bigg(\,g^{(1)}+{\cal F}^{(1)}\,\bigg)\,.
\eeq
The right-hand side of eq. (\ref{cpncdetX}) can be expanded in powers of
$X$ by means of eq. (\ref{detX-expansion}).
In order to write the terms resulting from the expansion in a neat 
form, let us introduce the auxiliary metric:
\beq
d\hat s^{\,2}\equiv{\cal G}_{mn}\,d\xi^m\,d\xi^n=
{\rho^2+L^2\over R^2}\,\big(-(dx^0)^2+\cdots+(dx^3)^2\big)+
{R^2\over \rho^2+L^2}\,\big((dy^1)^2+\cdots+(dy^4)^2\big)\,,
\label{cpncosmetric}
\eeq
where, as usual, $\rho^2=(y^1)^2+\cdots+(y^4)^2$, being the radial
coordinate of the subspace of the worldvolume of the D7-brane transverse
to the Minkowski directions.
Notice that the metric ${\cal G}$ is nothing but the induced metric 
for the unperturbed
embeddings in the commutative ($\Theta=0$) geometry. Moreover, it  turns
out that the matrix $\big(\,g^{(0)}+{\cal F}^{(0)}\,\big)^{-1}$ can be
written in  terms of the inverse
metric ${\cal G}^{-1}$ as:
\beq
\big(\,g^{(0)}+{\cal F}^{(0)}\,\big)^{-1}={\cal G}^{-1}+
{\cal J}\,,
\label{openmetric}
\eeq
where ${\cal J}$ is an antisymmetric matrix whose only non-vanishing 
values are:
\beq
{\cal J}^{x^2x^3}=-{\cal J}^{x^3x^2}=-\Theta^2\,R^2\,.
\eeq
We are going to use this representation of $\big(\,g^{(0)}+{\cal 
F}^{(0)}\,\big)^{-1}$
to obtain the traces appearing in the expansion
(\ref{detX-expansion}). Up to quadratic order in  the fluctuations, one
gets:
\bear
&&\tr X={\cal 
G}^{mn}\,g_{mn}^{(1)}+2\Theta^2\,R^2\,F_{x^2x^3}\,,\rc\rc
&&\tr X^2=F_{mn}\,F^{nm}+2\Theta^4\,R^4\,\big(F_{x^2x^3}\big)^{2}
\,,
\eear
where $F^{mn}={\cal G}^{mp}{\cal G}^{nq}F_{pq}$. Using this result 
one can prove that:
\beq
e^{-\phi}\,\sqrt{-\det(g+{\cal F})}=1+\Theta^2\,R^2\,F_{x^2x^3}+
{1\over 2}\,{\cal G}^{mn}\,g_{mn}^{(1)}+{1\over 4}\,F_{mn}\,F^{mn}\,.
\eeq
Notice that the quadratic terms are covariant with respect to the metric 
${\cal G}$. Dropping
the constant and linear terms (which do not contribute to the 
equations of motion), we
have:
\beq
{\cal L}_{BI}=-\sqrt{-\det {\cal G}}\,\left[{R^2\over 2(\rho^2+L^2)}\,
{\cal G}^{mn}\,\partial_{m}\vec y\cdot \partial_{n}\vec y+
{1\over 4}\,F_{mn}\,F^{mn}\right]\,,
\label{quadraticBI}
\eeq
where we have included the factor $\sqrt{-\det {\cal G}}$ (which is 
one in the coordinates
$(x^0,\cdots,x^3,y^1,\cdots,y^4)$). Remarkably, the Born-Infeld 
lagrangian written above is
exactly the same as that of the fluctuations in the $\Theta=0$ 
geometry and, in particular, is Lorentz invariant in the Minkowski
directions $x^0\cdots x^3$. This  is so because there is a conspiracy
between the terms of the metric (\ref{MRmetric}) that  break Lorentz
symmetry in the
$x^0\cdots x^3$ coordinates and the
$B$ field which results in the same quadratic Born-Infeld lagrangian 
for the fluctuations as in
the commutative theory. Actually, for $\rho>>L$ the effective metric 
${\cal G}$ approaches that
of the $AdS_5\times S^3$ space, which can be interpreted as the fact 
that the conformal
invariance, broken by the mass of the hypermultiplet and the 
non-commutativity, is restored at
asymptotic energies.

It is interesting at this  point to recall from  the analysis of ref.
\cite{SW} that the metric relevant for the non-commutative gauge theory
is the so-called open string metric, which is the effective metric seen
by the open strings and should not be confused with the closed string
metric.  By comparing the commutative formalism with the $B$ field and
the non-commutative description without the $B$ field one can find the
open string metric in terms of the closed string metric and the 
$B$ field. Moreover,  one can establish the so-called Seiberg-Witten
map between the non-commutative and commutative gauge fields \cite{SW}.

It has been proposed in ref. \cite{LiWu} that the metric
(\ref{MRmetric}) of the supergravity dual should be understood as the
closed string metric. We would like to argue here that  ${\cal G}$ is the
open string metric relevant for our problem. First of all,  notice that
we are identifying our fluctuations with open string degrees of freedom
and, thus, it is natural to think that the metric governing their
dynamics is not necessarily the same as that of the closed string
background. Moreover, ${\cal G}$ certainly contains the effect of the
coupling to the $B$ field and, actually (see  eq. (\ref{openmetric})), ${\cal G}^{-1}$
is the symmetric part of $(\,g^{(0)}\,-\,P[B]\,)^{-1}$, in  agreement
with the expression of the open string metric given in ref. \cite{SW}. Notice also that
we are keeping quadratic terms in our expansion, which is enough to study the mass
spectrum. Higher order terms, which represent interactions, are of course dependent on
$\Theta$. This fact is consistent with the Seiberg-Witten map since the
$\Theta$-dependent $*$-product can be replaced by the ordinary
multiplication in the quadratic terms of the action\footnote{It is also
interesting to consider the action of a D3-brane probe extended along
$x^0\cdots x^3$ in the (D1,D3) background. By expanding the Born-Infeld
action one can verify that the metric appearing in the quadratic
terms of the lagrangian for the fluctuations is also
$\Theta$-independent and Lorentz invariant in the directions $x^0\cdots
x^3$.}.

Let us consider next the Wess-Zumino term of the lagrangian, written 
in eq. (\ref{cpncWZ}). First of all, since $P[B]\wedge P[B]=P[\hat
C^{(4)}]\wedge P[B]=0$, one can write ${\cal  L}_{WZ}$, at
quadratic order in the fluctuations, as:
\beq
{\cal L}_{WZ}={1\over 2}\,P[\,{\cal C}^{(4)}\,]\wedge F\wedge F+
{1\over 2}\,P[\,{\tilde C}^{(4)}\,]\wedge F\wedge F-
\,P[\,{\tilde C}^{(4)}\,]\wedge F\wedge B\,,
\label{finalWZ}
\eeq
where the four-form ${\cal C}^{(4)}$ is defined as:
\beq
{\cal C}^{(4)}\equiv\hat C^{(4)}-C^{(2)}\wedge B\,.
\eeq
A straightforward calculation, using the expressions of $B$, $C^{(2)}$
and $\hat C^{(4)}$ written in eqs. (\ref{potentials}) and (\ref{C4}), 
gives the value of
${\cal C}^{(4)}$, namely:
\beq
{\cal C}^{(4)}={r^4\over R^4}\,dx^0\wedge\cdots \wedge dx^3\,.
\label{calC4}
\eeq
Notice that the first term  of ${\cal L}_{WZ}$ in eq. (\ref{finalWZ}) 
and the expression
of the potential ${\cal C}^{(4)}$ displayed in eq. (\ref{calC4}), are 
exactly the same as
those corresponding to the D7-brane probes in the $\Theta=0$ geometry.
In order to find out the contribution of the other two terms in eq. 
(\ref{finalWZ}),
let us determine the explicit form of ${\tilde C}^{(4)}$. With this 
purpose, let us
choose new coordinates, such that:
\beq
(dy^1)^2\,+\cdots+(dy^4)^2=d\rho^2+\rho^2\,d\Omega_3^2\,,
\eeq
where $d\Omega_3^2$ is the line element of a three-sphere of unit 
radius. In this system
of coordinates, one has:
\beq
{}^*\hat F^{(5)}={4\rho^4\,R^4\over (\,\rho^2+\vec
y^{\,2}\,)^3}\,\,\omega_3\wedge  dy^5\wedge dy^6+{4\rho^3\,R^4\over
(\,\rho^2+\vec y^{\,2}\,)^3}\, d\rho\wedge \omega_3\wedge
(\,y^5dy^6-y^6dy^5\,)\,,
\eeq
where $\omega_3$ is the volume form of the three-sphere. It is not 
difficult now to obtain
the expression of a four-form  ${\tilde C}^{(4)}$ such that ${}^*\hat
F^{(5)}=d{\tilde C}^{(4)}$, namely:
\beq
{\tilde C}^{(4)}=-\,R^4\,{2\rho^2+\vec y^{\,2}\over (\,\rho^2+\vec
y^{\,2}\,)^2}\,
\omega_3\wedge (\,y^5dy^6-y^6dy^5\,)\,.
\eeq
Let us consider next, without loss of generality,  the fluctuations 
around the configuration with
$\vec y=(0, L)$ and, following ref. \cite{KMMW}, let us write:
\beq
y^5=\varphi\,,\qquad\quad
y^6=L+\chi\,.
\eeq
Then, at first order in the fluctuations ${\tilde C}^{(4)}$ can be written as:
\beq
{\tilde C}^{(4)}=L\,R^4\,{2\rho^2+L^{2}\over 
(\,\rho^2+L^{2}\,)^2}\,
\omega_3\wedge d\varphi\,.
\label{tildeC4}
\eeq
It is immediate from this expression that the second term in 
(\ref{finalWZ}) is negligible
at second order. Let us denote by $\tilde{\cal L}_{WZ}$ the third 
term in ${\cal L}_{WZ}$. By
plugging the value of ${\tilde C}^{(4)}$ given in eq. (\ref{tildeC4}) 
one easily obtains
$\tilde{\cal L}_{WZ}$, namely:
\beq
\tilde{\cal L}_{WZ}=-\Theta^2\,\,R^2\,L\,(\,2\rho^2+L^2\,)\,h(\rho)\,
\sqrt{\det \tilde g}\,\big(\,\partial_{\rho}\varphi\,F_{x^0x^1}+
\partial_{x^0}\varphi\,F_{x^1\rho}+\partial_{x^1}\varphi\,F_{\rho 
x^0}\,\big)\,,
\eeq
where $\tilde g$ is the metric of the $S^3$ and
$h(\rho)^{-1}=1+\Theta^4\,(\,\rho^2+L^2\,)^2$. By 
integrating by parts,
and using the Bianchi identity, $\partial_{\rho}F_{x^0x^1}+
\partial_{x^0}\,F_{x^1\rho}+\partial_{x^1}\,F_{\rho 
\,x^0}=0$, one can write a more
simplified  expression for $\tilde{\cal L}_{WZ}$, namely:
\beq
\tilde{\cal L}_{WZ}=R^2\,f(\rho)\,\sqrt{\det \tilde 
g}\;\varphi\,F_{x^0x^1}\,,
\label{WZfluct}
\eeq
where we have defined the function:
\beq
f(\rho)\equiv
\Theta^2\,L\,\partial_{\rho}\left[(\,2\rho^2+L^2\,)\,h(\rho)\right]\,.
\eeq
Thus, the effect of the non-commutativity on the fluctuations is just 
the introduction of a
coupling between the scalar fluctuation $\varphi$ and the gauge field 
components $A_{x^0}$
and $A_{x^1}$. Notice that $f(\rho)\to 0$ as $\rho\to \infty$, which 
means that the scalar and
vector fluctuations decouple at the UV boundary and thus they 
behave as in the commutative
theory when $\rho$ is large.

The equations of motion for the other scalar fluctuation ($\chi$) and for
the remaining components of the gauge field are the same as in the 
$\Theta=0$ case. Here we will concentrate on the study of  the
fluctuations that depend on the non-commutativity parameter $\Theta$
which, as follows from the  above equations,  are
those corresponding to the fields $\varphi$, $A_{x^0}$ and $A_{x^1}$. 
As it happened in chapter \ref{cp2dpinter}, by imposing
the condition $\partial^{x^\mu}\,A_{x^\mu}=0$ one can consistently put
to zero the gauge field components along the three-sphere and the 
radial coordinate $\rho$.
These are the  modes called of type II in chapter \ref{cp2dpinter}, and
here they are mixed with the fluctuations along the scalar
$\varphi$. The corresponding  equations of motion are:
\bear
&& {R^4\over
(\rho^2+L^2)^2}\,\partial^{\mu}\partial_{\mu}\,A_{x^0}+{1\over
\rho^3}\,
\partial_{\rho}\big(\rho^3\partial_{\rho} A_{x^0}\big)+
{1\over \rho^2}\,\nabla^i\nabla_i\,A_{x^0}-R^2\,{f(\rho)\over
\rho^3}\,\partial_{x^1}\,\varphi=0\,,\rc\rc
&& {R^4\over
(\rho^2+L^2)^2}\,\partial^{\mu}\partial_{\mu}\,A_{x^1}+{1\over
\rho^3}\,
\partial_{\rho}\big(\rho^3\partial_{\rho} A_{x^1}\big)+
{1\over \rho^2}\,\nabla^i\nabla_i\,A_{x^1}-R^2\,{f(\rho)\over
\rho^3}\,\partial_{x^0}\,\varphi=0\,,\rc\rc
&& {R^4\over
(\rho^2+L^2)^2}\,\partial^{\mu}\partial_{\mu}\,\varphi+{1\over
\rho^3}\,
\partial_{\rho}\big(\rho^3\partial_{\rho} \varphi\big)+
{1\over \rho^2}\,\nabla^i\nabla_i\,\varphi+R^2\,{f(\rho)\over
\rho^3}\,F_{x^0x^1}=0\,.
\label{typeII}
\eear
Notice that, as the scalar $\varphi$ decouples from the equation of
$-\partial_{x^0}\,A_{x^0}+\partial_{x^1}\,A_{x^1}$,
the Lorentz condition $\partial^{x^\mu}\,A_{x^\mu}=0$  is consistent with the
equations written above.

\subsection{Decoupling of the equations}

The introduction of some non-commutativity parameter $[x_\nu,x_\nu]
\sim \theta_{\mu\nu}$ explicitly breaks the Lorentz group
$SO(1,3) \to SO(1,1) \times SO(2)$. Therefore, one can build two
Casimir operators related to $p_\mu p^\mu$ and $p_\mu \theta^{\mu\nu}
p_\nu$ \cite{gaume}. Accordingly, let us define the operator:
\beq
{\cal P}^2\equiv-\partial^2_{x^0}+\partial^2_{x^1}\,.
\eeq
Moreover, the squared mass $M^2$ is the eigenvalue of the 
operator
$\partial^{\mu}\partial_{\mu}=-\partial^2_{x^0}+\partial^2_{x^1}+\partial^2_{x^2}
+\partial^2_{x^3} \ $. Let us use these operators  to decouple the
equations (\ref{typeII}). First of all,  we can combine  the equations of
$A_{x^0}$ and $A_{x^1}$ to get an  equation for the field
strength $F_{x^0x^1}$. The equation for $F_{x^0x^1}$ becomes:
\beq
{R^4\over (\rho^2+L^2)^2}\,\partial^{\mu}\partial_{\mu}\,F_{x^0x^1}+
{1\over \rho^3}\partial_{\rho}\big(\rho^3\partial_{\rho} F_{x^0x^1}\big)+
{1\over \rho^2}\,\nabla^i\nabla_i\,F_{x^0x^1}+
R^2\,{f(\rho)\over \rho^3}\,{\cal P}^2
\,\varphi=0\,.
\label{eqF}
\eeq
This last equation, together with the equation of motion of $\varphi$ 
(the last equation in
(\ref{typeII})), constitute a system of coupled equations. In order 
to decouple them,
let us  define  the following combinations of $F_{x^0x^1}$ and $\varphi$:
\beq
\phi_{\pm}\equiv F_{x^0x^1}\pm {\cal P}\varphi\,.
\eeq
Notice that ${\cal P}=\sqrt{-\partial^2_{x^0}+\partial^2_{x^1}}$ 
makes sense acting on
a plane wave. It is straightforward to get the following system of decoupled
equations for $\phi_{\pm}$:
\beq
{R^4\over (\rho^2+L^2)^2}\,\partial^{\mu}\partial_{\mu}\,\phi_{\pm}+
{1\over \rho^3}\partial_{\rho}\big(\rho^3\partial_{\rho} \phi_{\pm}\big)+
{1\over \rho^2}\,\nabla^i\nabla_i\,\phi_{\pm}\,\pm\,
R^2\,{f(\rho)\over \rho^3}\,{\cal P}\phi_{\pm} =0\,.
\eeq
Let us  now expand $\phi_{\pm}$ in a basis of plane waves and 
spherical harmonics:
\beq
\phi_{\pm}=\xi_{\pm}(\rho)\,e^{ikx}\,Y^l(S^3)\,,
\eeq
where the product $kx$ in the plane wave is performed with the 
standard Minkowski metric.
Notice that:
\beq
{\cal P}\,\phi_{\pm}\, =\,k_{01}\,\phi_{\pm}\,,
\eeq
where $k_{01}$ is defined as:
\beq
k_{01}\,\equiv \sqrt{(k_0)^2-(k_1)^2}\,.
\eeq
Moreover, in order to get rid of the factors $R$ and $L$ in the 
differential equations,  let us
define the variable:
\beq
\varrho\equiv {\rho\over L}\,,
\eeq
and the following rescaled quantities:
\beq
\bar M^2\equiv-R^4\,L^{-2}\,k_{\mu}k^{\mu}\,,\qquad\quad
\bar k_{01}\equiv R^2\,L^{-1}\, k_{01}\,,\qquad\quad
\bar\Theta=\Theta\,L\,.
\eeq
Using these definitions it is an easy exercise to obtain the following
equation for the function $\xi_{\pm}$:
\beq
\partial_{\varrho}\left(\varrho^3\,\partial_{\varrho} \,\xi_{\pm}\right)+
\left[\bar M^2\,{\varrho^3\over(\varrho^2+1)^2}-l(l+2)\,\varrho\,\pm\,
f(\varrho)\,\bar k_{01}\right]\xi_{\pm}=0\,,
\label{decoupled}
\eeq
where $f(\varrho)$ is given by:
\beq
f(\varrho)=\bar\Theta^2\,\partial_{\varrho}\left[{2\varrho^2+1\over
1+\bar\Theta^4\,(\varrho^2+1)^2}\right]\,.
\eeq
Notice that $\bar M$ and $\bar k_{01}$ are related as:
\beq
\bar k_{01}=\sqrt{\bar M^2+\bar k_{23}^2}\,,
\label{Mkrelation}
\eeq
where $\bar k_{23}$ is defined as
$\bar k_{23}\equiv R^2L^{-1}\,\sqrt{k_2^2+k_3^2}$. The explicit 
appearance of the momentum
$\bar k_{01}$ in the fluctuation equation (\ref{decoupled}) is a 
reflection of the breaking of
Lorentz invariance induced by the non-commutativity. We will show 
below that the admissible
solutions of eq. (\ref{decoupled}) only occur for some particular 
values of  $\bar M$ and $\bar
k_{01}$. Actually, for a given value of the momentum $\bar k_{23}$ in 
the non-commutative plane,
$\bar k_{01}$ depends on $\bar M$ (see eq. (\ref{Mkrelation})). Thus, 
the study of eq. (\ref{decoupled}) will provide us with information
about the mass  spectrum of the theory and of
the dispersion relation satisfied by the corresponding modes.

When $\bar\Theta=0$, the equations (\ref{decoupled}) can be solved 
analytically in terms of the hypergeometric function. Indeed, they
reduce to the general equation (\ref{cp2fluc}) with $p_2=7$, $d=3$ and
$\gamma_1+\gamma_2=2$, which is a particular case of the equation solved
analytically in section \ref{cp2scexactly-solvable}. There, by requiring
that the solution is well behaved in the IR ($\varrho\to0$) and in the UV
($\varrho\to\infty$) we obtained a quantization condition for $\bar M$
given by eq. (\ref{cp2exactM}), which for $p_2=7$ and $d=3$ leads to:
\beq
\bar M^2(\bar \Theta=0)=4\,(n+l+1)\,(n+l+2)\,.
\label{commutativeM}
\eeq

\subsection{Study of the decoupled equations}
We are now going to study the general features of the fluctuation 
equations (\ref{decoupled})
for a  general value of the non-commutativity parameter $\Theta$. In 
what follows the
fluctuation modes corresponding to $\xi_+$ and $\xi_-$ will be 
referred to simply as $+$ or
$-$modes respectively. A general technique to analyze equations such 
as those written in (\ref{decoupled}) is the one presented in section
\ref{generalfluctuations} consisting in mapping the equation to the zero
energy Schr\"odinger equation. Indeed, by applying the change of
variables (\ref{Sch-variables}) with $p_2=7$ and $d=3$ in equations
(\ref{decoupled}) one gets an equation of the form (\ref{Sch}) with a
potential given by:
\beq
V_{\pm} (y)=-\bar M^2\,\,{e^{2y}\over 
(\,e^{2y}+1\,)^2}+(l+1)^2\,\mp\,
{4\bar\Theta^2 \bar k_{01}\over
[1+\bar\Theta^4\,(\,e^{2y}+1\,)^2\,]^2}\,\,
[1-\bar\Theta^4e^{2y}(\,e^{2y}+1\,)\,]\,,\rc
\label{ncpotential}
\eeq
where $V_+(y)$ ($V_-(y)$) corresponds to the Schr\"odinger equation for
the +($-$)modes.
Therefore, the problem of finding the values of $\bar M$, as a function
of $\bar k_{23}$, which give rise to admissible solutions of eq.
(\ref{decoupled}) can be  rephrased as that of finding the
values of  $\bar M$ such that a zero-energy level for the potential 
(\ref{ncpotential}) exists.
When $\bar\Theta=0$, the potential (\ref{ncpotential}) represents a 
well centered around the
point $y=0$, where it is negative, while it becomes strictly positive 
at $y=\pm\infty$ (see
figure \ref{cpNCpotenfig}).
\begin{figure}
\centerline{\hskip -.8in \epsffile{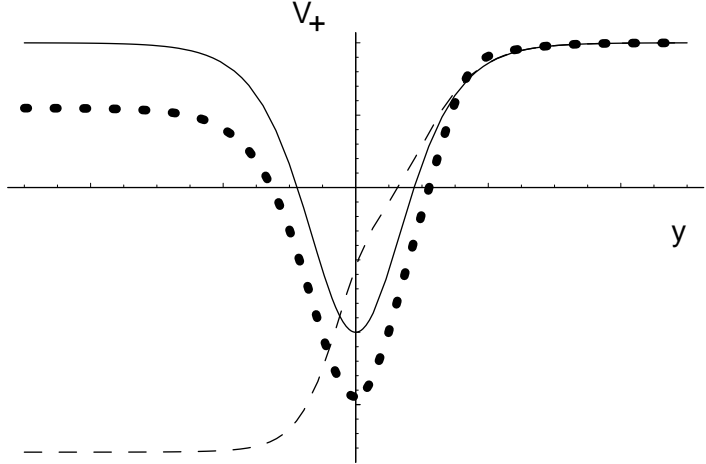}}
\caption{ The potential $V_+(y)$ for $\Theta=0$ (solid line) and for 
two non-vanishing values of
$\Theta$. The dashed line represents a potential such that there is 
only one turning point for
zero energy.}
\label{cpNCpotenfig}
\end{figure}
  The Schr\"odinger equation in this commutative case has a discrete 
spectrum of bound states and
when  $\bar M$  is of the form  (\ref{commutativeM}) one of these 
bound states has zero energy.
When
$\Theta\not=0$ the shape of $V_{\pm}$ is deformed (see figure 
\ref{cpNCpotenfig}) and it might happen
that there is no discrete bound state spectrum such that includes the 
zero-energy level. In
order to avoid this last possibility it is clear that one should have 
two turning points at
zero energy and, thus, the potential must be such that $\lim_{y\to 
\pm\infty}\,V_{\pm} (y)\ge
0$.  From the explicit expression of $V_{\pm} (y)$ given in eq. 
(\ref{ncpotential}) one readily
proves that:
\bear
&&\lim_{y\to +\infty}\,V_{\pm} (y)=(l+1)^2\,,\rc\rc
&&\lim_{y\to -\infty}\,V_{\pm} (y)=(l+1)^2\mp {4\bar\Theta^2 \bar 
k_{01}\over
(1+\bar\Theta^4)^2}\,.
\label{limits}
\eear
It is clear from (\ref{limits}) that the potential $V_{-}$ for the 
$-$modes is always positive
at $y=\pm\infty$. However, by inspecting the right-hand side of eq. 
(\ref{limits}) one easily
realizes this is not  the case for the
$+$modes. Actually,  from the condition $\lim_{y\to -\infty}\,V_{+} 
(y)\ge 0$ we get the
following upper bound on
$k_{01}$:
\beq
\bar k_{01}\,\le\,k_*(\bar\Theta)\,,
\qquad {\rm (+modes)}\,,
\label{bound}
\eeq
where the function $k_*(\bar\Theta)$ is defined as:
\beq
k_*(\bar\Theta)\equiv{(1+\bar\Theta^4)^2\over 
4\bar\Theta^2}\,\,(l+1)^2\,.
\label{k*}
\eeq
In  terms of the original unrescaled quantities, the above bound becomes:
\beq
k_{01}\,\le\,{L\over R^2}\,{(1+\Theta^4\,L^4)^2\over 4\Theta^2 
L^2}\,\,(l+1)^2\,.
\label{truebound}
\eeq

\begin{figure}
\centerline{\hskip -.8in \epsffile{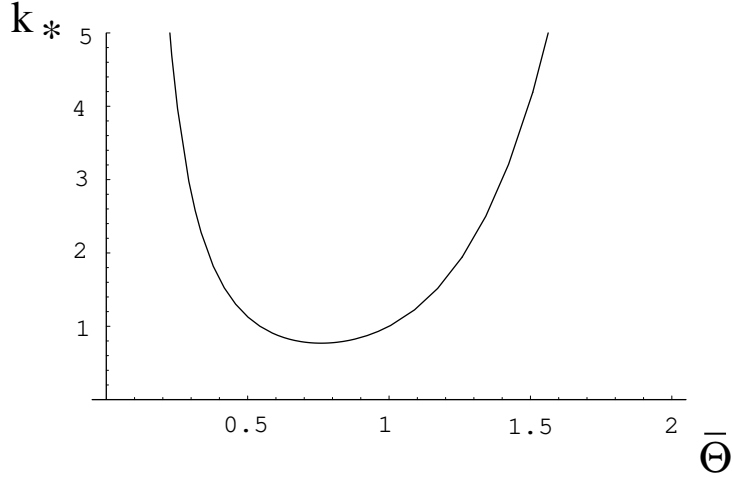}}
\caption{The function  $k_*(\bar\Theta)$  for $l=0$.}
\label{Theta}
\end{figure}

For a given value of the momentum $k_1$ along the $x^1$ direction, 
the above inequality
gives a bound on the energy $k_0$ of the $+$modes which resembles the 
stringy exclusion
principle. Indeed, as in string theory, we are dealing with a theory with a
fundamental length scale. In such theories one expects that it would be
impossible to explore distances smaller than the fundamental length which,
in turn, implies that some sort of upper bound on the energy and momentum
must hold. The function $k_*(\Theta)$ has been plotted in figure 
\ref{Theta}. Notice that this
upper bound is infinity when
$\Theta\to 0$ (as it should) but also grows infinitely when
$\Theta\to\infty$. Actually the function
$k_*(\bar\Theta)$ satisfies the following duality relation:
\beq
k_*(\bar\Theta)=\bar\Theta^4\,k_*({1\over \bar\Theta})\,.
\label{duality}
\eeq

A simple analysis of the function $k_*(\bar\Theta)$ reveals that it 
has a minimum
at a value of $\bar\Theta$ equal to:
\beq
\bar\Theta_0={1\over \root 4 \of 3}\approx 0.7598\,,
\eeq
where it reaches the value
$k_*(\bar\Theta_0)={4\over 3\sqrt{3}}\,(l+1)^2\approx 0.7698\,(l+1)^2$.

Notice that the upper bound (\ref{bound}) satisfied by the $+$modes 
implies, in particular, that
the spectrum of these modes is not an infinite tower as in the 
$\Theta=0$ case (see eq.
(\ref{commutativeM})). Instead, we will have a discrete set of values 
of $\bar M$ which, for a
given value of $l$, is parametrized by an integer $n$ such that $n\le 
n_*(\Theta)$, where
$n_*(\Theta)$ is a function of the non-commutativity parameter which 
diverges when
$\Theta=0,\infty$. Actually, we will find below that, for some 
intermediate values of $\Theta$,
there are no $+$modes  satisfying the bound (\ref{bound}), \ie\ they 
disappear from the
spectrum.

Another interesting conclusion that one can extract from the analysis 
of the potential
$V_{\pm} (y)$ in eq. (\ref{ncpotential}) is the fact that for 
$\bar\Theta$ sufficiently large the
potential and therefore, the spectrum, reduces to the one 
corresponding to the commutative
theory. Indeed, as shown in the plots of figure \ref{cpNCpotenfig},
$V_{\pm}  (y)$ is the same for all
values of $\bar\Theta$ in the region $y\to+\infty$. On the contrary, 
the limit of $V_{\pm}
(y)$ when $y\to-\infty$ does depend on $\bar\Theta$ (see eq. 
(\ref{limits})). However, it
follows from eq. (\ref{limits}) that when $\bar\Theta^6>>\bar k_{01}$ 
the $\bar\Theta$-dependent
term of $\lim_{y\to-\infty}\,V_{\pm} (y)$ can be neglected and, 
actually, when this condition
for $\bar\Theta$ is satisfied the form of
$V_{\pm} (y)$ in the whole range of $y$ is approximately the same as 
in the $\bar\Theta=0$ case.
We will verify below, both analytically and numerically,  that the spectrum for
$\bar\Theta\to\infty$ reduces to the one in the commutative theory. 
This fact, which might seem
surprising at first sight, is reminiscent of the Morita duality 
between irreducible modules
over the non-commutative torus (see \cite{NCFT} and references therein).

%
%
%
%
%
%
%
%
%

\setcounter{equation}{0}
\section{Meson spectrum}
\label{cpncscspectrum}
\medskip
In this section we will analyze in detail the solution of the 
differential equations
(\ref{decoupled}) for the $\pm$modes. The goal of this analysis is to 
determine the meson
spectrum of the corresponding non-commutative field theory at strong 
coupling. We will first study this spectrum in the framework of the
semiclassical WKB  approximation as it has already been done in section
\ref{cp2WKBssc} for the fluctuations of intersecting branes. The WKB 
approximation is only reliable for small $\Theta$ and large principal
quantum number $n$, although in  some cases it turns out to
give the exact result. In our case it will provide us of analytical 
expressions for the energy levels, allowing us to extract the main
characteristics of  the spectrum. We will confirm and enlarge the WKB
results by means of a numerical analysis of the  differential equations
(\ref{decoupled}).

\subsection{WKB quantization}
We shall follow the steps of the analysis performed in section
\ref{cp2WKBssc}. Thus, by comparing the equations (\ref{decoupled})
with the equation (\ref{WKBdiffeq}), one can easily compute the
coefficients $\alpha_i$ and $\beta_i$ defined in eqs.
(\ref{WKBalphabetauno}) and (\ref{alphabeta2}):
\beq
\alpha_1=\beta_1=2\;,\qquad\quad
\alpha_2=2(l+1)\sqrt{1\mp {\bar k_{01}\over
k_*(\bar\Theta)}}\;,\qquad\quad
\beta_2=2(l+1)\,,
\label{cpNCalphabeta}
\eeq
while for the function $\xi$ defined in (\ref{WKBxidef}) one gets:
\beq
\xi=\int_0^{\infty} {d\varrho\over 1+\varrho^2}={\pi\over 2}\,.
\label{cpNCwkbxi}
\eeq
Substituting these results into eq. (\ref{generalWKBlevels}) one arrives
at the following expression for $\bar M$:
\beq
\bar M^2_{WKB}=4(n+1)\,\Bigg(n+l+1+(l+1)\sqrt{1\mp {\bar k_{01}\over
k_*(\bar\Theta)}}\,\,\Bigg)\,.
\label{WKBlevels}
\eeq
Notice that this spectrum only makes sense if the condition 
(\ref{bound}) is satisfied. When
$\bar\Theta=0$ the above formula is exact for $l=0$ and for 
non-vanishing $l$ it reproduces
exactly the quadratic and linear terms in $n$. In general, as we will 
check by comparing  it
with the numerical results,  eq. (\ref{WKBlevels})  is a good
approximation for small $\bar\Theta$ and  $n>> 2(l+1)$. Notice also 
that, due to the property
(\ref{duality}) of $k_*(\bar\Theta)$, the spectrum for 
$\bar\Theta\to\infty$ is identical to
that for
$\bar\Theta\to 0$.

Let us now  analyze some of the consequences of eq. (\ref{WKBlevels}).
Notice first of all that, as $\bar k_{01}$ and $\bar M$ are related as in eq.
(\ref{Mkrelation}), eq. (\ref{WKBlevels}) is really an equation which must be
solved to obtain $\bar M_{WKB}$ as a function of
$\bar k_{23}$ for given quantum numbers $n$ and $l$. Let us 
illustrate this fact when $n=l=\bar
k_{23}=0$. In this case $\bar k_{01}=\bar M$ and by a simple 
manipulation of eq.
(\ref{WKBlevels}) one can verify that
$\bar M_{WKB}$ is obtained by solving the equation:
\beq
F_{\pm}(\bar M_{WKB})\equiv{\bar M_{WKB}^3\over 16}-{\bar 
M_{WKB}\over 2}\,\pm\,
{1\over k_*(\bar\Theta)}=0,
\eeq
where the two signs correspond to those in eq. (\ref{WKBlevels}).
The function $F_{\pm}(\bar M_{WKB})$ has a unique minimum at a value of
$\bar M_{WKB}=\bar M_*=\sqrt{{8\over 3}}$, where it takes the value:
\beq
F_{\pm}(\bar M_*)=-{\sqrt{8}\over 3\sqrt{3}}\pm {1\over 
k_*(\bar\Theta)}\,.
\eeq
Obviously, the equation $F_{\pm}(\bar M_{WKB})=0$ has a solution for 
positive $\bar M_{WKB}$
iff  $F_{\pm}(\bar M_*)\le 0$. This condition is satisfied for all 
values of $\bar
\Theta$ for $F_-$, whereas for  $F_+$ the non-commutativity parameter 
must be such
that:
\beq
k_*(\bar\Theta)\,\ge {3\sqrt{3}\over \sqrt{8}}\,,
\quad\quad {\rm (+modes)}\,.
\eeq
 From the form of the function $k_*(\bar\Theta)$ we conclude that the above
inequality is satisfied when $\bar\Theta\le \bar\Theta_1$ and
$\bar\Theta\ge \bar\Theta_2$, where $\bar\Theta_1$ and $\bar\Theta_2$ 
are the two
solutions of the equation $k_*(\bar\Theta)=3\sqrt{3}/\sqrt{8}$. By numerical
calculation one obtains the following values of  $\bar\Theta_1$ and
$\bar\Theta_2$:
\beq
\bar\Theta_1\approx0.376\,,
\qquad\quad
\bar\Theta_2\approx1.239\,,
\qquad({\rm WKB})\,.
\label{WKBrange}
\eeq
Therefore, the ground state for the $+$modes disappears from the spectrum if
$\bar\Theta_1<\bar\Theta<\bar\Theta_2$. One can check similarly that  the
modes with $n>0$ also disappear if 
$\bar\Theta_1<\bar\Theta<\bar\Theta_2$ . Thus $(\bar\Theta_1,
\bar\Theta_2)$ is a forbidden interval of the non-commutativity parameter
for the $+$modes. Moreover, notice that the condition (\ref{bound})
reduces in this 
$k_{23}=0$ case to the
inequality $\bar M\le k_*(\bar\Theta)$. One can check that when
$\bar\Theta\,=\bar\Theta_{1,2}$ the bound (\ref{bound}) is indeed satisfied,
since $\bar M=\sqrt{{8\over 3}}<k_*(\bar\Theta_{1,2})=3\sqrt{3}/\sqrt{8}$.

For a non-vanishing value of $\bar\Theta$ the energy levels of the 
$+$modes are cut off at some
maximal value $n_*(\bar \Theta)$ of the quantum number $n$. If we 
consider states with zero
momentum $\bar k_{23}$ in the non-commutative plane, for which $\bar 
M=\bar k_{01}$, the
function $n_*(\bar \Theta)$ can be obtained by solving the equation:
\beq
  \bar M^2_{WKB}\Big|_{n=n_*(\bar \Theta)}=k_*(\bar\Theta)^2\,,
\label{cutoff}
\eeq
where the left-hand side is given by the solution of eq. 
(\ref{WKBlevels}). One can get an
estimate of $n_*(\bar \Theta)$ for small and  large values of $\bar 
\Theta$ by putting
$k_*(\bar\Theta)\to\infty$ on the left-hand side of eq. 
(\ref{cutoff}). In this case eq.
(\ref{WKBlevels}) yields immediately the value of $\bar M$ and if, 
moreover,  we
consider states with $l=0$, one arrives at the approximate equation:
\beq
4(\,n_*+1\,)\,(\,n_*+2\,)\approx k_*(\bar\Theta)^2\,,
\eeq
which can be easily solved, namely:
\beq
n_*(\bar \Theta)\approx -{3\over 2}+{1\over 
2}\,\sqrt{1+k_*(\bar\Theta)^2}\,.
\eeq

\subsection{Numerical results}
We would like now to explore numerically the spectrum of values of 
$\bar M$ for the differential
equation (\ref{decoupled}). This can be done by applying the shooting
technique as it was described in section \ref{cp2ssgeneralnumeric} of
the second chapter. First, one needs to know the behavior of the
solutions of the equation (\ref{decoupled}) in the IR ($\varrho\sim0$)
and in the UV ($\varrho\to\infty$). For $\varrho\sim0$ the regular
solutions of eq. (\ref{decoupled}) are given by:
\beq
\xi_{\pm}\sim \varrho^{-1+(l+1)\,\sqrt{1\mp {\bar k_{01}\over 
k_*(\bar\Theta)}}}\,, \qquad (\varrho\sim 0)\,.
\label{IRxi}
\eeq
As $k_*(\bar\Theta)\to \infty$ when  $\bar\Theta \to 0$,
it is straightforward to verify that eq. (\ref{IRxi}) reduces to 
$\xi_{\pm}\sim \varrho^l$ when  $\bar\Theta = 0$  and thus, the IR
behavior of $\xi_{\pm}$ coincides with the one corresponding to the
commutative fluctuations, namely eq. (\ref{cp2generalIR}),  when the
non-commutative deformation is switched off. Moreover, it is interesting
to notice  that the condition  $\bar k_{01}\le k_*(\bar\Theta)$ for the
$+$modes (eq. (\ref{bound})) appears naturally if we require 
$\xi_{+}$ to be real in the IR.

Next, as in section \ref{cp2ssgeneralnumeric}, one has to
match the $\varrho\sim0$ behavior (\ref{IRxi}) with the asymptotic trend
of the solution for large $\varrho$. In this limit one can easily find a
solution of the form $\xi_{\pm}\sim \varrho^{\Lambda}$. 
An elementary calculation shows that there are two possible solutions for
the exponent $\Lambda$, namely $\Lambda=l, -(l+2)$, which coincide with
the ones of the commutative case, given by eq. (\ref{IRroots}) with
$p_2=7$ and $d=3$. Therefore, the general behavior of
$\xi_{\pm}$ for large
$\varrho$ will be of the form:
\beq
\xi_{\pm}\,\sim\,c_1\,\varrho^{l}+c_2\,\varrho^{-(l+2)}\;,
\qquad (\varrho\sim \infty)\,,
\label{UVxi}
\eeq
where $c_1$ and $c_2$ are constants. The allowed solutions are those 
which vanish at infinity,
\ie\ those for which the coefficient $c_1$ in (\ref{UVxi}) is zero. 
For a given value of the
momentum $\bar k_{23}$ in the non-commutative plane,
this condition only happens
for a discrete set of values of $\bar M$, which can be found 
numerically by solving the
differential equation (\ref{decoupled}) for a function which behaves 
as in eq. (\ref{IRxi}) near
$\varrho=0$ and then by applying the shooting technique to determine 
the values of $\bar M$ for
which $c_1=0$ in eq. (\ref{UVxi}). Proceeding in this way one gets a 
tower of values of $\bar
M$, which we will order according to the increasing value of $\bar 
M$. In agreement with the
general expectation for this type of boundary value problems, the fluctuation
$\xi_{\pm}$ corresponding to the $n^{th}$ mode has $n$ nodes. 
Moreover, as happened in the WKB
approximation, when $\bar \Theta\not =0,\infty$ the tower of states 
for the $+$modes terminates
at some maximal value $n=n_*$ and, for some values of $\bar\Theta$ 
and $\bar k_{23}$, there is
no solution for $\xi_{+}$ satisfying the boundary conditions and the 
bound (\ref{bound}) (see
below).

Let us now discuss the results of the numerical calculation when the 
momentum $\bar k_{23}$ in
the non-commutative direction is zero. In this case  one must put
$\bar k_{01}=\bar M$ in the differential equation (\ref{decoupled}) (see eq.
(\ref{Mkrelation})). For small $\bar \Theta$ the numerical results 
should be close to those
given by the WKB equation  (\ref{WKBlevels}). In order to check this 
fact we compare in table \ref{cpncnumtable} the values  of $\bar M^2$
obtained numerically and those  given by the WKB formula
(\ref{WKBlevels})  for the  $\pm$modes  for $l=0$ and $\bar \Theta=.1$

\begin{table}[!h]
\begin{tabular}[b]{|c|c|c|}
\hline
\multicolumn{3}{|c|}{
 \rule{0mm}{4.5mm}$\bar M^2$ for $+$modes for $l=0$, $\bar \Theta=.1$}\\
\hline
  $n$  &  Numerical & WKB \\
\hline
\ \ 0 & $7.37$  & $7.77$  \\
\ \ 1 & $22.24$  & $23.19$  \\
\ \ 2 & $44.67$  & $46.24$  \\
\ \ 3 & $74.76$  & $76.89$  \\
\ \ 4 & $112.52$  & $115.11$  \\
\ \ 5 & $157.94$  & $160.85$  \\
\hline
\end{tabular}
\qquad\qquad\qquad
\begin{tabular}[b]{|c|c|c|}
\hline
\multicolumn{3}{|c|}{\rule{0mm}{4.5mm}$\bar M^2$ for $-$modes for $l=0$,
$\bar \Theta=.1$}\\
\hline
  $n$  & Numerical   & WKB \\
\hline
\ \ 0 & $8.69$  & $8.22$  \\
\ \ 1 & $25.84$  & $24.76$  \\
\ \ 2 & $51.35$  & $49.58$  \\
\ \ 3 & $85.10$  & $82.68$  \\
\ \ 4 & $126.95$  & $124.05$  \\
\ \ 5 & $176.90$  & $173.66$  \\
\hline
\end{tabular}
\caption{Values of $\bar M^2$ obtained numerically and with the WKB method
for the +modes (left) and the $-$modes (right) when $\bar k_{23}=0$.}
\label{cpncnumtable}
\end{table}

We notice that, indeed,  the WKB values represent reasonably well the 
energy levels especially,
as it should, when the number $n$ is large.

It is also interesting to analyze the behavior of the ground state eigenvalue
$\bar M(n=0,l=0)$ with the non-commutativity parameter $\bar\Theta$. In 
figure \ref{spectrum}
we plot our numerical results for the $\pm$modes when the momentum 
$\bar k_{23}$ is zero.
As expected, when $\bar\Theta=0$ we recover the value corresponding 
to the mass of the lightest
meson of the commutative theory. Moreover, if we increase 
$\bar\Theta$ the value of $\bar M$
for the $+$modes decreases, until we reach a  point from which there is no
solution of the boundary value problem satisfying the bound $\bar 
M\le k_*(\bar\Theta)$ in a
given interval  $\bar\Theta_1\le \bar\Theta\le \bar\Theta_2$ of the 
non-commutativity parameter.
The numerical values of $\bar\Theta_{1,2}$  are slightly different 
from the WKB result
(\ref{WKBrange}), namely $\bar\Theta_{1}=0.41$, 
$\bar\Theta_{2}=1.37$. It is interesting to notice
the jump and the different behavior  of $\bar M$ at both sides of 
the forbidden region. Notice
also that, as expected, $\bar M$ approaches  the commutative result 
when $\bar \Theta$ is
very large.

\begin{figure}
\centerline{\hskip -.1in \epsffile{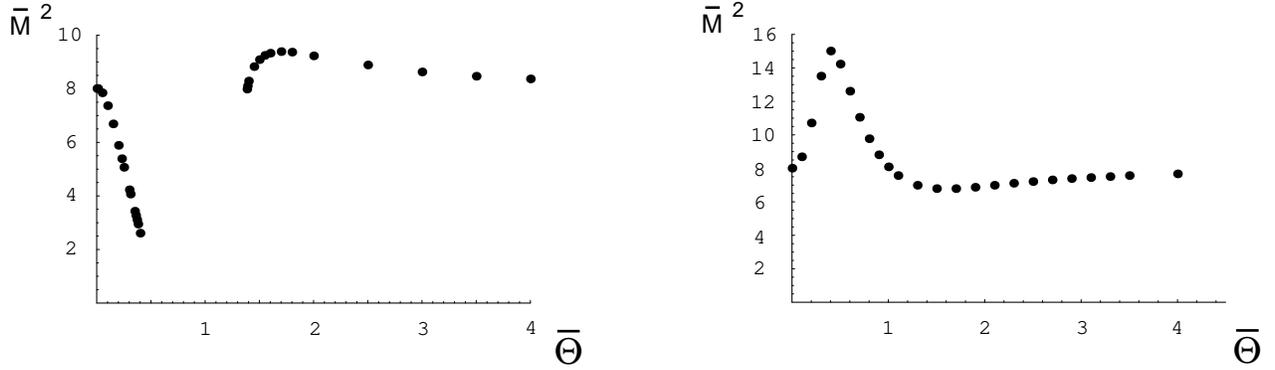}}
\caption{Numerical values of $\bar M^2$ for the ground state $n=l=0$ 
for $\bar k_{23}=0$ as a
function of $\bar \Theta$ for $\xi_+$ (left) and $\xi_-$(right). The 
corresponding value of the
commutative theory is $\bar M^2=8$.}
\label{spectrum}
\end{figure}

For the $-$modes there is always a solution for the ground state for 
all values of $\bar\Theta$
and, again, the corresponding value of $\bar M$ equals the 
commutative result when
$\Theta=0,\infty$. Interestingly, the range of values of $\bar\Theta$ 
for which $\bar M$ differs
significantly from its commutative value is approximately the same as 
the forbidden interval for
the $+$modes (see figure \ref{spectrum}).

The differential equations for our fluctuations break explicitly the 
Lorentz invariance among
the Minkowski coordinates $x^0,\cdots,x^3$. In order to find out how 
this breaking is reflected
in the spectrum it is interesting to study the values of $\bar M^2$ 
for $\bar k_{23}\not=0$. In
this case $\bar k_{01}\not= \bar M$ (see eq. (\ref{Mkrelation})) and 
one can regard $\bar
k_{23}$ as a external parameter in our boundary value problem. When 
$\bar \Theta\not=0$ the
values of $\bar M$ that one obtains by solving the differential 
equation (\ref{decoupled}) do
depend on  $\bar k_{23}$. This dependence encodes the modification of 
the relativistic
dispersion relation due to the non-commutative deformation. For 
illustrative purposes let us
consider the spectrum for the $+$modes. The values of $\bar M^2$ as a 
function of $\bar k_{23}$
for the ground state ($n=l=0$) and two different values of 
$\bar\Theta$ are shown in figure
\ref{spectrumk23}. Notice that, as a consequence of  (\ref{bound}), 
the momentum $\bar k_{23}$ is
bounded from above. Moreover, $\bar M^2$ increases (decreases) with 
$\bar k_{23}$ when
$\bar\Theta < \bar\Theta_1$ ($\bar\Theta > \bar\Theta_2$), while it 
becomes independent of
$\bar k_{23}$ as $\Theta\to 0,\infty$.

\begin{figure}[h!]
\centerline{\hskip -.1in \epsffile{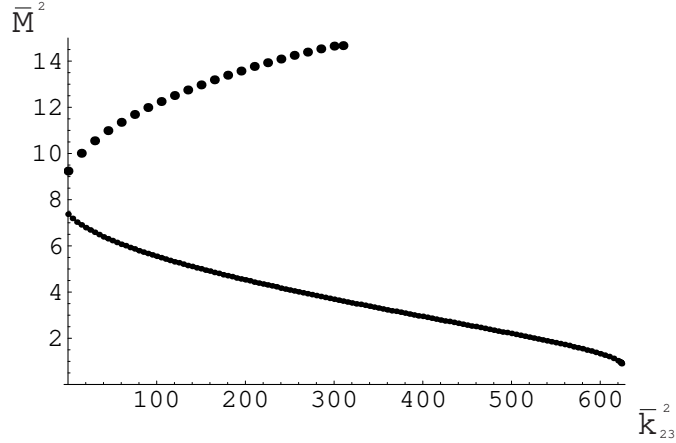}}
\caption{Numerical values of $\bar M^2$ for the ground state $n=l=0$ 
of the $+$modes  as a
function of $\bar k_{23}^2$ for two different values of $\bar 
\Theta$. The continuous line
corresponds to $\bar\Theta=0.1<\bar\Theta_1$, while the dotted line 
are the values of $\bar
M^2$ for $\bar\Theta=2> \bar\Theta_2 $.}
\label{spectrumk23}
\end{figure}

%
%
%
%
%
%
%
%
%

\setcounter{equation}{0}
\section{Semiclassical strings in the non-commutative background}
\label{cpncscstrings}
\medskip

Let us now study the meson spectrum for large four-dimensional spin. 
An exact calculation of
this spectrum would require the analysis of the fluctuations of open 
strings attached to the
D7-brane in the non-trivial gravitational background of section
\ref{cpncscback}.  This calculation is not
feasible, even for the case of D7-branes in the $AdS_5\times S^5$ 
geometry considered in  ref.
\cite{KMMW}. However, for large four-dimensional spin we can treat 
the open strings
semiclassically, as suggested in ref. \cite{rotating}. In this 
approach one solves
the classical equations of motion of rotating open strings with 
appropriate boundary conditions.
These equations are rather complicated and can only be solved 
numerically. However, from these
numerical solutions we will be able to extract the energy and angular 
momentum of the string and
study how they are correlated. The analysis for the commutative 
$\Theta=0$ case was performed in
ref. \cite{KMMW}. Here we would like to explore the effect of a 
non-commutative deformation on
the behavior found in ref. \cite{KMMW}.

\subsection{Strings rotating in the non-commutative plane}
\label{counterclockwise}
As explained above, the spectrum of mesons with
large four-dimensional spin  can be obtained from classical
rotating open strings. In order to study these configurations
let us consider strings attached to the
D7 flavor brane, extended in the radial direction, and rotating in
the $x^2x^3$ plane (i.e. in the non-commutative directions). The
relevant part of the metric is:
\beq
{r^2\over
R^2}\left[-dt^2+h\left(\left(dx^2\right)^2+\left(dx^3\right)^2\right)\right]+{R^2\over
r^2}dr^2.
\eeq
Defining the coordinate $z$ as:
\beq
z={R^2\over r}\,,
\eeq
and changing to polar
coordinates on the $x^2x^3$ plane, the  metric above becomes:
\beq
{R^2\over
z^2}\left[-dt^2+h\left(d\rho^2+\rho^2d\theta^2\right)+dz^2\right]\,,
\eeq
where $h$ and $B$, in terms of the new coordinate $z$, are:
\bear
h&=&\left(1+R^8{\Theta^4\over z^4}\right)^{-1},\rc\rc
B&=&-\Theta^2{r^4\over R^2}h\,\rho\,d\rho\wedge d\theta=-\Theta^2{R^6\over
z^4}h\,\rho\,d\rho\wedge d\theta.
\eear
Notice that the coordinate $\rho$ introduced above as radial 
coordinate in the $x^2x^3$ plane
has nothing to do with the one defined below eq. (\ref{cpncosmetric}). 
Actually, since the string is
rotating around one of its middle points, we must allow $\rho$ to 
have negative values.
Let us now consider an open
fundamental string moving in the above background. Let 
$(\tau,\sigma)$ be worldsheet
coordinates. The embedding of the string will be determined by the 
functions $X^M(\sigma,\tau)$,
where $X^M$ represent the target space coordinates. Moreover, the 
dynamics of the string is
governed by the standard Nambu-Goto action:
\beq
\label{rotact}
S=-{1\over 2\pi\alpha'}\int d\tau \,d\sigma \sqrt{-\det g}+{1\over
2\pi\alpha'}\int P[B]\,,
\eeq
where $g$ is the metric induced on the worldsheet of the string. Let 
us write the form of the
action (\ref{rotact}) for the following ansatz:
\beq
\label{rotans}
t=\tau \: ,\: \theta=\omega\tau \: ,\: \rho=\rho(\sigma) \: ,\:
z=z(\sigma),
\eeq
where $\omega$ is a constant angular velocity. If the prime denotes 
derivative with respect to
$\sigma$, the determinant of the induced metric and the pullback of 
the $B$ field for the ansatz
(\ref{rotans})  are given by:
\bear
\sqrt{-\det g}={R^2\over
z^2}\sqrt{\left(1-h\,\rho^2\omega^2\right)\left(h\,\rho'^2+z'^2\right)},\\
P[B]=\Theta^2{R^6\over z^4}h\,\rho\,\omega\,\rho'\,d\tau\wedge d\sigma\,.
\eear
By plugging this result in the  action (\ref{rotact}) one finds the 
following lagrangian
density:
\beq
\label{lgr}
{\cal L}={R^2\over
2\pi\alpha'}\left(-{1\over z^2}\sqrt{\left(1-h\,\rho^2\,\omega^2\right)
\left(h\,\rho'^2+z'^2\right)}+\Theta^2{R^4\over
z^4}h\,\rho\,\omega\,\rho'\right).
\eeq
The lagrangian (\ref{lgr}) does not depend explicitly on $t$ and 
$\theta$. Therefore, our system
has  two conserved quantities: the  energy $E$ and the (generalized) 
angular momentum $J$, whose
expressions are given by:
\bear
\label{Enrot}
E&=&\omega {\partial S \over \partial \omega} - S=
{R^2\over 2\pi\alpha'}\int d\sigma
{\sqrt{h \,\rho'^2+z'^2} \over z^2 \sqrt{1-h\,\rho^2\omega^2}} \, ,\rc
J&=&{\partial S \over \partial \omega}=
{R^2\over 2\pi\alpha'} \left[\int d\sigma \left({h\,\rho^2\omega \over
z^2} {\sqrt{h \,\rho'^2+z'^2} \over \sqrt{1-h\,\rho^2\omega^2}}+
\Theta^2{R^4\over
z^4}h\,\rho\,\rho'\right)\right]\,.
\eear
The physical angular momentum of the string is given by the first 
term on the second equation
in (\ref{Enrot}), namely:
\beq
J_1={R^2\over 2\pi\alpha'}\int
d\sigma
{h\rho^2\omega \over
z^2}{\sqrt{h\,\rho'^2+z'^2}\over\sqrt{1-h\,\rho^2\omega^2}}\,.
\label{J1}
\eeq

The equations of motion defining the time-independent profile of
the string can be obtained from (\ref{lgr}). Moreover, in addition one
must impose the boundary conditions that make the action stationary:
\beq
{\partial L\over \partial(X')^M}\delta X^M\Big|_{\partial\Sigma}=0\,.
\label{rotbc}
\eeq

As the endpoints of the string are
attached to the flavor brane placed at constant $z$, $\delta
z|_{\partial\Sigma}=0$. Moreover, since $\delta 
\rho\big|_{\partial\Sigma}$ is arbitrary, the
condition (\ref{rotbc}) reduces to:
\beq
{\partial {\cal  L}\over \partial \rho\,'}\Bigg|_{\partial\Sigma}=0\,.
\label{bcrho}
\eeq
Taking into account the explicit form of ${\cal  L}$ (eq. 
(\ref{lgr})), one can rewrite
eq. (\ref{bcrho}) as:
\beq
{\sqrt{1-h\,\rho^2\omega^2}\over \sqrt{h\,\rho'^2+z'^2}}\,\,\rho'
\,\Bigg|_{\partial\Sigma}={R^4\Theta^2\over 
z^2}\,\rho\omega\,\Bigg|_{\partial\Sigma}\,.
\label{bcrhodos}
\eeq
Eq. (\ref{bcrhodos}) can be used to find  the angle at which the 
string hits the flavor brane.
Indeed, let us suppose that the D7-brane is placed at $z=z_{D7}$ and 
that the string
intercepts the D7-brane at two points with coordinates 
$\rho=-\tilde\rho_{D7}<0$ and
$\rho=\rho_{D7}>0$. We will orient the string by considering the 
$\rho=-\tilde\rho_{D7}$
  ($\rho=\rho_{D7}$) end as its initial (final) point. It follows 
straightforwardly from eq.
(\ref{bcrhodos}) that the signs of $d\rho/d\sigma$ at the two ends of 
the string are:
\beq
{d\rho\over d\sigma}\,\Bigg|_{\rho=-\tilde\rho_{D7}}\,\le\,0\,,
\qquad\quad
{d\rho\over d\sigma}\,\Bigg|_{\rho=\rho_{D7}}\,\ge\,0\,.
\label{signhit}
\eeq
Notice that for $\Theta=0$ the right-hand side
of eq. (\ref{bcrhodos}) vanishes, which means that 
$\rho'_{|\partial\Sigma}=0$ and, therefore,
the string ends orthogonally on the
D7-brane, in agreement with the results of ref. \cite{KMMW}. Moreover,
$\tilde\rho_{D7}=\rho_{D7}$ in the commutative case  and the 
string configuration is
symmetric around the point
$\rho=0$ (see ref. \cite{KMMW}). On the contrary,
eq. (\ref{bcrhodos}) shows that $\rho'_{|\partial\Sigma}$ does not vanish
in the non-commutative theory  and thus
the string hits the D7-brane at a certain angle, which depends on the 
non-commutativity parameter
$\Theta$ and on the  $\rho$ and $z$ coordinates of $\partial\Sigma$. 
Actually, it follows from
the signs displayed in eq. (\ref{signhit}) that the string profile is 
not symmetric\footnote{Rotating strings which are not
symmetric with respect to a center of rotation were also considered
in \cite{talav} where the asymmetry came from considering
different quark masses.}
 around
$\rho=0$ when $\Theta\not= 0$ and that the string is tilted towards 
the region of negative $\rho$
\footnote{The strings rotating in the sense opposite to the one in
(\ref{rotans}), \ie\ with $\theta=-\omega\tau$, are tilted towards the region
of positive $\rho$. Apart from this, the other results in this section are not
modified if we change the sense of rotation.}. The actual values
of  $dz/d\rho$ at the two ends of the string can be obtained by 
solving the quadratic equation for
$z'/\rho'$ in (\ref{bcrhodos}). One gets:
\beq
\label{init}
{dz \over d\rho}\Bigg|_{\rho=-\tilde\rho_{D7}}=
-{\,z_{D7}^{2} \sqrt{1-\tilde\rho_{D7}^2 \,\omega^2}\over
\hat\Theta^2 \,\tilde\rho_{D7}\,\omega}\,,
\qquad\quad
{dz \over d\rho}\Bigg|_{\rho=\rho_{D7}}=
-{\,z_{D7}^{2} \sqrt{1-\rho_{D7}^2 \,\omega^2}\over
\hat\Theta^2 \,\rho_{D7}\,\omega}\,,
\eeq
where the sign of the right-hand side has been chosen to be in 
agreement with eq.
(\ref{signhit}) and we have defined $\hat \Theta=\Theta\,R^2$.

Setting $\sigma=\rho$, the equation of motion defining the string
profile $z(\rho)$ can be easily obtained from the lagrangian (\ref{lgr}):
\bear
h{z''\over h+z'^2}-{h\,\rho\,\omega^2\over 1-h\,\rho^2 \omega^2}z'
-{4\hat  \Theta^2\over z^3}
h^2\rho\,\omega\sqrt{{h+z'^2\over 1-h\,\rho^2
\omega^2}}+\rc +{2h^2\over z\left(1-h\,\rho^2
\omega^2\right)\left(h+z'^2\right)}\Bigg\{z'^2 \left[1-{\hat
\Theta^4\over z^4} +\rho^2\omega^2\left(1-2h+{\hat  \Theta^4\over
z^4}h\right)\right]+
\rc+h\left[1-\rho^2\omega^2h\left(1-{\hat \Theta^4\over z^4}\right)\right]
\Bigg\} =0,
\label{rotdiffeq}
\eear
where now $z'={dz\over d\rho}$. In order to solve the second order 
differential equation
(\ref{rotdiffeq}) we need to impose the value of $z$ and $z'$ at some 
value of $\rho$. Clearly,
the boundary condition (\ref{init}) fixes $z'$ at  $\rho=\rho_{D7}$. 
Since the string intersects
the D7-brane at this value of
$\rho$, it is evident that we have to impose also that:
\beq
z(\rho=\rho_{D7})=z_{D7}\,.
\label{zd7}
\eeq
By using the initial conditions (\ref{init}) at $\rho=\rho_{D7}$ and 
(\ref{zd7}), the equation of
motion (\ref{rotdiffeq}) can be numerically integrated for given 
values of $z_{D7}$, $\rho_{D7}$
and $\omega$. It turns out that these values are not uncorrelated. 
Indeed, we still have to
satisfy the condition written in eq. (\ref{init}) for
$z'(\rho=-\tilde\rho_{D7})$, which determines the angle at which the
$\rho=-\tilde\rho_{D7}$ end of the string hits the brane and it is 
not satisfied by arbitrary
values of $z_{D7}$, $\rho_{D7}$ and $\omega$. Actually, let us 
consider a fixed  value of
$z_{D7}$. Then,   by requiring  the fulfillment of eq. (\ref{init}), 
one gets a relation between
the quark-antiquark separation $\rho_{D7}+\tilde\rho_{D7}$ and the 
angular velocity $\omega$.
Before discussing this relation let us point out that, due to the 
tilting of the string, the
coordinate $\rho$ is not a good global  worldvolume coordinate in the 
region of negative  $\rho$,
since
$z(\rho)$ is a double-valued function in that region. In order to 
overcome this problem we will
solve eq. (\ref{rotdiffeq}) for $z(\rho)$ starting at 
$\rho=\rho_{D7}$ until a certain negative
value of $\rho$ and beyond that point we will continue the curve by 
parametrizing  the
string by means of a function
$\rho=\rho(z)$. The differential equation governing the function 
$\rho(z)$, which is similar to
the one written in (\ref{rotdiffeq}) for $z(\rho)$,   can be easily 
obtained from the lagrangian
density (\ref{lgr}) after taking $\sigma=z$. We have solved this 
equation by using as initial
conditions the values of the coordinate and slope of the last point 
of the $z=z(\rho)$ curve.
By performing the numerical integration in this way, the $\rho(z)$ 
curve is continued until
$z=z_{D7}$. Then $\tilde\rho_{D7}$ is determined as 
$-\tilde\rho_{D7}=\rho(z_{D7})$ and one can
check whether or not the string hits the flavor brane at 
$\rho=-\tilde\rho_{D7}$ with the angle
of eq. (\ref{init}). For a given value of the angular velocity 
$\omega$ this only happens for
some particular values of $\rho_{D7}$ and $\tilde\rho_{D7}$. Some of 
the profiles found by
numerical integration are shown in figure \ref{profile}. As explained 
above these curves are
tilted in general. This tilting increases with the non-commutativity
parameter and, for fixed
$\Theta\not=0$, it becomes more drastic as $\omega$ grows.

As mentioned above, for a given value of $z_{D7}$ the fulfillment of 
the conditions written in
eq.  (\ref{init}) determines a relation between $\tilde\rho_{D7}$, 
$\rho_{D7}$ and
$\omega$. In order to characterize this relation let us define 
$\bar\rho_{D7}$ as the half of the quark-antiquark separation, \ie\
$2\bar\rho_{D7}=\tilde\rho_{D7}+\rho_{D7}$. In figure \ref{numrot} we
have represented $\bar\rho_{D7}$ as a function of $\omega$ for some
non-vanishing  value of $\Theta$. From these
numerical results one concludes that $\bar\rho_{D7}\to \infty$ when 
$\omega\to 0$, while $\bar\rho_{D7}$ vanishes for large $\omega$.
Actually, the behaviors  found for small  and large $\omega$ can be
reproduced by a simple power law, namely:
\bear
&&\bar\rho_{D7}\,\sim \omega^{-{2\over 3}}\,,\qquad (\omega \to 
0)\,,\rc
&&\bar\rho_{D7}\,\sim \omega^{-{1}}\,,\qquad (\omega \to \infty)\,.
\label{powerrho}
\eear

\begin{figure}
\centerline{\epsffile{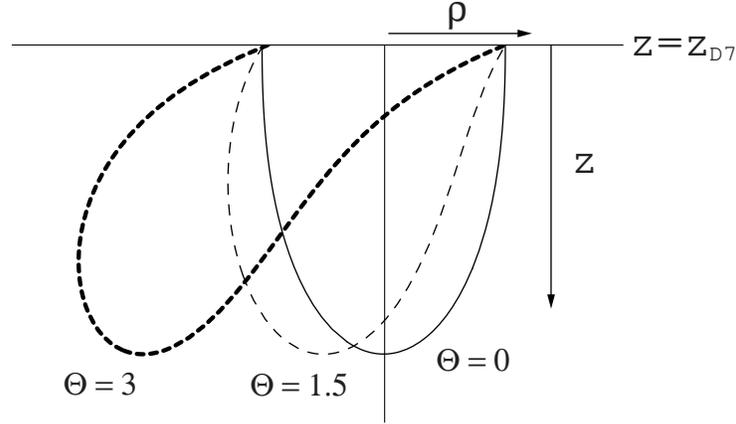}}
\caption{Rotating string profiles for three different values of the 
non-commutativity parameter
($\Theta=$0, 1.5 and 3). The three curves correspond to the same 
value of $\rho_{D7}$. As
$\Theta$ grows the tilting increases. However, the maximal value of 
$z$ is roughly the same for
the three curves. }
\label{profile}
\end{figure}

The power law behaviors displayed in eq. (\ref{powerrho})  coincide 
with the ones found in ref.
\cite{KMMW} for the $AdS_5\times S^5$ background. They imply, in 
particular that $\omega\to 0$
corresponds to having long strings, whereas for $\omega\to\infty$ we 
are dealing with very short
strings.

Once
the profile of the string  is known, we can plug it on the right-hand 
side of eqs. (\ref{Enrot})
and (\ref{J1}) and obtain the energy and the angular
momentum  of the rotating string by numerical integration. As happens for the
$AdS_5\times S^5$ background, large(small) angular velocity 
corresponds to small (large) values
of the angular momentum $J_1$. Actually, for small $\omega$ the
angular momentum $J_1$ diverges, while $J_1\to 0$ for large $\omega$. 
The spectrum $E(J_1)$ can
be obtained parametrically by integrating the equation of motion for 
different values of
$\omega$. The results  for some values of $\Theta$ have been plotted 
in figure \ref{numrot}. For
small  $J_1$ (or large
$\omega$) the meson energy follows a Regge trajectory since the 
energy $E$ grows linearly with
$\sqrt{J_1}$. The actual value of the Regge slope can be obtained 
analytically (see the next
subsection). This Regge slope, which is nothing but an effective 
tension for the rotating
string, is independent of $\Theta$ (see figure \ref{numrot}). An 
explanation of this fact is
provided in the next subsection.  For intermediate values of $J_1$ 
the $E(J_1)$ curve does
depend on the non-commutativity parameter while, on the
contrary, for large $J_1$ (or small $\omega$) one recovers again the 
$\Theta=0$ result, since the
energy becomes $2m_q$, with $m_q={R^2\over 2\pi\alpha'z_{D7}}$ being 
the mass of the quarks.
Actually this last result is quite natural since this large $J_1$ 
region corresponds to large
strings and one expects that the effects of the non-commutativity 
will disappear.
\begin{figure}[h!]
\centerline{\epsffile{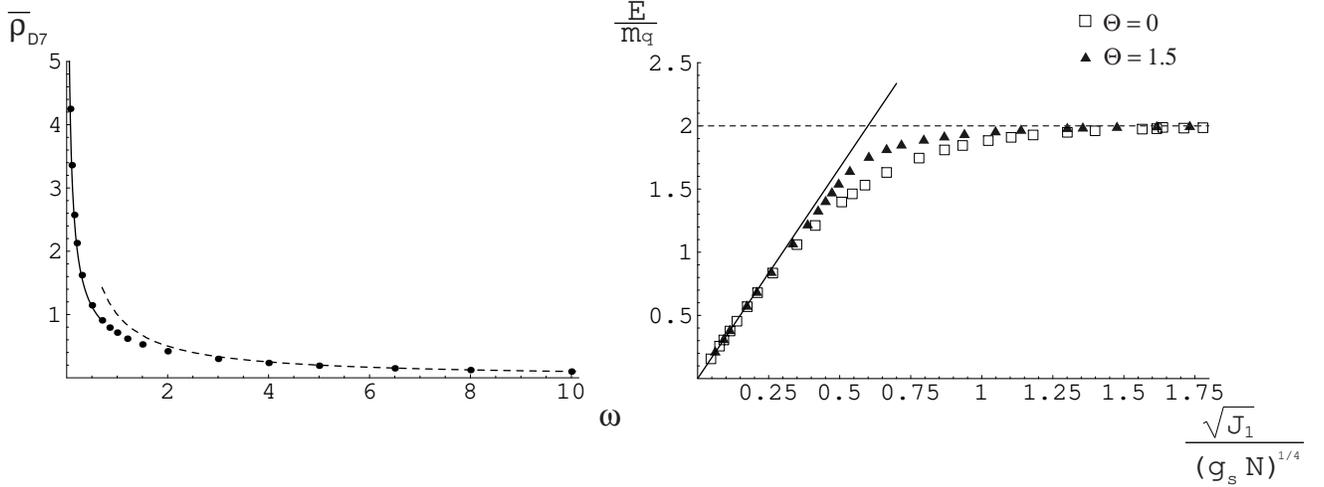}}
\caption{ On the left we plot the half of the quark-antiquark 
separation $\bar\rho_{D7}$ as a
function of the angular velocity $\omega$. For large $\omega$,
$\bar\rho_{D7}\approx \omega^{-1}$ (dashed line). In the $\omega\to 0$ region,
$\bar\rho_{D7}\sim \omega^{-2/3}$ (solid line). On the right plot we 
represent the
energy versus angular momentum of the rotating string for the 
commutative and non-commutative
theories. The meson masses follow Regge trajectories in the large $\omega$
(small $J_1$) region with an effective tension which does not depend on
the non-commutativity parameter $\Theta$ and is given by eq. 
(\ref{tauR}). The linear Regge
trajectory given by eq. (\ref{linearRegge}) is also plotted for 
comparison (solid line).
For large $J_1$ (small
$\omega$) which corresponds to long strings, the energy approaches 
the free value $2m_q$.
We are setting $z_{D7}=R^2=\alpha'=1$.}
\label{numrot}
\end{figure}

An interpretation of the behavior of the spectrum in the two 
limiting regimes (small and large
$J_1$) was given in ref. \cite{KMMW}. Let us recall, and adapt to our 
system, the arguments of
\cite{KMMW}. For small $J_1$ the string is very short and it is not 
much influenced by the
background geometry. As  a consequence the spectrum is similar to the 
one in flat space, \ie\ it
follows a Regge trajectory with a tension which is just the proper 
tension $1/2\pi\alpha'$
appropriately red-shifted. This effective tension is independent of 
$\Theta$. However, we
will verify in section \ref{cpncschangstrings} that this is not the case
when computing the static quark-antiquark potential energy.
The static and dynamic  tensions are different and they only 
coincide for $\Theta=0$,
where we recover the results of ref. \cite{KMMW}.

\begin{figure}
\centerline{\epsffile{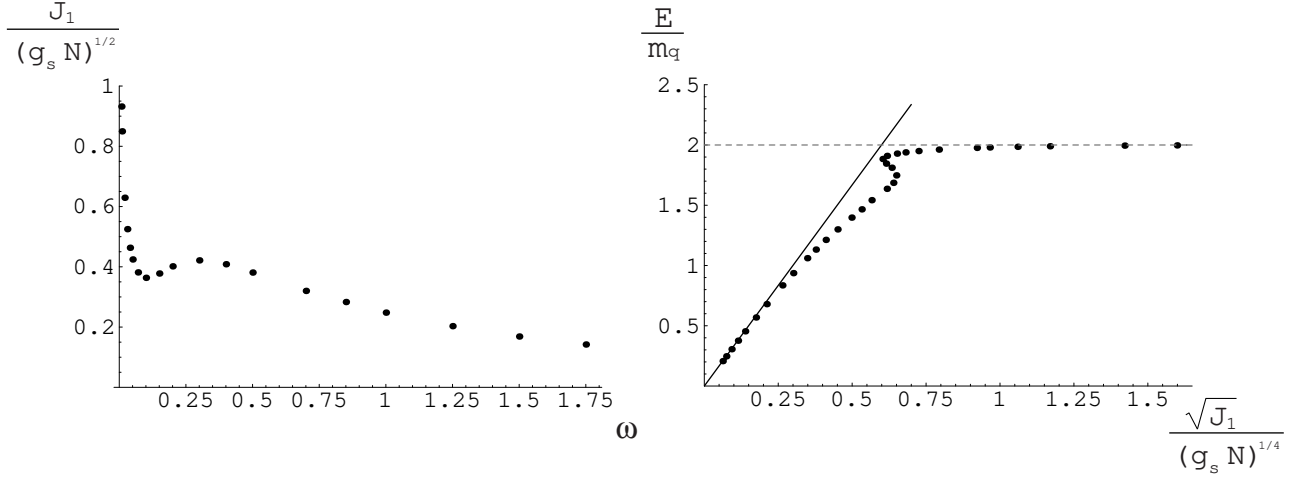}}
\caption{On the left we represent the angular momentum $J_1$ versus 
the angular velocity
$\omega$ for $\Theta=3$ and $z_{D7}=1$. The corresponding energy 
versus angular momentum plot is
shown on the right. The solid line corresponds to the linear Regge trajectory
(\ref{linearRegge}). }
\label{Largetheta}
\end{figure}

For large $J_1$ the spectrum corresponds to that of two non-relativistic
masses bound by a Coulomb potential. This is in agreement with the 
fact that in this long
distance regime the distance between the quark-antiquark pair is much 
larger than the inverse
mass of the lightest meson and one expects large screening 
corrections to the potential. We will
confirm this result by means of a static calculation, where we will 
compute the non-commutative
corrections to the large distance potential.

For $\Theta=0$ the $E(J_1)$ curve interpolates smoothly between the 
Regge behavior at small
$J_1$ and the Coulomb regime at large $J_1$. However, when  $\Theta$ 
is non-vanishing  and
large enough,  the crossover region is  more involved since, in some 
interval of $J_1$, the
energy decreases with increasing angular momentum. By analyzing the 
numerical results of $J_1$
as a function of $\omega$ (see figure \ref{Largetheta}), one can 
easily conclude that this effect
is due to the fact that, when $\Theta$ is large, the angular momentum 
$J_1$ has some local
extremum for some intermediate values of
$\omega$.

\subsubsection{Large angular velocity}

Let us now analyze the case $\omega\to \infty$. As we have checked by numerical
computation, the solution in this
limit consists of a very short string with $\rho_{D7}=1/\omega$ and
$z(\rho)\approx z_{D7}$. Actually, for large $\omega$ the shape of 
the string resembles  two
nearly parallel straight lines joined at some turning point (see 
figure \ref{parallel}). Let
$\rho=-\rho_{*}$ be the $\rho$ coordinate of the turning point. The 
value of $\rho_{*}$ can be
approximately obtained by noticing that $d^2z/d\rho^2$ should diverge 
at $\rho=-\rho_{*}$. By
inspecting the differential equation (\ref{rotdiffeq}) one readily 
realizes that this can only
happen if  $1-h\rho^2\omega^2$ vanishes. Taking into account that the 
coordinate $z$ is very
close to $z_{D7}$ for these short strings, one concludes that:
\beq
\rho_{*}\approx{1\over h_{D7}^{{1\over 2}}\,\, \omega}\,,
\label{rho*}
\eeq
where:
\beq
h_{D7}=\left(1+{\hat\Theta^4\over z_{D7}^4}\right)^{-1}\,.
\eeq
\begin{figure}
\centerline{\epsffile{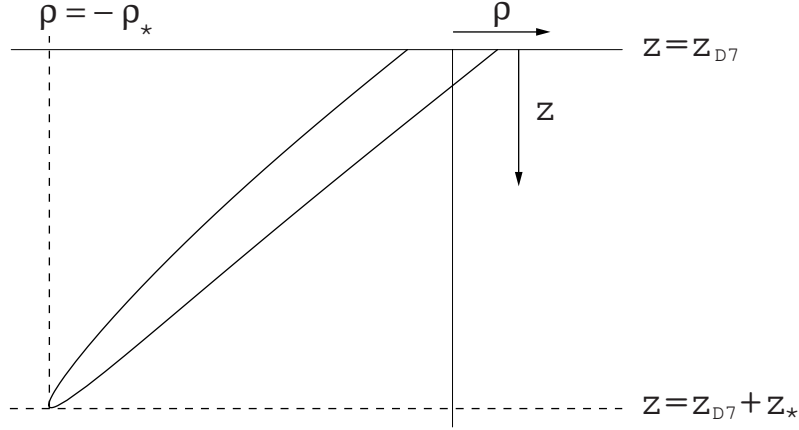}}
\caption{The string profile for large angular velocity. }
\label{parallel}
\end{figure}
Let $z_{D7}+z_{*}$ be the value of the function $z(\rho)$ at 
$\rho=-\rho_*$ (see figure
\ref{parallel}). It is clear that $z_{*}$ measures how far the rotating 
string is extended in the
holographic direction. Our numerical calculations show that, for a 
given value of the angular
velocity $\omega$, the tilting of the string grows as $\Theta$ is 
increased but $z_{*}$ remains
nearly the same. Let us check this fact from the above equations for 
large values of
$\hat\Theta$. Indeed, in this case we can obtain approximately 
$z_{*}$ by multiplying the slope
(\ref{init}) by $\rho_{*}$, namely:
\beq
z_{*}\approx-{dz\over d\rho}\Bigg|_{\rho=\rho_{D7}}\rho_{*}\,.
\label{z*}
\eeq
Moreover, the expression (\ref{rho*}) of $\rho_{*}$ for large 
$\hat\Theta$ reduces to:
\beq
\rho_{*}\approx{\hat\Theta^2\over z_{D7}^2\omega}\,,
\label{rho*appr}
\eeq
and using (\ref{init}) to evaluate the right-hand side of eq. 
(\ref{z*}), one gets:
\beq
z_{*}\approx{\sqrt{1-\rho_{D7}^2\omega^2}\over \rho_{D7}\omega^2}\,.
\label{z*appr}
\eeq
Notice that all the $\hat\Theta$ dependence has dropped out from the 
right-hand side of eq.
(\ref{z*appr}) and, thus,  $z_{*}$ is independent of $\hat\Theta$ as 
claimed. Moreover, by
comparing with our numerical calculations we have found that eq. 
(\ref{z*appr}) represents
reasonably well $z_{*}$.

Our numerical results indicate that the energy $E$ and the angular 
momentum $J_1$ do not depend
on $\Theta$. Again we can check this fact by computing $E$ and $J_1$ 
for large $\Theta$. Notice
that, as $\rho_{D7}\approx 1/\omega$, one gets from (\ref{rho*appr}) that
$\rho_{D7}/\rho_{*}\approx z_{D7}^2/\hat\Theta^2$ and, therefore,
$\rho_{D7}<<\rho_{*}$ if $\hat\Theta$ is large and, as a consequence, 
the profile of the string
degenerates into two coinciding straight lines. Therefore,
the energy and the angular momentum in this regime can be obtained
by taking  $z=z_{D7}={\rm constant}$ in eqs. (\ref{Enrot}) and 
(\ref{J1}) and performing the
integration of $\rho$ between $\rho=0$ and $\rho =\rho_*$:
\beq
E\approx 2\,{R^2\over 2\pi\alpha'}\,\,{h_{D7}^{{1\over 2}}\over 
z_{D7}^2}\,\,
\int_{0}^{\rho_*}\,\,{d\rho\over \sqrt{1-h_{D7}\,\rho^2\omega^2}}\,.
\eeq

By an elementary change of variables,
this integral can be done
analytically and, after taking into account the expression of 
$\rho_*$ (eq. (\ref{rho*})),  we
get the following expression of $E$:
\beq
E\approx {R^2\over 2\alpha'\omega\,z_{D7}^2}\,.
\label{ERegge}
\eeq
Notice that the dependence of $E$ on $h_{D7}$, and thus on $\Theta$, 
has disappeared.
Similarly, the angular momentum $J_1$ can be written as:
\beq
J_1\approx 2\,{R^2\over 2\pi\alpha'}\,\,
{h_{D7}^{{3\over 2}}\,\omega\over z_{D7}^2}\,\,
\int_{0}^{\rho_*}\,\,d\rho{\rho^2\over \sqrt{1-h_{D7}\,\rho^2\omega^2}}=
{R^2\over 4\alpha'\omega^2\,z_{D7}^2}\,,
\label{JRegge}
\eeq
and $J_1$ is also independent of the non-commutativity parameter.
Notice that $E$ and
$J_1$ depend on the angular velocity $\omega$ as $E\sim 1/\omega$ and 
$J_1\sim 1/\omega^2$. This
means that, indeed, $E\sim \sqrt{J_1}$ in this large $\omega$  regime 
and, actually, if we
define the effective tension for the rotating string as:
\beq
\tau_{eff}^R\equiv{E^2\over 2\pi J_1}\,,
\label{linearRegge}
\eeq
we have:
\bear
\tau_{eff}^R&=&{1\over 2\pi\alpha'_{eff}}\simeq
{R^2\over 2\pi\alpha'z_{D7}^2}\,.
\label{tauR}
\eear
As argued in ref. \cite{KMMW} for $\Theta=0$, the tension 
(\ref{tauR}) can be understood as the
proper tension ${1\over 2\pi\alpha'}$ at $z=z_{D7}$ which is then 
red-shifted  as
seen by a boundary observer.

\subsubsection{Small angular velocity}
 From the numerical computation displayed in figure \ref{numrot} one can
see that, in the large $J_1$ region (which corresponds to small $\omega$),
the energy becomes: $E\simeq 2m_q$, where $m_q={R^2\over
2\pi\alpha'z_{D7}}$ is the mass of the dynamical quarks. Furthermore, as
we can see in figure
\ref{numrot}, the separation between  the string endpoints $\bar\rho_{D7}$
when $\omega \to 0$ behaves as $\omega^{-2/3}$ (see eq. 
(\ref{powerrho})), which is the classical
result for two non-relativistic particles bound by a Coulomb 
potential, namely Kepler's law: the cube of the radius is proportional to
the square of the period. This is precisely what one gets in the
commutative case. In the next  section we will calculate
the static quark-antiquark potential for long strings and we will 
verify that the dominant term
of this potential is of the Coulomb form, with a strength which is 
independent of $\Theta$.
Therefore, as expected on general grounds,
in this limit
the rotating string behaves exactly as in the commutative theory, and
this can be understood taking into account that this $\omega\to 0$ limit
corresponds to long strings, much larger than the non-commutativity scale
of the theory.

%
%
%
%
%
%
%
%
%

\setcounter{equation}{0}
\section{Hanging strings}
\label{cpncschangstrings}
\medskip
In this section we will evaluate the potential energy for a static 
quark-antiquark pair and we
will compare the result with the calculation of the energy for a 
rotating string presented in
section \ref{cpncscstrings}. Following the analysis of ref. \cite{Wilson},
let us  consider a static configuration
consisting of a string stretched in the
$x^3$ direction with both ends attached to the $D7$-brane probe placed at
$r=r_{D7}$. The relevant part of the metric is:
\beq
{r^2\over R^2}\left[-dt^2+h\left(dx^3\right)^2\right]+{R^2\over
r^2}dr^2.
\eeq
Using $x^3$ as worldvolume coordinate and considering an ansatz of the form
$r=r(x^3)$, the Nambu-Goto action reads:
\beq
{\cal L}=-{1\over 2\pi\alpha'}\sqrt{-\det g}=-{1\over
2\pi\alpha'R^2}\sqrt{r^4h+R^4r'^2},
\eeq
where $r'$ denotes ${dr\over dx^3}$. As ${\cal L}$ does not depend 
explicitly on
$x^3$, the quantity
$r'{\partial {\cal L}
\over \partial r'} -{\cal L}$ is a constant of motion,\ \ie:
\beq
{r^4 h \over \sqrt{R^4r'^2 + r^4 h}}={\rm constant}\,.
\label{conservation}
\eeq
\begin{figure}
\centerline{\epsffile{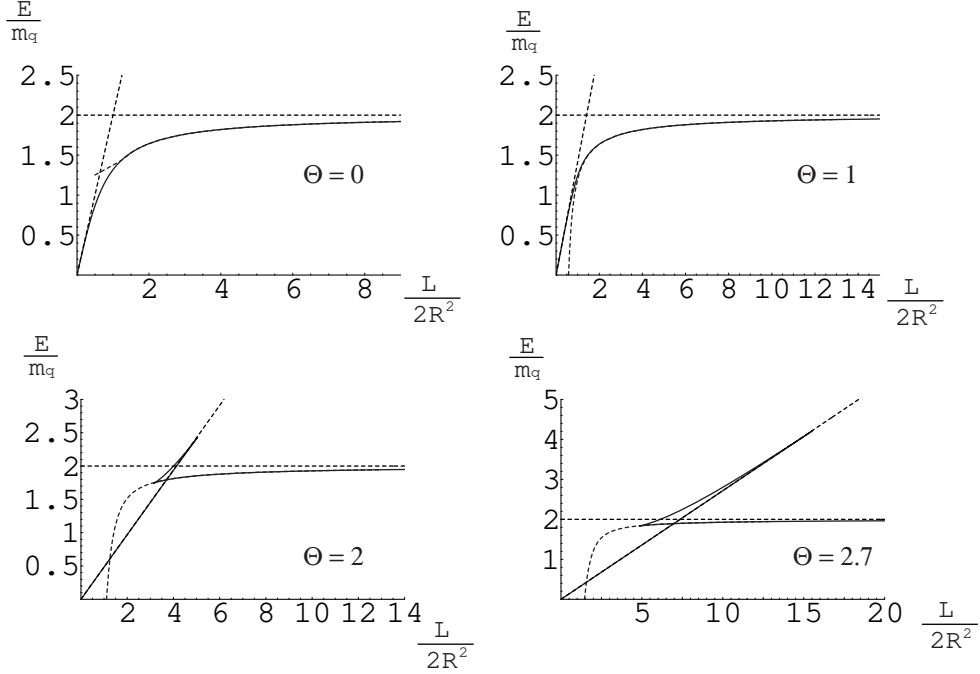}}
\caption{Energy of the static quark-antiquark configuration versus 
the string length $L$ for
$\Theta=0$ (commutative case), $\Theta=1$, $\Theta=2$, and 
$\Theta=2.7$ (setting
$z_{D7}=R^2=\alpha'=1$).
The solid line shows the result of the numeric computation, the straight
dashed line displays the Regge behavior obtained in section
\ref{shortwilson}, and the curved dashed line shows the Coulomb potential
computed in section \ref{longwilson}.}
\label{numwilson}
\end{figure}
The configurations we are interested in are those in which the string 
is hanging of the flavor
brane at $r=r_{D7}$ and reaching a minimum value $r_0$ of the 
coordinate $r$. Since
$r'=0$ when $r=r_0$, we can immediately evaluate the constant of 
motion and rewrite eq.
(\ref{conservation}) as:
\beq
{r^4 h \over \sqrt{R^4r'^2 + r^4 h}}= {r_0^2 \over \sqrt{1+\Theta^4
r_0^4}}\,\, .
\eeq
 From this
expression we get readily $r'$ in terms of $r$, from which we can 
compute the string
length (i.e. the quark-antiquark separation) with the result:
\bear
L&=&2\int_{r_0}^{r_{D7}}{dx^3\over dr} dr = 2R^2 \int_{r_0}^{r_{D7}}
{r_0^2(1+\Theta^4 r^4) \over r^2\sqrt{r^4-r_0^4}} dr =\rc
&=&{2R^2 \over r_0}\int_1^{\frac{r_{D7}}{r_0}}{dy \over y^2 \sqrt{y^4-1}}+
2R^2r_0^3\,\Theta^4\,\int_1^{\frac{r_{D7}}{r_0}}{y^2 \over
\sqrt{y^4-1}}dy\,.
\label{length}
\eear
The energy for this static configuration becomes:
\beq
\label{energy}
E=-\int{\cal L}\,d\sigma=
{1\over \pi\alpha'}r_0\sqrt{1+\Theta^4
\,r_0^4}\int_1^{\frac{r_{D7}}{r_0}} {y^2 \over  \sqrt{y^4-1}}\,dy\,.
\eeq
These last two expressions  for $E$ and $L$ can be evaluated
numerically for any $r_0$ between $0$ and $r_{D7}$. From this result we obtain
the energy of the static quark-antiquark configuration as a
function of the distance between the quark and the antiquark. These results
have been plotted in figure \ref{numwilson}, where  one can observe that the
potential is linear for small quark-antiquark separation, while for 
large separation the energy
becomes constant and equal to $2m_q$. Notice that this behavior is 
the same as the one we found
for the energy of the rotating strings. In the next two subsections 
we will analyze carefully the
small and long separation limits and we will obtain analytical 
expressions describing
the behavior of the configuration in those limits.

\subsection{Short strings limit}
\label{shortwilson}
The small separation limit $L\to 0$ is achieved when $r_0\to
r_{D7}$. Then, by defining
${r_{D7}\over r_0}=1+\epsilon$ and making the change of variable
$y=1+z$, it is easy to get, at leading order:
\beq
L={2R^2 \over r_{D7}} \left(1+\Theta^4
r_{D7}^4\right)\epsilon^{\frac{1}{2}}\,,
\qquad\quad
E={r_{D7}\over \pi\alpha'}\sqrt{1+\Theta^4
r_{D7}^4}\,\epsilon^{\frac{1}{2}}\,.
\eeq
 From the dependence of $L$ and $E$  on $\epsilon$  we notice that, indeed,
$E=\tau_{eff}^W L$, where $\tau_{eff}^W$ is the effective tension 
 and is given by:
\beq
\tau_{eff}^W={E\over L}={r_{D7}^2 \over 2\pi\alpha'R^2}\,{1 \over 
\sqrt{1+\Theta^4
r_{D7}^4}}={R^2\over 2\pi\alpha'z_{D7}^2}\,{1 \over \sqrt{1+\Theta^4
r_{D7}^4}}\, .
\eeq
Notice that $\tau_{eff}^W$ decreases with the non-commutativity parameter
$\Theta$. Actually, the maximal value of $\tau_{eff}^W$ is reached when
$\Theta=0$, where it  has the same value as the
tension $\tau_{eff}^R$  of the rapidly rotating string (see
eq. (\ref{tauR})). Therefore, for a general value of
$\Theta$ these two tensions have different values.

\subsection{Long strings limit}
\label{longwilson}
The limit $L\to\infty$ is achieved by taking the constant of
integration $r_0\to0$. It is easy to see from (\ref{length}),
(\ref{energy}) that the dependence on $\Theta$ cancels out
at leading order. Therefore, the Coulomb behavior of the commutative 
theory found in ref.
\cite{KMMW} is exactly recovered at long distances. This is expected on
general grounds since, at distances bigger than the scale set by the
non-commutativity parameter of the field theory, the commutative theory
must be obtained. However, let us see what we get beyond the leading
term. Since $\frac{r_{D7}}{r_0}=\frac{2\pi\alpha' m_q}{r_0}$,  the 
length $L$ in (\ref{length})
can be rewritten as:
\beq
\label{longl}
L={2R^2 \over r_0}\left(\int_1^{\infty}{dy \over y^2
\sqrt{y^4-1}} -\int_{\frac{2\pi\alpha' m_q}{r_0}}^{\infty}\,{dy \over y^2
\sqrt{y^4-1}}\right)+
2R^2r_0^3\,\Theta^4\,\int_1^{\frac{2\pi\alpha' m_q}{r_0}}dy{y^2
\over
\sqrt{y^4-1}}\,.
\eeq
The first integral in eq. (\ref{longl}) can be explicitly computed, namely:
\beq
{\cal C}\equiv\int_1^{\infty}{dy \over y^2
\sqrt{y^4-1}}={\sqrt{\pi}\,\Gamma\left({3\over4}\right)\over
\Gamma\left({1\over4}\right)}\approx 0.59907\,.
\eeq
Moreover, when $\frac{r_{D7}}{r_0}=\frac{2\pi\alpha' m_q}{r_0}$  is 
large, the second integral
in (\ref{longl}) can be approximated as:
\beq
\int_{\frac{2\pi\alpha'
m_q}{r_0}}^{\infty}\,{dy
\over y^2
\sqrt{y^4-1}}\simeq{r_0^3\over 3\left(2\pi\alpha' m_q\right)^3}\,,
\eeq
while the third integral in (\ref{longl}) can be rewritten as:
\bear
\int_1^{\frac{2\pi\alpha' m_q}{r_0}}
dy{y^2\over\sqrt{y^4-1}}&=&\frac{2\pi\alpha'
m_q}{r_0}+\left[-1+\int_1^\infty dy
\left({y^2\over \sqrt{y^4-1}} -1\right)\right]-\rc\rc &&-
\int_{\frac{2\pi\alpha' m_q}{r_0}}^\infty dy\left({y^2\over
\sqrt{y^4-1}} -1\right)\,.
\label{thirdint}
\eear
The quantity in square brackets on the right-hand side of 
(\ref{thirdint}) is equal to
$-{\cal C}$, and approximating the integrand in the last term by
${1\over 2y^4}$, $L$ reads:
\beq
L={2R^2{\cal C}\over r_0}-{2R^2r_0^2\over 3\left(2\pi\alpha' m_q\right)^3}
+2\Theta^4R^2\left[2\pi\alpha'm_qr_0^2-{\cal C}r_0^3-{r_0^6\over
6\left(2\pi\alpha' m_q\right)^3} \right]\, .
\eeq
By solving iteratively for $r_0(L)$, one gets:
\beq
\label{r0}
r_0= {2R^2{\cal C}\over L}\left[1-\left({8R^6{\cal C}^2\over
3\left(2\pi\alpha' m_q\right)^3}-16\Theta^4R^6{\cal
C}^2\pi\alpha'm_q\right){1\over L^3}\right]+o\left(L^{-5}\right).
\eeq
On the other hand, for small $r_0$  the energy (\ref{energy}) becomes:
\bear
\label{enapp}
E&\simeq&{1\over \pi\alpha'}\left(r_0+{1\over 2}\Theta^4r_0^5\right)
\int_1^{\frac{2\pi\alpha' m_q}{r_0}}dy{y^2\over\sqrt{y^4-1}}=\rc
&=&{1\over \pi\alpha'}\left[2\pi\alpha'm_q-r_0{\cal C}-{r_0^4\over
6\left(2\pi\alpha' m_q\right)^3}+\Theta^4\pi\alpha'm_qr_0^4\right]+
o\left(r_0^5\right).
\eear
Substituting (\ref{r0}) into (\ref{enapp}) we get:
\beq
E= 2m_q+{2R^2{\cal C}^2\over \pi\alpha'}\left[-{1\over L}+{4R^6{\cal
C}^2\over
L^4}\left({1\over3\left(2\pi\alpha'
m_q\right)^3}-2\Theta^4\pi\alpha'm_q\right)\right]+o\left(L^{-5}\right).
\label{longenergy}
\eeq
The first term in (\ref{longenergy}) is just the rest mass of the 
quark-antiquark pair. The
$1/L$ term is the Coulomb energy whose strength, as anticipated 
above, does not depend on
$\Theta$ and is given by the expression obtained in ref. 
\cite{Wilson}. The last term in
(\ref{longenergy}) contains the $1/L^4$ corrections to the Coulomb 
energy, which depend on the
non-commutativity parameter.

\subsection{Boosted hanging string}

As has already been pointed out, the non-commutativity parameter
explicitly breaks Lorentz invariance. Therefore, giving a velocity
to an object of the theory (and in particular to the string considered
in the previous subsections)
along $x_2$ or $x_3$ cannot be described as a trivial Lorentz boost.
Thus, it is worth taking a brief glance to a moving hanging string
in the supergravity dual.

In fact, a moving string  can be coupled to the background
$B$ field, adding to the action a term similar to that of a charged
particle moving in a magnetic field.
Let us consider the following (consistent) ansatz for the
embedding of the worldsheet of the string in the target space:
\beq
t=\tau\,,\qquad
x_2=v\tau\,,\qquad
x_3=x_3(\sigma)\,,\qquad
r=r(\sigma)\,.
\eeq
By inserting this in the Nambu-Goto action (\ref{rotact}),
one readily obtains the effective lagrangian (we use the static
gauge $\sigma =x_3$):
\beq
{\cal L}=-\frac{1}{2\pi\alpha'R^2}
\left(\sqrt{1-h\,v^2}\sqrt{r^4h+R^4 r'^2}+\Theta^2 r^4 h\, v\right)\,.
\label{movingaction}
\eeq
It is interesting to notice the presence of the $h$ factor inside
the first square root. In usual commutative spacetimes, the fact that
the speed of light is the limiting velocity gets reflected in this
expression because the square root should be real. On the contrary, in
this case we have $v<h^{-\frac12}$ so the limit is larger than
one since $h < 1$. This agrees with the known fact that in non-commutative
theories it is possible to travel faster than light \cite{hashi}.
In fact, the usual causality condition changes and the light-cone
is substituted by a light-wedge \cite{gaume} (there is no causal limit
along the non-commutative directions). Notice that $h\to 0$ when
$\Theta \to \infty$ and also when $r\to \infty$. This last fact is
connected to the observation in \cite{chu}
where it was proved that the appearance of the light-wedge is related
to the UV of the field theory. A study of causality in the field
theory from the string dual was performed in \cite{hubeny}.

Since (\ref{movingaction}) does not explicitly depend on $x_3$, one
can immediately obtain a first order condition that solves the
associated equations of motion $r'\frac{\partial {\cal L}}{\partial r'}
-{\cal L}={\rm constant}\,$:
\beq
\sqrt{1-h v^2}\frac{r^4 h}{\sqrt{r^4 h+ R^4 r'^2}}
+ \Theta^2 r^4 h v = \sqrt{1-h_0 v^2} r_0^2 h_0^\frac12
+\Theta^2r_0^4h_0 v\,,
\eeq
where $r_0$ is the minimal value of $r$, where the string turns back up and
$h_0=h(r_0)$. It is straightforward to obtain the energy of this configuration:
\beq
E=\frac{1}{\pi\alpha'}\int_{r_0}^{r_{D7}} dr \frac{r^2 h^\frac12}{
\sqrt{r^4h(1-hv^2)-\left[r_0^2 h_0^\frac12\sqrt{1-h_0 v^2}
+\Theta^2v(r_0^4 h_0 -r^4 h)\right]^2}} \,.
\eeq
Of course, in the commutative limit $\Theta=0$, $h=h_0=1$, the
$v$-dependence factors out of the integral and one recovers the
usual relation $E=\frac{1}{\sqrt{1-v^2}} m_0$, where $m_0$ denotes the
energy for $v=0$. In the general non-commutative case, this relation
is modified in a complicated fashion since $v$ cannot be factored
out and the integral above cannot be solved analytically.

%
%
%
%
%
%
%
%
%

\section{Discussion}
\label{cpncscconcl}
\medskip

In this chapter we have studied, from several points of view, the 
addition of flavor degrees of freedom to the supergravity dual of a
gauge theory with spatial non-commutativity.  This analysis is carried
out by considering flavor branes in the probe approximation and by
analyzing  the dynamics of open strings attached to them. First, by
using the Killing spinors of  the gravity background and the kappa
symmetry condition for the probes, we have explicitly found the stable,
supersymmetric embeddings for the flavor  branes. They turned out to be
the same as those of the commutative theory. Then,  by solving the
equations for the  fluctuations of the probe we have computed the spectrum
of scalar and vector mesons. We  have found that the
effective metric ${\cal G}$  appearing in the quadratic lagrangian 
(\ref{quadraticBI})  which
governs these fluctuations is exactly the same as in the commutative 
theory. Recall that  ${\cal
G}$  is Lorentz invariant in the Minkowski directions and is the 
result of the combined action
of the background metric and
$B$ field on the Born-Infeld lagrangian of the probe. We have interpreted
${\cal G}$ as the open string metric relevant for our problem.

The Wess-Zumino part of the
action gives rise to the term (\ref{WZfluct}) which depends on 
$\Theta$, breaks explicitly the
Lorentz symmetry and, in addition, couples the scalar and vector 
fluctuations. This Wess-Zumino
term vanishes in the UV limit and, as a consequence, the UV dynamics 
of the fluctuations
does not depend on the non-commutative deformation. This is in sharp 
contrast with what happens
to the metric of the background, for which the introduction of the
non-commutative deformation  changes drastically the UV behavior 
with respect to the
$AdS_5\times S^5$ geometry. By means of a change of variables we have 
been able to decouple the
differential equations of the fluctuations and we have studied the 
corresponding spectrum. By
looking at the $\Theta\to\infty$ limit of the equivalent 
Schr\"odinger problem we concluded that
the fluctuation spectrum for large $\Theta$ is the same as that 
corresponding to the $\Theta=0$
theory. We have also verified numerically that, for some intermediate 
values of $\Theta$, the
$+$modes are absent from the discrete spectrum.

For the first time in this work we have studied the spectrum of
mesons with large spin. This has been done by considering configurations
of open rotating strings with  their two ends attached to the D7-brane.
The presence of a $B$ field changes the boundary  conditions to be
imposed at the ends of the string and, as a consequence, the string is
tilted. Notice that this tilting also appears when the strings are
obtained as worldvolume solitons of D-branes in the presence of the
$B$ field (see, for example, \cite{NCbion}).  By numerical integration we
have obtained the profile of the string and determined its energy 
spectrum. Short strings present a Regge-like behavior with an asymptotic
slope which, despite of the  tilting of the string, is the same as in the
commutative theory. On the other hand, long strings  behave as two
non-relativistic particles bound by a Coulomb potential, which is 
also the way in which they behave in the commutative theory. Finally, in
section \ref{cpncschangstrings} we  have independently studied these two
limits for a static string configuration and briefly commented on the
modification of the causality relation.

Let us finally mention some aspects that it would be worth to study
further. First of all, it would be interesting to have a clearer
understanding of the fluctuation spectrum in the intermediate $\Theta$
region. Our results seem to indicate that new physics could show up there.
It is an open question whether or not this is an artifact of our approach
or  a real effect in the field theory dual. By expanding the flavor brane
action beyond second order we would get a hint on the meson interactions,
which we expect to depend on the non-commutativity parameter. Moreover, it
would also be worth studying the behavior of glueballs in 
non-commutative  backgrounds (see \cite{APT} for some work along this
line) in order to compare with the excitations coming from the flavor
sector.  Naively, since glueballs are dual to 
closed strings and feel the collapsing of the metric in the UV, one would 
expect a very different behavior from that found studying mesons, which, 
as explained above, are effectively embedded in the UV-finite open
string  metric. Nevertheless, from the field theory point of view we do
not expect a priori such different behaviors. For this reason it  would
be interesting to clarify this point  further.

An interesting  continuation of the analysis presented in this
chapter consists of moving on to less supersymmetric theories. Actually, in
ref. \cite{APR} it was studied the addition of flavor to the
non-commutative deformation of the Maldacena-N\'u\~nez (MN) background
\cite{MN, CV}. The non-commutative  version of this background was found
in ref. \cite{NCMN}. This solution, which corresponds to a  (D3,D5) bound
state  with the D3- and D5-branes wrapped on a two-cycle, preserves four 
supersymmetries and is dual to non-commutative ${\cal N}=1$ super
Yang-Mills in four dimensions. It is worth noting that, as it happens
for the maximally supersymmetric case, the embeddings of the flavor branes
(D5-branes in this case) are the same as in the commutative MN
background.

Another possible extension of this work would be trying to incorporate
dynamical baryons. It was suggested in ref. \cite{KMMW} that such
dynamical baryons could be constructed from the baryon vertex. Moreover,
according to the proposal of ref. \cite{Ncbaryon}, the baryon vertex for
the (D1,D3) background consists of a D7-brane wrapped on a five-sphere
and extended along the non-commutative plane. Within this approach the
dynamical baryons would be realized as bundles of fundamental strings
connecting the two types of D7-branes, namely the flavor brane and the
baryon vertex.

Non-commutative field theories have been formulated long time ago. They
display an intriguing UV/IR mixing whose implications are not fully
understood yet. The study of their non-perturbative structure might still
reserve us some surprises and we hope that our results could help to unveil
them.

%
%
%
%
%
%
%
%
%
%

\chapter{Supersymmetric probes on the conifold}
\label{cpSSprobes}
\setcounter{equation}{0}
\section{Introduction}

\medskip
A very fruitful extension of the work presented in the past chapters
would be to move on to less supersymmetric, and thus more realistic,
theories. String theory duals to four-dimensional gauge theories with
reduced SUSY can be constructed by placing D3-branes in curved
geometries instead of the maximally supersymmetric ten-dimensional flat
Minkowski space. Concretely, since the AdS/CFT correspondence involves
only the near-horizon region of the branes and every smooth manifold is
locally flat, backgrounds dual to theories with reduced supersymmetry
should be engineered by placing the branes at a singular point of the
transverse space. In \cite{KW} an ${\cal N}=1$ SUSY background was
constructed by placing $N$ D3-branes at the tip of the conifold: it is
known as the Klebanov-Witten (KW) model. The backreaction of the branes
causes the near-horizon geometry to be that of an $AdS_5\times T^{1,1}$
space with
$N$ units of Ramond-Ramond flux threading the $T^{1,1}$. The
corresponding dual field theory is a  four-dimensional ${\cal N}=1$
superconformal field theory with gauge group $SU(N)\times SU(N)$  coupled
to four chiral superfields in the bifundamental representation.

As we have pointed in the previous chapters, it will be interesting to
study the addition of flavors to the KW model. Fundamental
degrees of freedom living on defects of different codimension can be
introduced by following the same approach as for the case of D-branes in
flat space described in section \ref{cp1flavor}, that is to say, by
embedding D-branes in the supergravity dual. If the number of branes
embedded is much lower than the number of branes generating the
background, those can be treated as probes. Generically, the probe branes
will intersect the background branes along some common dimensions,
creating a defect where the fundamentals are confined. Motivated by this
application, and some more presented below, we will study in this chapter
the supersymmetric embeddings of different D-brane probes in the
$AdS_5\times T^{1,1}$ geometry.

Besides introducing fundamental degrees of freedom, the addition of
D-branes on the supergravity side permits to uncover stringy effects in
the Yang-Mills theory. Indeed, by embedding D-branes in the supergravity
dual, one can study interesting objects of the gauge theory which are not
represented by the supergravity fields. This approach
was pioneered by Witten in ref. \cite{Wittenbaryon} where he considered a
D3-brane wrapped over a topologically non-trivial cycle of the
$AdS_5\times {\bf \RR\IP}^5$ background and showed that this
configuration is dual to certain operators of dimension $N$ of the
$SO(N)$ gauge theory, namely the ``Pfaffians". Another  example along the
same lines is provided by the so-called giant gravitons, which are
rotating branes wrapped over a topologically trivial cycle \cite{GST}.
These branes are not topologically stable: they are stabilized
dynamically by their angular momentum. The corresponding field theory 
duals have been found in ref. \cite{BBNS}.

The effect of adding  different D-branes to the KW background
has already been studied in several places in the literature. In ref. 
\cite{GK} it was proposed that D3-branes wrapped over three-cycles of
$T^{1,1}$ are dual to dibaryon operators built out of products of $N$
chiral superfields (see also refs. \cite{BHK}-\cite{HMcK} for more
results on dibaryons in this model and in some orbifold theories).
Furthermore, it was also shown in ref. \cite{GK} that a D5-brane wrapped
over a two-cycle of $T^{1,1}$ behaves as a domain wall in $AdS_5$.
In addition, this configuration in which the D5-brane shares two spatial
dimensions with the background D3-branes is the natural candidate for
studying the addition of flavors living in a codimension one defect,
as we did for the ${\cal N}=4$ theory in chapter \ref{cp2dpinter}.
Similarly, one can add flavor living in a codimension zero defect,
\ie \  in the whole spacetime, by inserting spacetime filling D7-branes in
the $AdS_5\times T^{1,1}$ geometry \cite{KKW,Ouyang}. Actually, the meson
spectrum was computed in \cite{LeviOuy} by studying fluctuations of
a D7-brane probe and, recently, the backreacted solution corresponding to
having spacetime filling D7-branes wrapping a three-cycle of $T^{1,1}$ and
smeared along the transverse two-cycle has been obtained in
\cite{unqchKW}. A list of the stable D-branes in this background, obtained
with methods quite different from those employed here, has appeared in
ref. \cite{MS}. 

The main technique that we will employ to determine the supersymmetric
embeddings of the different D-brane probes in the $AdS_5\times T^{1,1}$
background is kappa symmetry. As we have already explained in section
\ref{cp1sskappasym}, one can look for embeddings of Dp-branes preserving
some fraction of the supersymmetry of the background by studying the
equation $\Gamma_\kappa\,\epsilon=\epsilon$, where $\epsilon$ is a
Killing spinor of the background and $\Gamma_\kappa$ is a matrix depending
on the induced metric on the worldvolume of the brane. Therefore, if the
Killing spinors are known, the equation gives rise to a set of first order
differential equations whose solutions (if they exist) determine the
supersymmetric embeddings of the probe. For these configurations the kappa
symmetry equation introduces some additional conditions on
$\epsilon$, which are only satisfied by some subset of the Killing
spinors. Thus, the probe only preserves some fraction of the original
supersymmetry of the background. For all the solutions we will find here
we will be able to identify  the supersymmetry that they preserve.
Moreover, we will verify that the corresponding embeddings satisfy the
equations of motion derived from the Dirac-Born-Infeld action of the
probe. Actually, in all the cases studied, we will establish a series of
BPS bounds for the energy, along the lines of those studied in ref.
\cite{GGT}, which are saturated by the kappa symmetric embeddings.

Clearly, to carry out the program sketched above we need to have a detailed
knowledge of the Killing spinors of the  $AdS_5\times T^{1,1}$ background.
In particular, it would be very useful to find a basis of frame one-forms
in which the spinors do not depend on the coordinates of the
$T^{1,1}$ space. It turns out that this frame is provided very naturally
when the conifold is obtained \cite{gaugedsugra} from an uplifting of
eight-dimensional gauged supergravity \cite{ss}. In this frame the Killing
spinors are characterized by simple algebraic conditions, and one can
systematically explore the solutions of the kappa symmetry equation. 

The first case we will study is that of the supersymmetric embeddings of D3-brane
probes. We will find a general family of three-cycles which contains, as a
particular case, the one used in ref. \cite{GK} to describe the dual of the
dibaryonic operator. We will be able to identify the field theory content of the
operators dual to our D3-brane embeddings. 

We will consider next D5-brane probes. We will be able to identify the
two-cycle on which the D5-brane must be wrapped to realize the domain wall
of the $AdS_5$ space.  

Our final case is that corresponding to D7-brane probes. We will study the
spacetime filling configurations. In this case we will be able
to find a two-parameter family of supersymmetric embeddings which, in
particular, include those proposed in refs. \cite{KKW,Ouyang} as suitable
to add flavor to this background. Our results confirm that these
configurations are kappa symmetric.

This chapter is organized as follows. In section \ref{cp5sKW} we review
the basic features of the Klebanov-Witten model. In particular, we give
the explicit form of the Killing spinors in the frame which is more
adequate to our purposes. We also introduce in this section the general
form of the kappa symmetry matrix $\Gamma_{\kappa}$ and discuss the
general strategy to solve the $\Gamma_{\kappa}\epsilon=\epsilon$
equation. 

Section \ref{cp5sD3} is devoted to the analysis of the supersymmetric
D3-brane embeddings. After choosing a set of convenient worldvolume
coordinates and an ansatz for the scalar fields that determine the
embedding, we will be able to find a pair of first order differential
equations whose solutions determine the supersymmetric wrappings of the
D3-brane over a three-cycle. This pair of equations can be solved in
general after a change of variables which converts them into the
Cauchy-Riemann equations. Similar analysis are carried out for the
D5-brane and D7-brane probes in sections \ref{cp5sD5} and \ref{cp5sD7}
respectively. In section \ref{cp5sdisc} we summarize our results and draw
some conclusions.  

%
%
%
%
%
%
%
%
%

\section{The Klebanov-Witten model}
\label{cp5sKW}
\setcounter{equation}{0}
\medskip
The conifold is a non-compact Calabi-Yau threefold with a conical singularity. Its
metric can be written as $ds^2_{6}=dr^2+r^2\,ds^2_{T^{1,1}}$, where 
$ds^2_{T^{1,1}}$ is the metric of the $T^{1,1}$ coset 
$(SU(2)\times SU(2))/U(1)$, which is the base of the cone. The $T^{1,1}$ space is
an Einstein manifold whose metric can be written \cite{Candelas}
explicitly by using the fact that
it is a $U(1)$ bundle over $S^2\times S^2$. Actually, if $(\theta_1,\phi_1)$ and 
$(\theta_2,\phi_2)$ are the standard coordinates of the $S^2$s and if
$\psi\in [0,4\pi)$ parametrizes the $U(1)$ fiber, the metric may be written as:
\beq
ds^2_{T^{1,1}}={1\over 6}\,\sum_{i=1}^{2}\,
\big(\,d\theta_i^2+\sin^2\theta_i\,d\phi_i^2\,)+
{1\over9}\,\big(\,d\psi+\sum_{i=1}^{2}\cos\theta_id\phi_i\,\big)^2\,.
\label{t11metric}
\eeq
This metric has an $SU(2)\times SU(2)\times U(1)$ isometry. The $ U(1)$
corresponds to shifts along the fiber, while the $SU(2)\times
SU(2)$ isometries act on the $S^2\times S^2$ base.
 
The conifold can also be described as the locus of points  
in $\CC^4$ which satisfy the equation:
\beq
z_1\,z_2-z_3\,z_4=0\,,
\label{conifold}
\eeq
which obviously has an isolated conical singularity at the origin of $\CC^4$. 
The relation between the holomorphic coordinates $z_i$ with the angles and
$r$ is:
\bear
&&z_1=r^{3/2}\,e^{{i\over 2}\,(\psi-\phi_1-\phi_2)}\,
\sin{\theta_1\over 2}\,\sin{\theta_2\over 2}\;,
\qquad\quad
z_2=r^{3/2}\,e^{{i\over 2}\,(\psi+\phi_1+\phi_2)}\,
\cos{\theta_1\over 2}\,\cos{\theta_2\over 2}\,,
\rc\rc
&&z_3=r^{3/2}\,e^{{i\over 2}\,(\psi+\phi_1-\phi_2)}\,
\cos{\theta_1\over 2}\,\sin{\theta_2\over 2}\;,
\qquad\quad
z_4=r^{3/2}\,e^{{i\over 2}\,(\psi-\phi_1+\phi_2)}\,
\sin{\theta_1\over 2}\,\cos{\theta_2\over 2}\,.\rc
\label{holomorphic}
\eear
It is also interesting to find some combinations of the $z_i\,$s which
only depend on the coordinates $(\theta_1,\phi_1)$ or
$(\theta_2,\phi_2)$. Actually, from the parametrization
(\ref{holomorphic}) it is straightforward to prove that:
\beq
{z_1\over z_3}={z_4\over z_2}=
e^{-i\phi_1}\,\tan{\theta_1\over 2}\;\qquad\quad
{z_1\over z_4}={z_3\over z_2}=
e^{-i\phi_2}\,\tan{\theta_2\over 2}\,.
\label{zratio}
\eeq

By adding four Minkowski coordinates to the conifold we obtain a Ricci flat
ten-dimensional metric. Let us now place in this geometry a stack of $N$ coincident
D3-branes extended along the Minkowski coordinates and located at the singular
point of the conifold. The resulting model is the so-called Klebanov-Witten (KW)
model. The corresponding near-horizon metric and Ramond-Ramond self-dual
five-form are given by:
\bear
ds^2_{10}&=&[h(r)]^{-{1\over 2}}\,dx^2_{1,3}+[h(r)]^{{1\over 2}}\,
\big(\,dr^2+r^2\,ds^2_{T^{1,1}}\,\big)\,,\rc\rc
h(r)&=&{L^4\over r^4}\,,\rc\rc
F^{(5)}&=&d^4x\,\wedge dh^{-1}+{\rm Hodge\,\,\, dual}\,,\rc\rc
L^4&=&{27\over 4}\,\pi g_s N\alpha'^{\,2}\,.
\label{KW}
\eear
The gauge theory dual to the supergravity background (\ref{KW}) is an ${\cal N}=1$
superconformal field theory with some matter multiplets.  Actually, the metric in
(\ref{KW}) can be written as:
\beq
ds^2_{10}={r^2\over L^2}\,dx^2_{1,3}+{L^2\over r^2}\,dr^2+
L^2\,ds^2_{T^{1,1}}\,,
\label{adspoincare}
\eeq
which corresponds to the $AdS_5\times T^{1,1}$ space. In order to exhibit the field
theory dual to this background, let us solve the conifold equation (\ref{conifold})
by introducing four homogeneous coordinates $A_1$, $A_2$, $B_1$ and $B_2$ as
follows:
\beq
z_1=A_1\,B_1\,,\qquad\quad
z_2=A_2\,B_2\,,\qquad\quad
z_3=A_1\,B_2\,,\qquad\quad
z_4=A_2\,B_1\,.
\label{homogeneous}
\eeq 
Following the analysis of ref. \cite{KW}, one can show that the dual superconformal
theory can be described as an ${\cal N}=1$ $SU(N)\times SU(N)$ gauge theory which
includes four ${\cal N}=1$ chiral multiplets, which can be identified with (the
matrix generalization of) the  homogeneous coordinates $A_1$, $A_2$, $B_1$ and
$B_2$.  The fields $A_1$ and $A_2$ transform in the $({\bf N}, {\bf \bar N})$
representation of the gauge group, while $B_1$ and $B_2$ transform in the 
$({\bf \bar  N}, {\bf  N})$ representation. These fields are coupled through an
exactly marginal superpotential $W$ of the form:
\beq
W=\lambda \,\epsilon^{ij}\,\,\epsilon^{kl}\,
tr(A_iB_kA_jB_l)\,,
\eeq
where $\lambda$ is a constant.
The $SU(2)\times SU(2)$ isometry of $T^{1,1}$ appears as a global symmetry
in the gauge theory. Under this symmetry the fields
$A_i$, $B_j$ transform as $({\bf 2}\,,\,{\bf 1})$ and $({\bf 1}\,,\,{\bf
2})$ respectively. In addition, the theory possesses an anomaly-free
$U(1)_{\cal R}$ symmetry, dual to the $U(1)$ isometry of the $T^{1,1}$.
Therefore, the fields $A_i$, $B_j$ have R-charge $1/2$ and their conformal
dimension is $\Delta=(3/2){\rm R}=3/4$.

\subsection{Killing spinors}

As argued in ref. \cite{KW}, the KW  model preserves eight supersymmetries
(see also ref. \cite{Kehagias}). Notice that this is in agreement with the
${\cal N}=1$ superconformal character of the corresponding dual field
theory, which has four ordinary supersymmetries and four superconformal
ones. 

To obtain the explicit form of the Killing spinors, one has to impose
the vanishing of the SUSY variations of the fermionic type IIB
supergravity fields, namely the dilatino and the gravitino.
For type IIB supergravity with constant dilaton and no three-forms
the variation of the dilatino vanishes trivially while the variation of
the gravitino reads \cite{SUSYIIB}:
\beq
\delta\psi_{\mu}=D_{\mu}\,\epsilon+{i\over 1920}\,
F_{\mu_1\cdots\mu_5}^{(5)}\,\Gamma^{\mu_1\cdots\mu_5}\Gamma_{\mu}\epsilon
\,,
\label{cp5susyvars}
\eeq
where $\psi$ is the gravitino. The Killing spinors of the KW background
are given by the spinors $\epsilon$ satisfying the equations
$\delta\psi_\mu=0$.

It turns out that the final result of the calculation is greatly
simplified if some particular basis of the frame one-forms for the
$T^{1,1}$ part of the metric is chosen. In order to specify this basis,
let us define three one-forms associated to a two-sphere:
\beq
\sigma^1=d\theta_1\,,\qquad\quad
\sigma^2=\sin\theta_1\,d\phi_1\,,\qquad\quad
\sigma^3=\cos\theta_1d\phi_1\,,
\label{sigmaoneforms}
\eeq
and three one-forms associated to a three-sphere:
\bear
w^1&=&\sin\psi\sin\theta_2 d\phi_2+\cos\psi d\theta_2\,,\rc
w^2&=&-\cos\psi\sin\theta_2 d\phi_2+\sin\psi d\theta_2\,,\rc
w^3&=&d\psi+\cos\theta_2 d\phi_2\,.
\label{woneforms}
\eear
After an straightforward calculation one can verify that
these forms satisfy:
\beq
d\sigma^i=-{1\over 2}\epsilon_{ijk}\,\sigma^j\wedge\sigma^k\,,\qquad\quad
dw^i={1\over 2}\epsilon_{ijk}\,w^j\wedge w^k\,.
\eeq
Moreover, the $T^{1,1}$ metric  (\ref{t11metric}) can be rewritten as:
\beq
ds^2_{T^{1,1}}={1\over 6}\,\big(\,(\sigma^1)^2+
(\sigma^2)^2+(w^1)^2+(w^2)^2\,\big)+
{1\over 9}\,\big(\,w^3+\sigma^3\,\big)^2\,.
\eeq
This form of writing the $T^{1,1}$ metric is the one that arises naturally when
the conifold geometry is obtained \cite{gaugedsugra} in the framework of the
eight-dimensional gauged supergravity obtained from a Scherk-Schwarz 
reduction of eleven-dimensional supergravity on an SU(2) group manifold
\cite{ss}. In this approach one starts with a domain wall problem in eight
dimensions and looks for BPS solutions of the equations of motion. These
solutions are subsequently uplifted to eleven dimensions, where they
represent gravity duals of branes wrapping non-trivial cycles. The
topological twist needed to realize supersymmetry with wrapped branes is
implemented in this approach in a very natural way. As shown in ref.
\cite{gaugedsugra}, the conifold metric is obtained as the gravity dual of
D6-branes wrapping a holomorphic $S^2$ inside a K3 manifold. Moreover,
from the consistency of the reduction, the Killing spinors should not
depend on the coordinates of the group manifold and, actually, in the
one-form basis we will use they do not depend on any angular coordinate of
the $T^{1,1}$ space. Accordingly, let us consider the following frame for
the ten-dimensional metric (\ref{KW}):
\bear
&&e^{x^{\alpha}}={r\over L}\,\,dx^{\alpha}
\;,\qquad(\alpha=0,1,2,3)\;,\qquad\quad
e^{r}={L\over r}\,\,dr\,,\rc\rc
&&e^{i}={L\over \sqrt{6}}\,\,\sigma^i\;,\qquad(i=1,2)\,,\rc\rc
&&e^{\hat i}=\,{L\over \sqrt{6}}\,\,w^i\;,\qquad(i=1,2)\,,\rc\rc
&&e^{\hat 3}={L\over 3}\,\,(\,w^3+\sigma^3\,)\,.
\label{frame}
\eear
Let us also define the matrix $\Gamma_{*}$ as:
\beq
\Gamma_{*}\equiv i\Gamma_{x^0x^1x^2x^3}\,.
\label{gamma*}
\eeq
Then, the Killing spinors for the type IIB background (\ref{KW}) take the following
form (see \cite{tesina} for the detailed computation):
\beq
\epsilon=r^{{\Gamma_{*}\over 2}}\,\,
\left(1+{\Gamma_r\over 2L^2}\,\,x^{\alpha}\,\Gamma_{x^{\alpha}}\,\,
(1-\Gamma_{*}\,)\right)\,\eta\,,
\label{adsspinor}
\eeq
where $\eta$ is a constant  spinor satisfying:
\beq
\Gamma_{12}\,\eta=i\eta\,,
\qquad\quad
\Gamma_{\hat 1\hat 2}\,\eta=-i\eta\,.
\label{tspinor}
\eeq
In eq. (\ref{adsspinor}) we are parametrizing the dependence of $\epsilon$ on the
$AdS_5$ coordinates as in ref. \cite{LPT}. Notice that, as the matrix multiplying
$\eta$ in eq. (\ref{adsspinor}) commutes with  $\Gamma_{12}$ and $\Gamma_{\hat1\hat 2}$,  the spinor $\epsilon$  also satisfies the  conditions (\ref{tspinor}),
namely:
\beq
\Gamma_{12}\,\epsilon=i\epsilon\,,
\qquad\quad
\Gamma_{\hat 1\hat 2}\,\epsilon=-i\epsilon\,.
\label{tsprojections}
\eeq
It is clear from eqs. (\ref{adsspinor}) and (\ref{tspinor}) that our system is $1/4$
supersymmetric, \ie\ it preserves 8 supersymmetries, as it corresponds to
the supergravity dual of an ${\cal N}=1$ superconformal field theory in
four dimensions. Moreover, let us decompose the constant spinor $\eta$ according to
the different eigenvalues of the matrix $\Gamma_{*}$:
\beq
\Gamma_{*}\,\eta_{\pm}=\pm\eta_{\pm}\,.
\label{etamasmenos}
\eeq
Using this decomposition in eq. (\ref{adsspinor}) we obtain two types
of Killing spinors:
\bear
\epsilon_{+}&=&r^{1/2}\,\eta_+\,,\rc\rc
\epsilon_{-}&=&r^{-1/2}\,\eta_-+{r^{1/2}\over L^2}\,\,
\Gamma_r\,x^{\alpha}\Gamma_{x^{\alpha}}\,\eta_-\,.
\label{chiraladsspinor}
\eear
Notice that the four spinors $\epsilon_{+}$ are independent of the
coordinates $x^{\alpha}$ and 
$\Gamma_{*}\,\epsilon_{+}=\epsilon_{+}$. On the contrary, the 
$\epsilon_{-}$s do depend on the $x^{\alpha}$s and are not eigenvectors
of $\Gamma_{*}$. The latter correspond to the four superconformal supersymmetries,
while the $\epsilon_{+}$s are the ones corresponding to the ordinary ones.

It is also interesting to write the form of the Killing spinors when
global coordinates are used for the $AdS_5$ part of the metric. In these
coordinates the ten-dimensional metric takes the form:
\beq
ds^2_{10}=L^2\,\Big[\,-\cosh^2\rho \,\,dt^2+d\rho^2+
\sinh^2\rho\,\,d\Omega_3^2\,\Big]+ L^2\,ds^2_{T^{1,1}}\,,
\label{globalads}
\eeq
where $d\Omega_3^2$ is the metric of a unit three-sphere parametrized by
three angles $(\alpha^1, \alpha^2,\alpha^3)$:
\beq
d\Omega_3^2=(d\alpha^1)^2+\sin^2\alpha^1\Big(\,
(d\alpha^2)^2+\sin^2\alpha^2\,(d\alpha^3)^2\,\Big)\,,
\eeq
with $0\le\alpha^1,\alpha^2\le \pi$ and $0\le\alpha^3\le 2\pi$. 
In order to write down the Killing spinors in these coordinates, let us
choose the following frame for the $AdS_5$ part of the metric:
\bear
&&e^{t}=L\cosh\rho\,dt\,,\qquad\quad
e^{\rho}=L\,d\rho\,,\rc\rc
&&e^{\alpha^1}=L\sinh\rho\,d\alpha^1\,,\rc\rc
&&e^{\alpha^2}=L\sinh\rho\,\sin\alpha^1\,d\alpha^2\,,\rc\rc
&&e^{\alpha^3}=L\sinh\rho\,\sin\alpha^1\,\sin\alpha^2\,d\alpha^3\,.
\eear
We will continue to use the same frame forms as in eq. (\ref{frame}) for
the 
$T^{1,1}$ part of the metric. If we now  define the matrix
\beq
\gamma_*\equiv\Gamma_{t}\,\Gamma_{\rho}\,
\Gamma_{\alpha^1\,\alpha^2\,\alpha^3}\,,
\eeq
then, the Killing spinors in these coordinates can be written as \cite{globalads}:
\beq
\epsilon=
e^{-i\,{\rho\over 2}\,\Gamma_{\rho}\gamma_*}\,
e^{-i\,{t\over 2}\,\Gamma_{t}\gamma_*}\,
e^{-{\alpha^1\over 2}\,\Gamma_{\alpha^1 \rho}}\,
e^{-{\alpha^2\over 2}\,\Gamma_{\alpha^2 \alpha^1}}\,
e^{-{\alpha^3\over 2}\,\Gamma_{\alpha^3 \alpha^2}}\,\eta\,,
\label{globalspinor}
\eeq
where $\eta$ is a constant spinor which satisfies the same conditions as
in eq. (\ref{tspinor}). 

\subsection{Supersymmetric probes}
\label{cp5sssusypr}
We have introduced in section \ref{cp1sskappasym} the use of kappa
symmetry as a tool to look for supersymmetric embeddings of brane probes
in a given background. Let us consider a Dp-brane probe in the KW
background (\ref{KW}) with worldvolume coordinates $\xi^{a}$ $(a=0,\cdots
,p)$. One  can decompose the complex spinor $\epsilon$ used up to now in
its real and imaginary parts as $\epsilon\,=\,\epsilon_1+i\epsilon_2$ and
arrange the two Majorana-Weyl spinors $\epsilon_1$ and $\epsilon_2$ as a
two-dimensional vector $\pmatrix{\epsilon_1\cr\epsilon_2}\,$. Acting on
this real two-component spinor the kappa symmetry matrix takes the form
given in eq. (\ref{kpGammakpIIB}) and after taking into
account that the NSNS $B$ field is zero for the KW background and assuming
that there are not worldvolume gauge fields on the Dp-brane reduces to:
\beq
\Gamma_{\kappa}\,=\,{1\over (p+1)!\sqrt{-g}}\,\epsilon^{a_1\cdots
a_{p+1}}\,(\tau_3)^{{p-3\over 2}}\,i\tau_2\,\otimes\,
\gamma_{a_1\cdots a_{p+1}}\,,
\label{cp5gammakappa}
\eeq
where now we are calling $\tau_i$ $(i=1,2,3)$ to the Pauli matrices
acting on the two-dimensional vector
$\pmatrix{\epsilon_1\cr\epsilon_2}\,$. As before, $g$ is the
determinant of the induced metric $g_{ab}=\partial_{a}
X^{M}\,\partial_{b} X^{N}\,G_{MN}$ ($G_{MN}$ is the ten-dimensional
metric of the background) and $\gamma_{a_1\cdots a_{p+1}}$ denotes the
antisymmetrized product of the induced gamma matrices
$\gamma_{a}=\partial_a\,X^{M}\,E_{M}^{\underline{N}}\,\Gamma_
{\underline{N}}$ (recall that $E_{M}^{\underline{N}}$ are the
coefficients appearing in the expression of the frame one-forms  in
terms of the differentials of the coordinates). In addition, let us write the
pullback of the frame one-forms
$e^{\underline M}$ as:
\beq
P\,\left[e^{\underline{M}}\right]=E_{N}^{\underline{M}}\,
\partial_a\,X^{N}\,d\xi^a\equiv C_a^{\underline{M}}
\,d\xi^a\,,
\eeq
with $C_a^{\underline{M}}\equiv E_{N}^{\underline{M}}\,\partial_a\,X^{N}$.
Then, the induced gamma matrices can be rewritten in the form:
\beq
\gamma_a=C_a^{\underline{M}}\,\Gamma_{\underline{M}}\,.
\label{cp5inducedgamma}
\eeq

The assumption that there are no worldvolume gauge fields is consistent
with the equations of motion of the probe if there are not source terms
for the worldvolume gauge field in the action. These source terms must be
linear in the gauge field and they can only come from the Wess-Zumino
term of the action of the Dp-brane (\ref{cp1WZ}), responsible for
the coupling of the probe to the RR fields of the background.
In our case we have only one of such  Ramond-Ramond fields, namely the
self-dual five-form $F^{(5)}$. If we denote by
$C^{(4)}$ its potential ($F^{(5)}=dC^{(4)}$), it is clear that the only
term linear in the worldvolume gauge field $A$ in the Wess-Zumino
lagrangian is:
\beq
\int F\wedge C^{(4)}=\int A\wedge F^{(5)}\,,
\label{source}
\eeq
where $F=dA$ and we have integrated by parts. In eq. (\ref{source}) it is
understood that  the pullback of the Ramond-Ramond fields to the
worldvolume is being taken. By counting the degree of the form under the integral
in eq. (\ref{source}), it is obvious that such a term can only exist for a
D5-brane and is zero if the  brane worldvolume
does not capture the  flux of the $F^{(5)}$. As it can be easily checked
by inspection, this happens in all the cases studied in this chapter.

Nevertheless, we could try to find embeddings with non-vanishing worldvolume gauge
fields even when the equations of motion allow to put them to zero.
Though for simplicity we would not try to do this in this chapter, in ref.
\cite{conifold} a supersymmetric embedding of a D5-brane with flux of
the worldvolume gauge field was obtained.

We have shown in section \ref{cp1sskappasym} that the supersymmetric
BPS configurations of the brane probe are obtained by requiring the
condition:
\beq
\Gamma_{\kappa}\,\epsilon=\epsilon\,,
\label{kappacondition}
\eeq
where $\epsilon$ is a Killing spinor of the background. It follows from
eq. (\ref{cp5gammakappa}) that $\Gamma_{\kappa}$ depends on the induced
metric and Dirac matrices, which in turn are determined by the D-brane
embedding 
$X^{M}(\xi^a)$. Actually, eq. (\ref{kappacondition}) should be regarded
as an equation whose unknowns are both the embedding $X^{M}(\xi^a)$
and the Killing spinors $\epsilon$. The number of solutions for
$\epsilon$ determines the amount of background supersymmetry that is
preserved by the probe. Notice that we have written $\Gamma_{\kappa}$ in
eq. (\ref{cp5gammakappa}) as a matrix acting on real two-component
spinors, while we have written  the Killing spinors of the background  in
complex notation. However, it is straightforward to find the following
rules to pass from complex to real spinors:
\beq
\epsilon^*\,\leftrightarrow\,\tau_3\,\epsilon\,,\qquad\quad
i\epsilon^*\,\leftrightarrow\,\tau_1\,\epsilon\,,\qquad\quad
i\epsilon\,\leftrightarrow\,-i\tau_2\,\epsilon\,.
\label{rule}
\eeq
As an example of the application of these rules, notice that the projections 
(\ref{tsprojections}) satisfied by the Killing spinors of the $AdS_5\times T^{1,1}$
background can be written as:
\beq
\Gamma_{12}\,\otimes\,i\tau_2\,\epsilon=-
\Gamma_{\hat 1\hat 2}\,\otimes\,i\tau_2\,\epsilon=\epsilon\,.
\label{tprojectionpauli}
\eeq

Let us now discuss the general strategy to solve the kappa symmetry equation 
(\ref{kappacondition}). First of all, notice that, by using the explicit form 
(\ref{cp5inducedgamma}) of the induced Dirac matrices in the expression of
$\Gamma_{\kappa}$ (eq. (\ref{cp5gammakappa})), eq. (\ref{kappacondition})
takes the form:
\beq
\sum_{i}\,c_i\,\Gamma_{AdS_5}^{(i)}\,
\Gamma_{T^{1,1}}^{(i)}\,\otimes\,
(\tau_3)^{{p-3\over 2}}\,i\tau_2\,\,\epsilon=\epsilon\,,
\label{kappaexplicit}
\eeq
where $\Gamma_{AdS_5}^{(i)}$ ($\Gamma_{T^{1,1}}^{(i)}$) are antisymmetrized products
of constant ten-dimensional Dirac matrices along the $AdS_5$
($T^{1,1}$) directions and the coefficients $c_i$ depend on the embedding 
$X^{M}(\xi^a)$ of the Dp-brane in the $AdS_5\times T^{1,1}$ space.
Actually, due to the relations (\ref{tprojectionpauli}) satisfied by the
Killing spinors
$\epsilon$, some of the terms in eq. (\ref{kappaexplicit}) are not independent.
After expressing eq. (\ref{kappaexplicit}) as a sum of independent contributions, we
obtain a new projection for the Killing spinor $\epsilon$. This projection is not,
in general, consistent with the conditions (\ref{tprojectionpauli}) since some of
the matrices  appearing on the left-hand side of eq. (\ref{kappaexplicit}) do not
commute with those appearing in (\ref{tprojectionpauli}). The only way of making
eqs. (\ref{tprojectionpauli}) and (\ref{kappaexplicit}) consistent with each other
is by requiring the vanishing of the coefficients $c_i$ of these non-commuting
matrices, which gives rise to a set of first order BPS differential
equations for the embedding $X^{M}(\xi^a)$. 

Notice that the kappa symmetry projection of the BPS configurations must be
satisfied at any point of the worldvolume of the brane probe. However, the Killing
spinors $\epsilon$ do depend on the coordinates 
(see eqs. (\ref{adsspinor}) or (\ref{globalspinor})). Thus, it is not obvious at
all that the $\Gamma_{\kappa}\epsilon=\epsilon$ condition can be imposed at all
points of the worldvolume. This fact would be guaranteed if we could recast eq. 
(\ref{kappacondition}) for BPS configurations as an algebraic condition on the
constant spinor $\eta$ of eqs.  (\ref{adsspinor}) or (\ref{globalspinor}). This
algebraic condition on $\eta$ must involve a constant matrix projector and its
fulfillment is generically achieved by imposing some extra conditions to the spinor 
$\epsilon$ (which reduces the amount of supersymmetry preserved by the
configuration) or by restricting appropriately the embedding. For example, when
working on the coordinates (\ref{adspoincare}), one should check whether the kappa
symmetry projector commutes with the matrix $\Gamma_{*}$ of eq.
(\ref{gamma*}). If this is the case, one can consider spinors such as the
$\epsilon_+$s of eq. (\ref{chiraladsspinor}), which are eigenvectors of 
$\Gamma_{*}$ and, apart from an irrelevant factor depending on the radial
coordinate, are constant. In case we use the parametrization (\ref{globalspinor}),
we should check that, for the BPS embeddings, the kappa symmetry projection
commutes with the matrix multiplying  the spinor $\eta$ on the right-hand side of
eq. (\ref{globalspinor}). 

If the BPS differential equations can be solved, one should verify that the
corresponding configuration also solves the equations of motion derived from the
Dirac-Born-Infeld action of the probe. In all the cases analyzed in this chapter the
solutions of the BPS equations also solve the equations of motion. Actually, we
will verify that these BPS configurations saturate a bound for the energy, as is
expected for a supersymmetric worldvolume soliton. 

%
%
%
%
%
%
%
%
%

\setcounter{equation}{0}
\section{Kappa symmetry for a D3-brane probe}
\label{cp5sD3}
\medskip
As our first example of D-brane probe in the Klebanov-Witten background, let us
consider a D3-brane. By particularizing eq. (\ref{cp5gammakappa}) to this
$p=3$ case, we obtain that $\Gamma_{\kappa}$ is given by: 
\beq
\Gamma_{\kappa}=-{i\over 4!\sqrt{-g}}\,\epsilon^{a_1\cdots a_4}\,
\gamma_{a_1\cdots a_4}\,,
\label{Gammad3}
\eeq
where we have used the dictionary (\ref{rule}) to obtain the expression 
of  $\Gamma_{\kappa}$ acting on complex spinors. 

One can consider several possible configurations with different  number of
dimensions on which the D3-brane is wrapped. Since the  $T^{1,1}$ space is
topologically $S^2\times S^3$, it is natural to consider branes wrapped
over three- and two-cycles. We will study the case of D3-branes wrapped
over a three-dimensional manifold, where we will find a rich set of BPS
configurations. The configurations where the D3-brane wraps a two-cycle of
the $T^{1,1}$ were studied in \cite{conifold} and no supersymmetric
embeddings were found.

\subsection{D3-branes wrapped on a three-cycle}
\label{cp5ssd33c}

Let us use global coordinates  as in eq. (\ref{globalads}) for the $AdS_5$ part of
the metric. We will search for supersymmetric configurations which are pointlike
from the $AdS_5$ point of view and wrap a compact three-manifold within $T^{1,1}$.
Accordingly, let us take the following set of 
worldvolume coordinates:
\beq
\xi^a=(\,t,\theta_1,\phi_1,\psi\,)\,,
\label{vwcoordinatesd3}
\eeq
and consider embeddings of the type:
\beq
\theta_2=\theta_2(\theta_1, \phi_1)\,,\qquad\quad
\phi_2=\phi_2(\theta_1,\phi_1)\,,
\label{ansatzd3}
\eeq
with the radial coordinate $\rho$ and the angles $\alpha^i$  being
constant. For these embeddings $\Gamma_{\kappa}$ in  eq. (\ref{Gammad3}) reduces to:
\beq
\Gamma_{\kappa}=-iL\,\,{\cosh\rho\over \sqrt{-g}}\,
\Gamma_{t}\,\,\gamma_{\theta_1\phi_1\psi}\,.
\label{gammad3s3}
\eeq
The induced gamma matrices along the worldvolume coordinates can be readily obtained
from the general expression (\ref{cp5inducedgamma}). The result is:
\bear
\gamma_{\theta_1}&=&{L\over \sqrt{6}}\,\Big[\,\Gamma_1\,
+\,(\cos\psi\,\partial_{\theta_1}\theta_2
+\sin\psi\sin\theta_2\partial_{\theta_1}\phi_2\,)
\Gamma_{\hat 1}+\rc\rc
&&+\,(\,\sin\psi\,\partial_{\theta_1}\theta_2-
\cos\psi\sin\theta_2\,\partial_{\theta_1}\phi_2\,)\,
\Gamma_{\hat 2}\,\,\Big]+
{L\over 3}\,\cos\theta_2\,\partial_{\theta_1}\phi_2
\Gamma_{\hat 3}\,,\rc\rc
\gamma_{\phi_1}&=&{L\over \sqrt{6}}\,\,\Big[\,\sin\theta_1
\,\Gamma_2+ (\,\sin\theta_2\,
\sin\psi\,\partial_{\phi_1}\phi_2+\cos\psi\,\partial_{\phi_1}\theta_2
\,)
\Gamma_{\hat 1}\,+\rc\rc
&&+\,(\sin\psi\,\partial_{\phi_1}\theta_2-
\cos\psi\,\sin\theta_2\,\partial_{\phi_1}\phi_2\,
)\Gamma_{\hat 2}\,\Big]\,
+\,{L\over 3}\,
(\,\cos\theta_1+\,\cos\theta_2\partial_{\phi_1}\phi_2\,)\,
\Gamma_{\hat 3}\,,\rc\rc
\gamma_{\psi}&=&{L\over 3}\,\,\Gamma_{\hat 3}\,.
\label{inducedgammasd3}
\eear
By using these expressions and the projections (\ref{tsprojections}), it
is easy to verify that:
\beq
{18\over L^3}\,\,
\gamma_{\theta_1\phi_1\psi}\,\epsilon=ic_1\Gamma_{\hat 3}\epsilon+
(c_2+ic_3)\,e^{-i\psi}\,\Gamma_{1\hat 2\hat 3}\,\epsilon\,,
\label{gammathetaphipsi}
\eeq
with the coefficients $c_1$, $c_2$ and $c_3$ being:
\bear
c_1&=&
\sin\theta_1+\sin\theta_2\,\Big(\,
\partial_{\theta_1}\theta_2\,\partial_{\phi_1}\phi_2-
\partial_{\theta_1}\phi_2\,\partial_{\phi_1}\theta_2\,
\Big)\,,\rc\rc
c_2&=&
\sin\theta_1\,\partial_{\theta_1}\theta_2-
\sin\theta_2\partial_{\phi_1}\phi_2\,,\rc\rc
c_3&=&
\partial_{\phi_1}\theta_2+
\sin\theta_1\, \sin\theta_2\,\partial_{\theta_1}\phi_2\,.
\label{d3cs}
\eear
Following the general strategy discussed at the end of section
\ref{cp5sssusypr}, we have to ensure that the kappa symmetry projection
$\Gamma_{\kappa}\epsilon=\epsilon$ is compatible with the conditions
(\ref{tsprojections}). By inspecting the right-hand side of eq.
(\ref{gammathetaphipsi}) it is fairly obvious that the terms containing
the matrix $\Gamma_{1\hat 2\hat 3}$ would give rise to contributions not
compatible with the projection (\ref{tsprojections}). Thus, it is clear
that to have 
$\Gamma_{\kappa}\,\epsilon=\epsilon$ we must impose the condition:
\beq
c_2=c_3=0\,,
\eeq
which yields the following differential equations for 
$\theta_2(\theta_1,\phi_1)$
and $\phi_2(\theta_1,\phi_1)$:
\bear
\sin\theta_1\,\partial_{\theta_1}\theta_2&=&
\sin\theta_2\,\partial_{\phi_1}\phi_2\,,\rc\rc
\partial_{\phi_1}\theta_2&=&-\sin\theta_1\sin\theta_2\,
\partial_{\theta_1}\phi_2\,.
\label{bpsd3}
\eear
We will prove below that the first order equations (\ref{bpsd3}),
together with some extra condition on the Killing spinor $\epsilon$, are
enough to ensure that 
$\Gamma_{\kappa}\epsilon\,=\,\epsilon$, \ie\ that our D3-brane probe configuration
preserves some fraction of supersymmetry. For this reason we will refer to 
(\ref{bpsd3}) as the BPS equations of the embedding. It is clear from eq.
(\ref{gammad3s3}) that, in order to compute $\Gamma_{\kappa}$, we need to calculate
the determinant $g$ of the induced metric.
For our ansatz (\ref{ansatzd3}), it is easy to verify that the non-vanishing elements of the
induced metric are:
\bear
&&g_{\tau\tau}=-L^2\cosh^2\rho\,,\rc\rc
&&g_{\theta_1\theta_1}={L^2\over 6}\,\,
\Big[\,1+(\partial_{\theta_1}\theta_2)^2+\sin^2\theta_2\,
(\partial_{\theta_1}\phi_2)^2
\Big]+{L^2\over 9}\,\cos^2\theta_2\,
(\partial_{\theta_1}\phi_2)^2
\,,\rc\rc
&&g_{\phi_1\phi_1}={L^2\over 6}\,\Big[\,
\sin^2\theta_1+(\partial_{\phi_1}\theta_2)^2+
\sin^2\theta_2\,(\partial_{\phi_1}\phi_2)^2\,
\Big]+{L^2\over 9}\,
\big(\,\cos\theta_1+\cos\theta_2\,\partial_{\phi_1}\phi_2\,
\big)^2\,,\rc\rc
&&g_{\psi\psi}={L^2\over 9}\,,\rc\rc
&&g_{\phi_1\theta_1}={L^2\over 6}\,\Big[\,
\partial_{\theta_1}\theta_2\,\partial_{\phi_1}\theta_2+
\sin^2\theta_2\,\partial_{\theta_1}\phi_2\,\partial_{\phi_1}\phi_2\,
\Big]+{L^2\over 9}\,\cos\theta_2\,\partial_{\theta_1}\phi_2\,
(\,\cos\theta_1+\cos\theta_2\,\partial_{\phi_1}\phi_2\,)\,,
\rc\rc
&&g_{\phi_1\psi}={L^2\over 9}\,
\big(\,\cos\theta_1+\cos\theta_2\,\partial_{\phi_1}\phi_2\,\big)
\,,\rc\rc
&&g_{\theta_1\psi}={L^2\over 9}\,
\cos\theta_2\,\partial_{\theta_1}\,\phi_2\,.
\label{inducedmetricd3}
\eear
Let us now define:
\bear
\alpha&\equiv&{L^2\over
6}\,\Big[\,1+(\partial_{\theta_1}\theta_2)^2+
\sin^2\theta_2\,(\partial_{\theta_1}\phi_2)^2\,\Big]\,,\rc\rc
\beta&\equiv&{L^2\over 6}\,\Big[\,
\sin^2\theta_1+(\partial_{\phi_1}\theta_2)^2+
\sin^2\theta_2\,(\partial_{\phi_1}\phi_2)^2\,\Big]\,,\rc\rc
\gamma&\equiv&{L^2\over 6}\,\Big[\,
\partial_{\theta_1}\theta_2\,\partial_{\phi_1}\theta_2+
\sin^2\theta_2\,\partial_{\theta_1}\phi_2\,\partial_{\phi_1}\phi_2
\,\Big]\,.
\label{cpcfalpha}
\eear
From these values one can prove  that:
\beq
\sqrt{-g}={L^2\cosh\rho\over 3}\,
\sqrt{\alpha\beta-\gamma^2}\,.
\label{detd3}
\eeq
Moreover, if the BPS equations (\ref{bpsd3}) are satisfied, the functions 
$\alpha$, $\beta$ and $\gamma$ take the values:
\beq
\alpha_{|_{BPS}}={L^2\over 6\sin\theta_1}\,{c_1}_{|_{BPS}}\,,\qquad\quad
\beta_{|_{BPS}}={L^2\sin\theta_1\over 6}\,\,{c_1}_{|_{BPS}}\,,\qquad\quad
\gamma_{|_{BPS}}=0\,,
\eeq
where $c_1$ is written in eq. (\ref{d3cs})
and the determinant of the induced metric is:
\beq
\sqrt{-g}_{|_{BPS}}={L^4\over 18}\,\cosh\rho\,{c_1}_{|_{BPS}}\,.
\eeq

From this expression of $\sqrt{-g}_{|_{BPS}}$ it is straightforward to
verify that, if the first order system (\ref{bpsd3}) holds, one has:
\beq
\Gamma_{\kappa}\,\epsilon=\Gamma_{t}\Gamma_{\hat 3}\,\epsilon\,.
\eeq
Thus, the condition $\Gamma_{\kappa}\,\epsilon=\epsilon$ is equivalent to:
\beq
\Gamma_{t}\Gamma_{\hat 3}\,\epsilon=\epsilon\,.
\eeq
Let us now plug in this equation the explicit form (\ref{globalspinor}) of
the Killing spinors. Notice that, except for $\Gamma_{\rho}\,\gamma_*$,
$\Gamma_{t}\Gamma_{\hat 3}$ commutes with all matrices appearing on
the right-hand side of eq. (\ref{globalspinor}). Actually, only for
$\rho=0$ the coefficient of $\Gamma_{\rho}\,\gamma_*$ in 
(\ref{globalspinor}) vanishes and, thus, only at this point of $AdS_5$
the equation  $\Gamma_{\kappa}\,\epsilon=\epsilon$ can be
satisfied. In this case, it reduces to the following condition on the
constant spinor $\eta$:
\beq
\Gamma_{t}\Gamma_{\hat 3}\,\eta=\eta\,.
\eeq
Then, in order to have a supersymmetric embedding, we must place our
D3-brane probe at $\rho=0$, \ie\ at the center of the $AdS_5$ space. The
resulting configuration is $1/8$ supersymmetric: it preserves four
Killing spinors of the type (\ref{globalspinor}) with
$\Gamma_{12}\,\eta=-\Gamma_{\hat 1\hat 2}\,\eta\,=\,i\eta$, 
$\,\,\,\Gamma_{t}\Gamma_{\hat 3}\,\eta\,=\,\eta$.

\subsubsection{Integration of the first order equations}

Let us now integrate the first order differential equations (\ref{bpsd3}).
Remarkably, this same set of equations has been obtained in ref.
\cite{flavoring} in the study of the supersymmetric embeddings of
D5-brane probes in the Maldacena-N\'u\~nez background \cite{MN,CV}. It was
shown in ref. \cite{flavoring}  that, after a change of variables, the
pair of eqs. in (\ref{bpsd3}) can be converted into the Cauchy-Riemann
equations. Indeed,  let us define two new variables $u_1$ and  $u_2$,
related to  
$\theta_1$ and $\theta_2$ as follows:
\beq
u_1=\log\Big(\tan\,{\theta_1\over 2}\Big)\;,
\qquad\quad
u_2=\log\Big(\tan\,{\theta_2\over 2}\Big)\,.
\label{change}
\eeq
Then, it is straightforward to demonstrate that the equations 
(\ref{bpsd3}) can be written as:
\beq
{\partial u_2\over \partial u_1}=
{\partial \phi_2\over \partial \phi_1}\;,
\qquad\quad
{\partial u_2\over \partial \phi_1}=-
{\partial \phi_2\over \partial u_1}\;,
\eeq
\ie\ as the Cauchy-Riemann equations for the variables $(u_1,\phi_1)$ and 
$(u_2,\phi_2)$. Since $u_1,u_2\in (-\infty,+\infty)$ and
$\phi_1,\phi_2\in (0,2\pi)$, the above equations are actually the
Cauchy-Riemann equations in a band. The general integral of these
equations is obtained by requiring that $u_2+i\phi_2$ be an arbitrary
function of the holomorphic variable $u_1+i\phi_1$:
\beq
u_2+i\phi_2=f(u_1+i\phi_1)\,.
\label{generalholo}
\eeq
Let us now consider the particular case in which $u_2+i\phi_2$ depends
linearly on $u_1+i\phi_1$, namely:
\beq
u_2+i\phi_2=m(u_1+i\phi_1)+{\rm constant}\,,
\label{holo}
\eeq
where $m$ is constant. Let us further assume that $m$ is real and integer. By
equating the imaginary parts of both sides of eq. (\ref{holo}), one gets:
\beq
\phi_2=m\,\phi_1+{\rm constant}\,.
\label{mwindingphi}
\eeq
Clearly, $m$ can be interpreted as a winding number \cite{flavoring}. Moreover, from
the real part of eq. (\ref{holo}) we immediately obtain $u_2$ as a function
of $u_1$ for this embedding. By using the change of variables of eq.
(\ref{change}) we can convert this $u_2=u_2(u_1)$ function in a relation between
the angles
$\theta_1$ and
$\theta_2$, namely:
\beq
\tan{\theta_2\over 2}=C\,\Bigg(\,\tan{\theta_1\over 2}\,\Bigg)^m\,,
\label{mwindingtheta}
\eeq
with $C$ constant. Following ref. \cite{flavoring} we will call  $m$-winding
embedding to the brane configuration corresponding to eqs. 
(\ref{mwindingphi}) and (\ref{mwindingtheta}).  Notice that for $m=0$ the above
solution reduces to $\theta_2={\rm constant}$, $\phi_2={\rm constant}$. This
zero-winding configuration of the D3-brane is just the one proposed in ref.
\cite{GK} as dual to the dibaryon operators of the $SU(N)\times SU(N)$ gauge
theory. Moreover, when $m=\pm 1$ we have the so-called unit-winding embeddings.
When the constant $C$ in eq. (\ref{mwindingtheta}) is equal to one, it is easy to
find the following form of these unit-winding configurations:
\bear
&&\theta_2=\theta_1\,,\qquad\qquad\;\;
\phi_2=\phi_1\,,\qquad
(m=1)\,,\rc\rc
&&\theta_2=\pi-\theta_1\,,\qquad\quad
\phi_2=2\pi-\phi_1\,,\qquad
(m=-1)\,,
\label{unitwinding}
\eear
where we have adjusted appropriately the constant of eq. (\ref{mwindingphi}). Notice
that the two possibilities in (\ref{unitwinding}) correspond to the
two possible identifications of the two $(\theta_1,\phi_1)$ and 
$(\theta_2,\phi_2)$ two-spheres.

\subsubsection{Holomorphic structure}

It is also interesting to write the $m$-winding embeddings just found in
terms of the holomorphic coordinates $z_1,\cdots,z_4$ of the conifold.
Actually, by inspecting eq. (\ref{zratio}), and comparing it with the functions 
$\theta_2=\theta_2(\theta_1)$ and $\phi_2=\phi_2(\phi_1)$ corresponding
to an $m$-winding embedding (eqs. (\ref{mwindingphi}) and
(\ref{mwindingtheta})),  one concludes that the latter can be written,
for example, as\footnote{If the function $f$ in eq. (\ref{generalholo}) satisfies
that $\bar f(z)=f(\bar z)$, then the general solution (\ref{generalholo}) can be
written as $\log{z_1\over z_4}=f\big[\log{z_1\over z_3}\big]$.
}:
\beq
{z_1\over z_4}=C\,
\Bigg(\,{z_1\over z_3}\,\Bigg)^m\,.
\label{d3holomorphic}
\eeq
Thus, the  $m$-winding embeddings of the D3-brane in the $T^{1,1}$ space can be
characterized as the vanishing locus of a polynomial in the $z_i$ coordinates of
$\CC^4$. In order to find this polynomial in its full generality, let us consider
the solutions of the following polynomial equation:
\beq
z_1^{m_1}\,z_2^{m_2}\,z_3^{m_3}\,z_4^{m_4}={\rm constant}\,,
\label{d3pol}
\eeq
where the $m_i\,$s are real constants and we will assume that:
\bear
&&m_1+m_2+m_3+m_4=0\,,\rc\rc
&&m_1+m_3\not=0\,.
\label{d3polconditions}
\eear
By plugging the representation (\ref{holomorphic}) of the $z_i$ coordinates in the
left-hand side of eq. (\ref{d3pol}), one readily proves that, due to the first
condition in eq. (\ref{d3polconditions}), $r$ and $\psi$ are not restricted by eq.
(\ref{d3pol}). Moreover, by looking at the phase of the 
left-hand side of eq. (\ref{d3pol}) one realizes that, if the second condition in 
(\ref{d3polconditions}) holds, $\phi_2$ is related to $\phi_1$ as in eq. 
(\ref{mwindingphi}), with $m$ being given by:
\beq
m={m_2+m_3\over m_1+m_3}={m_1+m_4\over m_2+m_4}\,.
\label{m-winding}
\eeq
Furthermore, from the modulus of eq. (\ref{d3pol}) we easily prove that 
$\theta_2(\theta_1)$ is indeed given by eq. (\ref{mwindingtheta}) with the winding
$m$ displayed in eq. (\ref{m-winding}). As a check of these identifications 
let us notice that, by using the
conifold equation (\ref{conifold}), the embedding (\ref{d3pol}) is
invariant under the change:
\beq
(m_1,m_2,m_3,m_4)\rightarrow (m_1-n,m_2-n,m_3+n,m_4+n)\,,
\label{mchange}
\eeq
for arbitrary $n$. Our relation (\ref{m-winding}) of the winding $m$ and the
exponents
$m_i$  is also invariant under the change (\ref{mchange}).

As particular cases notice that  the zero-winding
embedding can be described by the equation $z_1=Cz_4$, while the unit-winding
solution corresponds to $z_3=C\,z_4$ for $m=1$ and $z_1=C\,z_2$ for $m=-1$. 

For illustrative purposes, let us consider the holomorphic structure of the
solutions (\ref{generalholo}) with a non-linear function $f$. It is easy to see
that, if the function $f$ is not linear, the corresponding holomorphic equation is
non-polynomial. For example, the embedding 
$u_2+i\phi_2=(u_1+i\phi_1)^2$ corresponds to the equation 
${z_1\over z_4}=\exp[\log^2{z_1\over z_3}]$. Contrary to what happens to the
solutions (\ref{d3pol}) (see below), the field theory dual of these
non-polynomial embeddings is completely unclear for us and we will not pursue their
study here.

\subsubsection{Energy bound}

The Dirac-Born-Infeld lagrangian density for the D3-brane probe is given by
\beq
{\cal L}=-\sqrt{-g}\,,
\label{DBIlagrangian}
\eeq 
where we have taken the D3-brane tension equal to one and the 
 value of $\sqrt{-g}$ for a general embedding of the type 
(\ref{ansatzd3}) has been written in
eq. (\ref{detd3}). We have   checked by explicit calculation that any solution of
the first order equations (\ref{bpsd3}) also satisfies the Euler-Lagrange
equations of motion derived from the lagrangian  (\ref{DBIlagrangian}).
Moreover, the hamiltonian density for the static configurations we are
considering is just
${\cal H}=-{\cal L}$. We are going to prove that ${\cal H}$ satisfies a bound which
is saturated just when the embedding satisfies the BPS equations (\ref{bpsd3}).
To check this fact, let us consider 
arbitrary functions  $\theta_2(\theta_1,\phi_1)$ and
$\phi_2(\theta_1,\phi_1)$. For an embedding at
$\rho=0$, we can write:
\beq
{\cal H}={L^2\over 3}\,\,
\sqrt{\alpha\beta-\gamma^2}\,,
\eeq
where $\alpha$, $\beta$ and $\gamma$ are given in eq. (\ref{cpcfalpha})
and we have used the value of  $\sqrt{-g}$    given in eq.
(\ref{detd3}).   Let us  now rewrite ${\cal H}$  as
${\cal H}=|{\cal Z}|+{\cal S}$, where 
${\cal Z}={L^4\over 18}\,c_1$, with $c_1$ given in the first expression
in eq.  (\ref{d3cs}). It is easily checked that ${\cal Z}$  can be
written as a total derivative:
\beq
{\cal Z}=\partial_{\theta_1}\,{\cal Z}^{\theta_1}+
\partial_{\phi_1}\,{\cal Z}^{\phi_1}\,,
\eeq 
with:
\beq
{\cal Z}^{\theta_1}=-{L^4\over 18}\,\,
(\,\cos\theta_1+\cos\theta_2\,\partial_{\phi_1}\phi_2)\,,\qquad\quad
{\cal Z}^{\phi_1}={L^4\over
18}\,\cos\theta_2\,\partial_{\theta_1}\phi_2\,.
\eeq
Clearly, ${\cal S}$ is given by:
\beq
{\cal S}={L^2\over 3}\,\,
\sqrt{\alpha\beta-\gamma^2}-{L^4\over 18}\,|c_1|\,.
\eeq
Let us now prove that ${\cal S}\ge 0$ for an arbitrary embedding of the type
(\ref{ansatzd3}). Notice that this is equivalent to the following bound:
\beq
{\cal H}\,\ge\,|{\cal Z}|\,.
\eeq
In fact, it is easy to verify that the condition ${\cal S}\ge 0$ is
equivalent to:
\beq
\Big(\,\sin\theta_1\,\partial_{\theta_1}\theta_2-\sin\theta_2
\partial_{\phi_1}\phi_2\,\Big)^2\,+
\Big(\,\partial_{\phi_1}\theta_2+\sin\theta_1\sin\theta_2\,
\partial_{\theta_1}\phi_2\,\Big)^2
\,\ge\,\,0\,,
\label{d3inequality}
\eeq
which is obviously satisfied for arbitrary functions $\theta_2(\theta_1,\phi_1)$ and
$\phi_2(\theta_1,\phi_1)$. Moreover, by inspecting eq. (\ref{d3inequality}) one
easily concludes that the equality in (\ref{d3inequality}) is equivalent to the
first order BPS equations  (\ref{bpsd3}) and thus  ${\cal H}=|{\cal Z}|$
for the BPS embeddings (actually, ${\cal Z}\ge 0$ if the BPS equations
are satisfied).

\subsubsection{Field theory dual}
In this subsection we will give some hints on the field theory dual  of a general
D3-brane $m$-winding embedding. First of all, let us try to find the conformal
dimension $\Delta$ of the corresponding operator in the dual field theory. According
to the general AdS/CFT arguments, and taking into account the zero-mode corrections
as in ref. \cite{BHK}, one should have:
\beq
\Delta=LM\,,
\label{Delta}
\eeq
where $L$ is given in (\ref{KW}) and $M$ is the mass of the wrapped D3-brane. The
latter can be written simply as:
\beq
M=T_3\,V_3\,,
\eeq
where $V_3$ is the volume of the cycle computed with the induced metric 
(\ref{inducedmetricd3}) and
$T_3$ is the tension of  the D3-brane, given by:
\beq
T_3={1\over 8\pi^3(\alpha')^2 g_s}\,.
\eeq

Taking into account that, as we have seen above, for the $m$-winding embedding
$\sqrt{-g} = {L^4\over18}\,c_1$, one can easily compute
the value of $V_3$ for  the three-cycle ${\cal C}^{(m)}$ corresponding
to the $m$-winding embedding. The result is:
\beq
V_3={8\pi^2 L^3\over 9}\,(1+|m|)\,.
\eeq
Now, the mass of the wrapped D3-brane can be readily obtained, namely:
\beq
M={L^3\over 9\pi (\alpha')^2 g_s}\,(1+|m|)\,.
\eeq
By plugging this result in eq. (\ref{Delta}), and using the value of $L$
given in eq. (\ref{KW}), one obtains the following value of the conformal
dimension $\Delta$:
\beq
\Delta={3\over 4}\,(1+|m|)\,N\,.
\label{Dimension}
\eeq
Notice that for $m=0$ we recover from (\ref{Dimension}) the result 
$\Delta={3\over 4}N$ of ref. \cite{GK}, which is the conformal dimension
of  an operator of the form $A^N$, with $A$ being the ${\cal N}=1$
chiral multiplets introduced in section \ref{cp5sKW} and a double
antisymmetrization over the gauge indices of $A$ is performed in the
$N^{{\rm th}}$ power of $A$ (for details see ref. \cite{GK}). Remember
that the conformal dimensions of the
$A$ and
$B$ fields are 
$\Delta(A)=\Delta(B)=3/4$. Thus, in view of the result (\ref{Dimension}), it is
natural to think that our wrapped D3-branes
for general $m$ correspond to operators with a field content of the form:
\beq
(A^{a}B^{b})^N\,,\qquad
a+b=1+|m|\,.
\label{dualfield}
\eeq
To determine the values of $a$ and $b$ in (\ref{dualfield}) one has to find  the
baryon number of the operator. Recall \cite{KW} that the $U(1)$ baryon number
symmetry acts on the ${\cal N}=1$ matter multiplets as:
\beq
A_i\,\to \,e^{i\alpha}\,A_i\,,\qquad\quad
B_i\,\to \,e^{-i\alpha}\,B_i\,,
\label{baryonnumber}
\eeq
and, thus, the $A$ ($B$) field has baryon number $+1$($-1$). Notice that the
$z_i$ coordinates are invariant under the transformation (\ref{baryonnumber}) of
the homogeneous coordinates $A_i$ and $B_i$ (see eq. (\ref{homogeneous})). On the
gravity side of the AdS/CFT correspondence, the baryon number (in units of $N$)
can be identified with the third homology class of the three-cycle 
${\cal C}^{(m)}$ over which the D3-brane is wrapped. Indeed, the third homology
group of the $T^{1,1}$ space is 
$H_3(T^{1,1})={\mathbb Z}$. Moreover, the homology class of the cycle can
be determined by representing it as the zero-locus of a polynomial in the
$A$ and $B$ coordinates which transforms homogeneously under the $U(1)$
symmetry (\ref{baryonnumber}). Actually, the charge of the polynomial
under the baryon-number transformations  (\ref{baryonnumber}) is just the
class of the three-cycle ${\cal C}^{(m)}$ in 
$H_3(T^{1,1})={\mathbb Z}$.

It is easy to rewrite the results for the holomorphic structure of the $m$-winding
embeddings in terms of the homogeneous coordinates $A$ and $B$. Indeed,
one can prove that the three-cycle ${\cal C}^{(m)}$ corresponding to the $m$-winding
embedding can be written as:
\beq
A_1\,B_2^m=c\,A_2\,B_1^m\,.
\label{homogeneouspol}
\eeq
Notice that changing $m\to -m$ in eq. (\ref{homogeneouspol}) is equivalent to the
exchange $B_1\leftrightarrow B_2$. Therefore, we can always arrange eq.
(\ref{homogeneouspol}) in such a way that the exponents of $A$ and $B$ are
positive\footnote{There is an obvious asymmetry in our equations between the $A$
and $B$ coordinates. The origin of this asymmetry is the particular choice of
worldvolume coordinates we have made in (\ref{vwcoordinatesd3}). If we choose
instead $\xi^a=(\,t,\theta_2,\phi_2,\psi\,)$ the role of $A$ and $B$
is exchanged. Alternatively, the same effect is obtained with the
coordinates (\ref{vwcoordinatesd3}) by changing $m\to 1/m$.}.
Moreover, the polynomial representing
${\cal C}^{(m)}$ transforms homogeneously under the  symmetry (\ref{baryonnumber}) 
with charge
$1-|m|$, which is just the class of  ${\cal C}^{(m)}$ in 
$H_3(T^{1,1})={\mathbb Z}$. One can confirm this result by computing the
integral over ${\cal C}^{(m)}$ of the (pullback) of the three-form
$\omega_3$:
\beq
\omega_3={1\over 16\pi^2}\,d\psi\wedge(\,\sin\theta_1\,d\theta_1\wedge
d\phi_1-\sin\theta_2\,d\theta_2\wedge d\phi_2)\,,
\eeq
which has been suitably normalized. 
One can easily check that:
\beq
\int_{{\cal C}^{(m)}}\,\omega_3=1-|m|\,,
\eeq
which is the same result as that obtained by representing ${\cal C}^{(m)}$ as in
eq. (\ref{homogeneouspol}). Thus, the baryon number of the dual operator must be
$(1-|m|)N$ and we have a new equation for the exponents $a$ and $b$ in
(\ref{dualfield}), namely $a-b=1-|m|$, which allows to determine the actual values
of $a$ and $b$,  \ie\ $a=1$ and $b=|m|$. Thus, we are led to the conclusion that
the field theory operator dual to our $m$-winding embedding must be of the form:
\beq
\big(\,A\,B^{|m|}\,\big)^N\,.
\label{fieldcontent}
\eeq
We will not attempt to determine here the gauge-invariant index structure of the
operator with the field content (\ref{fieldcontent}) which is dual to the
$m$-winding configurations of the D3-branes. Notice that,   for generic
values of $m$, the absolute value of the baryon number is greater than $N$, whereas
for $m=\pm 1$ it vanishes. This last case resembles that of a giant graviton,
although it is interesting to remember that our unit-winding embeddings are
static, \ie\ time-independent. 

%
%
%
%
%
%
%
%
%

\setcounter{equation}{0}
\section{Kappa symmetry for a D5-brane probe}
\label{cp5sD5}
\medskip
In this section we will explore the possibility of having supersymmetric
configurations of D5-branes which wrap some cycle of the $T^{1,1}$ space. Notice
that, according to the general expression of $\Gamma_{\kappa}$ (eq.
(\ref{cp5gammakappa})) and to the dictionary of eq. (\ref{rule}), one has
in this case:
\beq
\Gamma_{\kappa}\,\epsilon={i\over 6!\, \sqrt{-g}}\,\,
\epsilon^{a_1\cdots a_6}\,\gamma_{a_1\cdots a_6}\,\epsilon^*\,.
\label{Gammakappad5}
\eeq
The complex conjugation of the right-hand side of eq. (\ref{Gammakappad5}) will be
of great importance in what follows. Recall that we want 
the D5-brane kappa symmetry projector to be compatible with the 
conditions $\Gamma_{12}\,\epsilon\,=\,i\epsilon$ and 
$\Gamma_{\hat 1\hat 2}\,\epsilon\,=\,-i\epsilon$ of eq. (\ref{tsprojections}).
Since in this D5-brane case the action of $\Gamma_{\kappa}$ on $\epsilon$ involves
the complex conjugation, this compatibility with the conditions
(\ref{tsprojections}) will force us to select embeddings for which the kappa
symmetry projector mixes the two 
$S^2$ spheres, which will allow us to find the differential equations to be
satisfied by the embedding. 

Actually, we will  only be able to carry out successfully this program for the case
of a D5-brane wrapped on a two-cycle. It has been proposed in ref.
\cite{GK} that this kind of configurations behave as domain walls in
$AdS_5$.  The simplest example would be a D3-brane which is not wrapped
over the compact manifold. Through an analysis of the five-form flux as in
\cite{Wittenbaryon} one can easily conclude that when crossing the
domain wall the gauge group of the dual field theory changes from
$SU(N)\times SU(N)$ to $SU(N+1)\times SU(N+1)$. The D5-brane wrapped over
a two-cycle of the $T^{1,1}$ is also an object of codimension one in
$AdS_5$ and it was shown in \cite{GK} that upon crossing it the gauge
theory group changes from $SU(N)\times SU(N)$ to $SU(N)\times SU(N+1)$,
which makes these configurations useful for constructing supergravity
duals to $SU(N_1)\times SU(N_2)$ gauge theories. In the next subsection
we will explain in detail how to find such supersymmetric embeddings and
we will analyze some of their properties.

\subsection{D5-branes wrapped on a two-cycle}
\label{cp5ssd52c}
Let us consider a D5-brane wrapped on a two-cycle of the $T^{1,1}$ space. In order to
preserve supersymmetry in a D3-D5 intersection the two branes must share two spatial
directions. Accordingly, we will place the D5-brane probe at some constant value of
one of the Minkowski coordinates (say $x^3$) and we will extend it along  the radial
direction. Following this discussion, let us take the following set of
worldvolume coordinates:
\beq
\xi^a=(x^0,x^1,x^2,r,\theta_1,\phi_1)\,,
\eeq
and consider embeddings with $x^3$ and $\psi$ constant in which:
\beq
\theta_2= \theta_2(\theta_1,\phi_1)\,,
\qquad\quad
\phi_2= \phi_2(\theta_1,\phi_1)\,.
\eeq
In this case eq. (\ref{Gammakappad5}) takes the form:
\beq
\Gamma_{\kappa}\,\epsilon={i\over \sqrt{-g}}\,{r^2\over L^2}\,\,
\Gamma_{x^0x^1x^2r}\,\,\gamma_{\theta_1\phi_1}\,\epsilon^*\,.
\eeq
The induced gamma matrices along the $\theta_1$ and $\phi_1$ directions are 
given by the same equations as in
the D3-brane embeddings of section \ref{cp5sD3} (see eq.
(\ref{inducedgammasd3})). Denoting by $\psi_0$ the constant value of the
$\psi$ coordinate, we obtain after an straightforward calculation:
\beq
{6\over L^2}\,\,
\gamma_{\theta_1\phi_1}\,\epsilon^*\,=-ic_1\,\epsilon^*+(c_2\,-ic_3)
e^{i\psi_0}\,\Gamma_{1\hat 2}\,\epsilon^*\,+
(\,c_4-ic_5)\Gamma_{1\hat 3}\epsilon^*+
(c_6-ic_7)\,e^{i\psi_0}\,\Gamma_{\hat 1\hat 3}\,\epsilon^*\,,
\label{gammakappad5}
\eeq
where the coefficients $c_1$, $c_2$ and $c_3$ are just the same as in the D3-brane 
(eq. (\ref{d3cs})) and
$c_4,\cdots,c_7$ are given by:
\bear
c_4&=&\sqrt{{2\over
3}}\left(\cos\theta_1+\cos\theta_2\partial_{\phi_1}\phi_2\right),\rc\rc
c_5&=&-\sqrt{{2\over 3}}\,\,
\sin\theta_1\cos\theta_2\partial_{\theta_1}\phi_2\,,\rc\rc
c_6&=&\sqrt{{2\over 3}}\,
\left(\cos\theta_1+\cos\theta_2\partial_{\phi_1}\phi_2\right)
\partial_{\theta_1}\theta_2\,,\rc\rc
c_7&=&\sqrt{{2\over 3}}\left(\cos\theta_1+
\cos\theta_2\partial_{\phi_1}\phi_2\right)
\partial_{\theta_1}\phi_2\,\sin\theta_2\,.
\label{c4c7}
\eear

To implement the $\Gamma_{\kappa}\epsilon=\epsilon$ condition one must impose some
differential (BPS) equations which make  some of the $c_i$ coefficients of eq. 
(\ref{gammakappad5}) vanish. The remaining terms give rise to an extra projection
which must commute with the ones already satisfied by the Killing spinors in
order to be compatible with them. By inspecting the different terms on the
right-hand side of eq. (\ref{gammakappad5}), one easily concludes that only the
terms with the 
$\Gamma_{1\hat 2}$ matrix lead to a projection which commutes with the ones corresponding
to the $T^{1,1}$ coset (eq. (\ref{tsprojections})). Therefore, it seems clear that
we must require the vanishing of all the coefficients $c_i$ different from $c_2$ and
$c_3$. Moreover, from the fact that $\Gamma^2_{\kappa}=1$ for any embedding, one
easily proves that:
\beq
\sqrt{-g}_{|_{BPS}}={r^2\over 6}\,\big|c_2-ic_3\big|_{_{BPS}}\,.
\eeq
Then, if $\delta(\theta_1,\phi_1)$ denotes the phase of $c_2-ic_3$, it is clear that, if
the BPS equations hold, the kappa symmetry condition (\ref{kappacondition}) is
equivalent to the projection:
\beq
ie^{i\delta(\theta_1,\phi_1)}\, e^{i\psi_0}\,
\Gamma_{x^0x^1x^2r}\,\,\Gamma_{1\hat 2}\,\epsilon^*\,=\epsilon\,.
\label{projdelta}
\eeq
We want to translate the above projection (\ref{projdelta}) into an algebraic condition
involving a constant matrix acting on a constant spinor. It is rather evident by
inspecting eq. (\ref{projdelta}) that this can only be achieved if the phase $\delta$ does
not depend on the worldvolume angles $\theta_1$ and $\phi_1$, which is ensured if
$c_2-ic_3$ is either real or purely imaginary, \ie\ when $c_2$ or $c_3$ is zero. We will
demonstrate later on in this section that by imposing  $c_2=0$ one does not arrive
at a consistent set of equations. Thus, let us consider the case $c_3=0$, \ie\ let
us require that all the coefficients except $c_2$ vanish:
\beq
c_1=c_3=c_4=c_5=c_6=c_7=0\,.
\label{cvanishingd5}
\eeq
Notice, first of all, that the condition
$c_5=0$ implies that $\partial_{\theta_1}\phi_2=0$. Substituting this result in the
equation $c_3=0$ one gets that the other crossed derivative $\partial_{\phi_1}\theta_2$
also vanishes  (see eq. (\ref{d3cs})) and, thus, one must have embeddings of the
type:
\beq
\theta_2= \theta_2(\theta_1)\;,
\,\,\,\,\,\,\,\,\,\,\,\,\,\,
\phi_2= \phi_2(\phi_1)\,.
\label{bpsdependences}
\eeq
For these embeddings $c_7$ is automatically zero and the  conditions 
$c_4=c_6=0$ give rise to the equation:
\beq
\cos\theta_1+\cos\theta_2\partial_{\phi_1}\phi_2=0\,,
\label{d5bps1}
\eeq
while the remaining condition $c_1=0$ yields another first order
equation, namely:
\beq
\sin\theta_1+\sin\theta_2\partial_{\theta_1}\theta_2\partial_{\phi_1}
\phi_2=0\,.
\label{d5bps2}
\eeq
Eqs. (\ref{d5bps1}) and (\ref{d5bps2}) are equivalent to the conditions 
(\ref{cvanishingd5}) and are the first order BPS differential
equations we were looking for in this case.

Let us now try to find the supersymmetry preserved by the BPS configurations. First of all
we notice that:
\beq
\sqrt{-g}_{\,\,|_{BPS}}={r^2\over 6}\,{|c_2|}_{\,\,_{BPS}}\,,
\eeq
and thus, the action of $\Gamma_{\kappa}$ on a Killing spinor $\epsilon$ when the
BPS conditions are satisfied is:
\beq
\Gamma_{\kappa}\,\epsilon_{\,\,|_{BPS}}=i\,{\rm sign} (c_2)\,e^{i\psi_0}\,
\Gamma_{x^0x^1x^2r}\,\,\Gamma_{1\hat 2}\,\epsilon^*\,.
\eeq
Therefore, we must require that:
\beq
i\,{\rm sign} (c_2)\,e^{i\psi_0}\,
\Gamma_{x^0x^1x^2r}\,\,\Gamma_{1\hat 2}\,\epsilon^*=\epsilon\,.
\label{d5projector}
\eeq
We want to convert eq. (\ref{d5projector}) into an algebraic condition on a
constant spinor. With this purpose in mind, let us write the general form of
$\epsilon$ as the sum of the two types of spinors written in eq.
(\ref{chiraladsspinor}), namely:
\beq
\epsilon=r^{-{1\over 2}}\,\eta_-+
r^{{1\over 2}}\,\Big(\,{\bar x^3\over L^2}\,
\Gamma_{rx^3}\,\eta_-+\eta_+\,\Big)\,+
\,{r^{{1\over 2}}\over L^2}\,x^p\,\Gamma_{rx^p}\,\eta_-\,,
\label{generalepsilon}
\eeq
where $\bar x^3$ is the constant value of the coordinate $x^3$ in the embedding,
$\eta_{\pm}$ are constant spinors satisfying eq. (\ref{etamasmenos}) and the index
$p$ runs over the set  $\{0,1,2\}$. In eq. (\ref{generalepsilon}) we have
explicitly displayed the dependence of $\epsilon$ on the coordinates $r$ and $x^p$.
By substituting eq.  (\ref{generalepsilon}) on both sides of eq.
(\ref{d5projector}), one can get the conditions that $\eta_{+}$  and
$\eta_{-}$ must satisfy\footnote{A similar analysis, for the D5-brane
configurations in the $AdS_5\times S^5$ background, was performed in ref. \cite{ST}.
}. In order to write these conditions, let us define ${\cal
P}$ as the operator that acts on any spinor $\epsilon$ as follows:
\beq
{\cal P}\,\epsilon\equiv{\rm sign} (c_2) e^{i\psi_0}\,\Gamma_{rx^3}\,
\Gamma_{1\hat 2}\,\epsilon^*\,.
\eeq
Then, eq. (\ref{d5projector}) is equivalent to
\bear
&&{\cal P}\,\eta_-=\eta_-\,,\rc\rc
&&(1+{\cal P}\,)\,\eta_+=-{2 \bar x^3\over
L^2}\,\Gamma_{rx^3}\,\eta_-\,.
\label{d5system}
\eear
Since ${\cal P}^2=1$, we can classify the four spinors $\eta_+$ according to their 
${\cal P}$-eigenvalue as:
\beq
{\cal P}\,\eta_+^{(\pm)}=\pm\eta_+^{(\pm)}\,.
\eeq
We can now solve the system (\ref{d5system}) by taking $\eta_-=0$ and
$\eta_+$ equal to one of the two spinors $\eta_+^{(-)}$ of negative
${\cal P}$-eigenvalue. Moreover, there are other two solutions which
correspond to taking a spinor $\eta_+^{(+)}$ of positive 
${\cal P}$-eigenvalue and a
spinor $\eta_-$  related to the former as:
\beq
\eta_{-}={L^2\over \bar x^3}\,\Gamma_{r x^3}\,\,\eta_+^{(+)}\,.
\label{secondspinord5}
\eeq
Notice that, according to the first equation in (\ref{d5system}), the spinor
$\eta_-$ must have positive ${\cal P}$-eigenvalue, in agreement with eq.
(\ref{secondspinord5}). All together this configuration preserves four
supersymmetries, \ie\ one half of the supersymmetries of the background, 
as expected for a domain wall.

\subsubsection{Integration of the first order equations}
Let us now integrate the BPS equations (\ref{d5bps1}) and (\ref{d5bps2}).
First of all, notice that eq. (\ref{d5bps1}) can be written as:
\beq
{\partial \phi_2\over \partial\phi_1}=-{\cos\theta_1\over
\cos\theta_2}= {\rm constant}\,,
\label{constancy}
\eeq
where we have already taken into account  the only way in which eq. (\ref{d5bps1})
can be consistent with the dependencies displayed in eq. (\ref{bpsdependences}).
Moreover, by combining eq. (\ref{d5bps2}) with eq. (\ref{d5bps1}), one can
eliminate 
$\partial_{\phi_1}\phi_2$ and obtain the following equation for 
$\partial_{\theta_1}\theta_2$:
\beq
{\partial\theta_2\over
\partial\theta_1}={\tan\theta_1\over\tan\theta_2}\,.
\label{thetaderivative}
\eeq
Eq. (\ref{thetaderivative}) is easily integrated with the result:
\beq
\cos\theta_2=k\cos\theta_1\,,
\label{cp5d5cos}
\eeq 
where $k$ is a constant such that $|k|\leq1$.
In addition, eq. (\ref{constancy}) can be easily solved as:
\beq
\phi_2=n\,\phi_1+{\rm constant}\,,
\label{cp5d5phisol}
\eeq
with $n$ constant. Furthermore, by plugging (\ref{cp5d5phisol}) and
(\ref{cp5d5cos}) into eq. (\ref{constancy}) one gets the following
relation between the constants $k$ and $n$:
\beq
k\,n=-1\,,
\label{cp5d5consrel}
\eeq
which implies that they never vanish and thus, contrary to what happened
in section \ref{cp5ssd33c} for the D3-branes, there is no zero-winding
embedding. The unit-winding embeddings corresponding to the values
$k=-n=\pm1$ are given by:
\bear
&&\theta_2=\theta_1\,,
\qquad\qquad\quad
\phi_2=2\pi-\phi_1\,,
\rc\rc
&&\theta_2\,=\pi-\theta_1\,,\qquad\quad
\phi_2=\phi_1\,.
\label{D5embeddings}
\eear
Notice the similarity of (\ref{D5embeddings}) and (\ref{unitwinding}), although the
two solutions are actually very different (see, for example, their different
holomorphic structure). Furthermore, it is interesting to point out that
$c_2=2\sin\theta_1$ for the solution with $\theta_2=\theta_1$ while 
$c_2=-2\sin\theta_1$ when  $\theta_2=\pi-\theta_1$. These two solutions correspond to the
two possible signs in the projection (\ref{d5projector}). Moreover, the two-cycles 
(\ref{D5embeddings}) mix the $(\theta_1,\phi_1)$ and $(\theta_2,\phi_2)$
two-spheres, in agreement with the results of \cite{DasMuk}.

To finish this subsection, let us discuss the possibility of requiring the
vanishing of all $c\,$s in eq. (\ref{gammakappad5})
except for $c_3$. From the vanishing of  $c_1$,  $c_4$ and  $c_5$ we
obtain again eqs.  (\ref{d5bps1}) and (\ref{d5bps2}). In addition, from
$c_2=0$ we get a new equation:
\beq
\partial_{\phi_1}\phi_2={\sin\theta_1\over \sin\theta_2}\,\,
\partial_{\theta_1}\,\theta_2\,.
\eeq
By combining this new equation with eq. (\ref{d5bps1}), we obtain:
\beq
\partial_{\theta_1}\,\theta_2=-{\tan\theta_2\over\tan\theta_1}\,.
\label{tantheta}
\eeq
It is easy to see that this equation is inconsistent with eq. (\ref{thetaderivative})
(which follows from eqs. (\ref{d5bps1}) and (\ref{d5bps2})). Thus, we conclude that this
way of proceeding does not lead to any new solution of the kappa symmetry condition 
$\Gamma_{\kappa}\,\epsilon=\epsilon$.

\subsubsection{Holomorphic structure}
Let us now write the embeddings just found in terms of the holomorphic coordinates
$z_1,\cdots,z_4$ of the conifold. From the expressions of the ratios of
the $z\,$s in terms of the angles $(\theta_i, \phi_i)$ (eq.
(\ref{zratio})), it is immediate to realize that for the
$\theta_1=\theta_2$ embedding of eq. (\ref{D5embeddings}) one  has:
\beq
{z_1\over z_3}={z_4\over z_2}=
{\bar z_1\over \bar z_4}={\bar z_3\over \bar z_2}\,,
\eeq
which can be written as a single quadratic equation such as:
\beq
z_1\bar z_4-\bar z_1\,z_3=0\,.
\eeq
(This embedding also satisfies that $|z_3|=|z_4|$). Notice that the equations found are
not holomorphic. This is in correspondence with the fact that the kappa symmetry projector
of a D5 involves a complex conjugation of the spinor. Similarly, for the 
$\theta_2=\pi-\theta_1$ embedding of eq. (\ref{D5embeddings}), one has:
\beq
{z_2\over z_3}={z_4\over z_1}=
{\bar z_4\over \bar z_2}={\bar z_1\over \bar z_3}\,,
\eeq
which, again,  can be recast as a single quadratic equation, which in this case can be
written as:
\beq
z_1\,\bar z_4- \bar z_2\, z_4=0\,,
\eeq
and one has that $|z_1|=|z_2|$ for this solution.

\subsubsection{Energy bound}
It can be easily verified by explicit calculation that any solution of the type
(\ref{bpsdependences}) of the first order equations (\ref{d5bps1}) and
(\ref{d5bps2}) is also a solution of the Euler-Lagrange equations derived from the
Dirac-Born-Infeld lagrangian density
${\cal L}=-\sqrt{-g}$. Actually, for a generic configuration 
with $\theta_2= \theta_2(\theta_1)$ and 
$\phi_2= \phi_2(\phi_1)$, the hamiltonian density ${\cal H}=-{\cal L}$ is:
\beq
{\cal H}={r^2\over 6}\,\,\sqrt{1+(\partial_{\theta_1}\theta_2)^2}\,
\sqrt{\sin^2\theta_1\,+\sin^2\theta_2\,(\partial_{\phi_1}\phi_2)^2
+\,{2\over
3}\,(\,\cos\theta_1+\cos\theta_2\,\partial_{\phi_1}\phi_2\,)^2}\,.
\eeq
Let us now show that ${\cal H}$ satisfies a BPS bound of the type
${\cal H}\ge\big|{\cal Z}\big|$, which is saturated precisely when 
(\ref{d5bps1}) and (\ref{d5bps2}) are satisfied. First of all, 
we rewrite ${\cal H}$ as ${\cal H}=\big|{\cal Z}\big|+{\cal S}$, where:
\beq
{\cal Z}={r^2\over 6}\,\Big(\sin\theta_1\,\partial_{\theta_1}\theta_2-
\sin\theta_2\,\partial_{\phi_1}\phi_2\,\Big)\,.
\eeq
It can be straightforwardly proven that ${\cal Z}$ can be written as a total
derivative, \ie:
\beq
{\cal Z}=\partial_{\theta_1}\,{\cal Z}^{\theta_1}+
\partial_{\phi_1}\,{\cal Z}^{\phi_1}\,,
\eeq
and it is not difficult to find the explicit expressions of ${\cal Z}^{\theta_1}$ and
${\cal Z}^{\phi_1}$, namely:
\beq
{\cal Z}^{\theta_1}={r^2\over 6}\,\theta_2\sin\theta_1\,,
\qquad\quad
{\cal Z}^{\phi_1}=-{r^2\over 6}\,\Big
(\phi_2\sin\theta_2+\theta_2\phi_1\cos\theta_1\,\Big)\,.
\eeq
Moreover, one can check that ${\cal S}\ge 0$ is equivalent to:
\beq
\Big(\,\sin\theta_1+\sin\theta_2\,\partial_{\theta_1}\theta_2
\partial_{\phi_1}\phi_2\,\Big)^2+{2\over 3}\,
\Big(\,1+(\partial_{\theta_1}\theta_2)^2\,\Big)\,
(\,\cos\theta_1+\cos\theta_2\,\partial_{\phi_1}\phi_2\,)^2\,
\ge\,0\,,
\eeq
which is obviously satisfied and reduces to an equality when the BPS equations 
(\ref{d5bps1}) and (\ref{d5bps2}) are
fulfilled. Clearly, in this case the  bound ${\cal H}\ge\big|{\cal
Z}\big|$ is saturated. Notice that ${\cal Z}\ge 0$ for the BPS embedding
of (\ref{D5embeddings}) with $\theta_2=\theta_1$, while  ${\cal Z}\le 0$
when
$\theta_2=\pi-\theta_1$ and $\phi_2=\phi_1$. 

%
%
%
%
%
%
%
%
%

\setcounter{equation}{0}
\section{Kappa symmetry for a D7-brane probe}
\label{cp5sD7}
\medskip
In this section we will try to find supersymmetric embeddings of a D7-brane probe
in the $AdS_5\times T^{1,1}$ geometry. The corresponding kappa symmetry 
 matrix can be obtained from eqs. (\ref{cp5gammakappa}) and (\ref{rule}),
namely:
\beq
\Gamma_{\kappa}=-{i\over 8!\sqrt{-g}}\,\epsilon^{a_1\cdots a_8}\,
\gamma_{a_1\cdots a_8}\,.
\label{kappad7}
\eeq
The main interest of studying D7-branes in the $AdS_5\times T^{1,1}$ background
comes from their use as flavor branes, \ie\ as branes whose fluctuations can be
identified with dynamical mesons of the corresponding gauge theory. These flavor
branes must be spacetime filling, \ie\ they must be extended along all the gauge
theory directions. Moreover, their worldvolume should include some holographic,
non-compact, direction. In this section we will determine some supersymmetric
configurations which fulfill these requirements.

\subsection{Spacetime filling D7-brane}
\label{cp5ssd7}

As explained above, we are interested in configurations where the D7-brane
is extended along the $x^0\cdots x^3$ coordinates. Notice that in this
case the D7-brane probe and the D3-branes of the background share four
dimensions, which is just what is needed for a supersymmetric D3-D7
intersection. Accordingly,  let us choose  the following set of
worldvolume coordinates: 
\beq
\xi^a=(x^0,\cdots,x^3,\theta_1,\phi_1,\theta_2,\phi_2)\,.
\eeq
The remaining ten-dimensional coordinates $r$ and $\psi$ will be considered as
scalars. Actually, we will restrict ourselves to those configurations in which the
dependence of $r$ and $\psi$ on the worldvolume coordinates is the following:
\beq
\psi=\psi(\phi_1,\phi_2)\,,\qquad\quad
r=r(\theta_1,\theta_2)\,.
\label{d7ansatz}
\eeq
In this case the kappa symmetry matrix (\ref{kappad7}) takes the form:
\beq
\Gamma_{\kappa}=-i\,{h^{-1}\over \sqrt{-g}}\,\,
\Gamma_{x^0\cdots x^3}\,\,\gamma_{\theta_1\phi_1\theta_2\phi_2}\,,
\label{d7kappa}
\eeq
where $h$ is the warp factor of eq. (\ref{KW}). 
The induced gamma matrices along the angular coordinates of the
worldvolume are:
\bear
&&h^{-{1/4}}\,\gamma_{\theta_1}={r\over \sqrt{6}}\,\Gamma_1+
\partial_{\theta_1}r\,\Gamma_r\,,\rc\rc
&&h^{-{1/4}}\,\gamma_{\phi_1}={r\over \sqrt{6}}\,
\sin\theta_1\,\Gamma_2+
{r\over 3}\,\,(\cos\theta_1+\partial_{\phi_1}\psi\,)
\,\Gamma_{\hat 3}\,,\rc\rc
&&h^{-{1/4}}\,\gamma_{\theta_2}={r\over \sqrt{6}}\,\Big(\,\cos\psi\,
\Gamma_{\hat 1}+\sin\psi\,\Gamma_{\hat
2}\,\Big)+\partial_{\theta_2}r\,\Gamma_r\,,\rc\rc
&&h^{-{1/4}}\,\gamma_{\phi_2}={r\over \sqrt{6}}\,\sin\theta_2\,
\Big(\,\sin\psi\,
\Gamma_{\hat 1}-\cos\psi\,\Gamma_{\hat 2}\,\Big)+
{r\over 3}\,(\cos\theta_2+\partial_{\phi_2}\psi\,)\,
\Gamma_{\hat 3}\,.
\label{gammasd7}
\eear
By inspecting the form of the kappa symmetry matrix in eq.
(\ref{d7kappa}), one readily concludes that $\epsilon$ must be a
eigenvector of $\Gamma_*=i\Gamma_{x^0x^1x^2x^3}$. Then, it has to be of
the form $\epsilon_+$ (see eq. (\ref{chiraladsspinor})) and we can
write:
\beq
\Gamma_{\kappa}\,\epsilon_+=-{h^{-1}\over \sqrt{-g}}\,\,
\gamma_{\theta_1\phi_1\theta_2\phi_2}\,\epsilon_+\,.
\label{d7gammakappa}
\eeq
Moreover, after taking into account that $\epsilon_+$ has fixed
ten-dimensional chirality:
\beq
\Gamma_{x^0x^1x^2x^3}\,\Gamma_{r 12 \hat 1\hat 2\hat 3}\,
\epsilon_+=-\epsilon_+\,,
\eeq
and using eq. (\ref{tsprojections}), one can easily verify that:
\beq
\Gamma_{r\hat 3}\,\epsilon_+=-i\epsilon_+\,.
\label{rhat3}
\eeq
By using these  projection conditions and the explicit form of
the $\gamma\,$s (eq. (\ref{gammasd7})), one can prove that:
\beq
h^{-1}\,\gamma_{\theta_1\phi_1\theta_2\phi_2}\,\epsilon_+=
d_1\,\epsilon_++ie^{-i\psi}\,d_2\,\Gamma_{\hat 1\hat
3}\,\epsilon_++ id_3\,\Gamma_{ 1\hat 3}\,\epsilon_++
e^{-i\psi}\,d_4\,\Gamma_{ 1\hat 1}\,\epsilon_+\,,
\label{gammakappad7}
\eeq
where the coefficients $d_i$ are given by:
\bear
&&d_1=-{r^4\over 36}\,\sin\theta_1\sin\theta_2+
{r^3\over 18}\Big[\,\sin\theta_1\partial_{\theta_2}r\left
(\cos\theta_2+\partial_{\phi_2}\psi\right)+
\sin\theta_2\partial_{\theta_1}r\left(\cos\theta_1+\partial_{\phi_1}\psi
\right)\Big]\,,\rc\rc
&&d_2={r^3\over 6\sqrt{6}}\,\sin\theta_1
\Big[\,\sin\theta_2\partial_{\theta_2}r+{r\over 3}\left(\cos\theta_2+
\partial_{\phi_2}\,\psi\right)\Big]\,,\rc\rc
&&d_3={r^3\over 6\sqrt{6}}\,\sin\theta_2
\Big[\,\sin\theta_1\partial_{\theta_1}r+{r\over 3}\left(\cos\theta_1+
\partial_{\phi_1}\,\psi\right)\Big]\,,\rc\rc
&&d_4={r^3\over 18}\Big[\,\sin\theta_1\
\partial_{\theta_1}r\left(\cos\theta_2+\partial_{\phi_2}\,\psi\right)-
\sin\theta_2\
\partial_{\theta_2}r\left(\cos\theta_1+\partial_{\phi_1}\,\psi\right)
\Big]\,.
\eear
Notice that the terms with $d_2$, $d_3$ and $d_4$ on the right-hand side of eq.
(\ref{gammakappad7}) give rise to projections which are not compatible with
those in eq.  (\ref{tsprojections}). Therefore, in order to have
$\Gamma_{\kappa}\,\epsilon_+=\epsilon_+$,  we impose:
\beq
d_2=d_3=d_4=0\,.
\eeq
The conditions $d_2=d_3=0$ lead to the following differential equations:
\beq
\partial_{\theta_1} r=-{r\over 3}\,
{\cos\theta_1+\partial_{\phi_1}\psi\over \sin\theta_1}\;,\qquad\quad
\partial_{\theta_2} r=-{r\over 3}\,
{\cos\theta_2+\partial_{\phi_2}\psi\over \sin\theta_2}\,,
\label{cpcfBPS}
\eeq
which, in turn, imply that $d_4=0$. The two differential equations in
(\ref{cpcfBPS}) are enough to guarantee the kappa symmetry condition
$\Gamma_{\kappa}\,\epsilon_+=\epsilon_+$. Actually (see eq.
(\ref{d7gammakappa})),  we have to check
that  $d_1=-\sqrt{-g}$ when the embedding satisfies eq. (\ref{cpcfBPS}).
This fact can be easily verified if one uses that, along the angular
directions,  the only non-vanishing elements of the induced metric are:
\bear
&&g_{\theta_i\theta_j}=h^{{1/2}}\,
\Big[\,\partial_{\theta_i}r\,\partial_{\theta_j}r+
{r^2\over 6}\,\delta_{ij}\,\Big]\,,\rc\rc
&&g_{\phi_i\phi_j}=h^{{1/2}}\,\Big[
\left(\cos\theta_i+\partial_{\phi_i}\psi\right)
\left(\cos\theta_j+\partial_{\phi_j}\psi\right){r^2\over 9}+
\sin^2\theta_i\,{r^2\over 6}\,\delta_{ij}\Big]\,.
\label{d7inducedmetric}
\eear

From the above analysis it is clear that any Killing spinor of the type 
$\epsilon=\epsilon_+$, with $\epsilon_+$ as in eq. (\ref{chiraladsspinor}),
satisfies the kappa symmetry condition $\Gamma_{\kappa}\epsilon=\epsilon$ if the BPS
equations  (\ref{cpcfBPS}) hold. Then, these embeddings preserve the four
ordinary  supersymmetries of the background and thus they are $1/8$
supersymmetric.

\subsubsection{Integration of the first order equations}

The BPS equations (\ref{cpcfBPS}) relate the different derivatives of $r$
and $\psi$. However, notice that according to our ansatz (\ref{d7ansatz}) 
the only dependence on $\phi_1$ and $\phi_2$ in  (\ref{cpcfBPS}) comes
from the derivatives of $\psi$. For consistency these derivatives must be
constant, \ie:
\beq
\partial_{\phi_1}\psi=n_1\,,\qquad\quad
\partial_{\phi_2}\psi=n_2\,,
\eeq
where $n_1$ and $n_2$  are two numbers which label the different solutions of the
BPS equations (\ref{cpcfBPS}). 
Thus, we can write:
\beq
\psi=n_1\phi_1+n_2\phi_2+{\rm constant}\,.
\label{winding}
\eeq
We will refer to a solution with given numbers $n_1$ and $n_2$ as an 
$(n_1, n_2)$-winding embedding. Below we will restrict ourselves to the case
in which $n_1$ and $n_2$ are integers. Plugging the function (\ref{winding}) on the
right-hand side of eq. (\ref{cpcfBPS}) one gets  an expression for the
partial derivatives of
$r(\theta_1,\theta_2)$ which is easy to integrate. The result is:
\beq
r^3={C\over
\left(\sin{\theta_1\over 2}\right)^{n_1+1}\,
\left(\cos{\theta_1\over 2}\right)^{1-n_1}\,
\left(\sin{\theta_2\over 2}\right)^{n_2+1}\,
\left(\cos{\theta_2\over 2}\right)^{1-n_2}}\,,
\label{rtheta}
\eeq
where $C$ is a constant of integration. 

Several remarks concerning the function
(\ref{rtheta}) are in order. First of all,
notice that it is impossible to choose $n_1$ and $n_2$ in such a way that $r$ is
constant \footnote{Except when the constant $C$ in eq. (\ref{rtheta}) is equal
to zero. In this case $r=0$, $\sqrt{-g}$ also vanishes and the matrix
$\Gamma_{\kappa}$ is not well-defined.} or even that $r$ depends only on one of the
angles $\theta_i$. Moreover, 
$r(\theta_1,\theta_2)$ is invariant under the change
$\theta_i\rightarrow\pi-\theta_i$ and
$n_i\rightarrow-n_i$, which means that, in the analysis of the function
(\ref{rtheta}),  we can restrict ourselves to the case $n_i\ge 0$. 

It is also
interesting to point out that the function $r(\theta_1,\theta_2)$ always
diverges for some particular values of the angles ($\theta_i=0,\pi$
for $n_i=0$, $\theta_i=0$ for $n_i\ge 1$ and $\theta_i=\pi$ for $n_i\le
-1$). Moreover, the probe brane reaches the origin $r=0$ of the $AdS_5$
space only when any of the $n_i\,$s is such that $|n_i|\ge 2$. When $n_1$
and $n_2$ are such that $|n_i|\le 1$ there is a minimum value $r_*$ of
the coordinate $r$, which is reached at some particular values of the
$\theta_i\,$s. For example, for the $(n_1,n_2)=(0,0)$ ($(n_1,n_2)=(1,1)$)
solution, $r_*^3=4C$ ($r_*^3=C$), and this value of $r$ is reached when
$\theta_1=\theta_2=\pi/2$ ($\theta_1=\theta_2=\pi$). The existence of
this minimal value of $r$ is important when one considers these D7-branes
as flavor branes. Indeed, in this case the minimal value of $r$ provides
us of an energy scale, which is naturally identified with the mass of the
dynamical quarks added to the gauge theory. Taking this fact into
account, we would be inclined to think that the above configurations with
$n_i=0,\pm 1$ are the most adequate to be considered as flavor branes for
the $AdS_5\times T^{1,1}$ background.

\subsubsection{Holomorphic structure}
\medskip

Let us now prove that the $(n_1,n_2)$-winding embeddings just discussed can be
described by means of a polynomial equation of the type of that written in eq. 
(\ref{d3pol}), where now the exponents $m_i$ must satisfy the following
equation:
\beq
m_1+m_2+m_3+m_4\not=0\,.
\eeq
Indeed, from the expression of the phase of both sides of eq. (\ref{d3pol}) in
terms of the angular coordinates, one finds that $\psi$ depends on $\phi_1$ and
$\phi_2$ as in eq. (\ref{winding}), with $n_1$ and $n_2$ being given by:
\beq
n_1={m_1-m_2-m_3+m_4\over m_1+ m_2+ m_3+ m_4}\,,\qquad\quad
n_2={m_1-m_2+m_3-m_4\over m_1+ m_2+ m_3+ m_4}\,.
\label{d7midentifications}
\eeq
We can confirm the identification (\ref{d7midentifications}) by extracting 
$r(\theta_1,\theta_2)$ from the modulus of eq. (\ref{d3pol}). Actually, one can
easily demonstrate that the function $r(\theta_1,\theta_2)$ obtained in this way is
just given by the right-hand side of eq. (\ref{rtheta}), with the $n_1$ and $n_2$ of
eq.  (\ref{d7midentifications}). As a further check of these identifications, notice
that $n_1$ and $n_2$ are left invariant under the transformation (\ref{mchange}).

The analysis performed above serves to identify some of our solutions with those
proposed in the literature as flavor branes for this background. In this sense,
notice that the unit-winding
case with $n_1=n_2=1$ corresponds to the  embedding $z_1=C$ proposed in ref.
\cite{Ouyang}, whose fluctuations where considered in \cite{LeviOuy} in
order to compute the meson spectrum of the gauge theory. While the
zero-winding case with
$n_1=n_2=0$ is the embedding 
$z_1z_2=C$ first considered in ref. \cite{KKW} .

\subsubsection {Energy bound}
\medskip
For a D7-brane embedding of  the type (\ref{d7ansatz}), the
Dirac-Born-Infeld lagrangian ${\cal L}\,=\,-\sqrt{-g}$ can be obtained easily from
the elements of the induced metric written in eq. 
(\ref{d7inducedmetric}). We have verified that the 
equations of motion derived from 
${\cal L}$ are satisfied if the BPS first order equations (\ref{cpcfBPS})
are fulfilled. Actually, as happened with the other supersymmetric
embeddings we have studied, the hamiltonian density ${\cal H}=-{\cal L}$
satisfies a bound of the type
${\cal H}\ge \big| {\cal Z}\big|$, which is saturated just when the BPS equations
(\ref{cpcfBPS}) are verified. In order to prove this statement let us
consider, as in our ansatz (\ref{d7ansatz}), arbitrary functions 
$\psi(\phi_1,\phi_2)$ and $r\,=\,r(\theta_1,\theta_2)$.
We define the functions $\Delta_1$ and
$\Delta_2$ as follows:
\beq
\Delta_i=-{r\over 3}\,{\cos\theta_i+\partial_{\phi_i}\psi
\over \sin\theta_i}\;,\qquad (i=1,2)\,.
\eeq
Notice that the BPS equations (\ref{cpcfBPS}) are just
$\partial_{\theta_i}r=\Delta_i$ and  the lagrangian 
${\cal L}=-\sqrt{-g}$ becomes:
\beq
{\cal L}=-r^2\sin\theta_1\sin\theta_2
\sqrt{\Big((\partial_{\theta_1}r)^2+(\partial_{\theta_2}r)^2+
{r^2\over 6}\Big)\Big(\Delta_1^2+\Delta_2^2+{r^2\over 6}\Big)}\,.
\eeq
Let us now rewrite the hamiltonian density ${\cal H}=-{\cal L}$  as
${\cal H}=\big|{\cal Z}\big|+{\cal S}$, where ${\cal Z}$ is given by:
\beq
{\cal Z}=r^2\sin\theta_1\sin\theta_2\,\Big(\,{r^2\over 6}+
\partial_{\theta_1}r\,\Delta_1+
\partial_{\theta_2}r\,\Delta_2\,\Big)\,.
\label{d7Z}
\eeq
When $\psi$ and $r$  are arbitrary functions 
of the type (\ref{d7ansatz}), it is straightforward to prove that 
${\cal Z}$ is a total divergence:
\beq
{\cal Z}=\partial_{\theta_1}{\cal Z}^{\theta_1}+
\partial_{\theta_2}{\cal Z}^{\theta_2}\,,
\eeq
with ${\cal Z}^{\theta_i}$ being given by:
\beq
{\cal Z}^{\theta_1}=-{r^4\over 12}\,(\cos\theta_1+\partial_{\phi_1}\psi)
\sin\theta_2\,,\qquad\quad
{\cal Z}^{\theta_2}=-{r^4\over 12}\,(\cos\theta_2+\partial_{\phi_2}\psi)
\sin\theta_1\,.
\eeq
Moreover, the function ${\cal S}={\cal H}-\big|{\cal Z}\big|$ is non-negative. 
Actually, the condition ${\cal S}\ge 0$ is equivalent to:
\beq
{r^2\over 6}\,\Big(\partial_{\theta_1}r-\Delta_1\Big)^2+
{r^2\over 6}\,\Big(\partial_{\theta_2}r-\Delta_2\Big)^2+
\Big(\partial_{\theta_1}r\,\Delta_2-\partial_{\theta_2}r\,\Delta_1
\Big)^2\,\ge\,0\,,
\eeq
which is obviously always satisfied and reduces to an equality when 
$\partial_{\theta_i}r=\Delta_i$. Thus, ${\cal H}\ge \big| {\cal Z}\big|$ and, as
previously claimed, the BPS conditions (\ref{cpcfBPS}) saturate the
bound. It is also clear from the expression of ${\cal Z}$ in (\ref{d7Z})
that  ${\cal Z}_{|BPS}\ge 0$. 

%
%
%
%
%
%
%
%
%

\section{Discussion}
\label{cp5sdisc}

Let us now summarize the main results of this chapter. We have used kappa
symmetry to explore in a systematic way the supersymmetric embeddings of 
D-brane probes in the $AdS_5\times T^{1,1}$ geometry. Our method is based
on a detailed knowledge of the Killing spinors of the background and
allows to determine the explicit form of the D-brane embedding, as well
as the fraction of supersymmetry preserved by the different
configurations. Generically, the supersymmetric embeddings are obtained
by integrating a system of first order BPS differential equations. We
have checked in all cases that the solutions of these BPS equations also
solve the equations of motion derived from the Dirac-Born-Infeld action
of the brane probe. Actually, we have verified that our embeddings
saturate an energy bound, as it is expected to occur for a worldvolume
soliton.

In the case of a D3-brane we have found a family of three-cycles which
generalizes the ones used in ref. \cite{GK} to construct the duals of the
dibaryon operators and we have determined the field content of the dual
operator for our cycles.  We  have also been able to find explicitly the
two-cycles over which one must wrap a D5-brane in order to preserve some
fraction of supersymmetry. It was shown in \cite{GK} that a D5-brane
wrapped over a two-cycle of the $T^{1,1}$ behaves as a domain wall in
$AdS_5$. Furthermore, in the configuration we have considered, the D5-brane
is a codimension one defect along the gauge theory directions. Therefore,
these embeddings become interesting in the context of the addition of
flavor, confined to a defect, to the dual gauge theory. The analysis of
spacetime filling configurations of D7-brane probes led us to determine a
two-parameter family of supersymmetric embeddings, some of them having
the right properties to be considered as flavor branes for the
Klebanov-Witten model. All the supersymmetric embeddings we have found can
be described by means of a simple polynomial equation in terms of the
holomorphic coordinates of the conifold. Notice, however, that our method
does not rely on this fact and can be applied to other backgrounds, as it
has been done in \cite{flavoring,CPR} for the Maldacena-N\'u\~nez
background \cite{MN,CV}, for which the algebraic-geometric techniques are
not available. Nevertheless, one might wonder if, for the background
studied here, it would not be more appropriate to use the holomorphic
coordinates from the beginning, instead of the angular and radial
variables that we have employed. To clarify this point, let us recall that
the $z$ coordinates are not independent, since they satisfy the constraint
(\ref{conifold}) and thus their use as worldvolume coordinates would be
rather cumbersome.

In addition to the configurations studied in this chapter, some other
possibilities were analyzed in ref. \cite{conifold}. First, it was
considered the case of D3-branes extended along one spacelike
Minkowski direction and wrapped over a two-cycle of the $T^{1,1}$. Though
it was not possible to find supersymmetric configurations, it was shown
that the  embeddings where the D3-branes wrap the two-cycle found in
section
\ref{cp5ssd52c} are a solution of the D3-brane equations of motion and
saturate an energy bound, thus being stable configurations.

Second, for the case of D5-brane probes it was found another two-cycle of
the $T^{1,1}$ such that the embeddings of D5-branes wrapped over this cycle
preserve the same supersymmetry as the configurations of section
\ref{cp5ssd52c}. Besides, it was studied the effect of adding flux to
the embeddings of section \ref{cp5ssd52c}, which leads to a supersymmetric
configuration where the D5-brane is bent along the Minkowski spacetime as
in the embeddings with flux of section \ref{cp3codim1sc}. The
baryon vertex of the KW model, a D5-brane wrapped over the entire
$T^{1,1}$, was also analyzed and it was argued that this cannot be
a supersymmetric configuration.

Finally, different supersymmetric embeddings of D7-brane probes were
also found in \cite{conifold}. Spacetime filling embeddings where the
D7-brane wraps a three-cycle of the $T^{1,1}$ and is extended along the
holographic coordinate were shown to preserve the same supersymmetries as
the configuration of section \ref{cp5ssd7}. In addition, it was found a
supersymmetric embedding of a D7-brane wrapping the $T^{1,1}$ and
extended along two spatial Minkowski directions and the holographic
coordinate.

Let us now discuss possible extensions of the work presented in this
chapter; some of which have already been worked out elsewhere, while some
others could be worth looking at. It would be interesting to compute the
spectrum of BPS excitations of the dibaryonic operators found in section
\ref{cp5ssd33c}: this can be achieved by studying the fluctuations of the
D3-brane probe around the embeddings found in that section, as it was done
in \cite{BHK} for the zero-winding embedding. Regarding the configurations
with D5-branes wrapping a two-cycle of the
$T^{1,1}$ considered in section \ref{cp5ssd52c}, notice that they
are the suitable  ones for the introduction of flavor\footnote{These would be massless flavors since the D5-branes are
extended along the holographic coordinate reaching the origin $r=0$ of
the $AdS_5$.} living on a codimension one defect in the dual field theory.
Again, by studying the fluctuations of those embeddings the mesonic
spectrum of the defect field theory could be determined. Furthermore, the
study of the fluctuations of the spacetime filling D7-branes dealt with in
section \ref{cp5ssd7} would allow one to extract the spectrum of
mesons for the codimension zero case (\ie \ no defect). This has already
been done in \cite{LeviOuy} for the unit-winding embedding of
\cite{Ouyang}. Another promising research line consists of the
application of the methodology used in this chapter to other
backgrounds, an interesting candidate being the so-called
Klebanov-Strassler (KS) background \cite{KS}. With this objective in mind,
it was computed in \cite{conifold} (see also \cite{tesina}) the explicit
form of the Killing spinors of that background in a basis where they are
independent of the compact coordinates. The mesonic spectrum for the KS
background has been studied in \cite{kuper} by computing fluctuations of a
D7-brane probe. As we have mentioned before, it would be interesting to
study the addition of unquenched flavors to both the KW and the KS
background. This has been done in
\cite{unqchKW,unqchKS} by following the smearing procedure of \cite{CNP}.
Finally, one could study the supersymmetric configurations of M2 and M5
brane probes in some backgrounds of eleven-dimensional supergravity. The
analogue of the case studied here would be considering a manifold of the
type $AdS_4\times X_7$, where $X_7$ is a seven-dimensional Einstein space
\cite{Kirsch}.
\newpage
\vspace{3cm}
\section*{Acknowledgements}
I would like to thank my supervisor Alfonso V. Ramallo, with whom it has been
a pleasure to work for these last four years. I am also deeply indebted to my collaborators \'Angel Paredes, David Crooks, Paolo Merlatti, Carlos N\'u\~nez and Diego Rodr\'\i guez-G\'omez. Finally, I wish to thank Felipe Canoura, Matteo Bertolini and Jonathan Shock for useful discussions and comments on the manuscript.

\vspace{0.8cm}

No meu caso a admiraci\'on pola beleza do universo e o interese por entender a s\'ua estrutura van da man dunha fascinaci\'on crecente polas insignificantes creaci\'ons
humanas. Unha delas, que \'e a esencia da nosa especie, \'e a lingua.
E cada un sente cari\~no por aquela que lle \'e propia. Por iso nestes tempos
escuros nos que \'e v\'itima de mentiras e desprezo quer\'\i a facer
unha pequena louvanza da nosa lingua. E aproveito esta tese, onde resumo a mi\~na min\'uscula contribuci\'on \'a comprensi\'on de algo tan fermoso como o universo, para facer un chamamento a defendermos unha marabilla que est\'a nos nosos cerebros e que calquera que goste da beleza non pode permitir que esmoreza.

%
%
%
%
%
%
%

\end{document}